\documentclass[review,authoryear,3p,times,10pt]{elsarticle}
\usepackage{multirow,setspace,amssymb,amsmath,graphicx,color,url,bm,pdflscape}
\usepackage{lineno}
\usepackage{natbib}
\usepackage{booktabs}
\usepackage{longtable}
\usepackage{lscape}  
\usepackage[normalem]{ulem}  
\usepackage{threeparttable}
\usepackage[table]{xcolor}
\usepackage{tabularx}
\usepackage{epstopdf}
\usepackage{mathrsfs}
\usepackage{makecell}
\usepackage[bookmarks=true,colorlinks,linkcolor=blue,anchorcolor=blue,citecolor=blue,unicode]{hyperref}
\usepackage{bookmark}
\usepackage{todonotes}
\usepackage{ragged2e}
\usepackage{dcolumn}
\usepackage{float}
\usepackage{dsfont}
\usepackage{listings}
\usepackage{relsize}
\usepackage{tikz}
\usepackage[percent]{overpic}
\usepackage{subcaption}  
\usepackage[font=small]{caption}
\usepackage{booktabs}
\usepackage{amsmath}
\usepackage{xstring}
\usepackage[all]{xy}
\usepackage{array}
\newcommand{\motif}[1]{%
	\IfStrEqCase{#1}{%
		{6}{\begin{xy}
				\POS (0,3) *\cir<2pt>{} ="a", (-3,-1)*\cir<2pt>{} ="b", (3,-1)*\cir<2pt>{} ="c"
				\POS "a" \ar @{->} "b" \POS "a" \ar @{->} "c" \end{xy}}
		{12}{\begin{xy}
				\POS (0,3) *\cir<2pt>{} ="a", (-3,-1)*\cir<2pt>{} ="b", (3,-1)*\cir<2pt>{} ="c"
				\POS "a" \ar @{<-} "b" \POS "a" \ar @{->} "c" \end{xy}}
		{14}{\begin{xy}
				\POS (0,3) *\cir<2pt>{} ="a", (-3,-1)*\cir<2pt>{} ="b", (3,-1)*\cir<2pt>{} ="c"
				\POS "a" \ar @{<->} "b" \POS "a" \ar @{->} "c" \end{xy}}
		{36}{\begin{xy}
				\POS (0,3) *\cir<2pt>{} ="a", (-3,-1)*\cir<2pt>{} ="b", (3,-1)*\cir<2pt>{} ="c"
				\POS "a" \ar @{->} "c" \POS "b" \ar @{->} "c" \end{xy}}
		{38}{\begin{xy}
				\POS (0,3) *\cir<2pt>{} ="a", (-3,-1)*\cir<2pt>{} ="b", (3,-1)*\cir<2pt>{} ="c"
				\POS "a" \ar @{->} "b" \POS "a" \ar @{->} "c" \POS "b" \ar @{->} "c" \end{xy}}
		{46}{\begin{xy}
				\POS (0,3) *\cir<2pt>{} ="a", (-3,-1)*\cir<2pt>{} ="b", (3,-1)*\cir<2pt>{} ="c"
				\POS "a" \ar @{<->} "b" \POS "a" \ar @{->} "c" \POS "b" \ar @{->} "c" \end{xy}}
		{74}{\begin{xy}
				\POS (0,3) *\cir<2pt>{} ="a", (-3,-1)*\cir<2pt>{} ="b", (3,-1)*\cir<2pt>{} ="c"
				\POS "a" \ar @{<->} "b" \POS "a" \ar @{<-} "c" \end{xy}}
		{78}{\begin{xy}
				\POS (0,3) *\cir<2pt>{} ="a", (-3,-1)*\cir<2pt>{} ="b", (3,-1)*\cir<2pt>{} ="c"
				\POS "a" \ar @{<->} "b" \POS "a" \ar @{<->} "c" \end{xy}}
		{98}{\begin{xy}
				\POS (0,3) *\cir<2pt>{} ="a", (-3,-1)*\cir<2pt>{} ="b", (3,-1)*\cir<2pt>{} ="c"
				\POS "a" \ar @{->} "b" \POS "a" \ar @{<-} "c" \POS "b" \ar @{->} "c" \end{xy}}
		{102}{\begin{xy}
				\POS (0,3) *\cir<2pt>{} ="a", (-3,-1)*\cir<2pt>{} ="b", (3,-1)*\cir<2pt>{} ="c"
				\POS "a" \ar @{->} "b" \POS "a" \ar @{<->} "c" \POS "b" \ar @{->} "c" \end{xy}}
		{108}{\begin{xy}
				\POS (0,3) *\cir<2pt>{} ="a", (-3,-1)*\cir<2pt>{} ="b", (3,-1)*\cir<2pt>{} ="c"
				\POS "a" \ar @{<-} "b" \POS "a" \ar @{<->} "c" \POS "b" \ar @{->} "c" \end{xy}}
		{110}{\begin{xy}
				\POS (0,3) *\cir<2pt>{} ="a", (-3,-1)*\cir<2pt>{} ="b", (3,-1)*\cir<2pt>{} ="c"
				\POS "a" \ar @{<->} "b" \POS "a" \ar @{<->} "c" \POS "b" \ar @{->} "c" \end{xy}}
		{238}{\begin{xy}
				\POS (0,3) *\cir<2pt>{} ="a", (-3,-1)*\cir<2pt>{} ="b", (3,-1)*\cir<2pt>{} ="c"
				\POS "a" \ar @{<->} "b" \POS "a" \ar @{<->} "c" \POS "b" \ar @{<->} "c" \end{xy}}
	}[\textbf{?}]
}

\graphicspath{{Figures/}}
\makeatletter
\def\ps@pprintTitle{%
	\let\@oddhead\@empty 
	\let\@evenhead\@empty
}
\newcolumntype{d}[1]{D{.}{.}{#1}} 

\definecolor{Commodity}{HTML}{FF69B4}
\definecolor{Equity}{HTML}{007D00}
\definecolor{Energy}{HTML}{FF0000}
\definecolor{Agriculture}{HTML}{8B0000}
\definecolor{Metal}{HTML}{0000FF}
\definecolor{GrainsOilseeds}{HTML}{8B4513}
\definecolor{Softs}{HTML}{FF69B4}
\definecolor{Livestock}{HTML}{800080}
\definecolor{Preciousmetals}{HTML}{DAA520}
\definecolor{Industrialmetals}{HTML}{1E90FF}
\begin{document}

\begin{frontmatter}
	\title{A Motif-Based Framework for Decomposing Risk Spillovers}
	
	\author[SUIBE]{Ying-Hui Shao}
	\ead{yhshao@suibe.edu.cn}
	
	\author[SILC]{Yan-Hong Yang \corref{cor1}}
	\ead{yangyh@shu.edu.cn}
	
	\author[SUIBE]{Yun Zhang}
	\ead{ zhangyunphd@163.com}
	\cortext[cor1]{Corresponding author.}
	
	\address[SUIBE]{School of Finance, Shanghai University of International Business and Economics, Shanghai 201620, China}
	\address[SILC]{SILC Business School, Shanghai University, Shanghai 201899, China}

\begin{abstract}
	Connectedness measures quantify aggregate risk spillovers but obscure the local interaction patterns that generate systemic risk. We develop a motif-based framework that first extracts multiscale backbones from quantile connectedness networks and then identifies directed triadic motifs whose frequencies exceed randomization baselines. To distinguish how assets' sectoral identities shape local spillover structures, we introduce colored motifs under sector partitions of increasing granularity. Using orbit positions that capture each node's structural role within directed triadic motifs, we construct portfolio strategies that exploit an asset's place in the spillover architecture. Applying the framework to 39 commodity and equity futures across lower, median, and upper conditional quantiles, we find that motif-based portfolios outperform minimum correlation and minimum connectedness benchmarks on risk-adjusted returns. We further show that in tail networks, assets with greater orbit-position diversity tend to act as net spillover transmitters rather than receivers, establishing positional diversity as a tail-specific marker of systemic influence. These findings demonstrate that local triadic topology carries portfolio-relevant information that aggregate connectedness measures miss.
\end{abstract}
\end{frontmatter}

\section{Introduction}
\label{S1:Introduction}

Financial markets are interconnected systems in which cross-asset linkages define the architecture of risk transmission. In such systems, risk spillovers are not solely determined by the magnitude of pairwise connections, but also by the structural organization of the transmission network. For risk analysis, the key challenge is therefore not merely to detect the presence of spillovers, but to understand how they propagate through an interconnected market system and which transmission pathways facilitate their amplification.

A substantial body of literature has emerged to quantify such cross-market risk spillovers, predominantly through variance decomposition and connectedness network approaches \citep{FJ-Diebold-Yilmaz-2014-JEconom,FJ-Ando-GreenwoodNimmo-Shin-2022-ManagSci}. However, the prevailing analytical paradigm largely reduces the estimated network to a limited set of summary statistics. These typically include node-level measures such as degree centrality, strength, and net directional spillovers, which characterize each asset's role in the system, and system-level aggregates such as the total connectedness index, which summarize overall market integration. While informative, such aggregate metrics tend to obscure the higher-order interaction patterns through which risk actually propagates, leaving a critical gap between the richness of the estimated network and the coarseness of the structural insights extracted from it.

This limitation is critical because systemic risk is an emergent property of network topology rather than a simple sum of bilateral ties \citep{FJ1-Haldane-May-2011-Nature, FJ-Acemoglu-Ozdaglar-TahbazSalehi-2015-AmEconRev, FJ-Battiston-Farmer-Flache-Garlaschelli-Haldane-Heesterbeek-Hommes-Jaeger-May-Scheffer-2016-Science}. Systems with similar aggregate connectedness may exhibit markedly different vulnerability profiles depending on their local transmission architectures. Shocks may dissipate in sparse configurations but amplify within interlocked triadic structures that facilitate feedback amplification. Without identifying these structural building blocks, risk analysis may overlook the precise pathways through which disturbances accumulate or diffuse across markets \citep{FJ-Squartini-vanLelyveld-Garlaschelli-2013-SciRep}. A more granular characterization of spillovers therefore requires explicit attention to recurrent subgraph patterns, particularly triadic configurations that encode how risk is amplified, redirected, or contained.

Network motif analysis provides a principled approach for recovering such subgraph-level structural information. In directed networks, triadic motifs, the 13 non-isomorphic three-node subgraph configurations, are especially informative because each encodes a topologically distinct transmission architecture
 \citep{Milo-ShenOrr-Itzkovitz-Kashtan-Chklovskii-Alon-2002-Science}. Moreover, the 30 structurally distinct node positions, or motif orbits, embedded in these configurations provide a richer and more position-sensitive characterization of an asset's systemic role than conventional scalar centrality measures.

Rigorous motif analysis of financial connectedness networks further requires the extraction of a sparse but statistically principled network backbone, because spillover matrices derived from variance decomposition methods are inherently dense and naive motif enumeration would be dominated by trivial subgraphs induced by network density rather than by salient structural patterns. We adopt the multiscale backbone approach of \cite{FJ-Serrano-Boguna-Vespignani-2009-PNAS}, which filters edges against a local null model while preserving statistically significant connections across scales. Despite growing use of backbone methods in trade and interbank networks \citep{FJ-Kobayashi-Takaguchi-2018-JBankFin, FJ-Xie-Li-Zhou-2020-JManagSciChina}, their integration with motif significance testing in financial risk spillover networks remains largely unexplored.

Against this background, we propose an integrated methodology that combines multiscale backbone extraction with motif significance analysis to identify salient local structures of risk transmission in connectedness networks. We further distinguish three categories of colored directed motifs at different levels of granularity and trace the 30 motif orbits embedded in the 13 directed triadic configurations. We apply this methodology to the commodity-equity futures nexus, a setting that is particularly suitable for systemic risk analysis because futures markets are highly interconnected, rapidly repriced, and central to both hedging and speculative activity. 
Empirically, we examine raw, decomposed, partial, and aggregate connectedness estimated via a quantile VAR (QVAR) specification under different conditional quantiles, and analyze the local structural features of the resulting networks through backbone extraction and directed triadic motif analysis.

Our results show that motif-level analysis reveals systematic differences in local propagation patterns that are obscured by conventional network summaries. Motif-informed portfolio strategies also outperform two traditional benchmarks, indicating that mesoscopic structural information is relevant not only for structural interpretation but also for risk-sensitive decision making.

This study contributes to the literature in three ways. First, it advances systemic risk analysis by introducing a mesoscopic topological perspective that treats spillovers as recurring neighborhood-level interaction patterns rather than merely aggregate network outcomes. Second, it develops an integrated framework that combines multiscale backbone extraction, motif significance testing, and colored motif design to identify salient building blocks of risk transmission. Third, it demonstrates empirically that these subgraph-level insights carry information relevant for both cross-market vulnerability assessment and portfolio design.

The rest of the paper is organized as follows. Section~\ref{S2:Literature} reviews the related literature. Section~\ref{S3:Methodology} outlines the research methodology, including the quantile connectedness framework and multiscale backbone extraction procedure. Section~\ref{S4:Data description} describes the data and summary statistics. Section~\ref{S5:Empirical results} presents the empirical results, covering risk spillover dynamics, backbone structures, directed triadic motif significance, colored motif taxonomy, and orbit-based positional diversity. Section~\ref{S6:Portfolio} evaluates portfolio performance. Section~\ref{S7:conclusion} concludes with key findings and implications.

\section{Literature review}
\label{S2:Literature}

The quantification of cross-market risk transmission has advanced considerably over the past decade, largely building on the generalized forecast error variance decomposition (GFEVD) approach developed by  \cite{FJ-Diebold-Yilmaz-2009-EconJ,FJ-Diebold-Yilmaz-2012-IntJForecast,FJ-Diebold-Yilmaz-2014-JEconom}. By decomposing the forecast error variance of each variable into contributions attributable to shocks from other variables in the system, this approach yields a complete matrix of directional spillovers from which aggregate connectedness measures can be constructed. Subsequent extensions have enriched it along several dimensions. \cite{FJ-Barunik-Krehlik-2018-JFinancEconom} decompose connectedness into frequency-specific components, allowing short-run and long-run transmission channels to be distinguished. \cite{FJ-Antonakakis-Chatziantoniou-Gabauer-2020-JRiskFinancialManag} embed the GFEVD in a time-varying parameter VAR setting, thereby avoiding rolling-window estimation and producing smoother measures of dynamic connectedness. More recently, \citet{FJ-Ando-GreenwoodNimmo-Shin-2022-ManagSci} and \citet{FJ-Chatziantoniou-Gabauer-Stenfors-2021-EconLett} extend connectedness analysis to conditional quantiles, making it possible to examine how spillover structures vary across market states and become asymmetric under extreme conditions. Taken together, these advances have substantially increased the informational richness of spillover matrices by incorporating heterogeneity over time, frequency, and distributional location.

The commodity-equity nexus provides a particularly important setting for the application of these tools. The financialization of commodity markets has been widely documented as a major force behind the deepening of cross-market linkages \citep{FJ-Tang-Xiong-2012-FinancAnalJ, FJ-Adams-Glueck-2015-JBankFinanc, FJ-Basak-Pavlova-2016-JFinanc}. Subsequent studies show that commodities increasingly behave as financial assets whose return dynamics are more strongly tied to broader capital market conditions than to commodity-specific fundamentals alone \citep{FJ-Basak-Pavlova-2016-JFinanc,FJ-Adams-Collot-Kartsakli-2020-EnergyEcon, FJ-Kang-Tang-Wang-2023-JCommodMark}. Empirical work using connectedness frameworks generally finds that commodity-equity risk transmission is strongly time varying, crisis sensitive, and often asymmetric, with spillovers intensifying during episodes of market stress such as the global financial crisis and the COVID-19 pandemic \citep{FJ-Mensi-Yousaf-Vo-Kang-2022-JIntFinMarkInstMoney, FJ-Jain-Maitra-McIver-Kang-2023-JCommodMark}.

In sum, the spillover literature has made substantial progress in 
measuring the magnitude, direction, and temporal variation of risk 
transmission \citep{FJ-Yang-Shao-Zhou-2025-FinanceResLett,Shi-Chen-2025-GlobFinJ}. Yet the structural analysis of the resulting 
connectedness networks has remained almost exclusively at the level 
of individual nodes or aggregate system indices. None of the studies 
reviewed above examines how spillovers are organized into recurrent 
higher-order subgraph patterns, nor whether specific triadic 
transmission configurations are over- or under-represented relative 
to what network density alone would predict. 
Network motif analysis offers a natural tool for filling this gap.

Since the foundational work of \cite{Milo-ShenOrr-Itzkovitz-Kashtan-Chklovskii-Alon-2002-Science, Milo-Itzkovitz-Kashtan-Levitt-ShenOrr-Ayzenshtat-Sheffer-Alon-2004-Science}, motif analysis has accumulated an extensive cross-disciplinary record. Methodological advances in orbit-level enumeration \citep{FJ-Yaveroglu-MalodDognin-Davis-Levnajic-Janic-Karapandza-Stojmirovic-Przulj-2014-SciRep,FJ-Ribeiro-Silva-2014-DataMinKnowlDisc} have made it computationally tractable to characterize each node's local embedding through its complete 30-orbit participation profile in directed networks. Empirically, motif-based approaches have revealed functionally important subgraph structures across a wide range of domains, including transcriptional regulation \citep{FJ-ShenOrr-Milo-Mangan-Alon-2002-NatGenet,FJ-Alon-2007-NatRevGenet}, ecological food webs \citep{FJ-Stouffer-Camacho-Jiang-Amaral-2007-ProcRSocB}, neural connectivity \citep{FJ-Sporns-Kotter-2004-PLoSBiol}, and large-scale social and information networks \citep{FJ-Benson-Gleich-Leskovec-2016-Science}.

In financial and economic networks, motif-based analysis has begun to 
attract attention but remains limited in scope.
 \cite{FJ-Ohnishi-Takayasu-Takayasu-2010-JEconInteractCoord} reported significant over-representation of certain triadic configurations in Japanese stock correlation networks but without connecting motif structure to economic mechanisms. \cite{FJ-Squartini-vanLelyveld-Garlaschelli-2013-SciRep} documented motif deviations in the Dutch interbank network but did not translate 
 their findings into risk measurement tools. In a socioeconomic setting,
 \cite{FJ-Xie-Yang-Li-Jiang-Zhou-2017-EPJDataSci} examined all 30 motif 
 orbit positions across the 13 directed triadic configurations in 
dependence networks, finding that positional diversity 
 across orbits is positively correlated with economic output. Their 
 work provides direct empirical evidence that orbit-level structural 
 information carries economically meaningful content. Turning to commodity and cross-market applications, \cite{FJ-Liu-Li-Guan-2019-JEconInteractCoord} combined GARCH-BEKK 
 spillover estimation with triadic motif significance testing in 
 international steel markets, though without a system-wide connectedness 
 approach, orbit-level analysis, or portfolio construction. \cite{FJ-Xie-Yong-Wei-Yue-Zhou-2021-NorthAmJEconFinance} used the 
 distribution of the 13 directed triadic motifs in transfer entropy 
 networks of 48 global stock indices to classify market states, showing 
 that information flow increases markedly during periods of financial 
 stress. \cite{FJ-Pagnottoni-Spelta-2024-StatMethodsAppl} extended motif 
 analysis to weighted commodity price networks by quantifying the 
 intensity and coherence of triadic subgraphs, documenting a shift from 
 simple to complex triadic structures in the aftermath of the global 
 financial crisis.

 These studies demonstrate the feasibility of motif-based analysis in economic and financial systems. However, none employs the QVAR connectedness approach, incorporates colored motifs to capture asset heterogeneity, or translates motif-level findings into portfolio construction strategies. This neglect of node heterogeneity is consequential because standard motif enumeration treats all vertices as interchangeable. In commodity--equity spillover networks, a feedforward loop among three energy commodities captures an intra-sector cascade, whereas the same motif spanning commodities and equity indices reflects a distinct cross-market transmission pattern. Colored motifs, which augment topology with categorical node 
labels, address this limitation and have seen application in biological 
networks \citep{FJ-Qian-Hintze-Adami-2011-PLoSOne}, but have received 
little attention in financial network research. The present study 
addresses these gaps by integrating backbone extraction, motif 
significance testing, colored motif taxonomy, and orbit-informed 
portfolio construction into a unified analytical framework.

\section{Methodology}
\label{S3:Methodology}

\subsection{Joint spillover metrics}

This study implements a novel framework that combines quantile vector autoregression with the extended joint connectedness approach, featuring a refined connectedness normalization technique that leads to more accurate results \citep{FJ-Lastrapes-Wiesen-2021-EconModel,FJ-Balcilar-Gabauer-Umar-2021-ResourPolicy,FJ-Cunado-Chatziantoniou-Gabauer-PerezdeGracia-Marfatia-2023-JCommodMark}. To analyze the risk spillovers between stock and commodity futures markets across various market states, we first estimate a $p$-order QVAR model at conditional quantiles $\tau \in [0, 1]$:
\begin{equation}
	\mathbf{Y}_t = \boldsymbol{\mu}(\tau) + \sum_{k=1}^p \boldsymbol{\Phi}_k(\tau) \mathbf{Y}_{t-k} + \mathbf{u}_t(\tau)
\end{equation}
where $\mathbf{Y}_t$ is an $N \times 1$ vector of returns for stocks and commodity futures, $\boldsymbol{\mu}(\tau)$ is the conditional mean vector,  $\boldsymbol{\Phi}_k(\tau)$ represents the $N \times N$ matrix of autoregressive coefficients at quantile $\tau$, and $\mathbf{u}_t(\tau)$ is the error vector with a variance–covariance matrix  $\boldsymbol{\Sigma}(\tau)$ \citep{FJ-Chatziantoniou-Gabauer-Stenfors-2021-EconLett}. To facilitate the decomposition of forecast error variance, the QVAR process is transformed into its infinite-order moving average representation via the Wold theorem: $\mathbf{Y}_t = \boldsymbol{\mu}(\tau) + \sum_{h=0}^\infty \boldsymbol{\Psi}_{h}(\tau) \mathbf{u}_{t-h}(\tau)$.

Based on this moving average representation, we employ the GFEVD to quantify the pairwise connectedness ($CT(\tau)$), which measures the impact a shock in variable $j$ has on the forecast error variance of variable $i$ at quantile $\tau$ \citep{FJ-Ando-GreenwoodNimmo-Shin-2022-ManagSci}. Following the extended joint framework, the unnormalized GFEVD element, denoted as $gSOV_{ij}(H)$ for a forecast horizon $H$, is defined as:

\begin{equation}
	gSOV_{ij}(H) = \frac{\boldsymbol{\Sigma}(\tau)_{ii}^{-1} \sum_{h=0}^{H-1} (\mathbf{e}_i' \boldsymbol{\Psi}_{h}(\tau) \boldsymbol{\Sigma}(\tau) \mathbf{e}_j)^2}{\sum_{h=0}^{H-1} (\mathbf{e}_i' \boldsymbol{\Psi}_{h}(\tau) \boldsymbol{\Sigma}(\tau) \boldsymbol{\Psi}_{h}(\tau)' \mathbf{e}_i)}
\end{equation}
where $\mathbf{e}_i$ is a selection vector. The standard pairwise connectedness is initially normalized as
\begin{equation}
	gSOT_{ij}(H)
	=
	\frac{gSOV_{ij}(H)}{\sum_{\ell=1}^N gSOV_{i\ell}(H)}.
\end{equation}
Under the extended joint framework, we first define the $H$-step ahead forecast error vector $\boldsymbol{\xi}_t(H)$ and its variance-covariance matrix:
\begin{equation}
	\boldsymbol{\xi}_t(H) = \mathbf{Y}_{t+H} - E(\mathbf{Y}_{t+H} | \mathbf{Y}_t, \mathbf{Y}_{t-1}, \dots) = \sum_{h=0}^{H-1} \boldsymbol{\Psi}_{h} \boldsymbol{\epsilon}_{t+H-h}
\end{equation}
\begin{equation}
	E(\boldsymbol{\xi}_t(H) \boldsymbol{\xi}_t'(H)) = \sum_{h=0}^{H-1} \boldsymbol{\Psi}_{h} \boldsymbol{\Sigma} \boldsymbol{\Psi}_{h}'
\end{equation}
The key novelty of the joint connectedness framework, relative to the original measure, is that its normalization is grounded in the familiar $R^2$ goodness-of-fit measure.
The joint contribution to the variance of variable $i$ from all other variables in the network ($FROM$) is formulated by excluding specific shocks. Let $\mathbf{M}_i$ be an $N \times (N-1)$ rectangular matrix that equals the identity matrix with its $i$-th column removed. The joint $FROM$ connectedness is defined as:

\begin{equation}
	\begin{aligned}
		S_{i\leftarrow \bullet}^{jnt,\mathrm{from}}(H)
		&=
		\frac{
			E\!\left(\xi_{i,t}^2(H)\right)
			-
			E\!\left[
			\xi_{i,t}(H)-E\!\left(\xi_{i,t}(H)\mid
			\boldsymbol{\epsilon}_{\forall \neq i, t+1},\dots,\boldsymbol{\epsilon}_{\forall \neq i, t+H}
			\right)
			\right]^2
		}{
			E\!\left(\xi_{i,t}^2(H)\right)
		}
		\\[4pt]
		&=
		\frac{
			\sum_{h=0}^{H-1}
			\mathbf{e}_i'
			\boldsymbol{\Psi}_{h,t}(\tau)\,
			\boldsymbol{\Sigma}_t(\tau)\,
			\mathbf{M}_i\,
			\big(\mathbf{M}_i'\boldsymbol{\Sigma}_t(\tau)\mathbf{M}_i\big)^{-1}\,
			\mathbf{M}_i'\boldsymbol{\Sigma}_t(\tau)\,
			\boldsymbol{\Psi}_{h,t}'(\tau)\,
			\mathbf{e}_i
		}{
			\sum_{h=0}^{H-1}
			\mathbf{e}_i'
			\boldsymbol{\Psi}_{h,t}(\tau)\,
			\boldsymbol{\Sigma}_t(\tau)\,
			\boldsymbol{\Psi}_{h,t}'(\tau)\,
			\mathbf{e}_i
		},
	\end{aligned}
\end{equation}
where $\boldsymbol{\epsilon}_{\forall \neq i,t+1}$ denotes the $(N-1)$-dimensional vector of shocks at time $t+1$ for all variables other than $i$. Unlike its generalized counterpart, this measure requires no additional normalization, as it is inherently bounded between $0$ and $1$. The joint spillover index ($TCI$) is defined as
\begin{equation}
	TCI=\frac{1}{N}\sum_{i=1}^{N} S_{i\leftarrow \bullet}^{jnt,\mathrm{from}}(H),
\end{equation}
which captures the overall connectedness of the system. Equivalently, $TCI$ represents the average relative contribution of shocks from all other variables to each variable's $H$-step forecast error variance.

Notably, an essential extension of \cite{FJ-Balcilar-Gabauer-Umar-2021-ResourPolicy} is to link $gSOT$ to $jSOT$ via variable-specific scaling factors:

\begin{equation}
	\lambda_i(H)
	= \frac{S_{i\leftarrow \bullet}^{jnt,\mathrm{from}}(H)}{S_{i\leftarrow \bullet}^{gen,\mathrm{from}}(H)},
\end{equation}
where $S_{i\leftarrow \bullet}^{gen,\mathrm{from}}(H)=\sum_{j=1,\,j\neq i}^{K} gSOT_{ij}(H)$.

\begin{equation}
	jSOT_{ij}(H)
	= \lambda_i(H)\, gSOT_{ij}(H).
\end{equation}

Accordingly, the joint total spillovers to all others from variable $i$ are further assessed through
$S_{i\to \bullet}^{jnt,to}(H)$, with the corresponding joint net total spillover given by
$S_i^{jnt,net}(H)$ and the joint net pairwise spillover between $i$ and $j$ denoted by
$S_{ij}^{jnt,net}(H)$:
\begin{equation}
	S_{i\to \bullet}^{jnt,\mathrm{to}}(H)
	=
	\sum_{j=1,\, j\neq i}^{N} jSOT_{ji}(H).
\end{equation}
\begin{equation}
	S_i^{jnt,\mathrm{net}}(H)
	=
	S_{i\to \bullet}^{jnt,\mathrm{to}}(H)
	-
	S_{i\leftarrow \bullet}^{jnt,\mathrm{from}}(H).
\end{equation}
\begin{equation}
	S_{ij}^{jnt,net}(H)
	=
	jSOT_{ji}(H)
	-
	jSOT_{ij}(H).
\end{equation}

If $S_i^{jnt,net}(H)>0$ $\big(S_i^{jnt,net}(H)<0\big)$, market $i$ is a net transmitter (receiver) of shocks. Besides, if $S_{ij}^{jnt,net}(H)>0$ $\big(S_{ij}^{jnt,net}(H)<0\big)$, market $i$ dominates (is dominated by) market $j$, implying that market $i$ exerts (receives) a stronger net influence on market $j$ than it receives (exerts) from it. Overall, this joint quantile framework allows the estimated spillovers to more faithfully reflect the joint dynamics of shocks across different conditional quantiles \citep{FJ-Lastrapes-Wiesen-2021-EconModel}.

\subsection{Decomposed, partial, and aggregate connectedness}

In addition, we employ the time-varying decomposed connectedness framework of \cite{FJ-Gabauer-Gupta-2018-EconLett} to examine contagion across market groups, where total connectedness is split into internal and external components. To isolate group-specific dynamics from system-wide effects, we further apply the partial connectedness approach of \cite{FJ-Chatziantoniou-Elsayed-Gabauer-Gozgor-2023-EnergyEcon}, contrasting inclusive and exclusive measures as a robustness check. We also follow \cite{FJ-Stenfors-Chatziantoniou-Gabauer-2022-JIntFinancMarkInstMoney} to report aggregate connectedness. The total connectedness indices for the internal, external, inclusive, exclusive, and aggregate frameworks are denoted by $TCI^{int}$, $TCI^{ext}$, $TCI^{inc}$, $TCI^{exc}$, and $TCI^{agg}$, respectively. Collectively, these frameworks help mitigate third-party influences and sharpen inference on group-level transmission \citep{FJ-Naifar-2025-JCommodMark}.

\subsection{Extraction of multiscale backbones}
\label{S5:Spillover analysis}

In financial risk spillover studies, connectedness networks are typically constructed via the forecast error variance decomposition framework \citep{FJ-Diebold-Yilmaz-2012-IntJForecast,FJ-Diebold-Yilmaz-2014-JEconom}. The resulting fully connected matrices blend meaningful systemic linkages with stochastic noise from micro-market fluctuations, complicating the identification of core contagion pathways 
\citep{FJ-Demirer-Diebold-Liu-Yilmaz-2018-JApplEconom,FJ-Xie-Li-Zhou-2020-JManagSciChina}. Conventional global thresholding often overlooks ``small but significant'' spillovers and risks compromising the network's topological integrity \citep{FJ-Serrano-Boguna-Vespignani-2009-PNAS,FJ-Masuda-Boyd-Garlaschelli-Mucha-2025-PhysRep,FJ-Bhattacharjee-Maiti-2025-Risks,FJ-Kirkley-2025-PhysRevX}. To isolate the robust structural ``skeleton'' of risk transmission while filtering out this noise, we implement a multiscale backbone extraction method based on the disparity filter algorithm \citep{FJ-Serrano-Boguna-Vespignani-2009-PNAS}.

\subsubsection{Dual-perspective local significance testing across market states}
To filter noise while preserving the multiscale architecture across different market states $\tau$, we evaluate each directed edge $i \to j$ at time $t$ from both the transmitter's and receiver's perspectives. Let $w_{ij,t}(\tau)$ be the spillover weight from node $i$ to node $j$ in matrix $CT_t(\tau)$. For the source node $i$ at quantile $\tau$, let $s_{i,t}^{\mathrm{out}}(\tau) = \sum_j w_{ij,t}(\tau)$ be its total out-strength and $k_{i,t}^{\mathrm{out}}(\tau)$ be its out-degree. The statistical significance of this output channel, $\alpha_{ij,t}^{\mathrm{out}}(\tau)$, is defined by the following integral form:

\begin{equation}
	\alpha_{ij,t}^{\mathrm{out}}(\tau) = 1 - (k_{i,t}^{\mathrm{out}}(\tau) - 1) \int_0^{w_{ij,t}(\tau)/s_{i,t}^{\mathrm{out}}(\tau)} (1 - x)^{k_{i,t}^{\mathrm{out}}(\tau) - 2} dx
\end{equation}
Similarly, for the target node $j$ at quantile $\tau$, the significance of the input channel $\alpha_{ij,t}^{\mathrm{in}}(\tau)$ is assessed relative to its total in-strength $s_{j,t}^{\mathrm{in}}(\tau) = \sum_i w_{ij,t}(\tau)$ and in-degree $k_{j,t}^{\mathrm{in}}(\tau)$:

\begin{equation}
	\alpha_{ij,t}^{\mathrm{in}}(\tau) = 1 - (k_{j,t}^{\mathrm{in}}(\tau) - 1) \int_0^{w_{ij,t}(\tau)/s_{j,t}^{\mathrm{in}}(\tau)} (1 - x)^{k_{j,t}^{\mathrm{in}}(\tau) - 2} dx 
\end{equation}

\subsubsection{Construction of the directed weighted backbone $A_t(\tau)$}

Following the dual-testing principle \citep{FJ-Serrano-Boguna-Vespignani-2009-PNAS}, an edge $i \to j$ is preserved in the daily directed weighted backbone network $A_t(\tau)$ if it satisfies the predefined significance level $\alpha$ from at least one perspective. The decision rule is formulated as:

\begin{equation}
	A_{ij,t}(\tau)=
	\begin{cases}
		w_{ij,t}(\tau), & \text{if } \min\bigl(\alpha_{ij,t}^{\mathrm{out}}(\tau), \alpha_{ij,t}^{\mathrm{in}}(\tau)\bigr) < \alpha,\\
		0, & \text{otherwise.}
	\end{cases}
\end{equation}
The choice of $\alpha$ involves a trade-off between noise suppression 
and structural coverage: smaller values yield sparser but more 
conservative backbones, whereas larger values retain more edges at 
the cost of admitting weaker links. We therefore report results at 
both $\alpha=0.05$, the standard convention in the disparity filter 
literature, and 
$\alpha=0.10$ as a sensitivity band, so that subsequent motif and 
portfolio results can be assessed against two filtering strictness 
levels.

By retaining the original spillover weights for significant edges while setting non-significant ones to zero, we effectively eliminate noise linkages generated by market micro-fluctuations. This procedure ensures that the backbone $A_t(\tau)$ captures the robust structural characteristics of risk transmission across extreme and median market states, providing a precise foundation for subsequent mesoscopic motif analysis.

\section{Data}
\label{S4:Data description}

We examine a comprehensive set of benchmark commodities that play key roles in international financial markets and the real economy. The commodity sample spans six sectors, comprising energy, grains and oilseeds, softs, livestock, precious metals, and industrial metals, with detailed contracts listed in Table~\ref{Tb:Statistics}. To represent major regions and economies, the equity side includes seven benchmarks, namely S\&P 500, Euro Stoxx 50, FTSE 100, Nikkei 225, ASX 200, Ibovespa, and CSI 300. 
The daily prices run from April 16, 2010 to September 5, 2025, 
with all series aligned on common business days and data sourced 
from \href{https://www.investing.com}{Investing.com}. The start 
date is set by the launch of CSI 300 futures, which ensures that 
all 39 futures contracts are simultaneously tradable throughout 
the sample.

\begin{table}[H]
	\centering
	\setlength{\abovecaptionskip}{0pt}
	\setlength{\belowcaptionskip}{10pt}
	\caption{Descriptive statistics of daily logarithmic returns.}
	\label{Tb:Statistics}
{\setlength{\tabcolsep}{10pt} 	
\resizebox{\textwidth}{!}{
\begin{tabular}{l l c*{4}{d{1.3}} r r @{}}            
 \toprule
	Category & Market & Symbol &
	\multicolumn{1}{c}{Mean ($\times 10^{-3}$)} &
    \multicolumn{1}{c}{S.D.} &
	\multicolumn{1}{c}{Skew} &
	\multicolumn{1}{c}{Ex.Kurt} &
	\multicolumn{1}{c}{JB} &
	\multicolumn{1}{c}{ADF} \\ 
  \midrule
   & Crude Oil & WTI & 0.076 & 0.025 & 0.061 & 25.353 & 107536.988*** & -18.028*** \\ 
   & Natural Gas & NGS & -0.070 & 0.035 & 0.257 & 7.413 & 9236.538*** & -19.993*** \\ 
  Energy & Heating Oil & HOL & 0.008 & 0.021 & -0.970 & 12.560 & 27019.603*** & -19.519*** \\ 
   & Gasoline & GAL & -0.037 & 0.026 & -1.479 & 28.120 & 133743.068*** & -18.628*** \\ 
   & Coal& COL & 0.057 & 0.023 & -3.036 & 110.780 & 2059198.299*** & -19.655*** \\ 
   \addlinespace
   & Corn & CRN & 0.023 & 0.017 & -1.717 & 24.317 & 100892.224*** & -17.993*** \\ 
   & Wheat & WHT & 0.005 & 0.020 & 0.424 & 4.481 & 3479.580*** & -19.762*** \\ 
   & Soybeans & SBN & 0.005 & 0.013 & -0.816 & 6.599 & 7731.529*** & -18.172*** \\ 
  Grains \& Oilseeds & Rice & RIC & -0.025 & 0.016 & -2.868 & 50.829 & 437717.601*** & -21.889*** \\ 
   & Oats & OAT & 0.081 & 0.023 & -1.469 & 20.216 & 69817.056*** & -19.836*** \\ 
   & Soybean Meal & SBM & -0.000 & 0.017 & -1.445 & 15.310 & 40610.409*** & -18.626*** \\ 
   & Soybean Oil & SBO & 0.060 & 0.015 & -0.145 & 2.587 & 1133.327*** & -18.562*** \\ 
   \addlinespace
   & Sugar & SGR & -0.006 & 0.019 & -0.132 & 3.256 & 1785.196*** & -19.292*** \\ 
   & Coffee & COF & 0.273 & 0.021 & 0.269 & 1.940 & 678.314*** & -19.254*** \\ 
  Softs & Cocoa & CCA & 0.227 & 0.021 & -0.746 & 12.027 & 24570.181*** & -19.093*** \\ 
   & Cotton & CTN & -0.054 & 0.018 & -1.335 & 18.027 & 55557.385*** & -19.977*** \\ 
   & Orange Juice & OJC & 0.134 & 0.023 & -0.263 & 3.070 & 1623.375*** & -20.742*** \\ 
   & Lumber & LUM & 0.132 & 0.026 & -0.411 & 29.226 & 143006.416*** & -18.603*** \\ 
   \addlinespace
   & Live Cattle & LCT & 0.217 & 0.012 & -1.484 & 18.083 & 56176.198*** & -21.520*** \\ 
  Livestock & Feeder Cattle & FCT & 0.291 & 0.011 & 0.635 & 12.825 & 27787.008*** & -20.211*** \\ 
   & Lean Hogs & HOG & 0.026 & 0.024 & -1.451 & 27.032 & 123649.708*** & -18.877*** \\ 
   & Milk & MLK & 0.079 & 0.019 & 1.518 & 208.419 & 7268419.063*** & -18.661*** \\ 
   \addlinespace
   & Gold & GLD & 0.288 & 0.010 & -0.560 & 5.865 & 5965.072*** & -19.730*** \\ 
  Precious metals & Silver & SLV & 0.210 & 0.019 & -0.780 & 6.978 & 8553.447*** & -19.351*** \\ 
   & Palladium & PLD & 0.184 & 0.022 & -0.235 & 9.231 & 14291.498*** & -21.675*** \\ 
   & Platinum & PLT & -0.050 & 0.016 & -0.236 & 3.866 & 2537.614*** & -20.151*** \\ 
   \addlinespace
   & Aluminium & ALM & 0.018 & 0.012 & -0.030 & 2.303 & 887.788*** & -20.738*** \\ 
   & Copper & CPR & 0.060 & 0.013 & -0.285 & 3.173 & 1738.722*** & -19.374*** \\ 
  Industrial metals & Zinc  & ZNC & 0.038 & 0.015 & -0.103 & 1.951 & 644.099*** & -19.230*** \\ 
   & Lead  & LED & -0.034 & 0.015 & -0.127 & 2.661 & 1195.534*** & -19.353*** \\ 
   & Nickel & NKL & -0.140 & 0.021 & 3.178 & 90.420 & 1374500.289*** & -22.338*** \\ 
   & Tin & TIN & 0.149 & 0.015 & -0.621 & 5.565 & 5438.661*** & -19.615*** \\ 
   \addlinespace
   & S\&P 500 & SPX & 0.422 & 0.011 & -0.679 & 13.169 & 29320.259*** & -19.144*** \\ 
   & Euro Stoxx 50 & ESX & 0.153 & 0.013 & -0.479 & 7.170 & 8754.575*** & -19.519*** \\ 
   & FTSE 100 & UKX & 0.120 & 0.010 & -0.691 & 8.540 & 12520.961*** & -20.170*** \\ 
  Equity & Nikkei 225 & NKI & 0.340 & 0.013 & -0.516 & 6.311 & 6841.955*** & -20.146*** \\ 
   & ASX 200 & ASX & 0.142 & 0.010 & -0.828 & 10.732 & 19726.521*** & -17.989*** \\ 
   & Ibovespa & IBV & 0.180 & 0.015 & -0.787 & 13.062 & 28955.868*** & -17.709*** \\ 
   & CSI 300 & CSI & 0.065 & 0.014 & -0.621 & 11.685 & 23101.331*** & -19.529*** \\ 
   \bottomrule
\end{tabular}}}
\begin{flushleft}
\footnotesize
\justifying Note: 
 JB is the Jarque-Bera normality statistic. ADF is the Augmented Dickey–Fuller unit root statistic. ***, **, and * denote significance at the 1\%, 5\%, and 10\% levels, respectively.
\end{flushleft} 
\end{table}

Table~\ref{Tb:Statistics} reports summary statistics and diagnostic tests for the logarithmic returns. Mean daily returns are close to zero for most series. Standard deviations vary considerably across asset classes, reflecting heterogeneous levels of return volatility. Most series display negative skewness, indicating a longer left tail consistent with occasional sharp price declines, although a few commodities such as nickel and milk exhibit notably positive skewness. Excess kurtosis is uniformly large and positive across all series, with particularly extreme values for milk, coal, nickel, and rice, confirming heavy-tailed distributions that assign substantially more probability to extreme returns than a Gaussian benchmark would imply. The Jarque-Bera test rejects normality at the 1\% level for every series without exception, underscoring the relevance of a quantile-based connectedness framework that can capture spillover dynamics across different parts of the return distribution rather than focusing exclusively on conditional mean behavior \citep{FJ-Ando-GreenwoodNimmo-Shin-2022-ManagSci}. The Augmented Dickey-Fuller test rejects the unit root null at the 1\% level throughout, confirming that all return series are stationary and suitable for the subsequent VAR-based estimation.

\section{Empirical results}
\label{S5:Empirical results}

\subsection{Risk spillover dynamics}

\begin{figure}[H]
	\centering
	\begin{overpic}[width=0.495\textwidth]{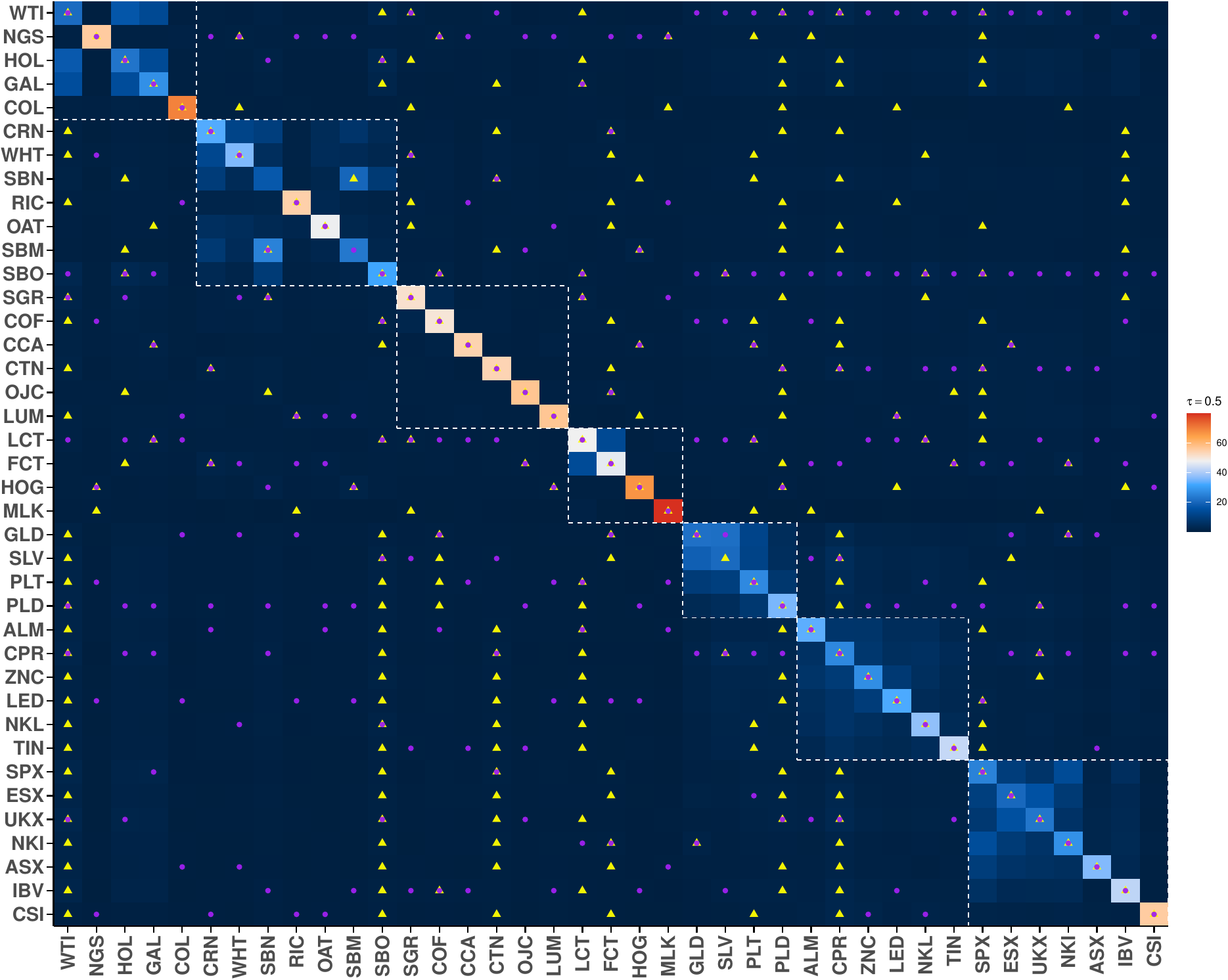}
		\put(96,76.5){\small\bfseries (a)} 
	\end{overpic}
	\begin{overpic}[width=0.499\textwidth]{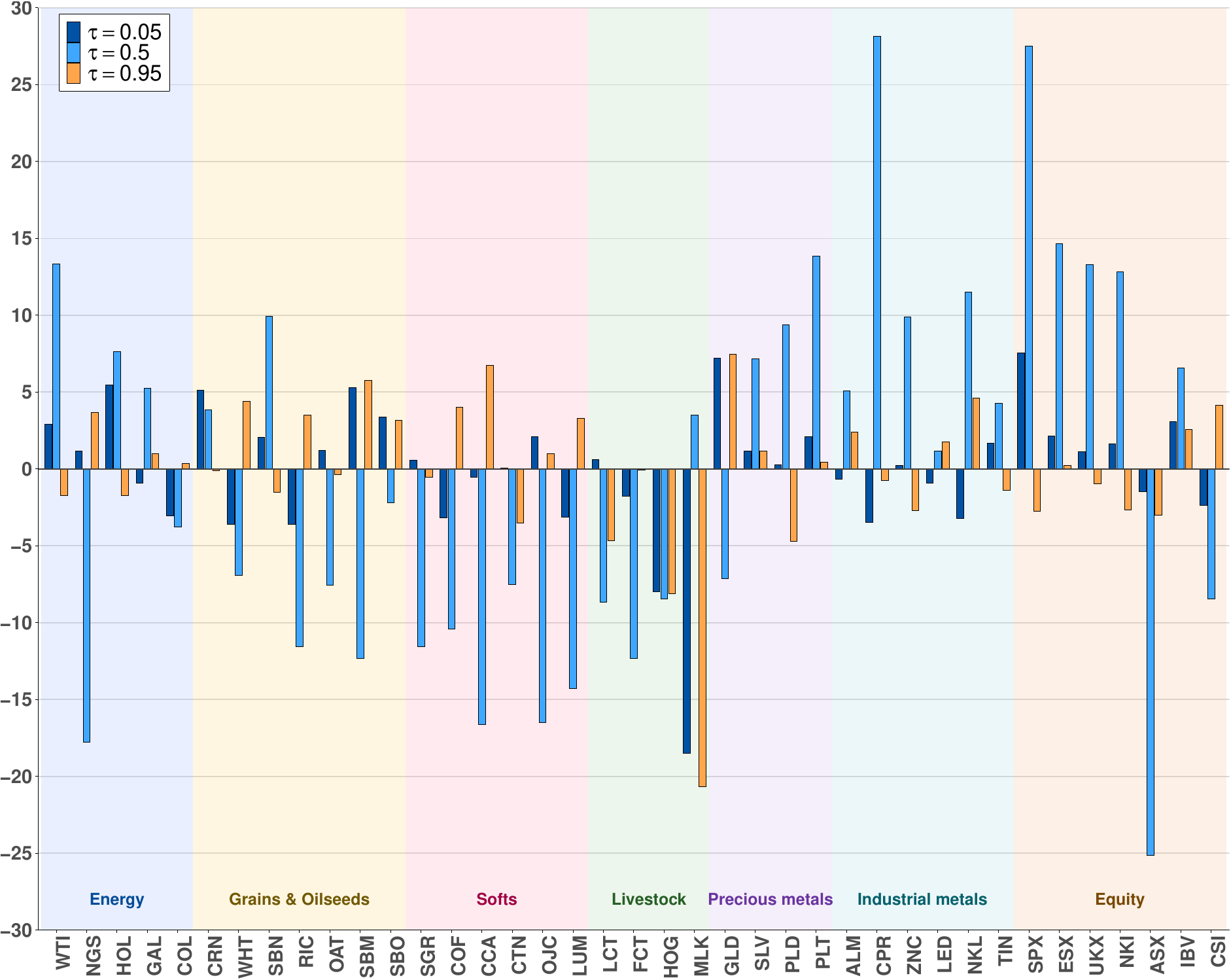}
		\put(94,75.5){\small\bfseries (b)}
	\end{overpic}
	\begin{overpic}[width=0.49\textwidth]{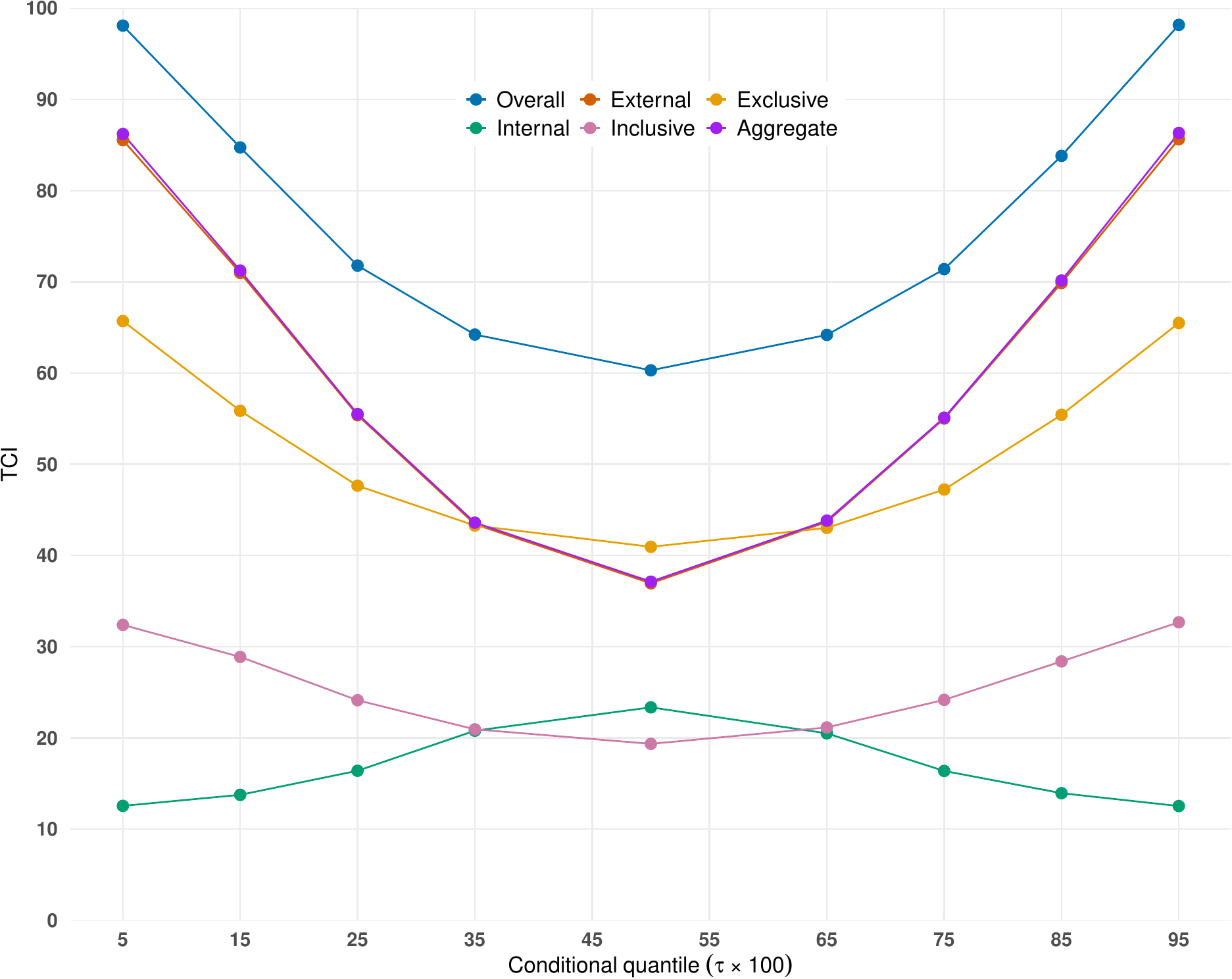}
		\put(97.5,76.5){\small\bfseries (c)} 
	\end{overpic}
	\hfill
	\begin{overpic}[width=0.499\textwidth]{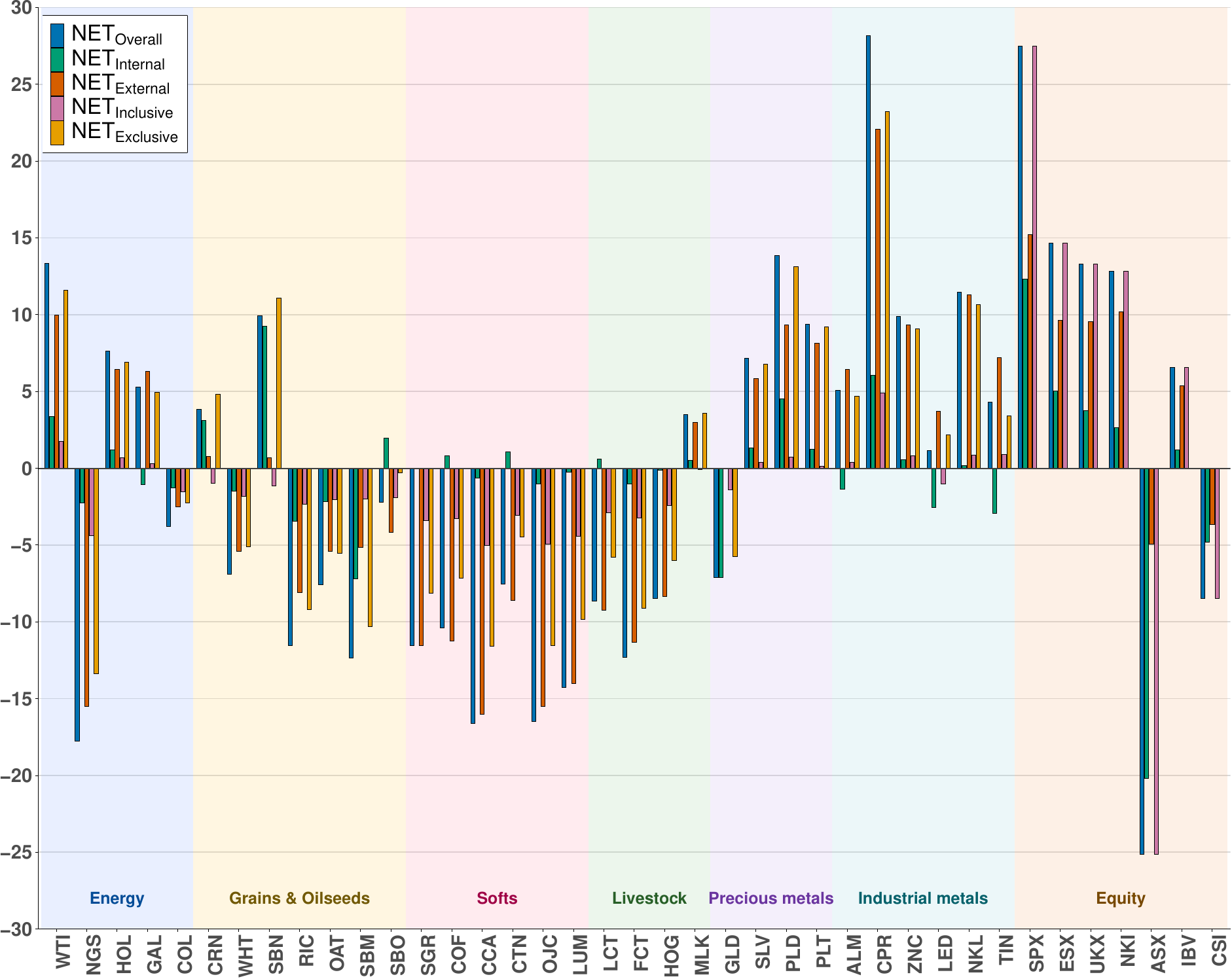}
		\put(94,75.5){\small\bfseries (d)}
	\end{overpic}
	\caption{Averaged dynamic risk spillovers. (a) Heatmap at the median quantile ($\tau=0.50$). Entries are spillovers from the transmitter (column) to the receiver (row), with dashed boxes marking sector blocks. Specifically, yellow triangles show, for each receiver, the largest incoming spillover within each block. Purple circles mark, for each transmitter, the largest outgoing spillover within each block. (b) Net spillovers by market at $\tau\in\{0.05,0.50,0.95\}$. Shaded bands indicate sectors. (c) TCI across conditional quantiles. (d) Net spillovers at the median quantile: overall, internal, external, inclusive, and exclusive.}
	\label{Fig:Static:Connectedness:M}
\end{figure}

We begin by examining the overall structure of risk spillovers across the commodity-equity system. Fig.~\ref{Fig:Static:Connectedness:M} (a) presents the averaged dynamic spillover heatmap at the median quantile, with corresponding heatmaps at the left and right tails provided in Fig.~\ref{Fig:Static:Connectedness:LR}. Spillovers are concentrated along the diagonal blocks, indicating that within-sector transmission dominates cross-sector transmission under normal market conditions. The yellow triangles and purple circles, which mark the largest incoming and outgoing spillovers within each sector block, reveal that a small number of assets serve as dominant within-sector transmitters and receivers simultaneously. Specifically, WTI in energy, SBO in grains, CTN in softs, LCT in livestock, PLD in precious metals, CPR in industrial metals, and SPX in equity emerge as the most active nodes within their respective sectors, acting as dual hubs that both absorb and propagate shocks within their sector blocks. In contrast, assets such as NGS, COL, MLK, and CSI retain high own-variance shares, suggesting weaker integration with the broader system. At the tails, the picture changes markedly. The heatmaps in Fig.~\ref{Fig:Static:Connectedness:LR} show that own-variance shares along the diagonal shrink considerably at both the left and right tails, indicating that assets become more exposed to external shocks under extreme market conditions.

Fig.~\ref{Fig:Static:Connectedness:M}(c) traces the total connectedness index across the full range of conditional quantiles for six decompositions: overall, internal, external, inclusive, exclusive, and aggregate. The overall TCI exhibits a pronounced U-shaped pattern, reaching 98.17\% at $\tau=0.05$ and 98.09\% at $\tau=0.95$, while falling to 60.30\% at the median. This indicates that, even under normal market conditions, the 39 futures contracts maintain a non-trivial degree of systemic connectedness. Under extreme conditions, whether bearish or bullish, risk spillovers intensify sharply and cross-market interdependence becomes markedly stronger across the system.

The decomposition into external and internal components reveals a clear contrast. External spillovers closely track the overall curve throughout the entire quantile range, whereas internal spillovers remain at the lowest level across all quantiles, ranging only from about 13\% to 23\%, and display a mild inverted-U profile. The persistent gap between the two suggests that cross-sector transmission, rather than within-sector linkages, constitutes the primary channel of systemic risk propagation. In other words, the sharp rise in overall connectedness at the tails is driven mainly by stronger transmission across sectors, rather than by a comparable escalation of within-sector contagion.

The exclusive and inclusive components further clarify the role of equity futures in the system. The exclusive TCI, which captures connectedness within the pure commodity network after excluding equity, remains substantially higher than the inclusive TCI, which reflects spillovers between equity and the six commodity sectors, across all quantiles. This gap conveys several implications. First, the commodity market itself forms a densely interconnected risk network even without the participation of equity indices. Second, the equity-commodity linkage is meaningful, as the inclusive component is non-negligible and also follows a U-shaped pattern. However, its magnitude remains well below that of the pure commodity network, suggesting that equity futures play a meaningful but secondary transmission role in the broader commodity risk system. Both components exhibit clear tail amplification, confirming that extreme conditions activate both channels simultaneously, although the pure commodity network consistently operates at a higher baseline.

The aggregate TCI also displays a pronounced U-shaped pattern, indicating that connectedness at the aggregated sector level intensifies markedly under extreme quantiles. Collectively, these results point to four main findings. First, risk spillovers exhibit strong nonlinear amplification under extreme market conditions. Second, cross-sector rather than within-sector transmission is the dominant source of systemic connectedness. Third, equity futures maintain a meaningful but non-dominant risk bridge with commodity markets. Fourth, the pure commodity network itself represents the more important structural basis of systemic risk in this system.

\begin{figure}[H]
	\centering
	\includegraphics[width=5.4cm]{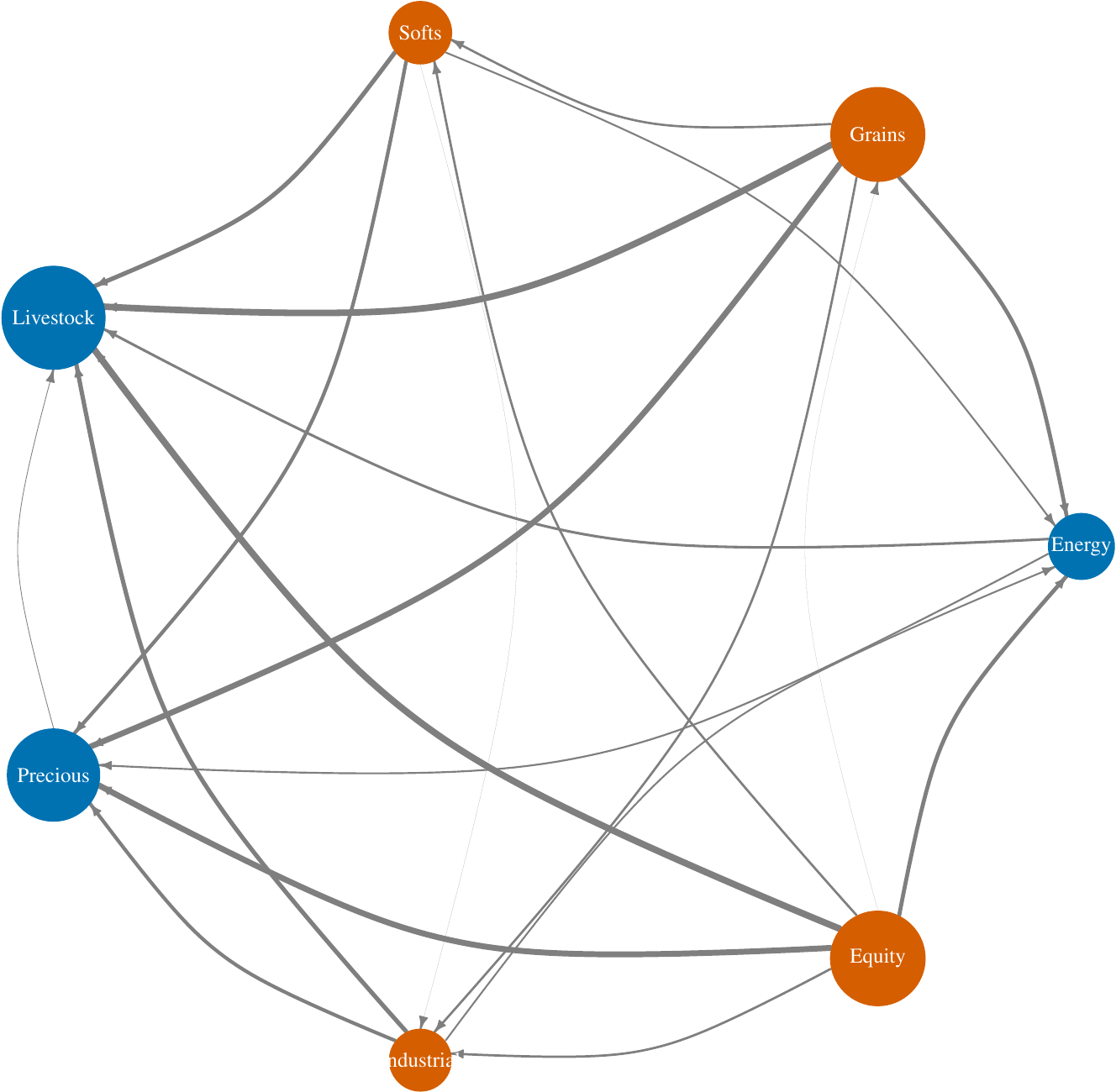}
	\includegraphics[width=5.4cm]{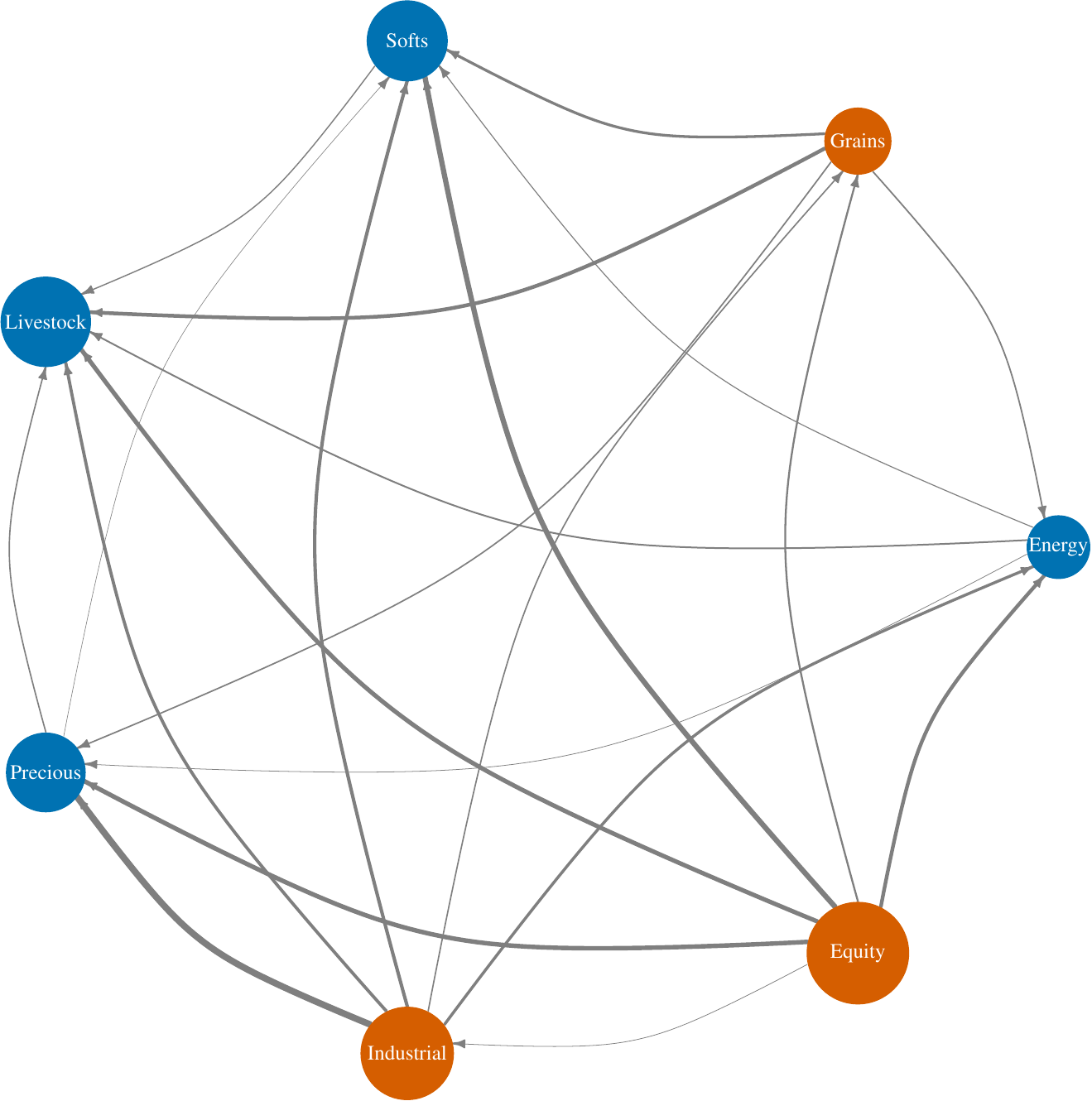}
	\includegraphics[width=5.4cm]{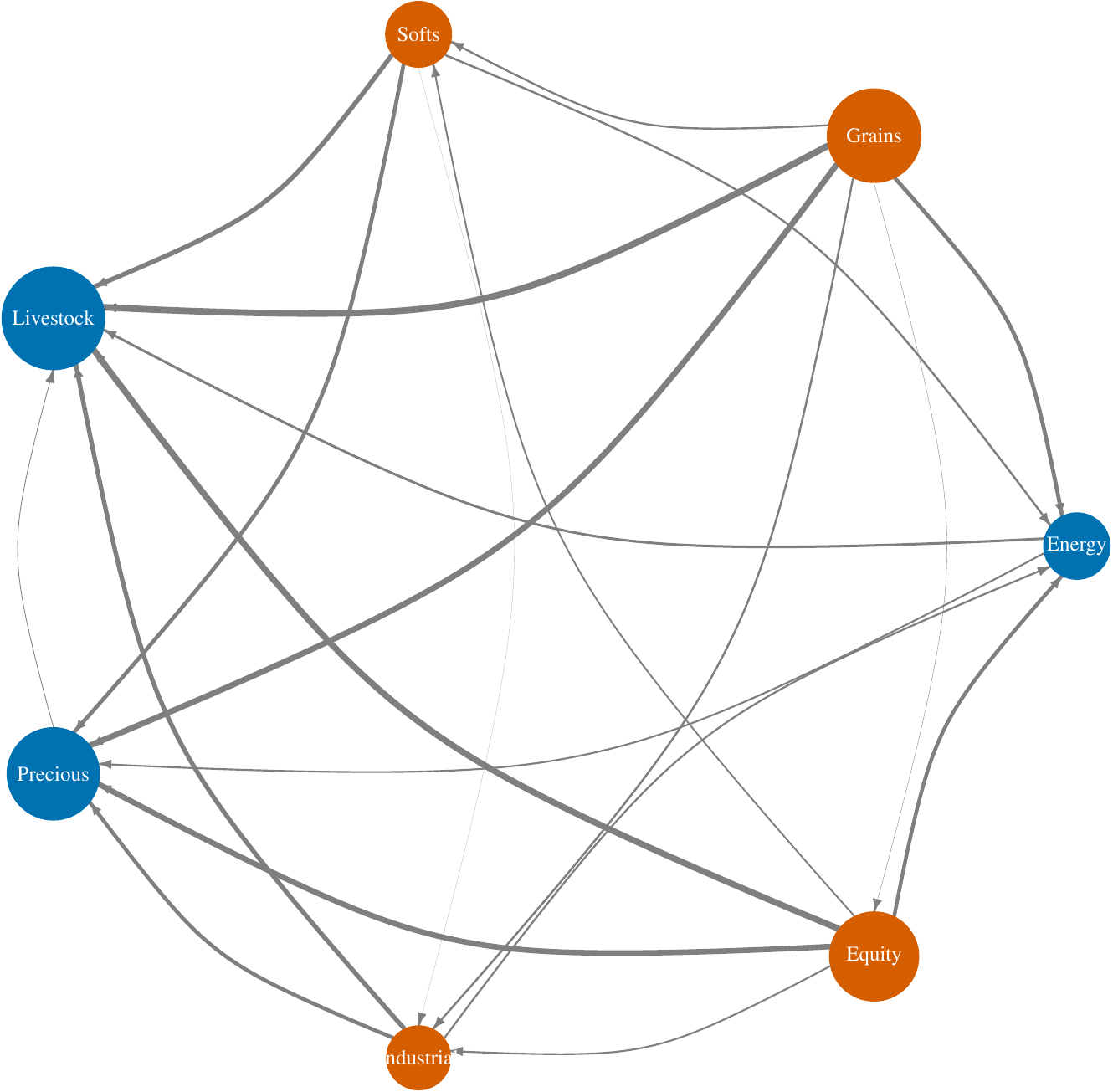}
	\caption{Net risk spillover networks derived from aggregate connectedness across seven sectors. Orange nodes denote net transmitters, while blue nodes denote net receivers. Node size reflects net total directional connectedness, and edge thickness indicates pairwise spillover strength. Left, middle, and right subfigures: $\tau=0.05, 0.50, 0.95$.}
	\label{Fig:Network:Aggregate}
\end{figure}

Fig.~\ref{Fig:Static:Connectedness:M}(b) reports net spillovers by
individual market at three quantiles ($\tau = 0.05, 0.50, 0.95$). At
the median, net spillover magnitudes are large and highly heterogeneous
across assets. CPR and SPX stand out as the dominant net transmitters at
28.17\% and 27.51\%, respectively, followed by PLT, ESX, UKX, NKI, and
WTI. On the receiving side, ASX ($-25.12\%$), NGS ($-17.75\%$), CCA
($-16.61\%$), OJC ($-16.49\%$), and LUM ($-14.28\%$) absorb the most
spillovers from the system. At the individual level, energy (except
NGS), industrial metals, and most developed-market equity indices tend to
be net transmitters, whereas softs, livestock, and most grains act as net
receivers. Precious metals present a mixed picture, with PLT, PLD, and
SLV transmitting on net while GLD absorbs.
Fig.~\ref{Fig:Network:Aggregate} provides a complementary perspective by
collapsing each sector into a single node. At the median, grains,
industrial metals, and equity emerge as net transmitting sectors, while
softs, livestock, precious metals, and energy are net receiving sectors.
The discrepancy for energy and precious metals between the individual and
aggregate views reflects the fact that much of the transmission by
individual assets in these sectors occurs within the sector rather than
across sector boundaries. At the tails, the net spillover profiles
compress markedly. Since the overall TCI approaches 98\% at both
extremes, nearly all forecast error variance is shared across the system,
leaving little room for large net directional imbalances. Several notable
role reversals occur at the individual level. CPR shifts from the
system's largest net transmitter at the median to a mild net receiver at
the left tail ($-3.46\%$). SPX turns from a strong net transmitter at
the median to a net receiver at the right tail ($-2.75\%$). GLD moves
from a net receiver at the median ($-7.12\%$) to the most prominent net
transmitter at both tails ($7.20\%$ at $\tau = 0.05$ and $7.45\%$ at
$\tau = 0.95$), consistent with gold's established role as a safe-haven
asset that transmits risk information more actively during market stress.
MLK turns from a modest net transmitter at the median ($3.52\%$) to the
largest net receiver at both tails ($-18.48\%$ and $-20.64\%$),
suggesting that it becomes a focal point for absorbing system-wide shocks
under extreme conditions. At the sector level, Fig.~\ref{Fig:Network:Aggregate} shows that
livestock, precious metals, and energy consistently act as net receiving
sectors across all three quantiles, while grains, industrial metals, and
equity remain net transmitting sectors throughout. The only sector-level
role reversal involves softs, which shifts from a net receiving sector at
the median to a net transmitting sector at both tails, suggesting that
soft commodities become a more active source of cross-sector risk
propagation under extreme market conditions.

Fig.~\ref{Fig:Static:Connectedness:M}(d) decomposes net spillovers at the median into overall, internal, external, inclusive, and exclusive components, with the corresponding tail-quantile decompositions shown in Fig.~\ref{Fig:Static:Connectedness:int:ext:NET:LR}. At the median, the overall, external, and exclusive bars move closely together for most assets, while the internal component remains near zero throughout, reinforcing the earlier finding that cross-sector transmission is the primary driver of net directional spillovers. The inclusive component is also small, confirming that the equity-commodity channel contributes modestly to the net transmission balance of individual assets. At the tails, all decomposition components compress in parallel, and the alignment between overall and external bars is preserved, indicating that the dominance of cross-sector transmission holds across the entire quantile spectrum.

\begin{figure}[H]
	\centering
	\includegraphics[width=10.4cm]{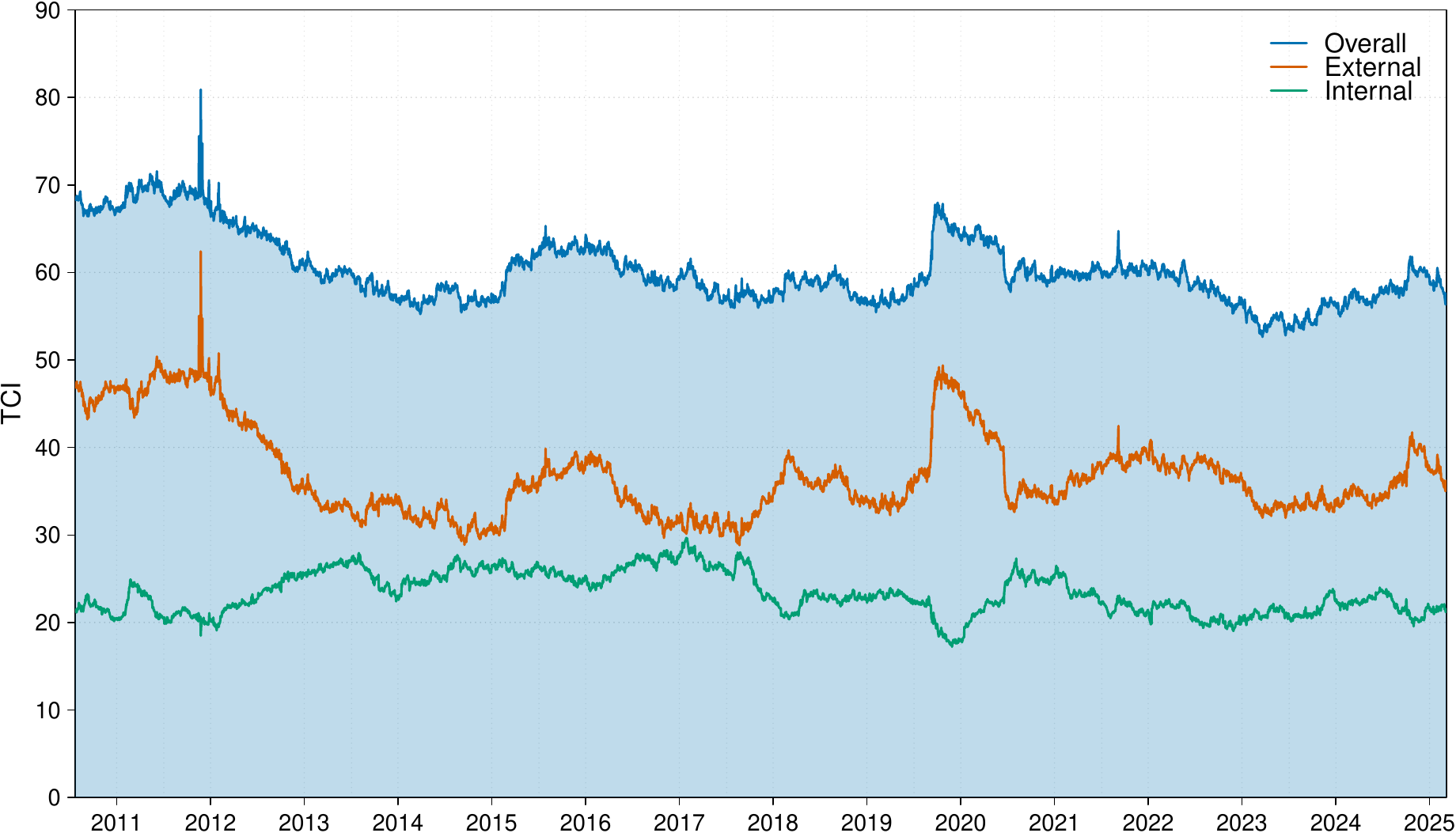}
	\includegraphics[width=10.4cm]{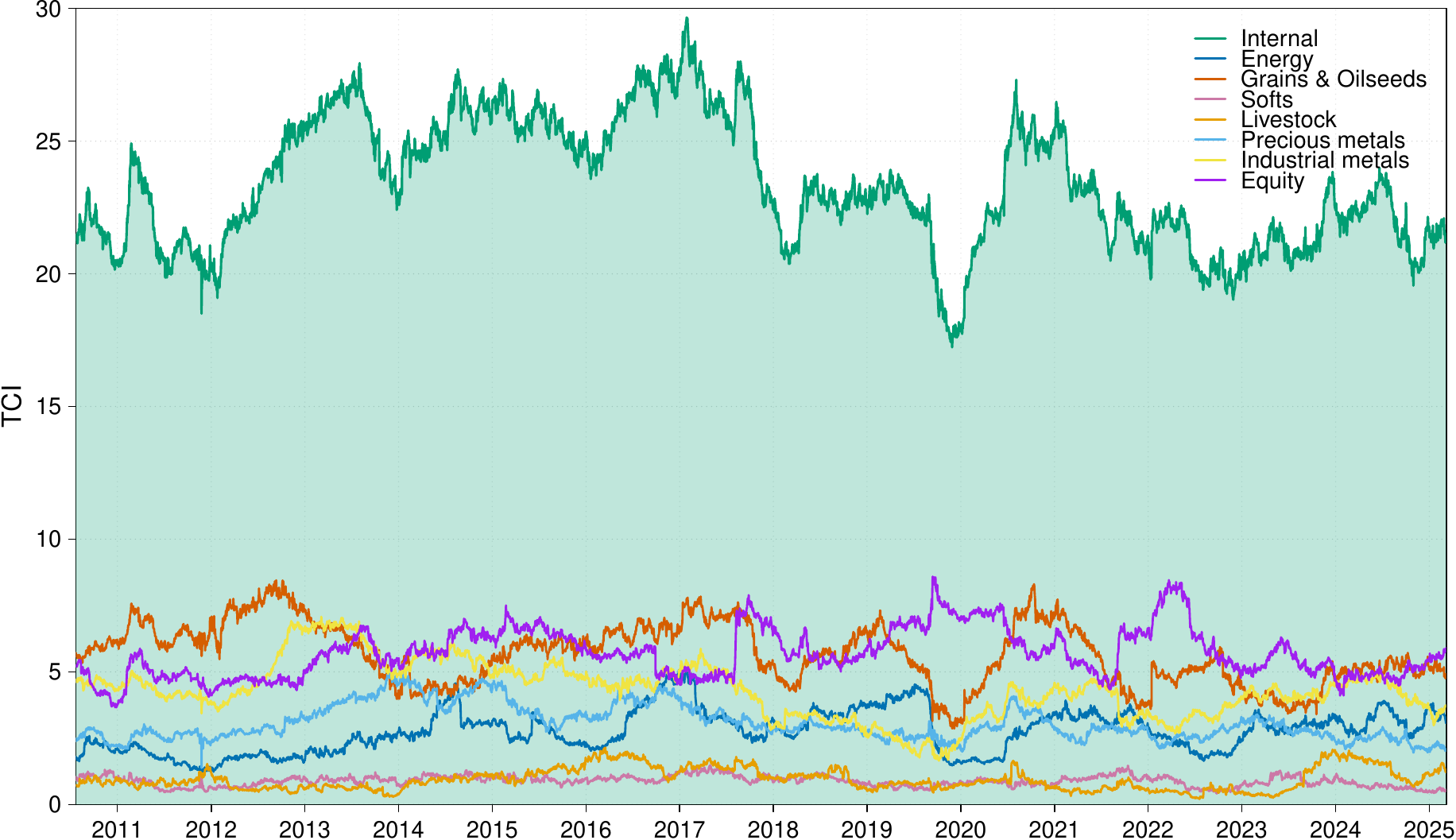}
	\includegraphics[width=10.4cm]{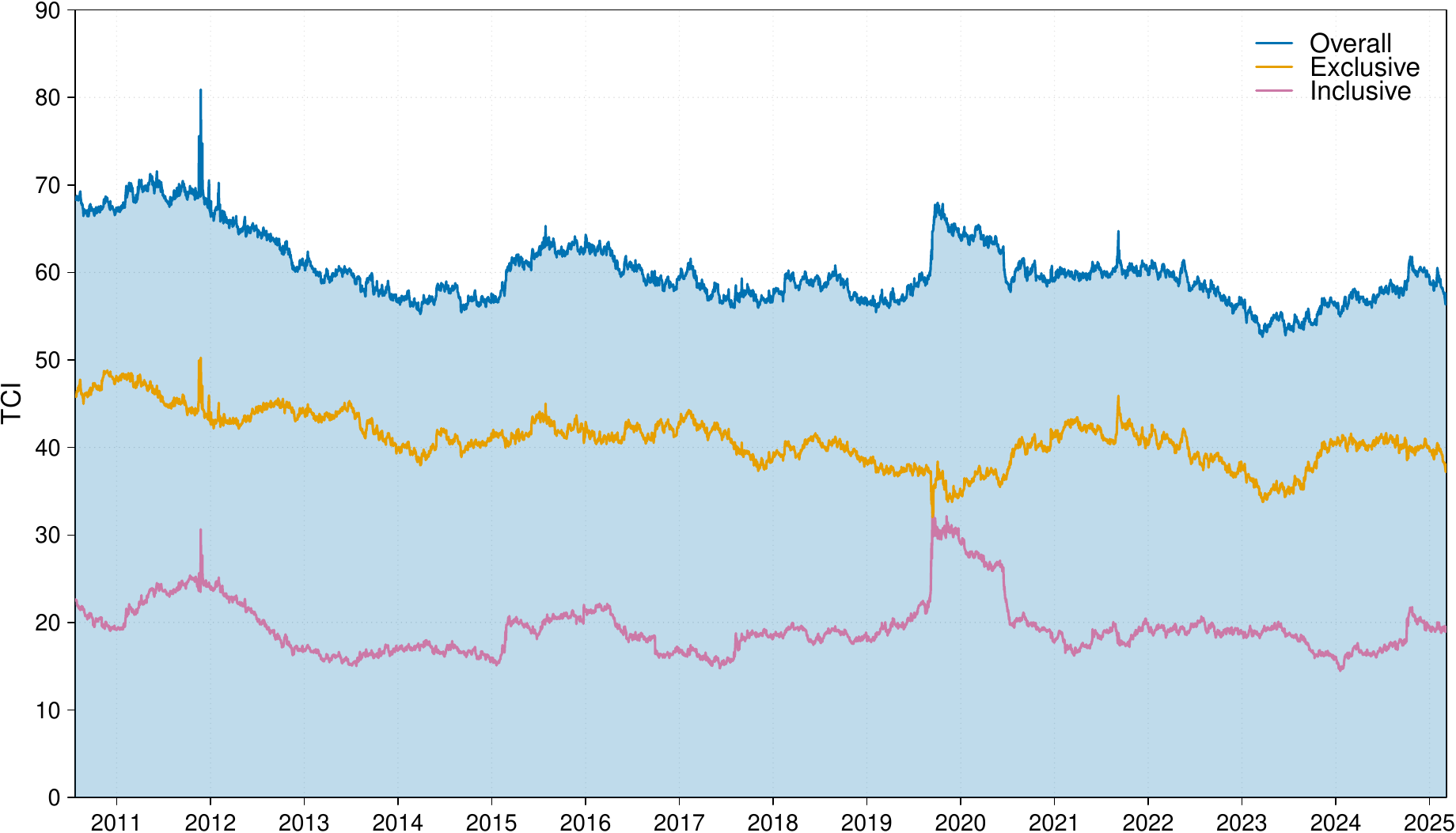}
	\caption{Time-varying TCI at the median quantile ($\tau=0.50$).
		Top row: overall, external, and internal. Middle row: sector-level decomposition of internal
		connectedness. Bottom row: overall, exclusive, and inclusive.}
	\label{Fig:Dynamic:TCI:M}
\end{figure}

Fig.~\ref{Fig:Dynamic:TCI:M} presents the time-varying TCI at the median quantile
($\tau=0.50$), with the corresponding tail-quantile dynamics shown in
Fig.~\ref{Fig:Dynamic:TCI:LR}. The temporal dynamics differ markedly across
quantiles.
At the median, the overall TCI fluctuates broadly between 55\% and 70\%,
exhibiting two clearly identifiable crisis-driven surges. The first and most prominent occurs during 2011 and 2012, when the overall TCI rises sharply, coinciding with the European sovereign debt crisis and heightened commodity price volatility. After a prolonged period of gradual decline through the mid-2010s, a second sharp surge emerges in early 2020, when the COVID-19 pandemic triggers a rapid jump in the overall TCI, followed by a gradual reversion. From 2021 onward, the overall TCI follows a downward trend. Notably, 
the Russia-Ukraine conflict does not generate a system-wide TCI surge 
comparable to the two preceding episodes, consistent with its more 
sector-specific, energy-centred profile rather than a broad-based 
financial-macro shock. These dynamics confirm that the aggregate risk transmission of the commodity-equity system is crisis-sensitive, but also that not all geopolitical or macroeconomic shocks generate comparable system-wide responses.

The external TCI closely mirrors the overall curve throughout, while the
internal TCI evolves within a narrower band of approximately 20\% to
27\%, confirming that the cross-sector dominance documented in the static
analysis persists dynamically. The sector-level decomposition of internal connectedness shows that the leading within-sector contributors alternate over time, with grains and oilseeds maintaining the most stable contribution and the equity sector exhibiting more episodic spikes in the second half of the sample. Energy, precious
metals, softs, and livestock consistently contribute less. Even so, these
within-sector components remain limited in magnitude relative to overall
and external connectedness, confirming that time variation in systemic
connectedness under normal conditions predominantly reflects the
evolution of inter-sector linkages.

The exclusive and inclusive components also exhibit clear time variation. The exclusive TCI fluctuates between approximately 35\% and 50\%,
consistently above the inclusive TCI which ranges from roughly 15\% to
30\%. A notable exception occurs during the COVID-19 pandemic, when the
inclusive component spikes from about 16\% to over 30\%, temporarily
narrowing the gap with the exclusive curve. This suggests that the
equity-commodity transmission channel is activated more strongly during
episodes of broad financial market distress, even though it remains
secondary under normal conditions.

At the tails, the dynamic picture changes fundamentally. As shown in
Fig.~\ref{Fig:Dynamic:TCI:LR}, the overall TCI at both $\tau = 0.05$ and
$\tau = 0.95$ is compressed near 98\% throughout the entire sample with
virtually no time variation, and all decomposition components remain
similarly flat. The left and right tail paths are highly similar. This
near-absence of temporal variation indicates that, under extreme
conditions, the risk transmission architecture reaches a saturated state
in which crisis-specific events no longer produce distinguishable
incremental effects on system-wide connectedness.

\begin{figure}[H]
	\centering
	\includegraphics[width=12cm]{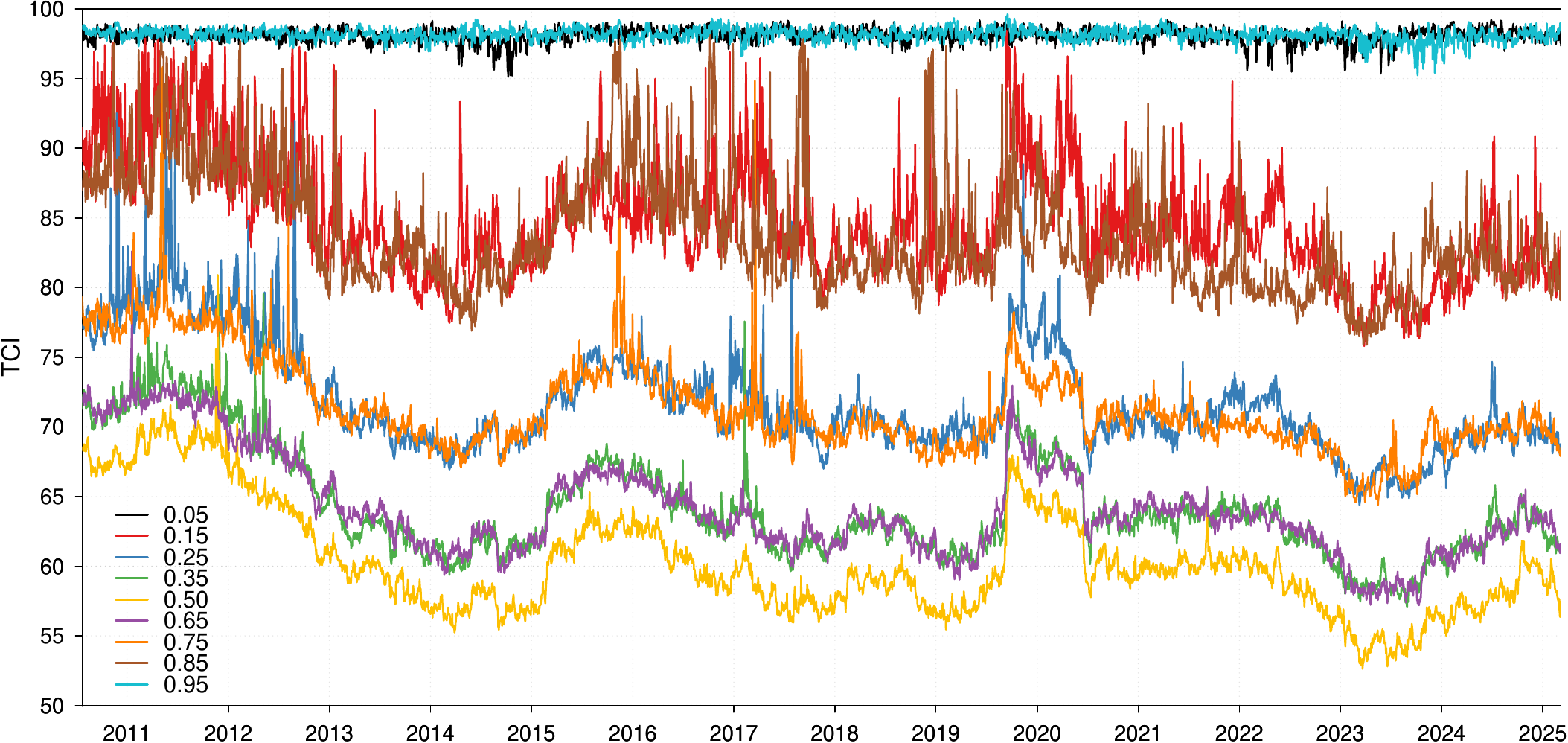}
	\caption{Time-varying overall TCI across conditional quantiles
		($\tau = 0.05, 0.15, 0.25, 0.35, 0.50, 0.65, 0.75, 0.85, 0.95$).}
	\label{Fig:TCI:Dynamic:quantiles}
\end{figure}

Fig.~\ref{Fig:TCI:Dynamic:quantiles} further shows that the
quantile ordering of connectedness is highly stable over time. The tail
quantiles ($\tau = 0.05$ and $\tau = 0.95$) hover near full
connectedness throughout the sample with minimal variation, while the
median quantile traces a substantially lower and more volatile path.
Intermediate quantiles lie smoothly between these extremes and exhibit
richer time variation, with visible responses to major crisis episodes.
This confirms that the U-shaped quantile dependence documented above is
not an artifact of time averaging but a persistent structural feature of
the system, and that the distinction between tail and normal states lies
not only in the level of connectedness but also in its temporal
stability.

\begin{figure}[H]
	\centering
	\begin{overpic}[width=0.495\textwidth]{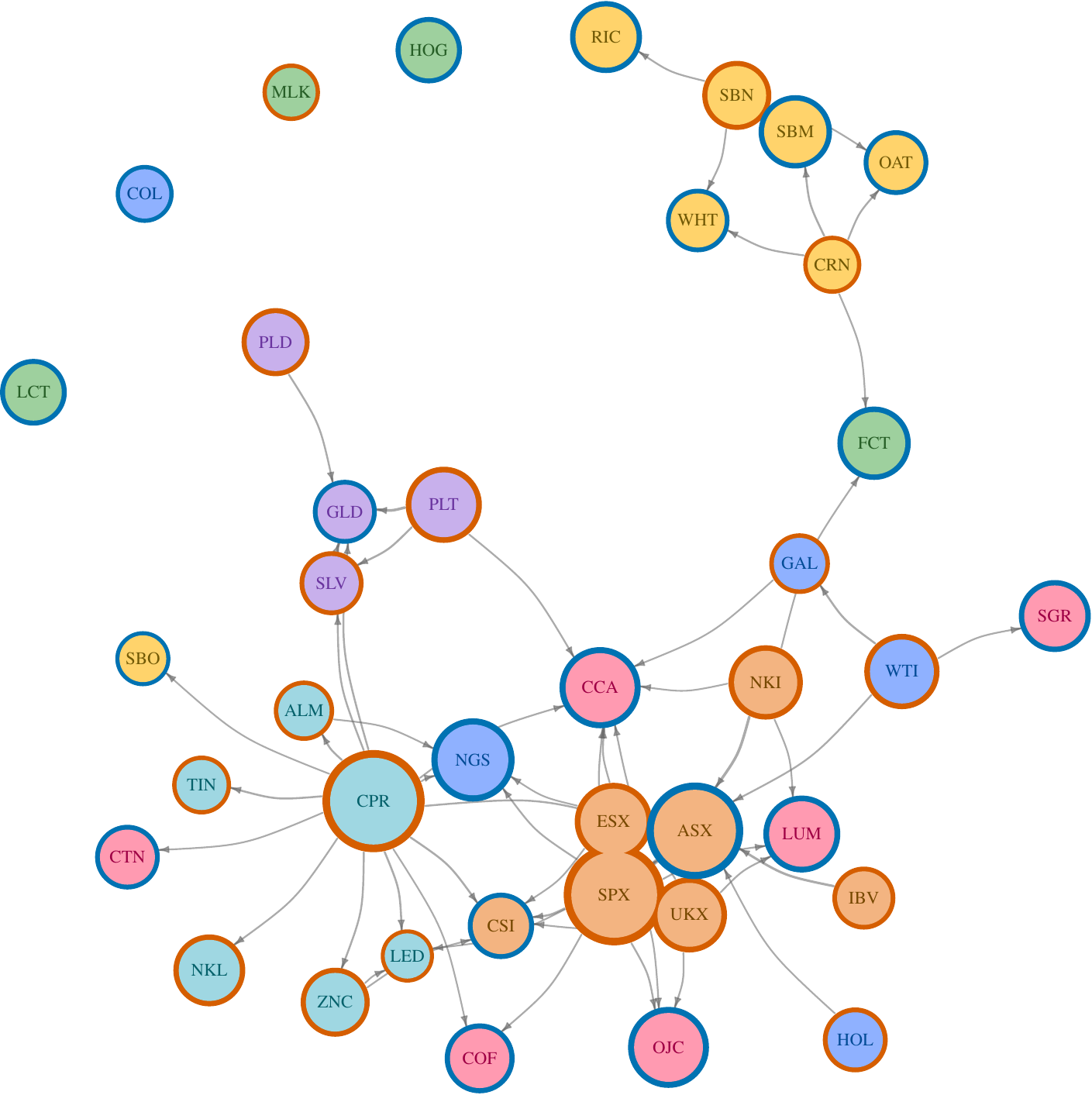}
		\put(1,90.5){\small\bfseries (a)} 
	\end{overpic}
	\hfill
	\begin{overpic}[width=0.499\textwidth]{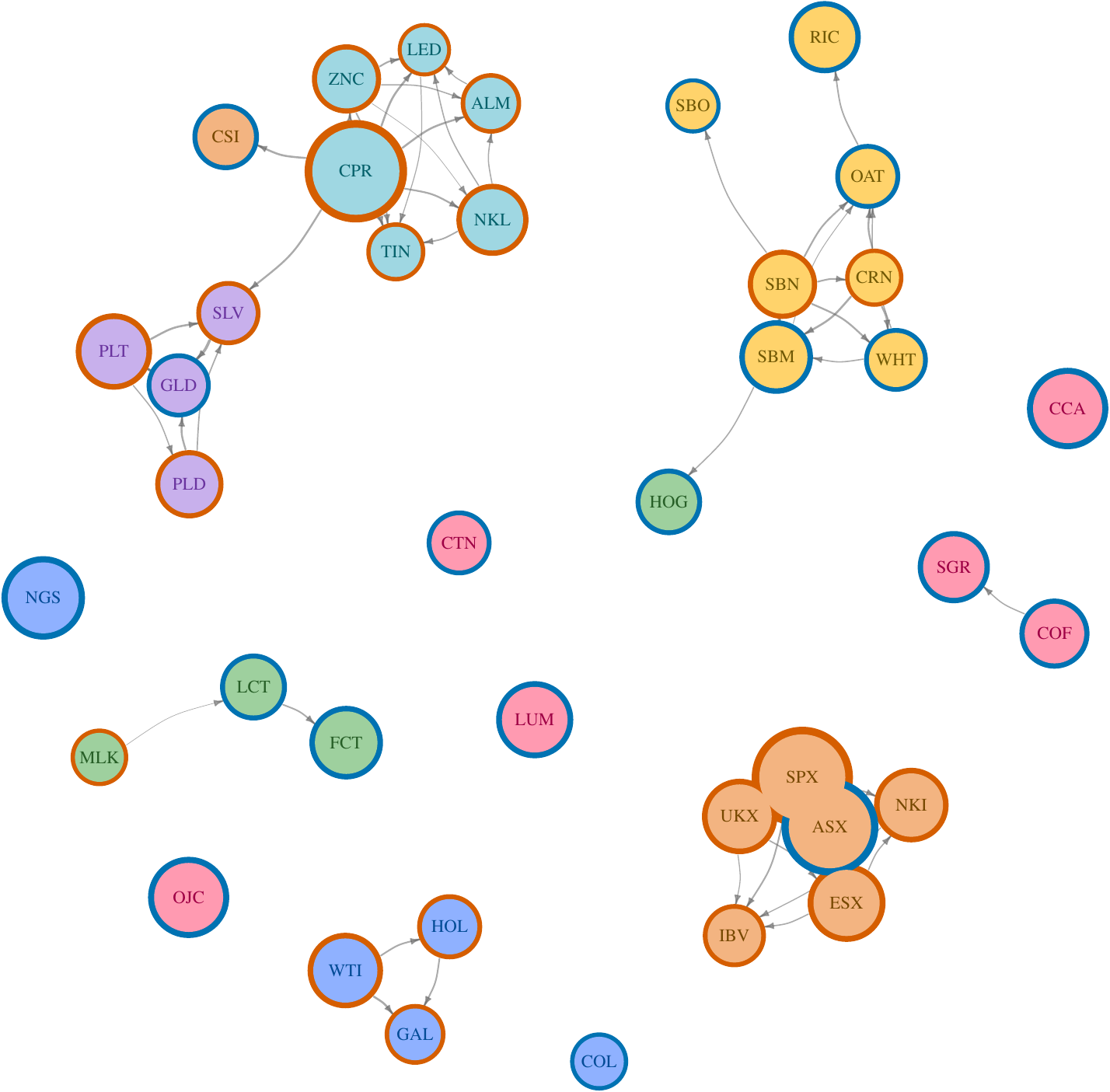}
		\put(1,90.5){\small\bfseries (b)}
	\end{overpic}
	\caption{Visualization of overall spillover risk networks ($\tau=0.50$). Left: Threshold-based network (cutoff = 0.1). Right: Backbone-based network ($\alpha = 0.1$). Outer rings distinguish net transmitters (orange) from net receivers (blue). Node size scales with net total directional connectedness; edge thickness represents pairwise spillover strength.}
	\label{Fig:NPDC:Backbone:Network:M}
\end{figure}

\subsection{Backbone extraction}

To distinguish persistent transmission channels from links that are
artifacts of excessive network density, we compare simple thresholding
with backbone extraction at both the averaged and daily network levels \citep{FJ-Serrano-Boguna-Vespignani-2009-PNAS}.
Figs.~\ref{Fig:NPDC:Backbone:Network:M} and \ref{Fig:NPDC:Backbone:Network:LR}
first present the contrast between threshold-based networks and
backbone-filtered networks across quantiles at the overall averaged
level. At the median quantile ($\tau=0.50$), the threshold-based network
remains relatively dense and preserves numerous cross-sector connections.
After backbone extraction, however, the topology becomes markedly sparser
and more modular, retaining only core within-sector clusters and
selective cross-sector bridges, while many weakly embedded nodes become
isolated. For instance, grains and oilseeds, industrial metals, precious
metals, and equity indices each form compact local communities. CPR
connects the industrial metals cluster to equity and precious metals, and
SBN links grains to softs. Peripheral assets such as CTN, OJC, COL, and
LUM retain few or no surviving edges, confirming that their bilateral
linkages in the full matrix are largely artifacts of network density.
This suggests that under normal market conditions the spillover network
contains a relatively stable structural core, whereas part of the density
observed under simple thresholding is in fact generated by weak or
redundant links. By contrast, at the tail quantiles ($\tau=0.05$ and
$\tau=0.95$), the threshold-based averaged networks are nearly complete,
whereas the corresponding backbone-filtered networks are reduced to
largely disconnected nodes. In other words, although aggregate connectedness is extremely high 
in tail states, the underlying pairwise transmission channels are 
not stable enough to form a persistent average backbone. Mechanically, 
the disparity filter compares each edge to a local null in which a 
node's total strength is uniformly distributed across its neighbours, 
so as tail-state weights become more evenly spread, edges deviate 
less from the null and fewer pass the significance threshold. Tail connectedness is thus broad in scope but locally diffuse, without a persistent structural core.

\begin{figure}[H]
	\centering
	\begin{overpic}[width=0.495\textwidth]{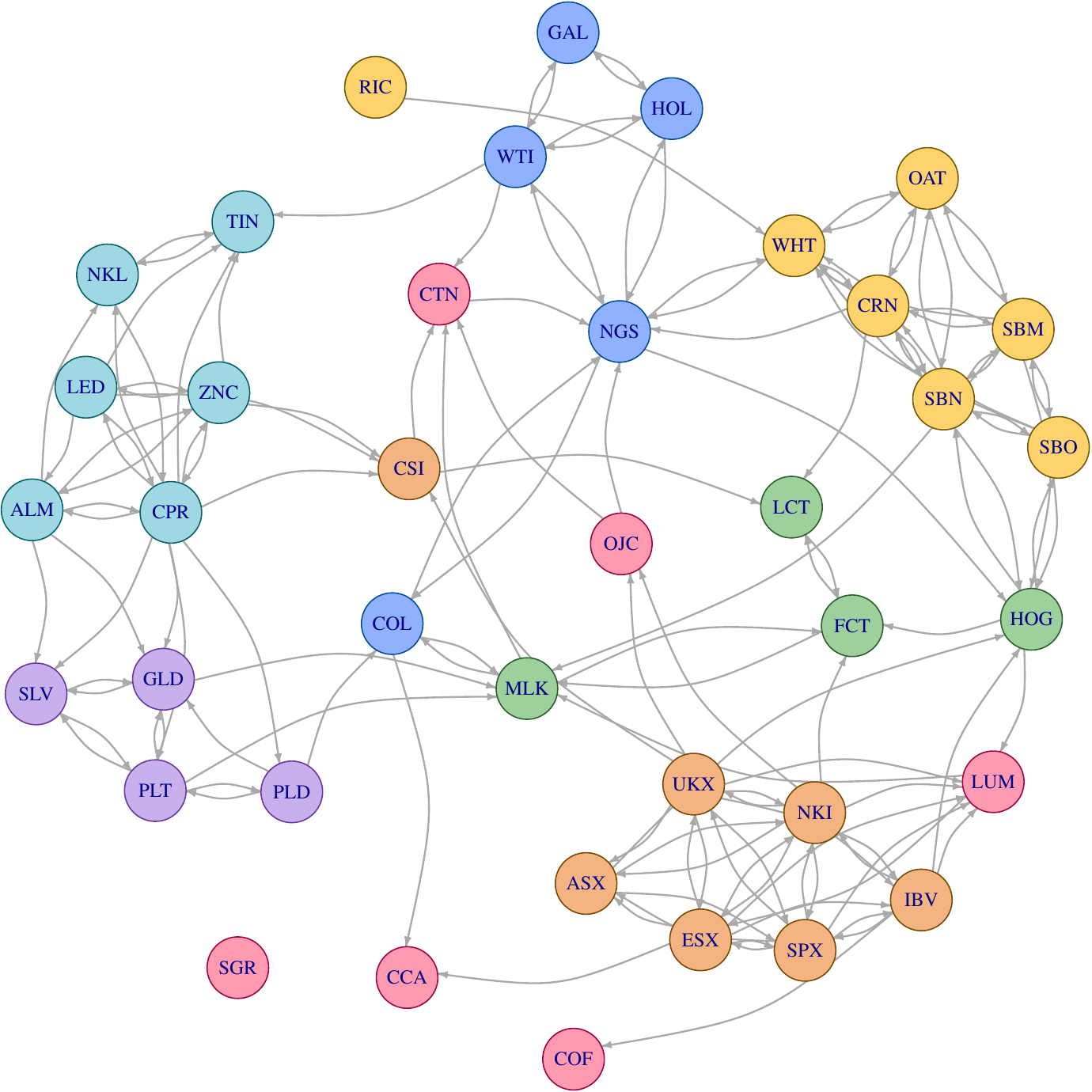}
		\put(1,94.5){\small\bfseries (a)} 
	\end{overpic}
	\begin{overpic}[width=0.499\textwidth]{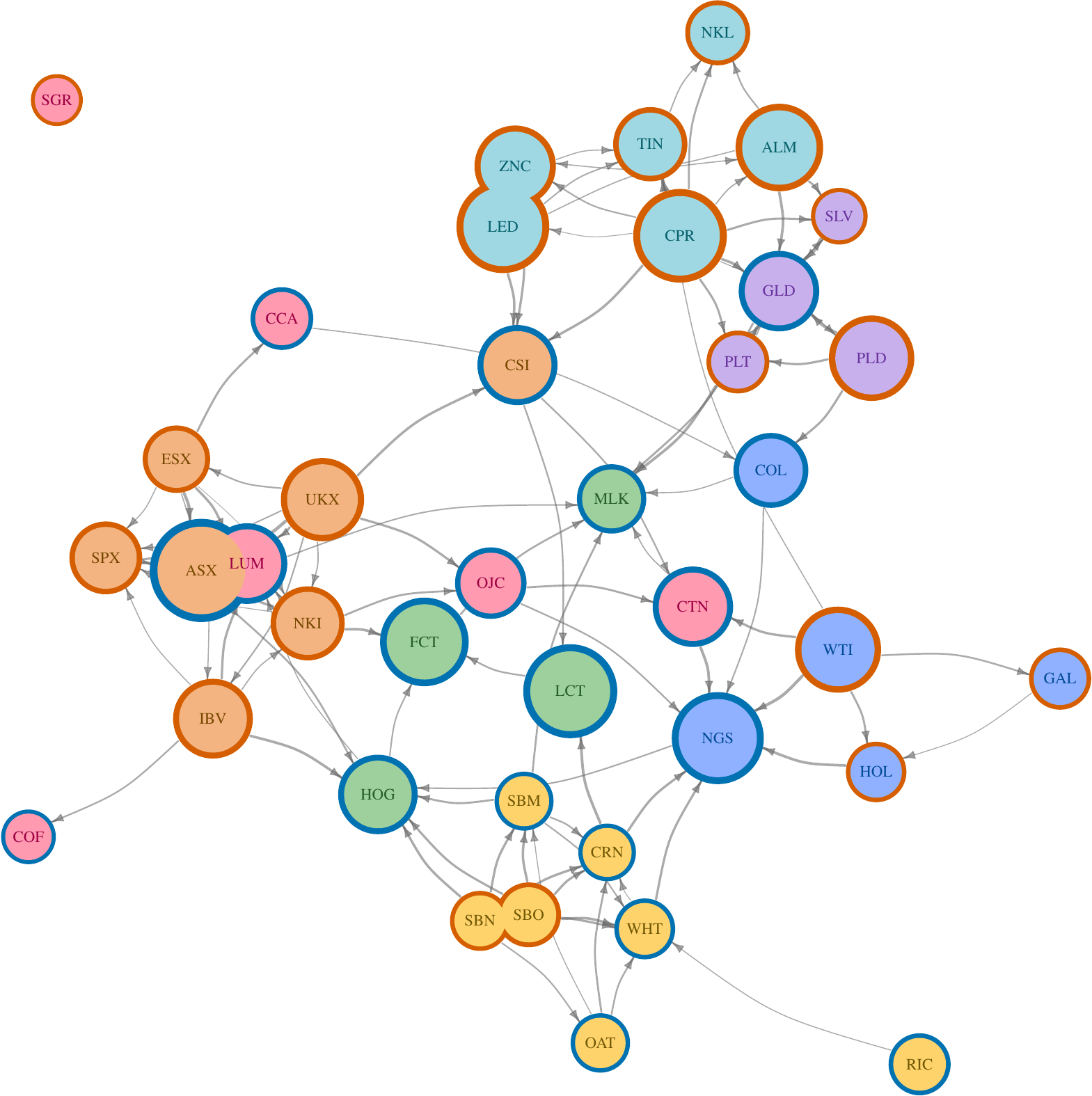}
		\put(1,94.5){\small\bfseries (b)}
	\end{overpic}
	\caption{Network backbones of a representative daily connectedness structure ($\alpha = 0.1$, $\tau=0.50$).
		Left: CT-based topology. Right: NPDC-based topology. In the NPDC plot, outer rings denote net transmitters (orange) and net receivers (blue). Node size reflects net total directional connectedness; edge thickness indicates pairwise spillover strength.}
	\label{Fig:CT:NPDC:Backbone:20120120:M}
\end{figure}

The representative daily structures in
Figs.~\ref{Fig:CT:NPDC:Backbone:20120120:M} and \ref{Fig:CT:NPDC:Backbone:20120120:LR}
further support this interpretation. At the median quantile, the daily
backbone extracted from pairwise connectedness $CT(\tau)$ remains sparse
and modular, while the NPDC representation further shows that the
surviving key links are directionally asymmetric and organized around a
few net transmitting and net receiving nodes. By contrast, even within
the same trading day, the backbone structures associated with the lower
tail ($\tau=0.05$) and the upper tail ($\tau=0.95$) are not identical:
the former is sparser and more selective, whereas the latter exhibits
stronger cross-sector concentration. This indicates that, although
connectedness in extreme states is highly pronounced at the aggregate
level, the underlying transmission topology varies substantially across
quantile states. Consistent with the earlier evidence on elevated tail
connectedness, these results further suggest that extreme market
conditions are associated not only with higher levels of connectedness,
but also with network structures that are less stable and more readily
reconfigured. Overall, backbone extraction shows that persistent
spillover architecture is most readily identifiable under normal market
conditions, whereas extreme states are characterized by high
connectedness but structurally unstable network configurations.

\subsection{Motif significance profiles}

\begin{figure}[H]
	\centering
	\includegraphics[width=0.85\linewidth]{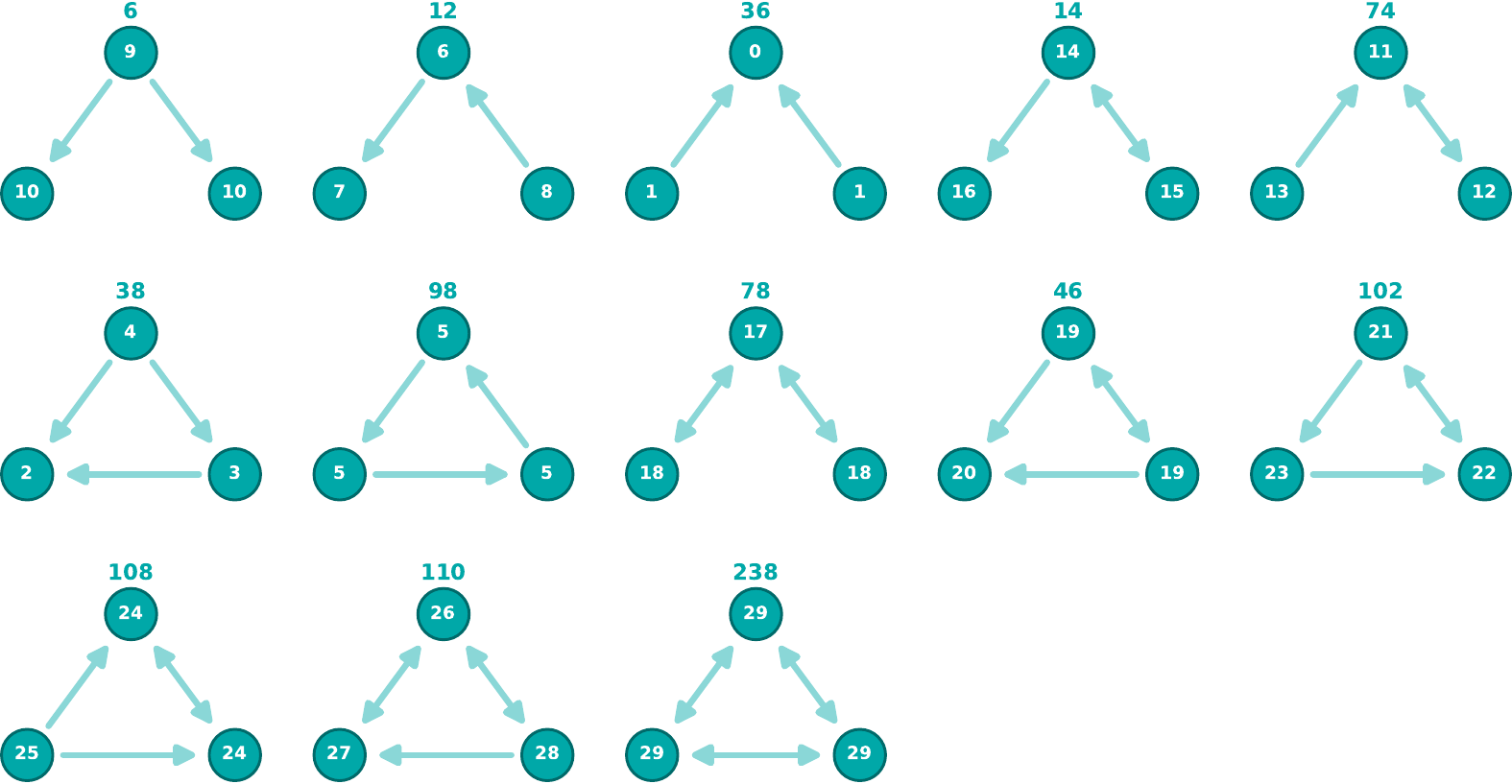}
	\caption{Schematic representation of the 13 directed triadic motifs and the 30 unique orbits.}
	\label{Fig:motif:schematic}
\end{figure}

As shown in Fig.~\ref{Fig:motif:schematic}, this study follows the three-node directed motif numbering scheme proposed by \cite{Milo-ShenOrr-Itzkovitz-Kashtan-Chklovskii-Alon-2002-Science} and further distinguishes node positions based on automorphism orbits, thereby identifying 13 directed triadic motifs and their corresponding 30 unique node orbits.
Tables~\ref{Tab:motif:M} and \ref{Tab:motif:LR} report the
daily directed motif statistics at the median and tail quantiles under
two backbone significance levels ($\alpha = 0.05$ and $\alpha = 0.10$).
Both levels are reported to assess the sensitivity of the results to the
backbone filtering threshold. A clear contrast emerges between the
median and the tails.

At the median quantile ($\tau = 0.50$), the motif significance profile
is strongly structured and highly robust across the two filtering levels.
Several motifs are markedly over-represented relative to the randomized
benchmark, most notably motif 238, followed by motifs 110, 78, and 46,
all of which remain significantly positive under both specifications.
These over-represented motifs share a common structural feature in that
they are dominated by reciprocal (bidirectional) edges. Motif 238, the
fully reciprocal triad in which all three pairs exchange risk
bilaterally, is by far the most over-represented configuration. Motifs
110 and 78 retain two reciprocal pairs, while motif 46 combines one
reciprocal pair with coordinated unidirectional transmission. The
prevalence of these configurations indicates that mutual risk exchange
between asset pairs is the statistically preferred mode of local
transmission under normal market conditions. By contrast, motifs 6, 12,
36, and 38 are strongly under-represented, and these configurations are
composed entirely of unidirectional edges, encoding purely one-way
transmission patterns such as fan-out broadcasting, fan-in convergence,
simple intermediation chains, and feedforward cascades. Their systematic
suppression indicates that purely asymmetric, one-way risk propagation is
significantly rarer than expected in the median-state spillover network.
The coexistence of strongly favored reciprocal motifs and strongly
suppressed unidirectional motifs confirms that normal-state spillovers
exhibit persistent local organization built around bilateral risk
exchange rather than diffuse or purely hierarchical linkage patterns.

This motif structure largely disappears at the tail quantiles. At both
the left tail ($\tau = 0.05$) and the right tail ($\tau = 0.95$),
z-scores of all motifs compress toward zero under both $\alpha$ levels,
and no motif displays the strong over-representation or
under-representation observed at the median. The ordering of motifs also
shifts. Motif 36, a fan-in structure that is significantly
under-represented at the median, becomes the motif with the highest
(though still modest) z-score at both tails, while motif 238 drops to
near zero. In terms of raw frequency, motif 6, a fan-out structure in
which a single node broadcasts risk unidirectionally to two others,
becomes by far the most prevalent configuration at both tails despite
being significantly under-represented at the median. Its z-score remains
modest, however, because the degree-preserving null model also generates
high counts for this configuration under dense tail-state backbones.
This reversal in both significance ranking and raw prevalence suggests
that the local transmission architecture is not merely weakened at the
tails but qualitatively reorganized. Extreme states thus intensify
spillovers at the aggregate level but do not produce a stable local
triadic architecture. The consistency of this result across the two tails
and across the two $\alpha$ levels confirms that the absence of strong
motif significance in extreme states is not a threshold artifact but
reflects the structural instability of tail-state spillover
organization.

\begin{table}[H]
	\centering
	\setlength{\abovecaptionskip}{0pt}
	\setlength{\belowcaptionskip}{10pt}
	\caption{Daily directed motif statistics at the median quantile.}
	\label{Tab:motif:M}
	\resizebox{\textwidth}{!}{
		\begin{tabular}{llrrrrr|llrrrrr}
			\toprule
			\multicolumn{7}{c|}{$\alpha=0.05$} & \multicolumn{7}{c}{$\alpha=0.10$}\\
			\cmidrule(lr){1-7} \cmidrule(lr){8-14}
			id & motif & $\mu$ & $\sigma$ & $\mu_{\text{rnd}}$ & $\sigma_{\text{rnd}}$ & $z$ &
			id & motif & $\mu$ & $\sigma$ & $\mu_{\text{rnd}}$ & $\sigma_{\text{rnd}}$ & $z$\\
			\midrule
			238 & \motif{238} & 13.928 & 6.335 & 0.005 & 0.073 & 189.951&
			238 & \motif{238} & 28.370 & 7.930 & 0.042 & 0.207 & 136.697\\
			110 & \motif{110} & 8.674 & 3.830 & 0.250 & 0.523 & 16.108&
			78 & \motif{78} & 52.938 & 18.435 & 5.107 & 3.356 & 14.253\\
			78 & \motif{78} & 21.128 & 8.409 & 1.332 & 1.503 & 13.173&
			110 & \motif{110} & 20.316 & 6.460 & 1.458 & 1.343 & 14.043\\
			46 & \motif{46} & 5.505 & 3.577 & 1.024 & 1.023 & 4.382&
			46 & \motif{46} & 13.822 & 10.698 & 4.832 & 2.296 & 3.916\\
			14 & \motif{14} & 42.664 & 13.585 & 25.874 & 9.627 & 1.744&
			14 & \motif{14} & 117.974 & 30.832 & 73.116 & 17.267 & 2.598\\
			108 & \motif{108} & 1.473 & 1.516 & 0.562 & 0.742 & 1.228&
			74 & \motif{74} & 82.485 & 19.979 & 60.175 & 14.351 & 1.555\\
			74 & \motif{74} & 24.913 & 8.285 & 21.032 & 7.775 & 0.499&
			108 & \motif{108} & 5.467 & 3.820 & 3.788 & 1.946 & 0.863\\
			102 & \motif{102} & 1.340 & 1.362 & 1.596 & 1.323 & -0.194&
			102 & \motif{102} & 4.762 & 2.884 & 9.301 & 3.216 & -1.411\\
			98 & \motif{98} & 0.162 & 0.415 & 1.273 & 1.106 & -1.005&
			98 & \motif{98} & 0.902 & 1.066 & 12.699 & 3.752 & -3.144\\
			38 & \motif{38} & 1.376 & 1.419 & 11.040 & 3.569 & -2.708&
			38 & \motif{38} & 6.074 & 4.061 & 54.270 & 9.353 & -5.153\\
			36 & \motif{36} & 11.404 & 5.517 & 83.765 & 8.409 & -8.605&
			6 & \motif{6} & 47.112 & 34.692 & 228.087 & 18.356 & -9.859\\
			6 & \motif{6} & 14.453 & 10.576 & 103.671 & 10.308 & -8.655&
			36 & \motif{36} & 36.830 & 11.594 & 187.854 & 15.287 & -9.879\\
			12 & \motif{12} & 23.073 & 9.856 & 232.110 & 21.276 & -9.825&
			12 & \motif{12} & 70.210 & 22.567 & 460.111 & 36.331 & -10.732\\
			\bottomrule
	\end{tabular}}
\end{table}

\subsection{Colored motif structures}

To examine whether the local spillover architecture is also organized by
market type, we extend the motif analysis from uncolored triads to
colored motifs under progressively finer sector partitions: two classes
(commodity versus equity), four classes (agriculture, energy, metals, and
equity), and seven classes (corresponding to the full sector taxonomy).
Tables~\ref{tab:colored_triads_2class}--\ref{tab:colored_triads_7class}
report the top-ranked colored motifs at the median quantile, while
Tables~\ref{tab:colored_triads_2class_L}--\ref{tab:colored_triads_7class_R}
present the corresponding results at the two tails. All results are
reported under both $\alpha = 0.05$ and $\alpha = 0.10$ and remain
qualitatively stable across the two backbone thresholds.

\begin{table}[H]
	\centering
	\setlength{\abovecaptionskip}{0pt}
	\setlength{\belowcaptionskip}{10pt}
	\caption{Colored triad motifs at the median quantile (two classes). }
	\label{tab:colored_triads_2class}
	\resizebox{\textwidth}{!}{
		\begin{tabular}{llrrrrr|llrrrrr}
			\toprule
			\multicolumn{7}{c|}{$\alpha=0.05$} & \multicolumn{7}{c}{$\alpha=0.10$}\\
			\cmidrule(lr){1-7} \cmidrule(lr){8-14}
			id & motif & $\mu$ & $\sigma$ & $\mu_{\text{rnd}}$ & $\sigma_{\text{rnd}}$ & $z$ &
			id & motif & $\mu$ & $\sigma$ & $\mu_{\text{rnd}}$ & $\sigma_{\text{rnd}}$ & $z$\\
			\midrule
			238 & \begin{xy}
				\POS (0,3) *{\textcolor{Equity}{\medbullet}} ="a",
				\POS (-3.5,-1.5) *{\textcolor{Equity}{\medbullet}} ="b",
				\POS (3.5,-1.5) *{\textcolor{Equity}{\medbullet}} ="c"
				\POS "a" \ar @{<->} "b"
				\POS "a" \ar @{<->} "c"
				\POS "b" \ar @{<->} "c"
			\end{xy} & 6.378 & 4.253 & 0.101 & 0.351 & 17.873 & 238 & \begin{xy}
				\POS (0,3) *{\textcolor{Equity}{\medbullet}} ="a",
				\POS (-3.5,-1.5) *{\textcolor{Equity}{\medbullet}} ="b",
				\POS (3.5,-1.5) *{\textcolor{Equity}{\medbullet}} ="c"
				\POS "a" \ar @{<->} "b"
				\POS "a" \ar @{<->} "c"
				\POS "b" \ar @{<->} "c"
			\end{xy} & 6.659 & 4.096 & 0.107 & 0.367 & 17.862\\
			110 & \begin{xy}
				\POS (0,3) *{\textcolor{Equity}{\medbullet}} ="a",
				\POS (-3.5,-1.5) *{\textcolor{Equity}{\medbullet}} ="b",
				\POS (3.5,-1.5) *{\textcolor{Equity}{\medbullet}} ="c"
				\POS "a" \ar @{<->} "b"
				\POS "a" \ar @{<->} "c"
				\POS "b" \ar @{->} "c"
			\end{xy} & 3.792 & 2.497 & 0.074 & 0.281 & 13.254 & 110 & \begin{xy}
				\POS (0,3) *{\textcolor{Equity}{\medbullet}} ="a",
				\POS (-3.5,-1.5) *{\textcolor{Equity}{\medbullet}} ="b",
				\POS (3.5,-1.5) *{\textcolor{Equity}{\medbullet}} ="c"
				\POS "a" \ar @{<->} "b"
				\POS "a" \ar @{<->} "c"
				\POS "b" \ar @{->} "c"
			\end{xy} & 3.921 & 2.542 & 0.078 & 0.291 & 13.212\\
			46 & \begin{xy}
				\POS (0,3) *{\textcolor{Equity}{\medbullet}} ="a",
				\POS (-3.5,-1.5) *{\textcolor{Equity}{\medbullet}} ="b",
				\POS (3.5,-1.5) *{\textcolor{Equity}{\medbullet}} ="c"
				\POS "a" \ar @{<->} "b"
				\POS "a" \ar @{->} "c"
				\POS "b" \ar @{->} "c"
			\end{xy} & 2.541 & 2.626 & 0.050 & 0.220 & 11.320 & 46 & \begin{xy}
				\POS (0,3) *{\textcolor{Equity}{\medbullet}} ="a",
				\POS (-3.5,-1.5) *{\textcolor{Equity}{\medbullet}} ="b",
				\POS (3.5,-1.5) *{\textcolor{Equity}{\medbullet}} ="c"
				\POS "a" \ar @{<->} "b"
				\POS "a" \ar @{->} "c"
				\POS "b" \ar @{->} "c"
			\end{xy} & 2.666 & 2.687 & 0.053 & 0.228 & 11.480\\
			14 & \begin{xy}
				\POS (0,3) *{\textcolor{Equity}{\medbullet}} ="a",
				\POS (-3.5,-1.5) *{\textcolor{Equity}{\medbullet}} ="b",
				\POS (3.5,-1.5) *{\textcolor{Commodity}{\medbullet}} ="c"
				\POS "a" \ar @{<->} "b"
				\POS "a" \ar @{->} "c"
			\end{xy} & 21.496 & 22.448 & 2.789 & 3.061 & 6.112 & 14 & \begin{xy}
				\POS (0,3) *{\textcolor{Equity}{\medbullet}} ="a",
				\POS (-3.5,-1.5) *{\textcolor{Equity}{\medbullet}} ="b",
				\POS (3.5,-1.5) *{\textcolor{Commodity}{\medbullet}} ="c"
				\POS "a" \ar @{<->} "b"
				\POS "a" \ar @{->} "c"
			\end{xy} & 22.160 & 22.062 & 2.884 & 3.149 & 6.122\\
			78 & \begin{xy}
				\POS (0,3) *{\textcolor{Equity}{\medbullet}} ="a",
				\POS (-3.5,-1.5) *{\textcolor{Equity}{\medbullet}} ="b",
				\POS (3.5,-1.5) *{\textcolor{Equity}{\medbullet}} ="c"
				\POS "a" \ar @{<->} "b"
				\POS "a" \ar @{<->} "c"
			\end{xy} & 2.949 & 2.784 & 0.199 & 0.509 & 5.402 & 78 & \begin{xy}
				\POS (0,3) *{\textcolor{Equity}{\medbullet}} ="a",
				\POS (-3.5,-1.5) *{\textcolor{Equity}{\medbullet}} ="b",
				\POS (3.5,-1.5) *{\textcolor{Equity}{\medbullet}} ="c"
				\POS "a" \ar @{<->} "b"
				\POS "a" \ar @{<->} "c"
			\end{xy} & 2.924 & 2.796 & 0.203 & 0.519 & 5.244\\
			14 & \begin{xy}
				\POS (0,3) *{\textcolor{Equity}{\medbullet}} ="a",
				\POS (-3.5,-1.5) *{\textcolor{Equity}{\medbullet}} ="b",
				\POS (3.5,-1.5) *{\textcolor{Equity}{\medbullet}} ="c"
				\POS "a" \ar @{<->} "b"
				\POS "a" \ar @{->} "c"
			\end{xy} & 4.054 & 2.631 & 0.435 & 0.752 & 4.815 & 14 & \begin{xy}
				\POS (0,3) *{\textcolor{Equity}{\medbullet}} ="a",
				\POS (-3.5,-1.5) *{\textcolor{Equity}{\medbullet}} ="b",
				\POS (3.5,-1.5) *{\textcolor{Equity}{\medbullet}} ="c"
				\POS "a" \ar @{<->} "b"
				\POS "a" \ar @{->} "c"
			\end{xy} & 4.051 & 2.685 & 0.450 & 0.774 & 4.652\\
			102 & \begin{xy}
				\POS (0,3) *{\textcolor{Equity}{\medbullet}} ="a",
				\POS (-3.5,-1.5) *{\textcolor{Equity}{\medbullet}} ="b",
				\POS (3.5,-1.5) *{\textcolor{Equity}{\medbullet}} ="c"
				\POS "a" \ar @{->} "b"
				\POS "a" \ar @{<->} "c"
				\POS "b" \ar @{->} "c"
			\end{xy} & 0.525 & 0.860 & 0.023 & 0.144 & 3.483 & 102 & \begin{xy}
				\POS (0,3) *{\textcolor{Equity}{\medbullet}} ="a",
				\POS (-3.5,-1.5) *{\textcolor{Equity}{\medbullet}} ="b",
				\POS (3.5,-1.5) *{\textcolor{Equity}{\medbullet}} ="c"
				\POS "a" \ar @{->} "b"
				\POS "a" \ar @{<->} "c"
				\POS "b" \ar @{->} "c"
			\end{xy} & 0.519 & 0.859 & 0.023 & 0.144 & 3.440\\
			6 & \begin{xy}
				\POS (0,3) *{\textcolor{Equity}{\medbullet}} ="a",
				\POS (-3.5,-1.5) *{\textcolor{Commodity}{\medbullet}} ="b",
				\POS (3.5,-1.5) *{\textcolor{Equity}{\medbullet}} ="c"
				\POS "a" \ar @{->} "b"
				\POS "a" \ar @{->} "c"
			\end{xy} & 6.186 & 7.867 & 1.116 & 1.587 & 3.196 & 6 & \begin{xy}
				\POS (0,3) *{\textcolor{Equity}{\medbullet}} ="a",
				\POS (-3.5,-1.5) *{\textcolor{Commodity}{\medbullet}} ="b",
				\POS (3.5,-1.5) *{\textcolor{Equity}{\medbullet}} ="c"
				\POS "a" \ar @{->} "b"
				\POS "a" \ar @{->} "c"
			\end{xy} & 6.406 & 7.800 & 1.151 & 1.635 & 3.214\\
			46 & \begin{xy}
				\POS (0,3) *{\textcolor{Equity}{\medbullet}} ="a",
				\POS (-3.5,-1.5) *{\textcolor{Commodity}{\medbullet}} ="b",
				\POS (3.5,-1.5) *{\textcolor{Equity}{\medbullet}} ="c"
				\POS "a" \ar @{<->} "b"
				\POS "a" \ar @{->} "c"
				\POS "b" \ar @{->} "c"
			\end{xy} & 3.705 & 9.374 & 0.644 & 0.965 & 3.173 & 46 & \begin{xy}
				\POS (0,3) *{\textcolor{Equity}{\medbullet}} ="a",
				\POS (-3.5,-1.5) *{\textcolor{Commodity}{\medbullet}} ="b",
				\POS (3.5,-1.5) *{\textcolor{Equity}{\medbullet}} ="c"
				\POS "a" \ar @{<->} "b"
				\POS "a" \ar @{->} "c"
				\POS "b" \ar @{->} "c"
			\end{xy} & 3.785 & 9.357 & 0.677 & 0.998 & 3.114\\
			78 & \begin{xy}
				\POS (0,3) *{\textcolor{Equity}{\medbullet}} ="a",
				\POS (-3.5,-1.5) *{\textcolor{Commodity}{\medbullet}} ="b",
				\POS (3.5,-1.5) *{\textcolor{Equity}{\medbullet}} ="c"
				\POS "a" \ar @{<->} "b"
				\POS "a" \ar @{<->} "c"
			\end{xy} & 5.237 & 4.255 & 1.269 & 1.586 & 2.502 & 78 & \begin{xy}
				\POS (0,3) *{\textcolor{Equity}{\medbullet}} ="a",
				\POS (-3.5,-1.5) *{\textcolor{Commodity}{\medbullet}} ="b",
				\POS (3.5,-1.5) *{\textcolor{Equity}{\medbullet}} ="c"
				\POS "a" \ar @{<->} "b"
				\POS "a" \ar @{<->} "c"
			\end{xy} & 5.360 & 4.187 & 1.294 & 1.615 & 2.517\\
			38 & \begin{xy}
				\POS (0,3) *{\textcolor{Equity}{\medbullet}} ="a",
				\POS (-3.5,-1.5) *{\textcolor{Equity}{\medbullet}} ="b",
				\POS (3.5,-1.5) *{\textcolor{Equity}{\medbullet}} ="c"
				\POS "a" \ar @{->} "b"
				\POS "a" \ar @{->} "c"
				\POS "b" \ar @{->} "c"
			\end{xy} & 0.414 & 0.826 & 0.029 & 0.162 & 2.377 & 74 & \begin{xy}
				\POS (0,3) *{\textcolor{Equity}{\medbullet}} ="a",
				\POS (-3.5,-1.5) *{\textcolor{Equity}{\medbullet}} ="b",
				\POS (3.5,-1.5) *{\textcolor{Commodity}{\medbullet}} ="c"
				\POS "a" \ar @{<->} "b"
				\POS "c" \ar @{->} "a"
			\end{xy} & 7.560 & 5.184 & 2.019 & 2.358 & 2.350\\
			74 & \begin{xy}
				\POS (0,3) *{\textcolor{Equity}{\medbullet}} ="a",
				\POS (-3.5,-1.5) *{\textcolor{Equity}{\medbullet}} ="b",
				\POS (3.5,-1.5) *{\textcolor{Commodity}{\medbullet}} ="c"
				\POS "a" \ar @{<->} "b"
				\POS "c" \ar @{->} "a"
			\end{xy} & 7.291 & 5.362 & 1.957 & 2.273 & 2.347 & 38 & \begin{xy}
				\POS (0,3) *{\textcolor{Equity}{\medbullet}} ="a",
				\POS (-3.5,-1.5) *{\textcolor{Equity}{\medbullet}} ="b",
				\POS (3.5,-1.5) *{\textcolor{Equity}{\medbullet}} ="c"
				\POS "a" \ar @{->} "b"
				\POS "a" \ar @{->} "c"
				\POS "b" \ar @{->} "c"
			\end{xy} & 0.405 & 0.828 & 0.029 & 0.164 & 2.286\\
			12 & \begin{xy}
				\POS (0,3) *{\textcolor{Equity}{\medbullet}} ="a",
				\POS (-3.5,-1.5) *{\textcolor{Commodity}{\medbullet}} ="b",
				\POS (3.5,-1.5) *{\textcolor{Equity}{\medbullet}} ="c"
				\POS "b" \ar @{->} "a"
				\POS "a" \ar @{->} "c"
			\end{xy} & 5.060 & 4.387 & 1.664 & 1.812 & 1.875 & 12 & \begin{xy}
				\POS (0,3) *{\textcolor{Equity}{\medbullet}} ="a",
				\POS (-3.5,-1.5) *{\textcolor{Commodity}{\medbullet}} ="b",
				\POS (3.5,-1.5) *{\textcolor{Equity}{\medbullet}} ="c"
				\POS "b" \ar @{->} "a"
				\POS "a" \ar @{->} "c"
			\end{xy} & 5.309 & 4.357 & 1.718 & 1.866 & 1.925\\
			78 & \begin{xy}
				\POS (0,3) *{\textcolor{Commodity}{\medbullet}} ="a",
				\POS (-3.5,-1.5) *{\textcolor{Commodity}{\medbullet}} ="b",
				\POS (3.5,-1.5) *{\textcolor{Commodity}{\medbullet}} ="c"
				\POS "a" \ar @{<->} "b"
				\POS "a" \ar @{<->} "c"
			\end{xy} & 37.613 & 16.009 & 28.167 & 6.188 & 1.526 & 78 & \begin{xy}
				\POS (0,3) *{\textcolor{Commodity}{\medbullet}} ="a",
				\POS (-3.5,-1.5) *{\textcolor{Commodity}{\medbullet}} ="b",
				\POS (3.5,-1.5) *{\textcolor{Commodity}{\medbullet}} ="c"
				\POS "a" \ar @{<->} "b"
				\POS "a" \ar @{<->} "c"
			\end{xy} & 38.313 & 14.857 & 28.716 & 6.309 & 1.521\\
			6 & \begin{xy}
				\POS (0,3) *{\textcolor{Commodity}{\medbullet}} ="a",
				\POS (-3.5,-1.5) *{\textcolor{Commodity}{\medbullet}} ="b",
				\POS (3.5,-1.5) *{\textcolor{Equity}{\medbullet}} ="c"
				\POS "a" \ar @{->} "b"
				\POS "a" \ar @{->} "c"
			\end{xy} & 18.326 & 28.126 & 11.568 & 5.228 & 1.293 & 108 & \begin{xy}
				\POS (0,3) *{\textcolor{Equity}{\medbullet}} ="a",
				\POS (-3.5,-1.5) *{\textcolor{Equity}{\medbullet}} ="b",
				\POS (3.5,-1.5) *{\textcolor{Equity}{\medbullet}} ="c"
				\POS "b" \ar @{->} "a"
				\POS "a" \ar @{<->} "c"
				\POS "b" \ar @{->} "c"
			\end{xy} & 0.230 & 0.547 & 0.027 & 0.158 & 1.284\\
			108 & \begin{xy}
				\POS (0,3) *{\textcolor{Equity}{\medbullet}} ="a",
				\POS (-3.5,-1.5) *{\textcolor{Equity}{\medbullet}} ="b",
				\POS (3.5,-1.5) *{\textcolor{Equity}{\medbullet}} ="c"
				\POS "b" \ar @{->} "a"
				\POS "a" \ar @{<->} "c"
				\POS "b" \ar @{->} "c"
			\end{xy} & 0.225 & 0.539 & 0.027 & 0.156 & 1.268 & 6 & \begin{xy}
				\POS (0,3) *{\textcolor{Commodity}{\medbullet}} ="a",
				\POS (-3.5,-1.5) *{\textcolor{Commodity}{\medbullet}} ="b",
				\POS (3.5,-1.5) *{\textcolor{Equity}{\medbullet}} ="c"
				\POS "a" \ar @{->} "b"
				\POS "a" \ar @{->} "c"
			\end{xy} & 18.836 & 27.890 & 11.934 & 5.383 & 1.282\\
			74 & \begin{xy}
				\POS (0,3) *{\textcolor{Commodity}{\medbullet}} ="a",
				\POS (-3.5,-1.5) *{\textcolor{Commodity}{\medbullet}} ="b",
				\POS (3.5,-1.5) *{\textcolor{Commodity}{\medbullet}} ="c"
				\POS "a" \ar @{<->} "b"
				\POS "c" \ar @{->} "a"
			\end{xy} & 51.740 & 17.436 & 43.372 & 6.599 & 1.268 & 74 & \begin{xy}
				\POS (0,3) *{\textcolor{Commodity}{\medbullet}} ="a",
				\POS (-3.5,-1.5) *{\textcolor{Commodity}{\medbullet}} ="b",
				\POS (3.5,-1.5) *{\textcolor{Commodity}{\medbullet}} ="c"
				\POS "a" \ar @{<->} "b"
				\POS "c" \ar @{->} "a"
			\end{xy} & 53.383 & 14.539 & 44.746 & 6.875 & 1.256\\
			238 & \begin{xy}
				\POS (0,3) *{\textcolor{Commodity}{\medbullet}} ="a",
				\POS (-3.5,-1.5) *{\textcolor{Commodity}{\medbullet}} ="b",
				\POS (3.5,-1.5) *{\textcolor{Commodity}{\medbullet}} ="c"
				\POS "a" \ar @{<->} "b"
				\POS "a" \ar @{<->} "c"
				\POS "b" \ar @{<->} "c"
			\end{xy} & 19.345 & 8.211 & 14.504 & 4.241 & 1.141 & 238 & \begin{xy}
				\POS (0,3) *{\textcolor{Commodity}{\medbullet}} ="a",
				\POS (-3.5,-1.5) *{\textcolor{Commodity}{\medbullet}} ="b",
				\POS (3.5,-1.5) *{\textcolor{Commodity}{\medbullet}} ="c"
				\POS "a" \ar @{<->} "b"
				\POS "a" \ar @{<->} "c"
				\POS "b" \ar @{<->} "c"
			\end{xy} & 20.636 & 7.208 & 15.384 & 4.479 & 1.173\\
			6 & \begin{xy}
				\POS (0,3) *{\textcolor{Equity}{\medbullet}} ="a",
				\POS (-3.5,-1.5) *{\textcolor{Equity}{\medbullet}} ="b",
				\POS (3.5,-1.5) *{\textcolor{Equity}{\medbullet}} ="c"
				\POS "a" \ar @{->} "b"
				\POS "a" \ar @{->} "c"
			\end{xy} & 0.673 & 1.065 & 0.175 & 0.462 & 1.077 & 6 & \begin{xy}
				\POS (0,3) *{\textcolor{Equity}{\medbullet}} ="a",
				\POS (-3.5,-1.5) *{\textcolor{Equity}{\medbullet}} ="b",
				\POS (3.5,-1.5) *{\textcolor{Equity}{\medbullet}} ="c"
				\POS "a" \ar @{->} "b"
				\POS "a" \ar @{->} "c"
			\end{xy} & 0.701 & 1.112 & 0.180 & 0.474 & 1.099\\
			110 & \begin{xy}
				\POS (0,3) *{\textcolor{Equity}{\medbullet}} ="a",
				\POS (-3.5,-1.5) *{\textcolor{Commodity}{\medbullet}} ="b",
				\POS (3.5,-1.5) *{\textcolor{Equity}{\medbullet}} ="c"
				\POS "a" \ar @{<->} "b"
				\POS "a" \ar @{<->} "c"
				\POS "b" \ar @{->} "c"
			\end{xy} & 1.320 & 2.246 & 0.471 & 0.807 & 1.051 & 110 & \begin{xy}
				\POS (0,3) *{\textcolor{Equity}{\medbullet}} ="a",
				\POS (-3.5,-1.5) *{\textcolor{Commodity}{\medbullet}} ="b",
				\POS (3.5,-1.5) *{\textcolor{Equity}{\medbullet}} ="c"
				\POS "a" \ar @{<->} "b"
				\POS "a" \ar @{<->} "c"
				\POS "b" \ar @{->} "c"
			\end{xy} & 1.343 & 2.245 & 0.497 & 0.840 & 1.006\\
			102 & \begin{xy}
				\POS (0,3) *{\textcolor{Equity}{\medbullet}} ="a",
				\POS (-3.5,-1.5) *{\textcolor{Equity}{\medbullet}} ="b",
				\POS (3.5,-1.5) *{\textcolor{Commodity}{\medbullet}} ="c"
				\POS "a" \ar @{->} "b"
				\POS "a" \ar @{<->} "c"
				\POS "b" \ar @{->} "c"
			\end{xy} & 0.458 & 0.832 & 0.121 & 0.342 & 0.988 & 102 & \begin{xy}
				\POS (0,3) *{\textcolor{Equity}{\medbullet}} ="a",
				\POS (-3.5,-1.5) *{\textcolor{Equity}{\medbullet}} ="b",
				\POS (3.5,-1.5) *{\textcolor{Commodity}{\medbullet}} ="c"
				\POS "a" \ar @{->} "b"
				\POS "a" \ar @{<->} "c"
				\POS "b" \ar @{->} "c"
			\end{xy} & 0.447 & 0.826 & 0.121 & 0.342 & 0.955\\
			74 & \begin{xy}
				\POS (0,3) *{\textcolor{Equity}{\medbullet}} ="a",
				\POS (-3.5,-1.5) *{\textcolor{Equity}{\medbullet}} ="b",
				\POS (3.5,-1.5) *{\textcolor{Equity}{\medbullet}} ="c"
				\POS "a" \ar @{<->} "b"
				\POS "c" \ar @{->} "a"
			\end{xy} & 0.856 & 1.385 & 0.305 & 0.605 & 0.911 & 12 & \begin{xy}
				\POS (0,3) *{\textcolor{Commodity}{\medbullet}} ="a",
				\POS (-3.5,-1.5) *{\textcolor{Commodity}{\medbullet}} ="b",
				\POS (3.5,-1.5) *{\textcolor{Equity}{\medbullet}} ="c"
				\POS "b" \ar @{->} "a"
				\POS "a" \ar @{->} "c"
			\end{xy} & 12.406 & 10.517 & 8.899 & 3.989 & 0.879\\
			12 & \begin{xy}
				\POS (0,3) *{\textcolor{Commodity}{\medbullet}} ="a",
				\POS (-3.5,-1.5) *{\textcolor{Commodity}{\medbullet}} ="b",
				\POS (3.5,-1.5) *{\textcolor{Equity}{\medbullet}} ="c"
				\POS "b" \ar @{->} "a"
				\POS "a" \ar @{->} "c"
			\end{xy} & 12.084 & 10.732 & 8.622 & 3.873 & 0.894 & 74 & \begin{xy}
				\POS (0,3) *{\textcolor{Equity}{\medbullet}} ="a",
				\POS (-3.5,-1.5) *{\textcolor{Equity}{\medbullet}} ="b",
				\POS (3.5,-1.5) *{\textcolor{Equity}{\medbullet}} ="c"
				\POS "a" \ar @{<->} "b"
				\POS "c" \ar @{->} "a"
			\end{xy} & 0.828 & 1.370 & 0.314 & 0.623 & 0.826\\
			12 & \begin{xy}
				\POS (0,3) *{\textcolor{Equity}{\medbullet}} ="a",
				\POS (-3.5,-1.5) *{\textcolor{Equity}{\medbullet}} ="b",
				\POS (3.5,-1.5) *{\textcolor{Equity}{\medbullet}} ="c"
				\POS "b" \ar @{->} "a"
				\POS "a" \ar @{->} "c"
			\end{xy} & 0.747 & 1.131 & 0.261 & 0.598 & 0.814 & 12 & \begin{xy}
				\POS (0,3) *{\textcolor{Equity}{\medbullet}} ="a",
				\POS (-3.5,-1.5) *{\textcolor{Equity}{\medbullet}} ="b",
				\POS (3.5,-1.5) *{\textcolor{Equity}{\medbullet}} ="c"
				\POS "b" \ar @{->} "a"
				\POS "a" \ar @{->} "c"
			\end{xy} & 0.748 & 1.136 & 0.268 & 0.613 & 0.783\\
			98 & \begin{xy}
				\POS (0,3) *{\textcolor{Equity}{\medbullet}} ="a",
				\POS (-3.5,-1.5) *{\textcolor{Equity}{\medbullet}} ="b",
				\POS (3.5,-1.5) *{\textcolor{Equity}{\medbullet}} ="c"
				\POS "a" \ar @{->} "b"
				\POS "c" \ar @{->} "a"
				\POS "b" \ar @{->} "c"
			\end{xy} & 0.098 & 0.331 & 0.014 & 0.112 & 0.747 & 110 & \begin{xy}
				\POS (0,3) *{\textcolor{Commodity}{\medbullet}} ="a",
				\POS (-3.5,-1.5) *{\textcolor{Commodity}{\medbullet}} ="b",
				\POS (3.5,-1.5) *{\textcolor{Commodity}{\medbullet}} ="c"
				\POS "a" \ar @{<->} "b"
				\POS "a" \ar @{<->} "c"
				\POS "b" \ar @{->} "c"
			\end{xy} & 13.310 & 5.328 & 11.017 & 3.094 & 0.741\\
			74 & \begin{xy}
				\POS (0,3) *{\textcolor{Commodity}{\medbullet}} ="a",
				\POS (-3.5,-1.5) *{\textcolor{Commodity}{\medbullet}} ="b",
				\POS (3.5,-1.5) *{\textcolor{Equity}{\medbullet}} ="c"
				\POS "a" \ar @{<->} "b"
				\POS "c" \ar @{->} "a"
			\end{xy} & 13.656 & 10.934 & 10.139 & 4.708 & 0.747 & 38 & \begin{xy}
				\POS (0,3) *{\textcolor{Commodity}{\medbullet}} ="a",
				\POS (-3.5,-1.5) *{\textcolor{Equity}{\medbullet}} ="b",
				\POS (3.5,-1.5) *{\textcolor{Equity}{\medbullet}} ="c"
				\POS "a" \ar @{->} "b"
				\POS "a" \ar @{->} "c"
				\POS "b" \ar @{->} "c"
			\end{xy} & 0.441 & 0.941 & 0.153 & 0.393 & 0.734\\
			110 & \begin{xy}
				\POS (0,3) *{\textcolor{Commodity}{\medbullet}} ="a",
				\POS (-3.5,-1.5) *{\textcolor{Commodity}{\medbullet}} ="b",
				\POS (3.5,-1.5) *{\textcolor{Commodity}{\medbullet}} ="c"
				\POS "a" \ar @{<->} "b"
				\POS "a" \ar @{<->} "c"
				\POS "b" \ar @{->} "c"
			\end{xy} & 12.516 & 5.681 & 10.431 & 2.966 & 0.703 & 74 & \begin{xy}
				\POS (0,3) *{\textcolor{Commodity}{\medbullet}} ="a",
				\POS (-3.5,-1.5) *{\textcolor{Commodity}{\medbullet}} ="b",
				\POS (3.5,-1.5) *{\textcolor{Equity}{\medbullet}} ="c"
				\POS "a" \ar @{<->} "b"
				\POS "c" \ar @{->} "a"
			\end{xy} & 14.003 & 10.627 & 10.461 & 4.842 & 0.732\\
			38 & \begin{xy}
				\POS (0,3) *{\textcolor{Commodity}{\medbullet}} ="a",
				\POS (-3.5,-1.5) *{\textcolor{Equity}{\medbullet}} ="b",
				\POS (3.5,-1.5) *{\textcolor{Equity}{\medbullet}} ="c"
				\POS "a" \ar @{->} "b"
				\POS "a" \ar @{->} "c"
				\POS "b" \ar @{->} "c"
			\end{xy} & 0.418 & 0.937 & 0.150 & 0.388 & 0.689 & 98 & \begin{xy}
				\POS (0,3) *{\textcolor{Equity}{\medbullet}} ="a",
				\POS (-3.5,-1.5) *{\textcolor{Equity}{\medbullet}} ="b",
				\POS (3.5,-1.5) *{\textcolor{Equity}{\medbullet}} ="c"
				\POS "a" \ar @{->} "b"
				\POS "c" \ar @{->} "a"
				\POS "b" \ar @{->} "c"
			\end{xy} & 0.096 & 0.328 & 0.014 & 0.112 & 0.729\\
			14 & \begin{xy}
				\POS (0,3) *{\textcolor{Commodity}{\medbullet}} ="a",
				\POS (-3.5,-1.5) *{\textcolor{Commodity}{\medbullet}} ="b",
				\POS (3.5,-1.5) *{\textcolor{Commodity}{\medbullet}} ="c"
				\POS "a" \ar @{<->} "b"
				\POS "a" \ar @{->} "c"
			\end{xy} & 67.400 & 24.258 & 61.899 & 9.229 & 0.596 & 14 & \begin{xy}
				\POS (0,3) *{\textcolor{Commodity}{\medbullet}} ="a",
				\POS (-3.5,-1.5) *{\textcolor{Commodity}{\medbullet}} ="b",
				\POS (3.5,-1.5) *{\textcolor{Commodity}{\medbullet}} ="c"
				\POS "a" \ar @{<->} "b"
				\POS "a" \ar @{->} "c"
			\end{xy} & 69.759 & 20.275 & 63.996 & 9.490 & 0.607\\
			36 & \begin{xy}
				\POS (0,3) *{\textcolor{Equity}{\medbullet}} ="a",
				\POS (-3.5,-1.5) *{\textcolor{Commodity}{\medbullet}} ="b",
				\POS (3.5,-1.5) *{\textcolor{Equity}{\medbullet}} ="c"
				\POS "a" \ar @{->} "c"
				\POS "b" \ar @{->} "c"
			\end{xy} & 2.804 & 2.550 & 1.733 & 1.886 & 0.568 & 36 & \begin{xy}
				\POS (0,3) *{\textcolor{Equity}{\medbullet}} ="a",
				\POS (-3.5,-1.5) *{\textcolor{Commodity}{\medbullet}} ="b",
				\POS (3.5,-1.5) *{\textcolor{Equity}{\medbullet}} ="c"
				\POS "a" \ar @{->} "c"
				\POS "b" \ar @{->} "c"
			\end{xy} & 2.979 & 2.523 & 1.806 & 1.960 & 0.598\\
			
			\bottomrule
	\end{tabular}}
	\begin{flushleft}
		\footnotesize
		\textit{Legend: }
		\textcolor{Equity}{{\normalsize\textbullet}}\, Equity;
		\textcolor{Commodity}{{\normalsize\textbullet}}\,  Commodity.
	\end{flushleft} 
\end{table}


\begin{table}[H]
	\centering
	\setlength{\abovecaptionskip}{0pt}
	\setlength{\belowcaptionskip}{10pt}
	\caption{Colored triad motifs at the median quantile (four classes).}
	\label{tab:colored_triads_4class}
	\resizebox{\textwidth}{!}{
		\begin{tabular}{llrrrrr|llrrrrr}
			\toprule
			\multicolumn{7}{c|}{$\alpha=0.05$} & \multicolumn{7}{c}{$\alpha=0.10$}\\
			\cmidrule(lr){1-7} \cmidrule(lr){8-14}
			id & motif & $\mu$ & $\sigma$ & $\mu_{\text{rnd}}$ & $\sigma_{\text{rnd}}$ & $z$ &
			id & motif & $\mu$ & $\sigma$ & $\mu_{\text{rnd}}$ & $\sigma_{\text{rnd}}$ & $z$\\
			\midrule
			238 & \begin{xy}
				\POS (0,3) *{\textcolor{Equity}{\medbullet}} ="a",
				\POS (-3.5,-1.5) *{\textcolor{Equity}{\medbullet}} ="b",
				\POS (3.5,-1.5) *{\textcolor{Equity}{\medbullet}} ="c"
				\POS "a" \ar @{<->} "b"
				\POS "a" \ar @{<->} "c"
				\POS "b" \ar @{<->} "c"
			\end{xy} & 6.386 & 4.249 & 0.105 & 0.360 & 17.455 & 238 & \begin{xy}
				\POS (0,3) *{\textcolor{Equity}{\medbullet}} ="a",
				\POS (-3.5,-1.5) *{\textcolor{Equity}{\medbullet}} ="b",
				\POS (3.5,-1.5) *{\textcolor{Equity}{\medbullet}} ="c"
				\POS "a" \ar @{<->} "b"
				\POS "a" \ar @{<->} "c"
				\POS "b" \ar @{<->} "c"
			\end{xy} & 6.659 & 4.096 & 0.111 & 0.376 & 17.423\\
			238 & \begin{xy}
				\POS (0,3) *{\textcolor{Metal}{\medbullet}} ="a",
				\POS (-3.5,-1.5) *{\textcolor{Metal}{\medbullet}} ="b",
				\POS (3.5,-1.5) *{\textcolor{Metal}{\medbullet}} ="c"
				\POS "a" \ar @{<->} "b"
				\POS "a" \ar @{<->} "c"
				\POS "b" \ar @{<->} "c"
			\end{xy} & 10.893 & 6.102 & 0.347 & 0.731 & 14.421 & 238 & \begin{xy}
				\POS (0,3) *{\textcolor{Metal}{\medbullet}} ="a",
				\POS (-3.5,-1.5) *{\textcolor{Metal}{\medbullet}} ="b",
				\POS (3.5,-1.5) *{\textcolor{Metal}{\medbullet}} ="c"
				\POS "a" \ar @{<->} "b"
				\POS "a" \ar @{<->} "c"
				\POS "b" \ar @{<->} "c"
			\end{xy} & 11.555 & 5.603 & 0.368 & 0.767 & 14.577\\
			110 & \begin{xy}
				\POS (0,3) *{\textcolor{Equity}{\medbullet}} ="a",
				\POS (-3.5,-1.5) *{\textcolor{Equity}{\medbullet}} ="b",
				\POS (3.5,-1.5) *{\textcolor{Equity}{\medbullet}} ="c"
				\POS "a" \ar @{<->} "b"
				\POS "a" \ar @{<->} "c"
				\POS "b" \ar @{->} "c"
			\end{xy} & 3.791 & 2.497 & 0.074 & 0.281 & 13.220 & 110 & \begin{xy}
				\POS (0,3) *{\textcolor{Equity}{\medbullet}} ="a",
				\POS (-3.5,-1.5) *{\textcolor{Equity}{\medbullet}} ="b",
				\POS (3.5,-1.5) *{\textcolor{Equity}{\medbullet}} ="c"
				\POS "a" \ar @{<->} "b"
				\POS "a" \ar @{<->} "c"
				\POS "b" \ar @{->} "c"
			\end{xy} & 3.927 & 2.539 & 0.078 & 0.292 & 13.179\\
			46 & \begin{xy}
				\POS (0,3) *{\textcolor{Equity}{\medbullet}} ="a",
				\POS (-3.5,-1.5) *{\textcolor{Equity}{\medbullet}} ="b",
				\POS (3.5,-1.5) *{\textcolor{Equity}{\medbullet}} ="c"
				\POS "a" \ar @{<->} "b"
				\POS "a" \ar @{->} "c"
				\POS "b" \ar @{->} "c"
			\end{xy} & 2.536 & 2.625 & 0.051 & 0.224 & 11.087 & 46 & \begin{xy}
				\POS (0,3) *{\textcolor{Equity}{\medbullet}} ="a",
				\POS (-3.5,-1.5) *{\textcolor{Equity}{\medbullet}} ="b",
				\POS (3.5,-1.5) *{\textcolor{Equity}{\medbullet}} ="c"
				\POS "a" \ar @{<->} "b"
				\POS "a" \ar @{->} "c"
				\POS "b" \ar @{->} "c"
			\end{xy} & 2.662 & 2.687 & 0.054 & 0.232 & 11.242\\
			110 & \begin{xy}
				\POS (0,3) *{\textcolor{Metal}{\medbullet}} ="a",
				\POS (-3.5,-1.5) *{\textcolor{Metal}{\medbullet}} ="b",
				\POS (3.5,-1.5) *{\textcolor{Metal}{\medbullet}} ="c"
				\POS "a" \ar @{<->} "b"
				\POS "a" \ar @{<->} "c"
				\POS "b" \ar @{->} "c"
			\end{xy} & 5.470 & 3.966 & 0.252 & 0.555 & 9.408 & 110 & \begin{xy}
				\POS (0,3) *{\textcolor{Metal}{\medbullet}} ="a",
				\POS (-3.5,-1.5) *{\textcolor{Metal}{\medbullet}} ="b",
				\POS (3.5,-1.5) *{\textcolor{Metal}{\medbullet}} ="c"
				\POS "a" \ar @{<->} "b"
				\POS "a" \ar @{<->} "c"
				\POS "b" \ar @{->} "c"
			\end{xy} & 5.775 & 3.868 & 0.266 & 0.577 & 9.550\\
			78 & \begin{xy}
				\POS (0,3) *{\textcolor{Metal}{\medbullet}} ="a",
				\POS (-3.5,-1.5) *{\textcolor{Metal}{\medbullet}} ="b",
				\POS (3.5,-1.5) *{\textcolor{Metal}{\medbullet}} ="c"
				\POS "a" \ar @{<->} "b"
				\POS "a" \ar @{<->} "c"
			\end{xy} & 8.600 & 7.706 & 0.676 & 1.042 & 7.605 & 78 & \begin{xy}
				\POS (0,3) *{\textcolor{Metal}{\medbullet}} ="a",
				\POS (-3.5,-1.5) *{\textcolor{Metal}{\medbullet}} ="b",
				\POS (3.5,-1.5) *{\textcolor{Metal}{\medbullet}} ="c"
				\POS "a" \ar @{<->} "b"
				\POS "a" \ar @{<->} "c"
			\end{xy} & 8.697 & 7.634 & 0.689 & 1.062 & 7.539\\
			14 & \begin{xy}
				\POS (0,3) *{\textcolor{Equity}{\medbullet}} ="a",
				\POS (-3.5,-1.5) *{\textcolor{Equity}{\medbullet}} ="b",
				\POS (3.5,-1.5) *{\textcolor{Agriculture}{\medbullet}} ="c"
				\POS "a" \ar @{<->} "b"
				\POS "a" \ar @{->} "c"
			\end{xy} & 13.550 & 11.203 & 1.487 & 1.865 & 6.468 & 14 & \begin{xy}
				\POS (0,3) *{\textcolor{Equity}{\medbullet}} ="a",
				\POS (-3.5,-1.5) *{\textcolor{Equity}{\medbullet}} ="b",
				\POS (3.5,-1.5) *{\textcolor{Agriculture}{\medbullet}} ="c"
				\POS "a" \ar @{<->} "b"
				\POS "a" \ar @{->} "c"
			\end{xy} & 14.136 & 10.900 & 1.538 & 1.919 & 6.565\\
			238 & \begin{xy}
				\POS (0,3) *{\textcolor{Energy}{\medbullet}} ="a",
				\POS (-3.5,-1.5) *{\textcolor{Energy}{\medbullet}} ="b",
				\POS (3.5,-1.5) *{\textcolor{Energy}{\medbullet}} ="c"
				\POS "a" \ar @{<->} "b"
				\POS "a" \ar @{<->} "c"
				\POS "b" \ar @{<->} "c"
			\end{xy} & 1.073 & 0.413 & 0.031 & 0.172 & 6.047 & 238 & \begin{xy}
				\POS (0,3) *{\textcolor{Energy}{\medbullet}} ="a",
				\POS (-3.5,-1.5) *{\textcolor{Energy}{\medbullet}} ="b",
				\POS (3.5,-1.5) *{\textcolor{Energy}{\medbullet}} ="c"
				\POS "a" \ar @{<->} "b"
				\POS "a" \ar @{<->} "c"
				\POS "b" \ar @{<->} "c"
			\end{xy} & 1.078 & 0.402 & 0.033 & 0.179 & 5.852\\
			78 & \begin{xy}
				\POS (0,3) *{\textcolor{Equity}{\medbullet}} ="a",
				\POS (-3.5,-1.5) *{\textcolor{Equity}{\medbullet}} ="b",
				\POS (3.5,-1.5) *{\textcolor{Equity}{\medbullet}} ="c"
				\POS "a" \ar @{<->} "b"
				\POS "a" \ar @{<->} "c"
			\end{xy} & 2.949 & 2.784 & 0.200 & 0.513 & 5.354 & 78 & \begin{xy}
				\POS (0,3) *{\textcolor{Equity}{\medbullet}} ="a",
				\POS (-3.5,-1.5) *{\textcolor{Equity}{\medbullet}} ="b",
				\POS (3.5,-1.5) *{\textcolor{Equity}{\medbullet}} ="c"
				\POS "a" \ar @{<->} "b"
				\POS "a" \ar @{<->} "c"
			\end{xy} & 2.925 & 2.796 & 0.204 & 0.523 & 5.201\\
			14 & \begin{xy}
				\POS (0,3) *{\textcolor{Metal}{\medbullet}} ="a",
				\POS (-3.5,-1.5) *{\textcolor{Metal}{\medbullet}} ="b",
				\POS (3.5,-1.5) *{\textcolor{Metal}{\medbullet}} ="c"
				\POS "a" \ar @{<->} "b"
				\POS "a" \ar @{->} "c"
			\end{xy} & 9.363 & 6.201 & 1.499 & 1.534 & 5.126 & 14 & \begin{xy}
				\POS (0,3) *{\textcolor{Metal}{\medbullet}} ="a",
				\POS (-3.5,-1.5) *{\textcolor{Metal}{\medbullet}} ="b",
				\POS (3.5,-1.5) *{\textcolor{Metal}{\medbullet}} ="c"
				\POS "a" \ar @{<->} "b"
				\POS "a" \ar @{->} "c"
			\end{xy} & 9.554 & 6.222 & 1.550 & 1.579 & 5.068\\
			14 & \begin{xy}
				\POS (0,3) *{\textcolor{Equity}{\medbullet}} ="a",
				\POS (-3.5,-1.5) *{\textcolor{Equity}{\medbullet}} ="b",
				\POS (3.5,-1.5) *{\textcolor{Equity}{\medbullet}} ="c"
				\POS "a" \ar @{<->} "b"
				\POS "a" \ar @{->} "c"
			\end{xy} & 4.054 & 2.631 & 0.437 & 0.757 & 4.779 & 14 & \begin{xy}
				\POS (0,3) *{\textcolor{Metal}{\medbullet}} ="a",
				\POS (-3.5,-1.5) *{\textcolor{Metal}{\medbullet}} ="b",
				\POS (3.5,-1.5) *{\textcolor{Agriculture}{\medbullet}} ="c"
				\POS "a" \ar @{<->} "b"
				\POS "a" \ar @{->} "c"
			\end{xy} & 17.334 & 9.665 & 3.280 & 2.941 & 4.778\\
			14 & \begin{xy}
				\POS (0,3) *{\textcolor{Metal}{\medbullet}} ="a",
				\POS (-3.5,-1.5) *{\textcolor{Metal}{\medbullet}} ="b",
				\POS (3.5,-1.5) *{\textcolor{Agriculture}{\medbullet}} ="c"
				\POS "a" \ar @{<->} "b"
				\POS "a" \ar @{->} "c"
			\end{xy} & 16.537 & 10.278 & 3.171 & 2.857 & 4.678 & 14 & \begin{xy}
				\POS (0,3) *{\textcolor{Equity}{\medbullet}} ="a",
				\POS (-3.5,-1.5) *{\textcolor{Equity}{\medbullet}} ="b",
				\POS (3.5,-1.5) *{\textcolor{Equity}{\medbullet}} ="c"
				\POS "a" \ar @{<->} "b"
				\POS "a" \ar @{->} "c"
			\end{xy} & 4.051 & 2.685 & 0.452 & 0.779 & 4.622\\
			78 & \begin{xy}
				\POS (0,3) *{\textcolor{Agriculture}{\medbullet}} ="a",
				\POS (-3.5,-1.5) *{\textcolor{Agriculture}{\medbullet}} ="b",
				\POS (3.5,-1.5) *{\textcolor{Agriculture}{\medbullet}} ="c"
				\POS "a" \ar @{<->} "b"
				\POS "a" \ar @{<->} "c"
			\end{xy} & 16.399 & 7.551 & 3.860 & 2.814 & 4.456 & 78 & \begin{xy}
				\POS (0,3) *{\textcolor{Agriculture}{\medbullet}} ="a",
				\POS (-3.5,-1.5) *{\textcolor{Agriculture}{\medbullet}} ="b",
				\POS (3.5,-1.5) *{\textcolor{Agriculture}{\medbullet}} ="c"
				\POS "a" \ar @{<->} "b"
				\POS "a" \ar @{<->} "c"
			\end{xy} & 16.680 & 7.252 & 3.935 & 2.868 & 4.444\\
			108 & \begin{xy}
				\POS (0,3) *{\textcolor{Metal}{\medbullet}} ="a",
				\POS (-3.5,-1.5) *{\textcolor{Metal}{\medbullet}} ="b",
				\POS (3.5,-1.5) *{\textcolor{Metal}{\medbullet}} ="c"
				\POS "b" \ar @{->} "a"
				\POS "a" \ar @{<->} "c"
				\POS "b" \ar @{->} "c"
			\end{xy} & 1.060 & 1.550 & 0.073 & 0.264 & 3.735 & 14 & \begin{xy}
				\POS (0,3) *{\textcolor{Metal}{\medbullet}} ="a",
				\POS (-3.5,-1.5) *{\textcolor{Metal}{\medbullet}} ="b",
				\POS (3.5,-1.5) *{\textcolor{Equity}{\medbullet}} ="c"
				\POS "a" \ar @{<->} "b"
				\POS "a" \ar @{->} "c"
			\end{xy} & 7.440 & 6.204 & 1.350 & 1.609 & 3.784\\
			74 & \begin{xy}
				\POS (0,3) *{\textcolor{Agriculture}{\medbullet}} ="a",
				\POS (-3.5,-1.5) *{\textcolor{Agriculture}{\medbullet}} ="b",
				\POS (3.5,-1.5) *{\textcolor{Agriculture}{\medbullet}} ="c"
				\POS "a" \ar @{<->} "b"
				\POS "c" \ar @{->} "a"
			\end{xy} & 17.403 & 7.215 & 5.955 & 3.117 & 3.673 & 108 & \begin{xy}
				\POS (0,3) *{\textcolor{Metal}{\medbullet}} ="a",
				\POS (-3.5,-1.5) *{\textcolor{Metal}{\medbullet}} ="b",
				\POS (3.5,-1.5) *{\textcolor{Metal}{\medbullet}} ="c"
				\POS "b" \ar @{->} "a"
				\POS "a" \ar @{<->} "c"
				\POS "b" \ar @{->} "c"
			\end{xy} & 1.091 & 1.604 & 0.076 & 0.270 & 3.762\\
			14 & \begin{xy}
				\POS (0,3) *{\textcolor{Metal}{\medbullet}} ="a",
				\POS (-3.5,-1.5) *{\textcolor{Metal}{\medbullet}} ="b",
				\POS (3.5,-1.5) *{\textcolor{Equity}{\medbullet}} ="c"
				\POS "a" \ar @{<->} "b"
				\POS "a" \ar @{->} "c"
			\end{xy} & 6.970 & 6.243 & 1.305 & 1.562 & 3.628 & 74 & \begin{xy}
				\POS (0,3) *{\textcolor{Agriculture}{\medbullet}} ="a",
				\POS (-3.5,-1.5) *{\textcolor{Agriculture}{\medbullet}} ="b",
				\POS (3.5,-1.5) *{\textcolor{Agriculture}{\medbullet}} ="c"
				\POS "a" \ar @{<->} "b"
				\POS "c" \ar @{->} "a"
			\end{xy} & 17.941 & 6.665 & 6.144 & 3.229 & 3.654\\
			102 & \begin{xy}
				\POS (0,3) *{\textcolor{Equity}{\medbullet}} ="a",
				\POS (-3.5,-1.5) *{\textcolor{Equity}{\medbullet}} ="b",
				\POS (3.5,-1.5) *{\textcolor{Equity}{\medbullet}} ="c"
				\POS "a" \ar @{->} "b"
				\POS "a" \ar @{<->} "c"
				\POS "b" \ar @{->} "c"
			\end{xy} & 0.529 & 0.862 & 0.024 & 0.146 & 3.455 & 102 & \begin{xy}
				\POS (0,3) *{\textcolor{Equity}{\medbullet}} ="a",
				\POS (-3.5,-1.5) *{\textcolor{Equity}{\medbullet}} ="b",
				\POS (3.5,-1.5) *{\textcolor{Equity}{\medbullet}} ="c"
				\POS "a" \ar @{->} "b"
				\POS "a" \ar @{<->} "c"
				\POS "b" \ar @{->} "c"
			\end{xy} & 0.523 & 0.861 & 0.024 & 0.147 & 3.405\\
			14 & \begin{xy}
				\POS (0,3) *{\textcolor{Equity}{\medbullet}} ="a",
				\POS (-3.5,-1.5) *{\textcolor{Equity}{\medbullet}} ="b",
				\POS (3.5,-1.5) *{\textcolor{Metal}{\medbullet}} ="c"
				\POS "a" \ar @{<->} "b"
				\POS "a" \ar @{->} "c"
			\end{xy} & 5.044 & 10.754 & 0.879 & 1.279 & 3.257 & 6 & \begin{xy}
				\POS (0,3) *{\textcolor{Equity}{\medbullet}} ="a",
				\POS (-3.5,-1.5) *{\textcolor{Agriculture}{\medbullet}} ="b",
				\POS (3.5,-1.5) *{\textcolor{Equity}{\medbullet}} ="c"
				\POS "a" \ar @{->} "b"
				\POS "a" \ar @{->} "c"
			\end{xy} & 4.018 & 4.089 & 0.614 & 1.033 & 3.297\\
			6 & \begin{xy}
				\POS (0,3) *{\textcolor{Equity}{\medbullet}} ="a",
				\POS (-3.5,-1.5) *{\textcolor{Agriculture}{\medbullet}} ="b",
				\POS (3.5,-1.5) *{\textcolor{Equity}{\medbullet}} ="c"
				\POS "a" \ar @{->} "b"
				\POS "a" \ar @{->} "c"
			\end{xy} & 3.830 & 4.103 & 0.596 & 1.004 & 3.220 & 46 & \begin{xy}
				\POS (0,3) *{\textcolor{Metal}{\medbullet}} ="a",
				\POS (-3.5,-1.5) *{\textcolor{Metal}{\medbullet}} ="b",
				\POS (3.5,-1.5) *{\textcolor{Metal}{\medbullet}} ="c"
				\POS "a" \ar @{<->} "b"
				\POS "a" \ar @{->} "c"
				\POS "b" \ar @{->} "c"
			\end{xy} & 1.664 & 2.054 & 0.180 & 0.461 & 3.216\\
			46 & \begin{xy}
				\POS (0,3) *{\textcolor{Metal}{\medbullet}} ="a",
				\POS (-3.5,-1.5) *{\textcolor{Metal}{\medbullet}} ="b",
				\POS (3.5,-1.5) *{\textcolor{Metal}{\medbullet}} ="c"
				\POS "a" \ar @{<->} "b"
				\POS "a" \ar @{->} "c"
				\POS "b" \ar @{->} "c"
			\end{xy} & 1.586 & 2.012 & 0.171 & 0.446 & 3.173 & 14 & \begin{xy}
				\POS (0,3) *{\textcolor{Equity}{\medbullet}} ="a",
				\POS (-3.5,-1.5) *{\textcolor{Equity}{\medbullet}} ="b",
				\POS (3.5,-1.5) *{\textcolor{Metal}{\medbullet}} ="c"
				\POS "a" \ar @{<->} "b"
				\POS "a" \ar @{->} "c"
			\end{xy} & 5.052 & 10.751 & 0.909 & 1.317 & 3.147\\
			46 & \begin{xy}
				\POS (0,3) *{\textcolor{Equity}{\medbullet}} ="a",
				\POS (-3.5,-1.5) *{\textcolor{Agriculture}{\medbullet}} ="b",
				\POS (3.5,-1.5) *{\textcolor{Equity}{\medbullet}} ="c"
				\POS "a" \ar @{<->} "b"
				\POS "a" \ar @{->} "c"
				\POS "b" \ar @{->} "c"
			\end{xy} & 2.375 & 5.768 & 0.344 & 0.654 & 3.107 & 46 & \begin{xy}
				\POS (0,3) *{\textcolor{Equity}{\medbullet}} ="a",
				\POS (-3.5,-1.5) *{\textcolor{Agriculture}{\medbullet}} ="b",
				\POS (3.5,-1.5) *{\textcolor{Equity}{\medbullet}} ="c"
				\POS "a" \ar @{<->} "b"
				\POS "a" \ar @{->} "c"
				\POS "b" \ar @{->} "c"
			\end{xy} & 2.452 & 5.758 & 0.361 & 0.677 & 3.087\\
			14 & \begin{xy}
				\POS (0,3) *{\textcolor{Equity}{\medbullet}} ="a",
				\POS (-3.5,-1.5) *{\textcolor{Equity}{\medbullet}} ="b",
				\POS (3.5,-1.5) *{\textcolor{Energy}{\medbullet}} ="c"
				\POS "a" \ar @{<->} "b"
				\POS "a" \ar @{->} "c"
			\end{xy} & 2.904 & 3.743 & 0.436 & 0.808 & 3.053 & 14 & \begin{xy}
				\POS (0,3) *{\textcolor{Equity}{\medbullet}} ="a",
				\POS (-3.5,-1.5) *{\textcolor{Equity}{\medbullet}} ="b",
				\POS (3.5,-1.5) *{\textcolor{Energy}{\medbullet}} ="c"
				\POS "a" \ar @{<->} "b"
				\POS "a" \ar @{->} "c"
			\end{xy} & 2.973 & 3.736 & 0.451 & 0.832 & 3.031\\
			14 & \begin{xy}
				\POS (0,3) *{\textcolor{Energy}{\medbullet}} ="a",
				\POS (-3.5,-1.5) *{\textcolor{Energy}{\medbullet}} ="b",
				\POS (3.5,-1.5) *{\textcolor{Agriculture}{\medbullet}} ="c"
				\POS "a" \ar @{<->} "b"
				\POS "a" \ar @{->} "c"
			\end{xy} & 4.239 & 3.242 & 0.715 & 1.246 & 2.829 & 14 & \begin{xy}
				\POS (0,3) *{\textcolor{Energy}{\medbullet}} ="a",
				\POS (-3.5,-1.5) *{\textcolor{Energy}{\medbullet}} ="b",
				\POS (3.5,-1.5) *{\textcolor{Agriculture}{\medbullet}} ="c"
				\POS "a" \ar @{<->} "b"
				\POS "a" \ar @{->} "c"
			\end{xy} & 4.398 & 3.170 & 0.739 & 1.281 & 2.856\\
			74 & \begin{xy}
				\POS (0,3) *{\textcolor{Metal}{\medbullet}} ="a",
				\POS (-3.5,-1.5) *{\textcolor{Metal}{\medbullet}} ="b",
				\POS (3.5,-1.5) *{\textcolor{Metal}{\medbullet}} ="c"
				\POS "a" \ar @{<->} "b"
				\POS "c" \ar @{->} "a"
			\end{xy} & 4.211 & 3.226 & 1.046 & 1.204 & 2.628 & 74 & \begin{xy}
				\POS (0,3) *{\textcolor{Metal}{\medbullet}} ="a",
				\POS (-3.5,-1.5) *{\textcolor{Metal}{\medbullet}} ="b",
				\POS (3.5,-1.5) *{\textcolor{Metal}{\medbullet}} ="c"
				\POS "a" \ar @{<->} "b"
				\POS "c" \ar @{->} "a"
			\end{xy} & 4.313 & 3.197 & 1.079 & 1.243 & 2.602\\
			14 & \begin{xy}
				\POS (0,3) *{\textcolor{Metal}{\medbullet}} ="a",
				\POS (-3.5,-1.5) *{\textcolor{Metal}{\medbullet}} ="b",
				\POS (3.5,-1.5) *{\textcolor{Energy}{\medbullet}} ="c"
				\POS "a" \ar @{<->} "b"
				\POS "a" \ar @{->} "c"
			\end{xy} & 4.175 & 4.126 & 0.930 & 1.260 & 2.576 & 14 & \begin{xy}
				\POS (0,3) *{\textcolor{Metal}{\medbullet}} ="a",
				\POS (-3.5,-1.5) *{\textcolor{Metal}{\medbullet}} ="b",
				\POS (3.5,-1.5) *{\textcolor{Energy}{\medbullet}} ="c"
				\POS "a" \ar @{<->} "b"
				\POS "a" \ar @{->} "c"
			\end{xy} & 4.308 & 4.072 & 0.961 & 1.298 & 2.580\\
			78 & \begin{xy}
				\POS (0,3) *{\textcolor{Equity}{\medbullet}} ="a",
				\POS (-3.5,-1.5) *{\textcolor{Energy}{\medbullet}} ="b",
				\POS (3.5,-1.5) *{\textcolor{Equity}{\medbullet}} ="c"
				\POS "a" \ar @{<->} "b"
				\POS "a" \ar @{<->} "c"
			\end{xy} & 1.457 & 2.256 & 0.200 & 0.488 & 2.574 & 78 & \begin{xy}
				\POS (0,3) *{\textcolor{Equity}{\medbullet}} ="a",
				\POS (-3.5,-1.5) *{\textcolor{Energy}{\medbullet}} ="b",
				\POS (3.5,-1.5) *{\textcolor{Equity}{\medbullet}} ="c"
				\POS "a" \ar @{<->} "b"
				\POS "a" \ar @{<->} "c"
			\end{xy} & 1.469 & 2.253 & 0.203 & 0.496 & 2.553\\
			74 & \begin{xy}
				\POS (0,3) *{\textcolor{Equity}{\medbullet}} ="a",
				\POS (-3.5,-1.5) *{\textcolor{Equity}{\medbullet}} ="b",
				\POS (3.5,-1.5) *{\textcolor{Energy}{\medbullet}} ="c"
				\POS "a" \ar @{<->} "b"
				\POS "c" \ar @{->} "a"
			\end{xy} & 1.897 & 2.680 & 0.303 & 0.628 & 2.538 & 74 & \begin{xy}
				\POS (0,3) *{\textcolor{Equity}{\medbullet}} ="a",
				\POS (-3.5,-1.5) *{\textcolor{Equity}{\medbullet}} ="b",
				\POS (3.5,-1.5) *{\textcolor{Energy}{\medbullet}} ="c"
				\POS "a" \ar @{<->} "b"
				\POS "c" \ar @{->} "a"
			\end{xy} & 1.924 & 2.686 & 0.313 & 0.647 & 2.489\\
			74 & \begin{xy}
				\POS (0,3) *{\textcolor{Metal}{\medbullet}} ="a",
				\POS (-3.5,-1.5) *{\textcolor{Metal}{\medbullet}} ="b",
				\POS (3.5,-1.5) *{\textcolor{Equity}{\medbullet}} ="c"
				\POS "a" \ar @{<->} "b"
				\POS "c" \ar @{->} "a"
			\end{xy} & 3.901 & 6.483 & 0.918 & 1.177 & 2.535 & 74 & \begin{xy}
				\POS (0,3) *{\textcolor{Metal}{\medbullet}} ="a",
				\POS (-3.5,-1.5) *{\textcolor{Metal}{\medbullet}} ="b",
				\POS (3.5,-1.5) *{\textcolor{Equity}{\medbullet}} ="c"
				\POS "a" \ar @{<->} "b"
				\POS "c" \ar @{->} "a"
			\end{xy} & 3.900 & 6.483 & 0.947 & 1.215 & 2.431\\
			38 & \begin{xy}
				\POS (0,3) *{\textcolor{Equity}{\medbullet}} ="a",
				\POS (-3.5,-1.5) *{\textcolor{Equity}{\medbullet}} ="b",
				\POS (3.5,-1.5) *{\textcolor{Equity}{\medbullet}} ="c"
				\POS "a" \ar @{->} "b"
				\POS "a" \ar @{->} "c"
				\POS "b" \ar @{->} "c"
			\end{xy} & 0.410 & 0.823 & 0.028 & 0.162 & 2.358 & 238 & \begin{xy}
				\POS (0,3) *{\textcolor{Agriculture}{\medbullet}} ="a",
				\POS (-3.5,-1.5) *{\textcolor{Agriculture}{\medbullet}} ="b",
				\POS (3.5,-1.5) *{\textcolor{Agriculture}{\medbullet}} ="c"
				\POS "a" \ar @{<->} "b"
				\POS "a" \ar @{<->} "c"
				\POS "b" \ar @{<->} "c"
			\end{xy} & 7.024 & 4.745 & 2.107 & 2.060 & 2.388\\
			238 & \begin{xy}
				\POS (0,3) *{\textcolor{Agriculture}{\medbullet}} ="a",
				\POS (-3.5,-1.5) *{\textcolor{Agriculture}{\medbullet}} ="b",
				\POS (3.5,-1.5) *{\textcolor{Agriculture}{\medbullet}} ="c"
				\POS "a" \ar @{<->} "b"
				\POS "a" \ar @{<->} "c"
				\POS "b" \ar @{<->} "c"
			\end{xy} & 6.430 & 4.623 & 1.986 & 1.956 & 2.272 & 110 & \begin{xy}
				\POS (0,3) *{\textcolor{Agriculture}{\medbullet}} ="a",
				\POS (-3.5,-1.5) *{\textcolor{Agriculture}{\medbullet}} ="b",
				\POS (3.5,-1.5) *{\textcolor{Agriculture}{\medbullet}} ="c"
				\POS "a" \ar @{<->} "b"
				\POS "a" \ar @{<->} "c"
				\POS "b" \ar @{->} "c"
			\end{xy} & 4.951 & 3.053 & 1.511 & 1.477 & 2.328\\
			\bottomrule
	\end{tabular}}
	\begin{flushleft}
		\footnotesize
		\textit{Legend: }
		\textcolor{Equity}{{\normalsize\textbullet}}\, Equity;
		\textcolor{Metal}{{\normalsize\textbullet}}\, Metal;
		\textcolor{Agriculture}{{\normalsize\textbullet}}\, Agriculture;
		\textcolor{Energy}{{\normalsize\textbullet}}\,  Energy.
	\end{flushleft} 
\end{table}


\begin{table}[htp]
	\centering
	\setlength{\abovecaptionskip}{0pt}
	\setlength{\belowcaptionskip}{10pt}
	\caption{Colored triad motifs at the median quantile (seven classes).}
	\label{tab:colored_triads_7class}
	\resizebox{\textwidth}{!}{
\begin{tabular}{llrrrrr|llrrrrr}
			\toprule
			\multicolumn{7}{c|}{$\alpha=0.05$} & \multicolumn{7}{c}{$\alpha=0.10$}\\
			\cmidrule(lr){1-7} \cmidrule(lr){8-14}
			id & motif & $\mu$ & $\sigma$ & $\mu_{\text{rnd}}$ & $\sigma_{\text{rnd}}$ & $z$ &
			id & motif & $\mu$ & $\sigma$ & $\mu_{\text{rnd}}$ & $\sigma_{\text{rnd}}$ & $z$\\
			\midrule
			238 & \begin{xy}
				\POS (0,3) *{\textcolor{Preciousmetals}{\medbullet}} ="a",
				\POS (-3.5,-1.5) *{\textcolor{Preciousmetals}{\medbullet}} ="b",
				\POS (3.5,-1.5) *{\textcolor{Preciousmetals}{\medbullet}} ="c"
				\POS "a" \ar @{<->} "b"
				\POS "a" \ar @{<->} "c"
				\POS "b" \ar @{<->} "c"
			\end{xy} & 2.555 & 1.315 & 0.011 & 0.083 & 30.644 & 238 & \begin{xy}
				\POS (0,3) *{\textcolor{Preciousmetals}{\medbullet}} ="a",
				\POS (-3.5,-1.5) *{\textcolor{Preciousmetals}{\medbullet}} ="b",
				\POS (3.5,-1.5) *{\textcolor{Preciousmetals}{\medbullet}} ="c"
				\POS "a" \ar @{<->} "b"
				\POS "a" \ar @{<->} "c"
				\POS "b" \ar @{<->} "c"
			\end{xy} & 2.588 & 1.295 & 0.012 & 0.086 & 29.818\\
			238 & \begin{xy}
				\POS (0,3) *{\textcolor{Industrialmetals}{\medbullet}} ="a",
				\POS (-3.5,-1.5) *{\textcolor{Industrialmetals}{\medbullet}} ="b",
				\POS (3.5,-1.5) *{\textcolor{Industrialmetals}{\medbullet}} ="c"
				\POS "a" \ar @{<->} "b"
				\POS "a" \ar @{<->} "c"
				\POS "b" \ar @{<->} "c"
			\end{xy} & 7.385 & 4.910 & 0.061 & 0.263 & 27.847 & 238 & \begin{xy}
				\POS (0,3) *{\textcolor{Industrialmetals}{\medbullet}} ="a",
				\POS (-3.5,-1.5) *{\textcolor{Industrialmetals}{\medbullet}} ="b",
				\POS (3.5,-1.5) *{\textcolor{Industrialmetals}{\medbullet}} ="c"
				\POS "a" \ar @{<->} "b"
				\POS "a" \ar @{<->} "c"
				\POS "b" \ar @{<->} "c"
			\end{xy} & 7.852 & 4.711 & 0.064 & 0.271 & 28.707\\
\bottomrule
	\end{tabular}}
	\begin{flushleft}
		\footnotesize
		\textit{Legend: }
		\textcolor{Equity}{{\normalsize\textbullet}}\, Equity;
		\textcolor{Industrialmetals}{{\normalsize\textbullet}}\, Industrial metals;
		\textcolor{Preciousmetals}{{\normalsize\textbullet}}\, Precious metals;
		\textcolor{Energy}{{\normalsize\textbullet}}\,  Energy;
		\textcolor{GrainsOilseeds}{{\normalsize\textbullet}}\, Grains \& oilseeds;
		\textcolor{Softs}{{\normalsize\textbullet}}\, Softs;
		\textcolor{Livestock}{{\normalsize\textbullet}}\, Livestock.
	\end{flushleft} 
\end{table}

At the median quantile ($\tau = 0.50$), colored motifs exhibit a clear
and highly structured significance profile. Under the two-class
partition, the most strongly over-represented motifs are dominated by
equity-only triads, especially the fully reciprocal triangle (motif 238,
$z = 17.87$), followed by partially reciprocal equity triangles such as
motifs 110 ($z = 13.25$) and 46 ($z = 11.32$). All-commodity
monochromatic triads, by contrast, register much weaker z-scores,
typically below 1.5. This indicates that the strong reciprocal motif
signature identified in the uncolored analysis is disproportionately
concentrated within the equity sector. Under the four-class scheme, the
monochromatic dominance extends beyond equity: all-metal motif 238
achieves a z-score of 14.42, almost as high as its equity counterpart,
indicating that the median-state network is locally organized around
tightly interconnected sector-specific cliques rather than a single
dominant hub sector. Under the seven-class decomposition, this sectoral
differentiation sharpens further. Motif 238 composed exclusively of precious metals reaches the highest z-score in the entire analysis at 30.64, followed by the industrial-metals-only version at 27.85, the equity-only version at 17.71, and the grains-only version at 14.64. This ordering reveals
that precious metals and industrial metals form the most cohesive
reciprocal clusters, followed by equity and grains.

Beyond the monochromatic patterns, two types of cross-sector colored
motifs emerge with moderate but significant z-scores. First, motif 14
frequently appears in configurations where a reciprocal pair from one
sector sends a unidirectional edge to a node in a different sector. At
the seven-class level, prominent examples include equity pairs
transmitting to softs ($z = 8.06$), industrial metals pairs transmitting
to precious metals ($z = 7.71$), and industrial metals pairs
transmitting to equity ($z = 5.66$). This suggests that within-sector
reciprocal cores serve as platforms from which risk is broadcast
unidirectionally to other sectors. Second, motif 78, a configuration
with two reciprocal edges sharing a hub node, appears significantly in
the industrial-precious-industrial combination ($z = 9.06$ at seven
classes), revealing a particularly strong bilateral exchange channel
between the two metal sub-sectors that goes beyond their individual
within-sector clustering. Putting these patterns together, the median-state network contains a distinctly layered
local structure: dense reciprocal triads arise most frequently within a
few specific sectors, while cross-sector motifs survive mainly when
anchored by a strongly integrated same-sector core.

At the tail quantiles, the colored motif significance profile collapses
in parallel with the uncolored results. Under the two-class partition,
commodity-heavy fan-out motifs (especially motif 6) dominate in raw
frequency at both tails, but their z-scores remain close to zero,
confirming that high triadic counts under extreme conditions do not
translate into statistically distinctive sector-colored building blocks.
Under the four-class and seven-class schemes, a similar pattern holds:
even when specific color combinations appear frequently, they are only
weakly differentiated from the degree-preserving null. One modest
exception at the right tail involves motif 78 in the
equity-energy-equity combination ($z = 2.30$ under seven classes),
suggesting that a bilateral exchange channel between equity and energy
may be selectively activated under extreme bullish conditions. This
exception aside, the general absence of significant colored motifs at the
tails confirms that extreme states generate high but locally
unstructured connectedness. Once market types are taken into account,
tail-state spillovers do not organize into stable and statistically
distinctive sector-colored patterns, and the layered within-sector
architecture observed at the median dissolves into a more diffuse and
sector-indifferent transmission regime.

\subsection{Positional diversity and risk transmission}

To examine whether an asset's structural role in local directed triadic configurations relates to its risk transmission behavior, we construct a positional profile for each asset on every trading day. Specifically, on each day $t$, we count the number of times asset $i$ occupies each of the 30 orbit positions in the directed triadic census and normalize these counts into a probability vector $\mathbf{p}_{i,t}=(p_{i1,t},\ldots,p_{i30,t})$ with $\sum_{j=1}^{30} p_{ij,t}=1$, which summarizes the asset's role distribution across local triadic structures. These probability vectors serve two purposes. First, we compute pairwise correlations of the position ratio vectors, yielding a daily structural similarity matrix $\mathbf{H}_t$ whose entries capture how closely two assets resemble each other in their local triadic roles rather than in their returns or volatilities. Second, we measure the evenness of each asset's role distribution via the Shannon entropy $d_{i,t}=-\sum_{j=1}^{30} p_{ij,t}\ln p_{ij,t}$, where higher values indicate more diversified structural participation \citep{FJ-Eagle-Macy-Claxton-2010-Science}. To link positional diversity to risk transmission, we compute the cross-sectional correlation between $d_{i,t}$ and each asset-level directional spillover measure (TO, FROM, and NET) on every trading day, producing three daily correlation series.

\begin{figure}[H]
	\centering
	\includegraphics[width=0.415\linewidth]{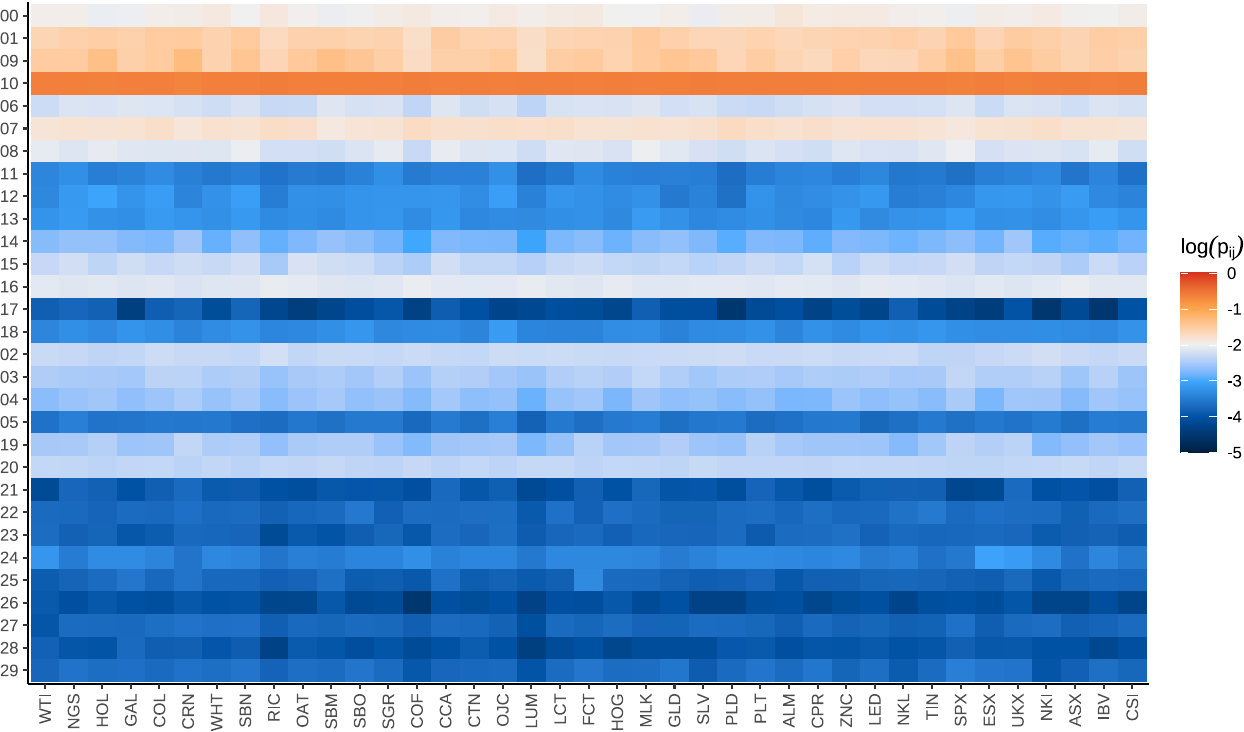}
	\includegraphics[width=0.415\linewidth]{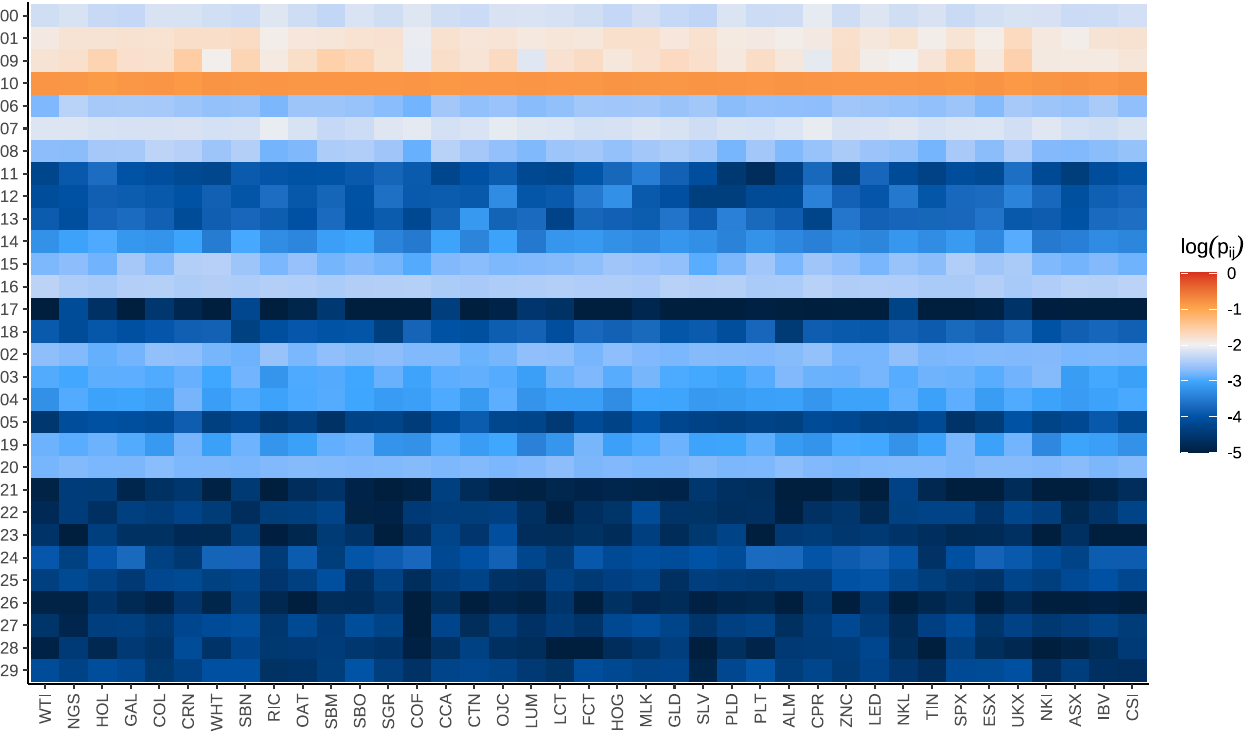}
	\includegraphics[width=0.415\linewidth]{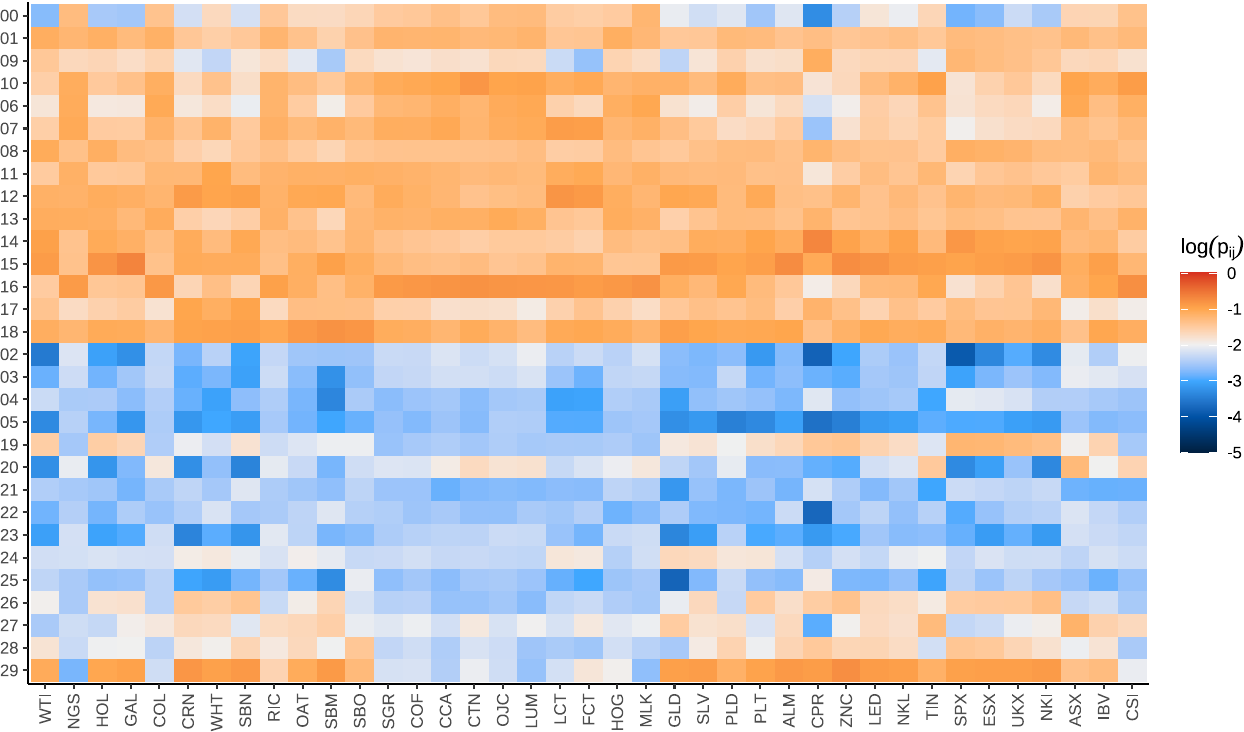}
	\includegraphics[width=0.415\linewidth]{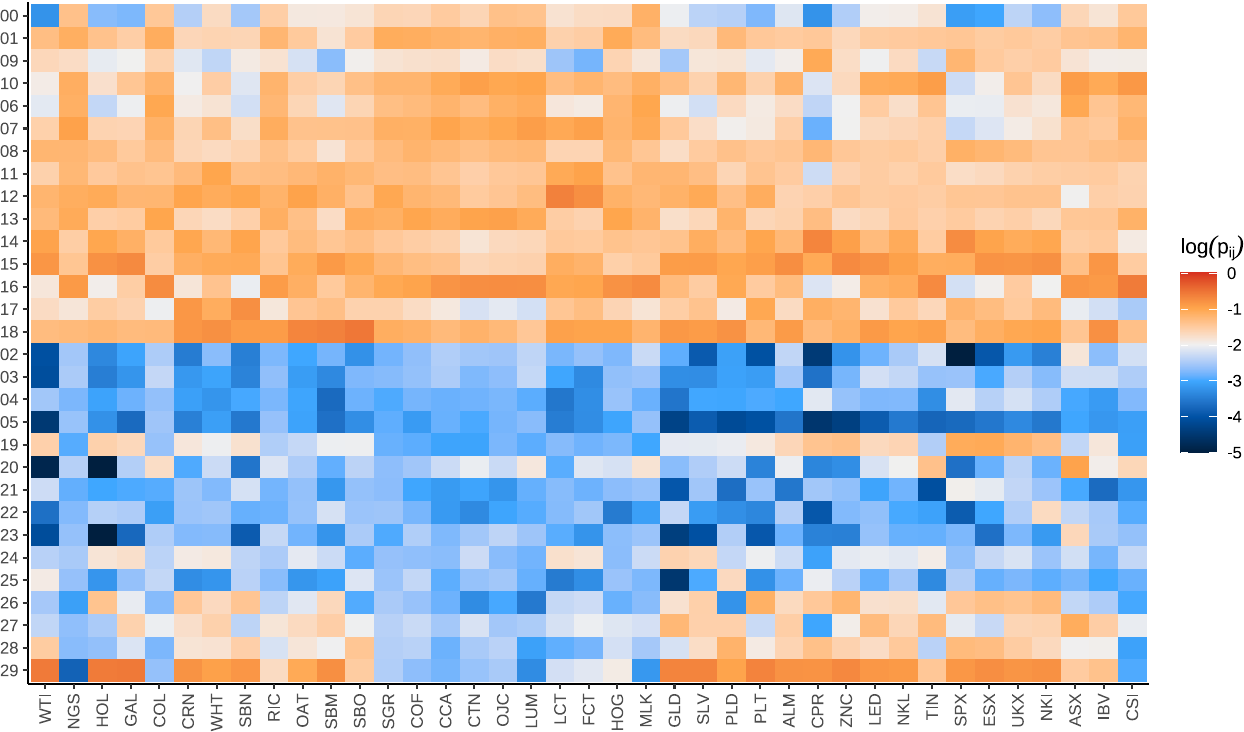}
	\includegraphics[width=0.415\linewidth]{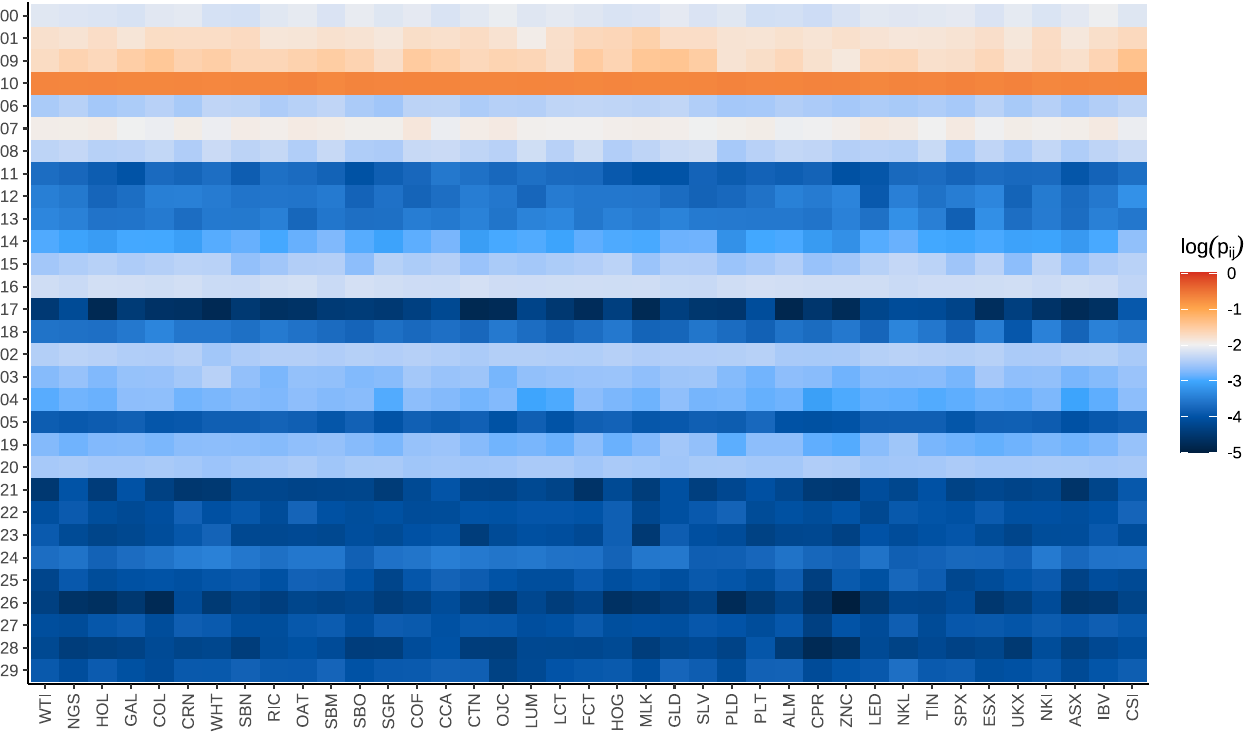}
	\includegraphics[width=0.415\linewidth]{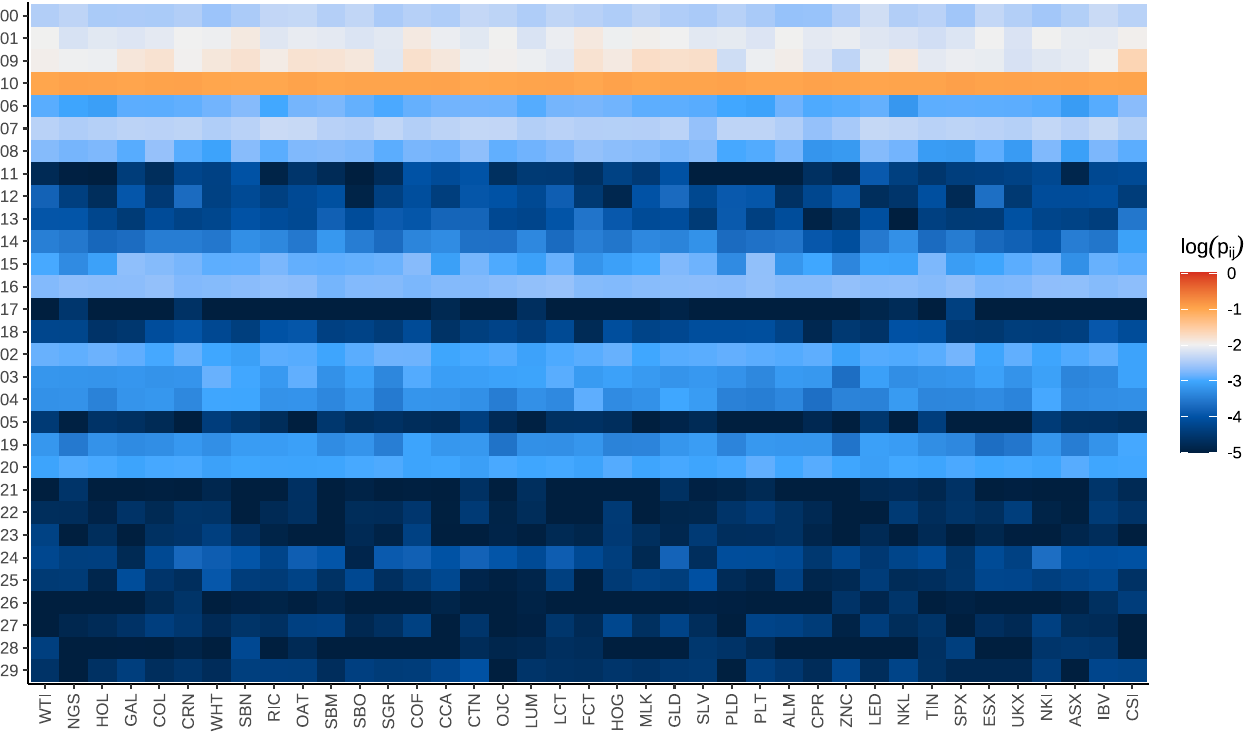}
	\caption{Average orbit-position ratio profiles across assets. For each asset $i$ and orbit position $j$, the daily relative frequency $p_{ij,t}$ is computed from normalized orbit counts and then averaged over the full sample to obtain $\bar{p}_{ij}$. Cell colors represent $\log_{10}(\bar{p}_{ij})$. Rows (top to bottom): $\tau=0.05$, $\tau=0.50$, $\tau=0.95$. 
		Columns (left to right): $\alpha=0.10$, $\alpha=0.05$.}
	\label{Fig:mean:occurrence:frequency}
\end{figure}

Fig.~\ref{Fig:mean:occurrence:frequency} displays the time-averaged orbit-position ratio profiles on a logarithmic scale. Across all six panels, the profiles reveal a clear hierarchy in orbit participation, although this hierarchy is much stronger at the tail quantiles. In the $\tau=0.05$ and $\tau=0.95$ panels, the probability mass is concentrated on a small subset of positions, most notably position 10, whereas several other positions, such as 17, 21, and 05, remain in the low-probability range throughout most assets. The resulting heatmap is heavily bimodal, with a few high-frequency rows in warm colors and the remainder in cold tones. By contrast, at $\tau=0.50$ the profiles become much more diffuse: many positions that are cold in the tail panels shift toward warmer colors, and cross-asset heterogeneity becomes substantially more visible, with 

\begin{figure}[H]
	\centering
	\includegraphics[width=0.395\linewidth]{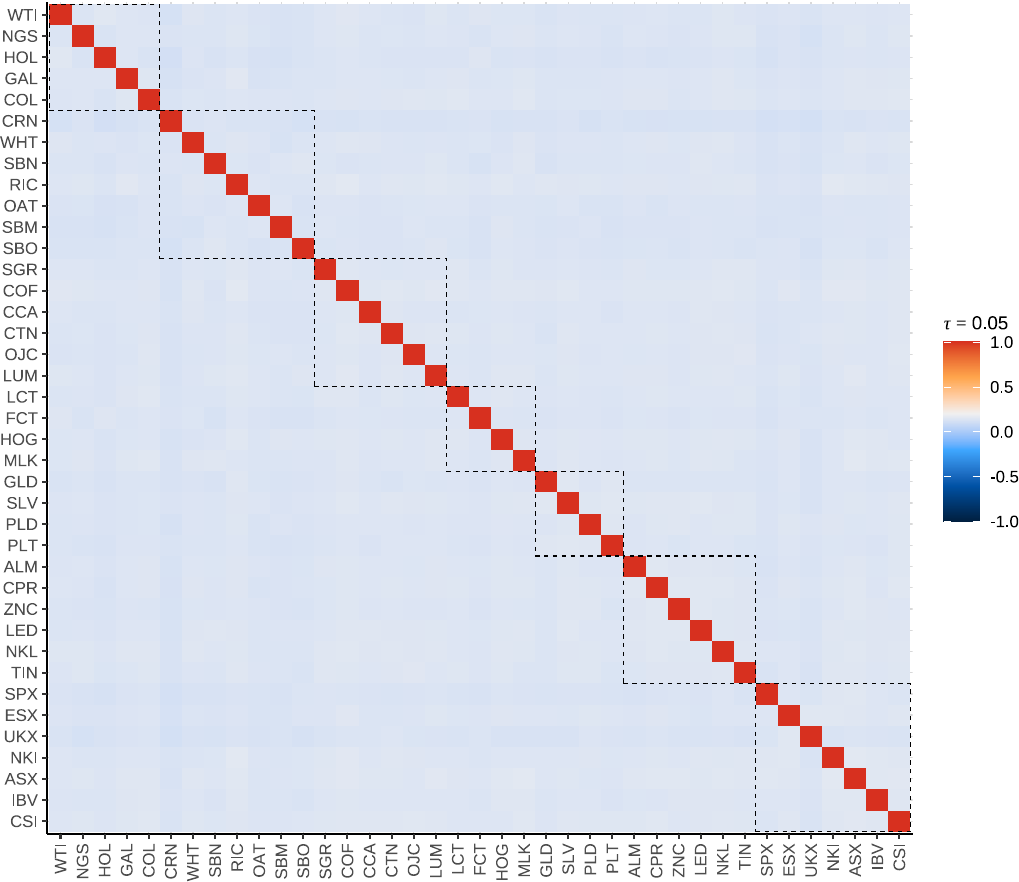}
	\includegraphics[width=0.395\linewidth]{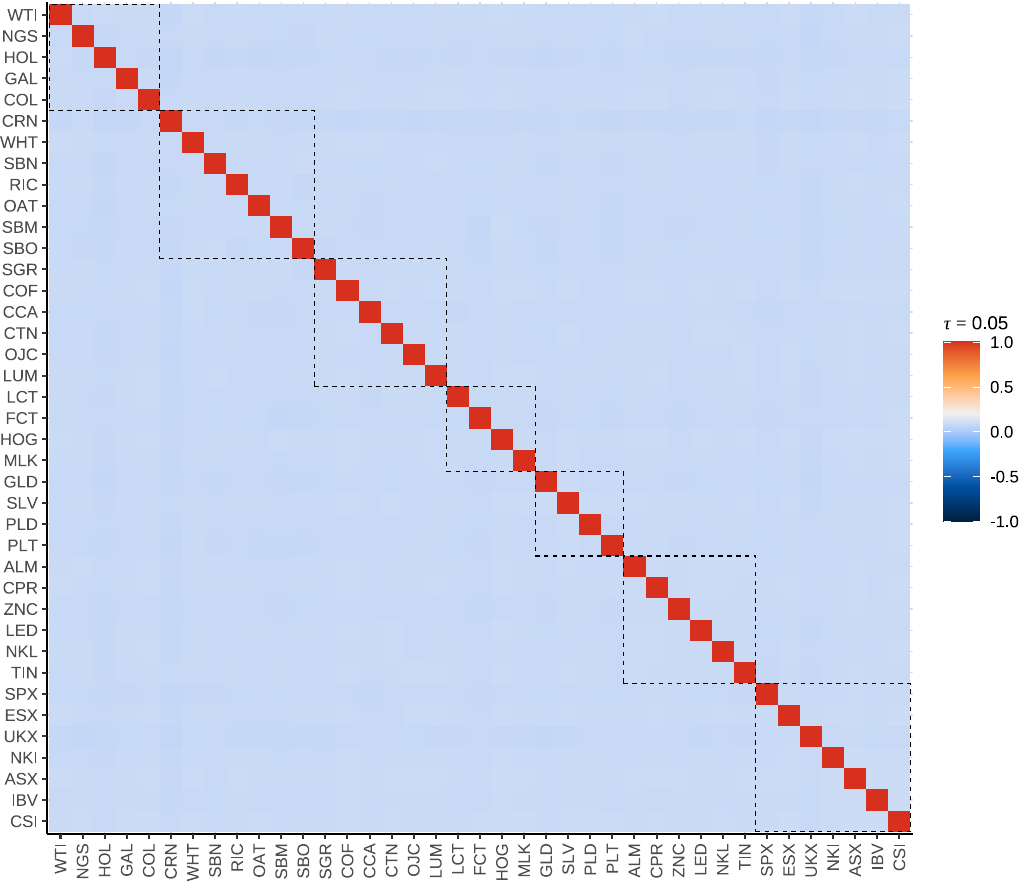}
	\includegraphics[width=0.395\linewidth]{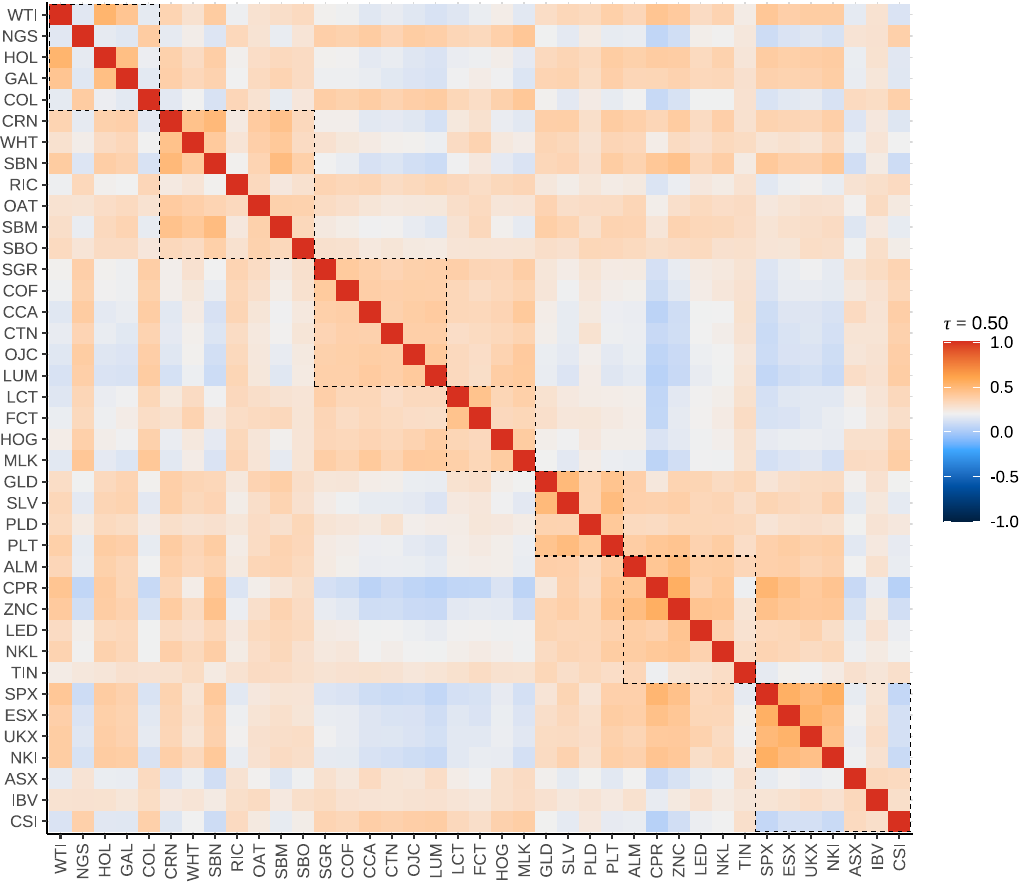}
	\includegraphics[width=0.395\linewidth]{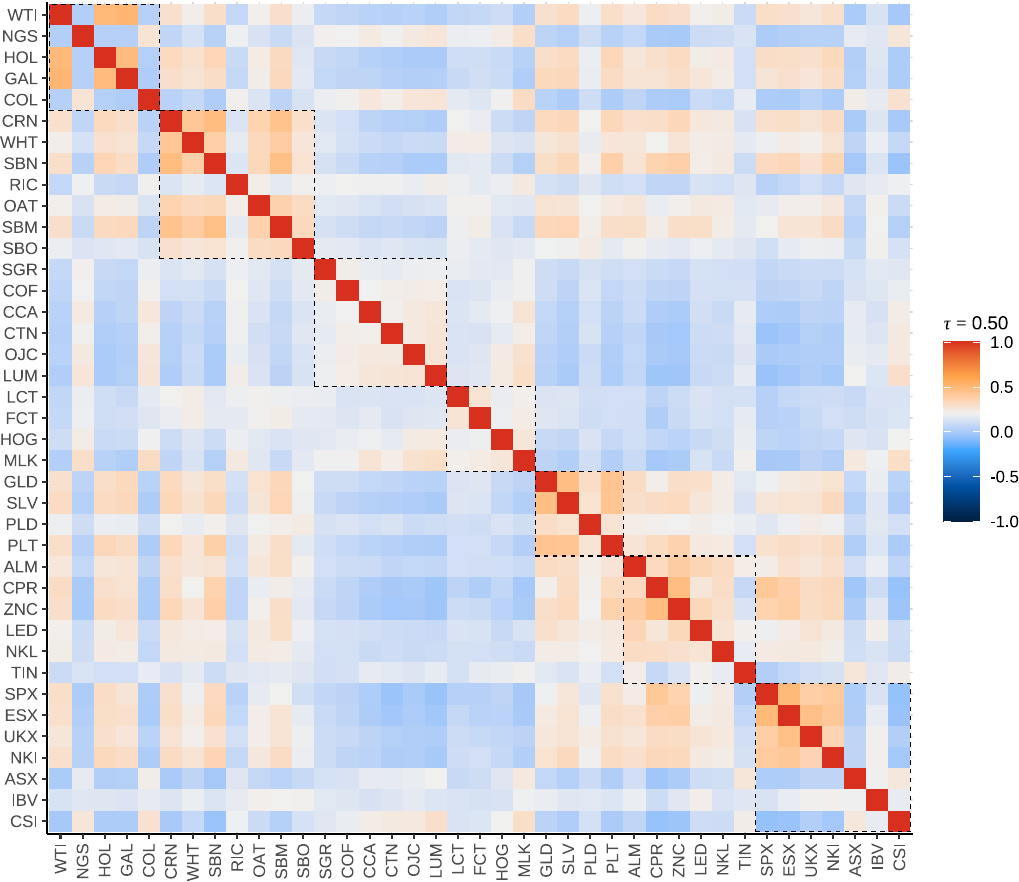}
	\includegraphics[width=0.395\linewidth]{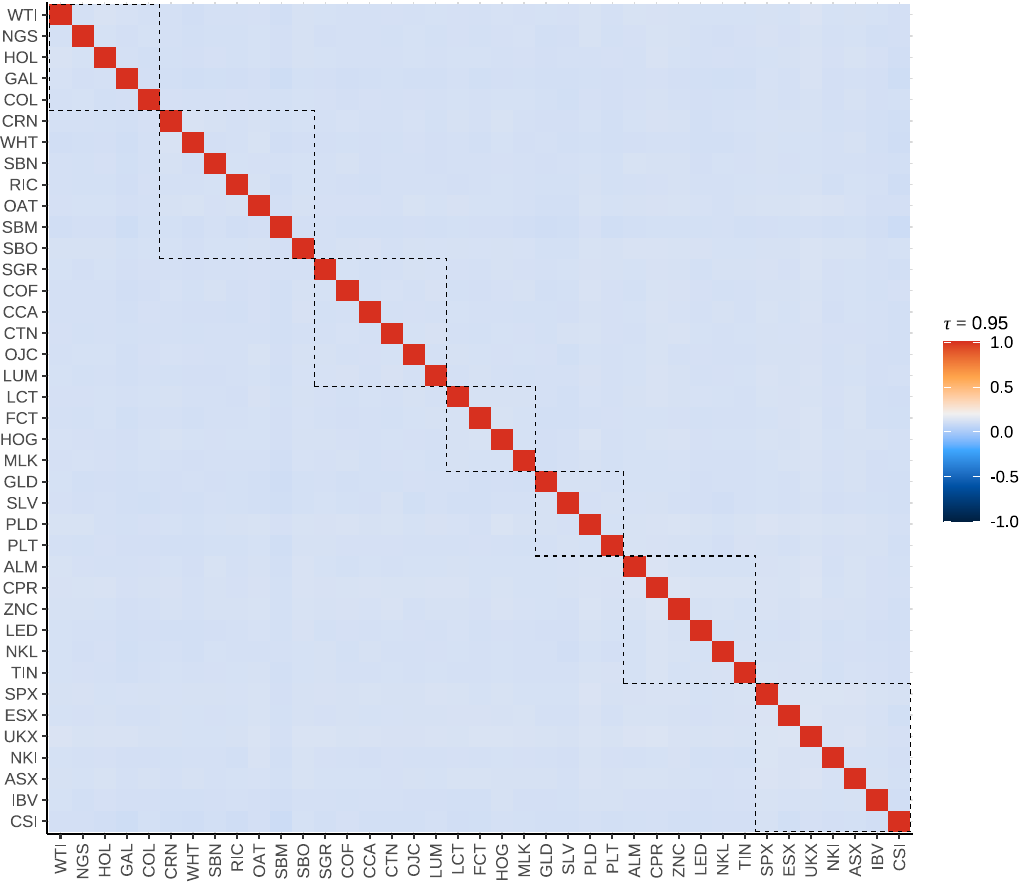}
	\includegraphics[width=0.395\linewidth]{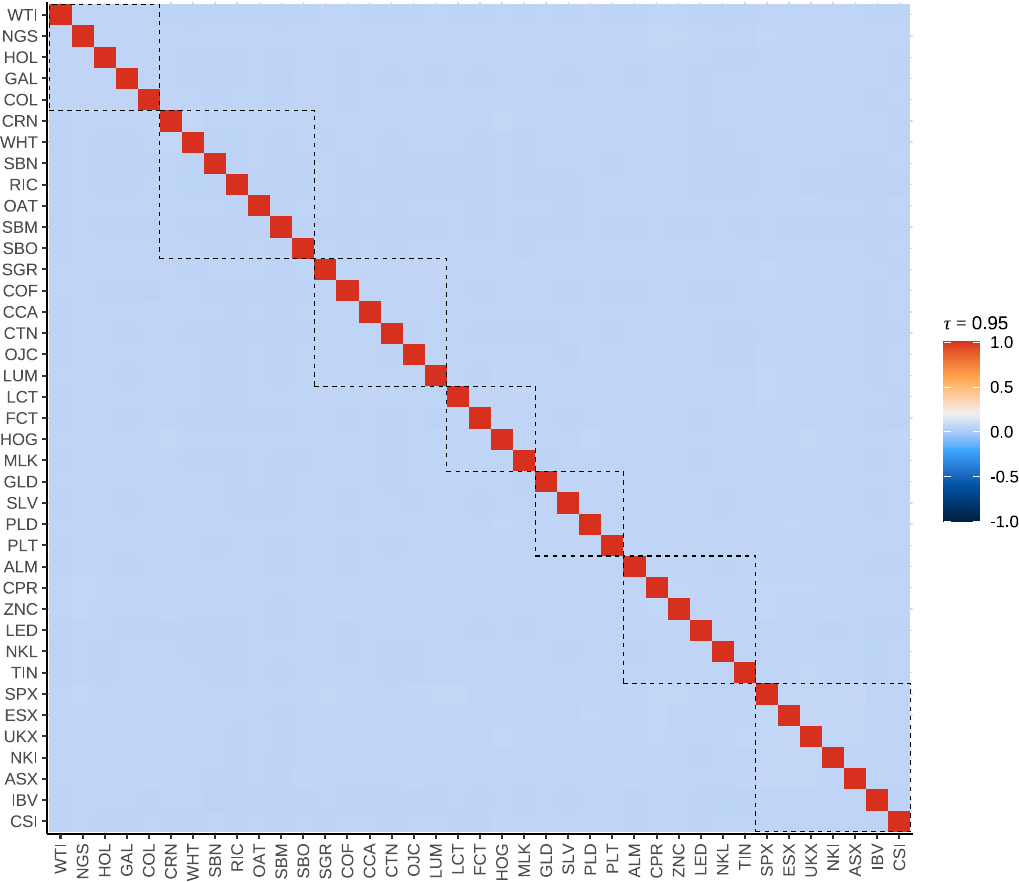}
	\caption{Time-averaged structural similarity matrix $\bar{\mathbf{H}}$. 
		Each daily matrix $\mathbf{H}_t$ is computed as the pairwise correlation 
		of orbit-position probability profiles across assets and then averaged 
		over the full sample. Rows from top to bottom correspond to $\tau=0.05$, 
		$\tau=0.50$, and $\tau=0.95$; left and right columns correspond to 
		$\alpha=0.10$ and $\alpha=0.05$, respectively. Dashed boxes indicate 
		predefined asset groups.}
	\label{Fig:Similarity}
\end{figure}

\noindent
some group-level differentiation emerging across asset columns. This indicates that under the median quantile, assets participate in a broader set of local motif roles, whereas the tail-quantile profiles are considerably more concentrated. Within each quantile level, the left and right columns remain visually similar, suggesting that the quantile dimension plays a stronger role than the backbone significance level in shaping the overall position profiles.

Fig.~\ref{Fig:Similarity} maps these position patterns into the time-averaged structural similarity matrix $\bar{\mathbf{H}}$. At $\tau=0.50$, the off-diagonal entries display a visibly richer structure than in the tail panels: several within-group blocks become moderately warmer than the surrounding entries, suggesting that assets within some economically related groups tend to share more similar orbit-position compositions. This pattern appears under both $\alpha=0.10$ and $\alpha=0.05$, although the contrast is somewhat sharper under $\alpha=0.10$. Cross-group entries remain mixed, with many values close to zero and some mildly negative patches, so the block structure is only partial rather than uniform across all predefined groups. By contrast, at $\tau=0.05$ and $\tau=0.95$, the matrices are dominated by near-zero off-diagonal entries and the dashed group blocks become much less distinct. Overall, the middle-quantile panels exhibit the clearest group-level differentiation in structural roles, whereas the tail-quantile panels imply much weaker average pairwise similarity in orbit-position composition. This is consistent with the position-ratio heatmaps, where the $\tau=0.50$ profiles are visibly more differentiated across assets while the tail-quantile profiles are much more concentrated.

\begin{figure}[H]
	\centering
	\includegraphics[width=0.495\linewidth]{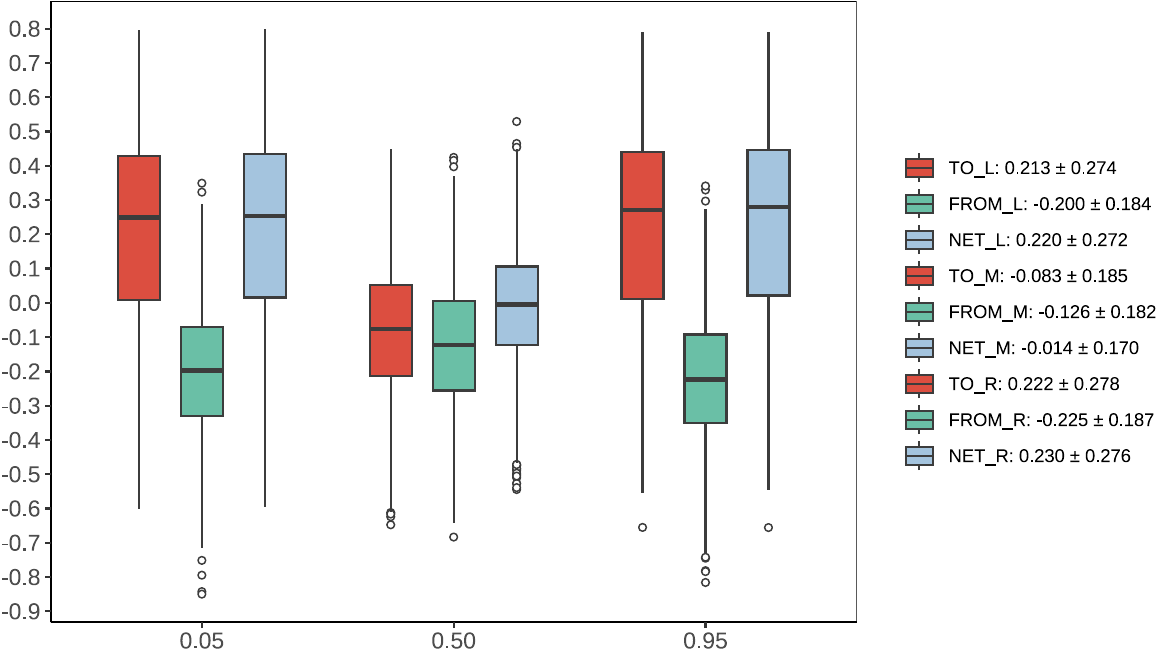}
	\includegraphics[width=0.495\linewidth]{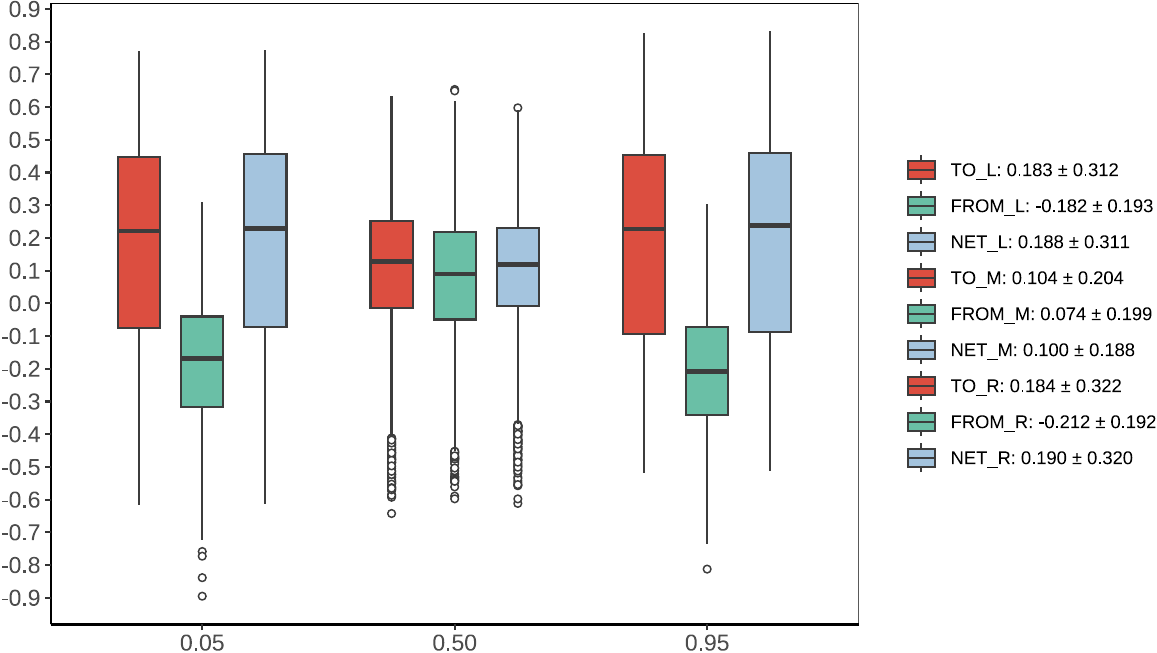}
	\caption{Box plots of the daily cross-sectional correlations between 
		orbit-position diversity and directional connectedness (TO, FROM, and 
		NET). Each box summarizes the distribution of one correlation series 
		over all trading days, grouped by quantile level $\tau$. Left and right 
		panels correspond to $\alpha=0.10$ and $\alpha=0.05$, respectively.}
	\label{Fig:corr:boxplot}
\end{figure}

Fig.~\ref{Fig:corr:boxplot} reports the distribution of the daily cross-sectional correlations between orbit-position diversity and directional connectedness measures. A clear sign separation emerges at the tail quantiles. For both $\tau=0.05$ and $\tau=0.95$, the correlations of diversity with TO and NET are predominantly positive, whereas the correlation with FROM is mostly negative. This indicates that, in tail-network states, assets characterized by greater positional diversity tend to be more closely associated with outward and net spillover roles than with inward spillover exposure. In contrast, at $\tau=0.50$ this separation is substantially attenuated: the three distributions compress toward zero, although the precise signs vary slightly across the two significance levels. The association between positional diversity and directional connectedness is therefore strongest in the lower- and upper-tail networks and materially weaker around the median quantile.

\begin{figure}[H]
	\centering
	\includegraphics[width=0.495\linewidth]{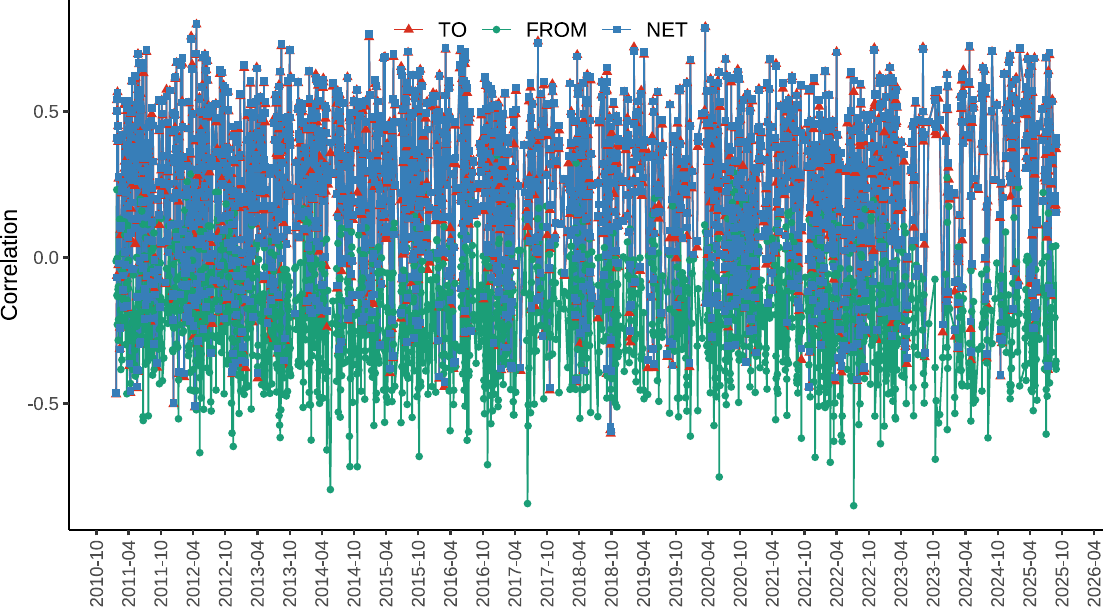}
	\includegraphics[width=0.495\linewidth]{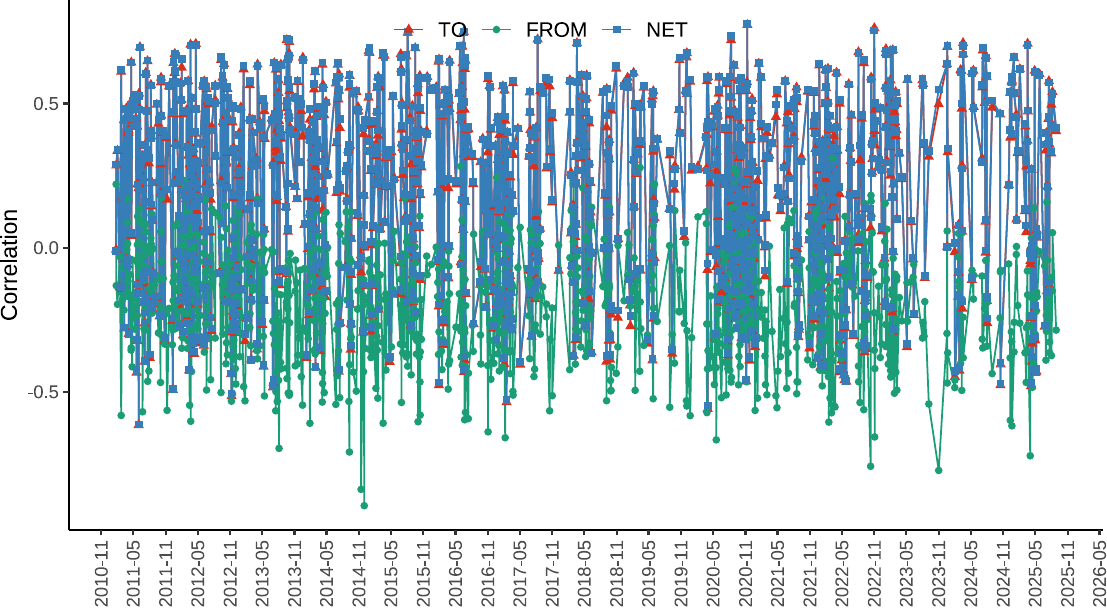}
	\includegraphics[width=0.495\linewidth]{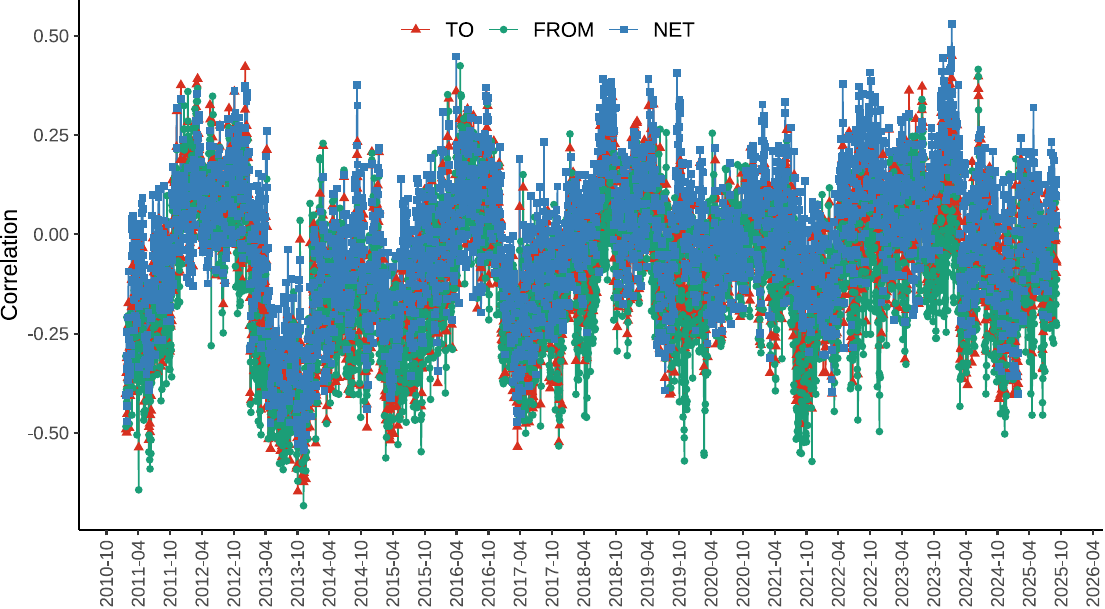}
	\includegraphics[width=0.495\linewidth]{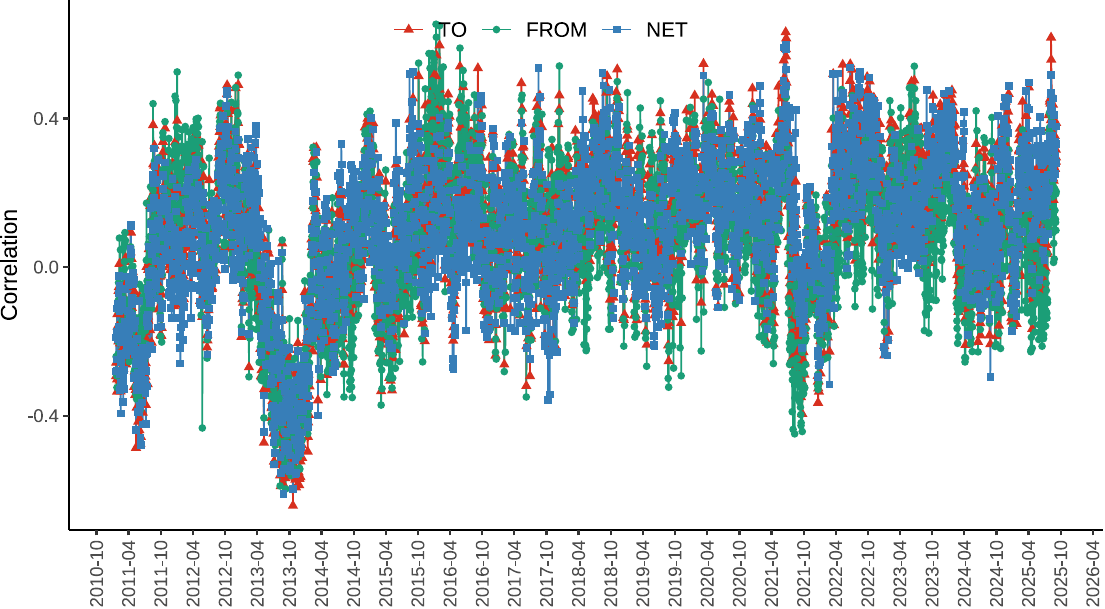}
	\includegraphics[width=0.495\linewidth]{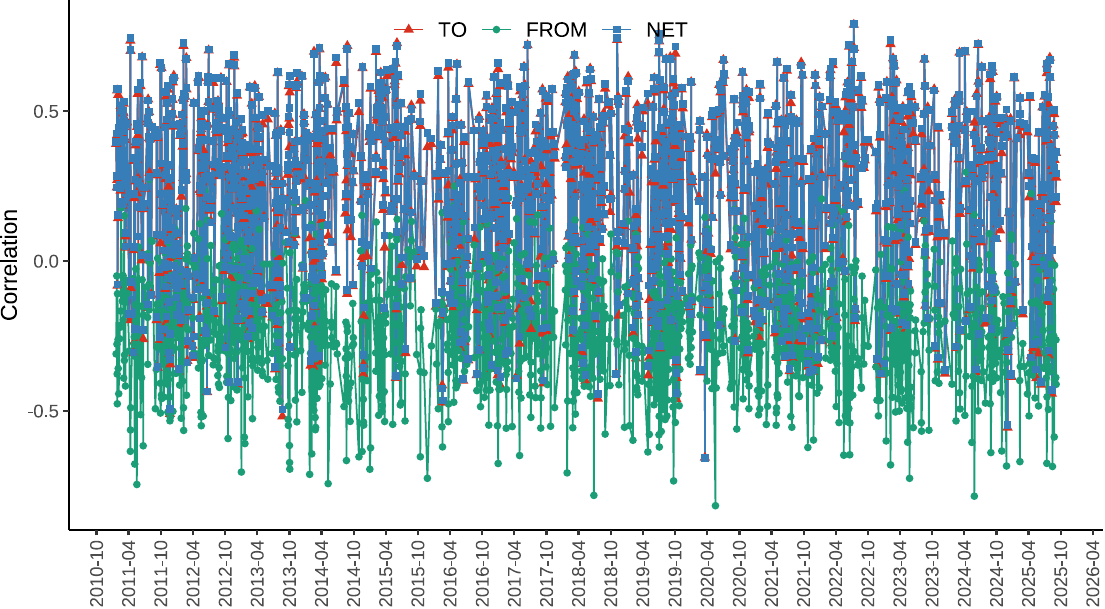}
	\includegraphics[width=0.495\linewidth]{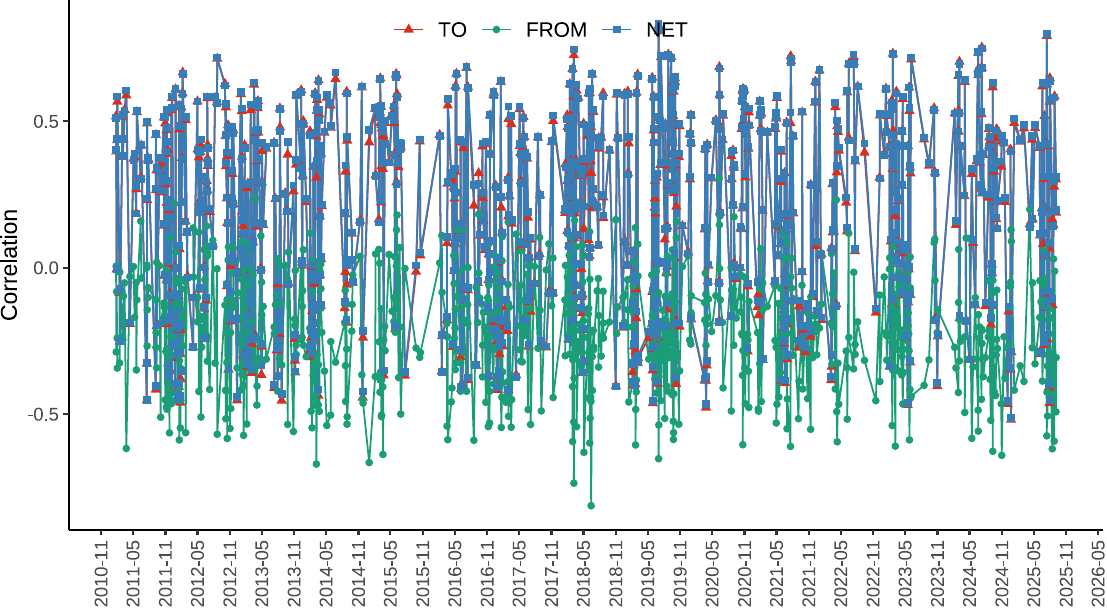}
	\caption{Daily cross-sectional correlations between orbit-position 
		diversity and directional connectedness (TO, FROM, and NET). Each point 
		represents the correlation computed across all assets on a single trading 
		day. Rows from top to bottom correspond to $\tau=0.05$, $\tau=0.50$, 
		and $\tau=0.95$; left and right columns correspond to $\alpha=0.10$ and 
		$\alpha=0.05$, respectively.}
	\label{Fig:corr:entropy:connectedness}
\end{figure}

Fig.~\ref{Fig:corr:entropy:connectedness} further shows that the tail-quantile pattern is persistent over time rather than being driven by time aggregation. In the $\tau=0.05$ and $\tau=0.95$ panels, the diversity--NET correlation remains positive for the vast majority of trading days, the diversity--TO correlation is likewise positive on most days, and the diversity--FROM correlation stays in negative territory throughout. Although the three series are not exact mirror images, their relative ordering is stable across the full period. By contrast, the $\tau=0.50$ panels display tighter co-movement among the three series and more pronounced regime dependence, with notable negative episodes around 2013 and again around 2017, followed by periods of mildly positive association. Near the median quantile, the link between positional diversity and directional connectedness is thus weaker and more state-contingent.

These results jointly show that motif-level positional diversity is systematically related to risk transmission roles in the spillover network, especially in tail states. It is worth noting that this evidence should be interpreted as complementary to, rather than identical with, the similarity structure used in portfolio construction. The matrix $\mathbf{H}_t$ is constructed from pairwise similarity in the full orbit-position probability profiles, whereas diversity is a scalar summary of profile dispersion. The diversity-based results therefore support the broader claim that local motif-position structure contains information relevant for portfolio construction, while $\mathbf{H}_t$ captures the finer cross-asset structural resemblance that enters the optimization step.

\subsection{Robustness checks}

\begin{figure}[H]
	\centering
	\includegraphics[width=0.85\linewidth]{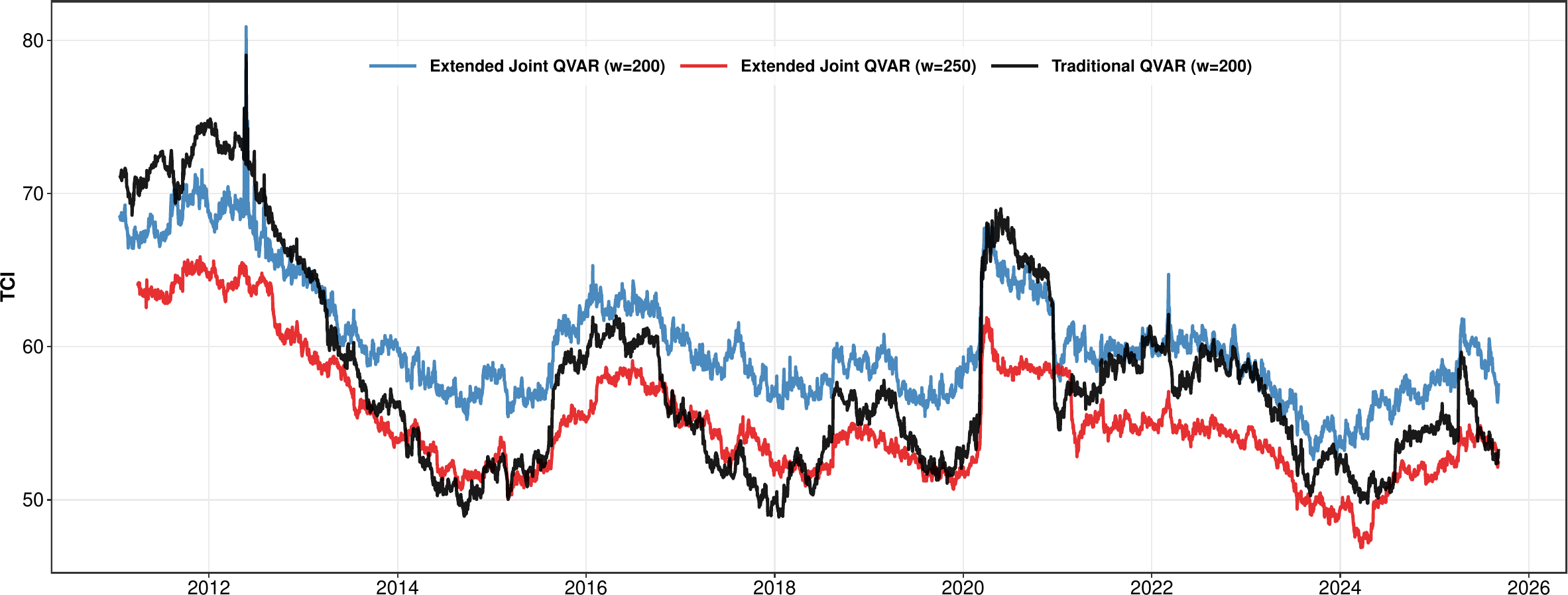}
	\caption{TCI under the traditional QVAR ($w=200$, black) and the 
		extended joint QVAR ($w=200$, blue; $w=250$, red).}
	\label{Fig:Robustness:Check}
\end{figure}

As a robustness check, Fig.~\ref{Fig:Robustness:Check} compares 
the time-varying TCI generated by the traditional QVAR ($w=200$) 
with that from the extended joint QVAR under two rolling-window 
settings ($w=200$ and $w=250$). The three series share highly 
consistent dynamic patterns: the timing of major peaks, troughs, 
and turning points remains broadly unchanged across specifications. 
Under the same window length, the traditional and extended joint 
QVAR produce closely aligned TCI paths, indicating that the 
extended modeling framework does not materially alter the 
connectedness dynamics. The more visible difference arises from 
window length: the longer window ($w=250$) yields lower and 
smoother TCI estimates, but the qualitative evolution is preserved. 
These results suggest that the main conclusions regarding the 
dynamics of systemic risk spillovers are robust to both the model 
specification and the choice of rolling-window length.

\section{Portfolio implications}
\label{S6:Portfolio}

The preceding analysis has shown that risk spillovers among commodity and equity futures exhibit rich local network structures that vary across quantile levels. We now turn to the question of whether this structural information can be exploited in portfolio construction. To this end, we consider four allocation strategies. The first two are well-established benchmarks: the minimum variance portfolio (MVP) of \citet{Markowitz-1952-JFinance} and the minimum connectedness portfolio (MCoP) of \citet{Broadstock-Chatziantoniou-Gabauer-2022-ApplEnergyFinance}. To these we add the minimum correlation portfolio (MCP) of \citet{Christoffersen-Errunza-Jacobs-Jin-2014-IntJForecast1}, which replaces the covariance matrix with a correlation-based analogue, and a new strategy proposed in this paper, the minimum structural similarity portfolio (MSP), which substitutes the motif-based structural similarity matrix $\mathbf{H}_t$ for the covariance or connectedness matrices used in the other approaches.

All four strategies share the same algebraic form, differing only in 
the input matrix that defines the risk notion being minimized. On each 
trading day $t$, the MVP weights are obtained from the conditional 
covariance matrix $\boldsymbol{\Sigma}_t$ as
\begin{equation}
	\boldsymbol{\omega}_{V,t}
	= \frac{\boldsymbol{\Sigma}_t^{-1}\mathbf{1}}
	{\mathbf{1}^{\prime}\boldsymbol{\Sigma}_t^{-1}\mathbf{1}},
\end{equation}
where $\mathbf{1}$ is a conformable vector of ones. The MCP replaces 
$\boldsymbol{\Sigma}_t$ with the conditional correlation matrix 
$\mathbf{R}_t$:
\begin{equation}
	\boldsymbol{\omega}_{R,t}
	= \frac{\mathbf{R}_t^{-1}\mathbf{1}}
	{\mathbf{1}^{\prime}\mathbf{R}_t^{-1}\mathbf{1}}.
\end{equation}
The MCoP constructs weights by minimising the pairwise connectedness 
of asset returns:
\begin{equation}
	\boldsymbol{\omega}_{C,t}
	= \frac{PCI_t^{-1}\mathbf{1}}
	{\mathbf{1}^{\prime}PCI_t^{-1}\mathbf{1}},
\end{equation}
where $PCI_t$ is the pairwise connectedness index matrix. The MSP 
follows the same algebraic form but replaces $PCI_t$ with the 
motif-based structural similarity matrix $\mathbf{H}_t$:
\begin{equation}
	\boldsymbol{\omega}_{S,t}
	= \frac{\mathbf{H}_t^{-1}\mathbf{1}}
	{\mathbf{1}^{\prime}\mathbf{H}_t^{-1}\mathbf{1}},
\end{equation}
so that assets whose motif roles are least similar to those of other 
assets receive higher weights, thereby diversifying across local 
network structures rather than across returns or spillover intensities.

To evaluate performance, we report portfolio returns, standard deviations, and Sharpe ratios based on both volatility and conditional value at risk (CVaR). Because investors may respond differently across the return distribution, portfolios built from connectedness or structural information estimated at different conditional quantiles are likely to exhibit distinct performance profiles \citep{Shi-Chen-2025-GlobFinJ}. We therefore construct all four strategies at $\tau=0.05$, $0.50$, and $0.95$ and compare their behaviour under normal and tail conditions.

\begin{table}[ht]
	\centering
	\setlength{\abovecaptionskip}{0pt}
	\setlength{\belowcaptionskip}{10pt}
	\caption{Performance of the MVP, MCP, MCoP, and MSP portfolios at the 5th, 50th, and 95th conditional quantiles.}
	\label{Tb:portfolio:performance}
	\resizebox{\textwidth}{!}{%
		\begin{tabular}{l|rrrrr|rrrrr|rrrrr}
			\toprule
			& \multicolumn{5}{c|}{$\tau = 0.05$} & \multicolumn{5}{c|}{$\tau = 0.50$} & \multicolumn{5}{c}{$\tau = 0.95$} \\
			\cmidrule(lr){2-6}\cmidrule(lr){7-11}\cmidrule(lr){12-16}
			& MVP & MCP & MCoP & MSP1 & MSP2 & MVP & MCP & MCoP & MSP1 & MSP2 & MVP & MCP & MCoP & MSP1 & MSP2 \\
			\midrule
			Return & 0.044 & 0.049 & -0.009 & -0.001& 0.001 & 0.039 & 0.014 & 0.004 & 0.002 & 0.024 & 0.030 & -0.021 & -0.006 & 0.011 & 0.027 \\
			StdDev & 0.075 & 0.107 & 0.134 & 0.112 & 0.121 & 0.065 & 0.089 & 0.107 & 0.120 & 0.120 & 0.077 & 0.115 & 0.128 & 0.113 & 0.130 \\
			Sharpe Ratio (StdDev) & 0.591 & 0.453 & -0.064&-0.012 & 0.007& 0.594 & 0.162 & 0.039 & 0.013& 0.202& 0.390 & -0.182 & -0.051 & 0.096& 0.206 \\
			Sharpe Ratio (CVaR) & 2.887 & 3.943 & -0.269&-0.049 & 0.022 & 2.983 & 0.943 & 0.180 & 0.060 & 1.017 & 1.723 & -0.858 & -0.260 & 2.250 & 0.093 \\
			\bottomrule
	\end{tabular}}
	\begin{flushleft}
		\footnotesize
		\justifying Note: The MSP1 and MSP2 refer to $\alpha=0.05$ and $\alpha=0.10$, respectively.
	\end{flushleft} 
\end{table}

\begin{figure}[H]
	\centering
	\includegraphics[width=0.85\linewidth]{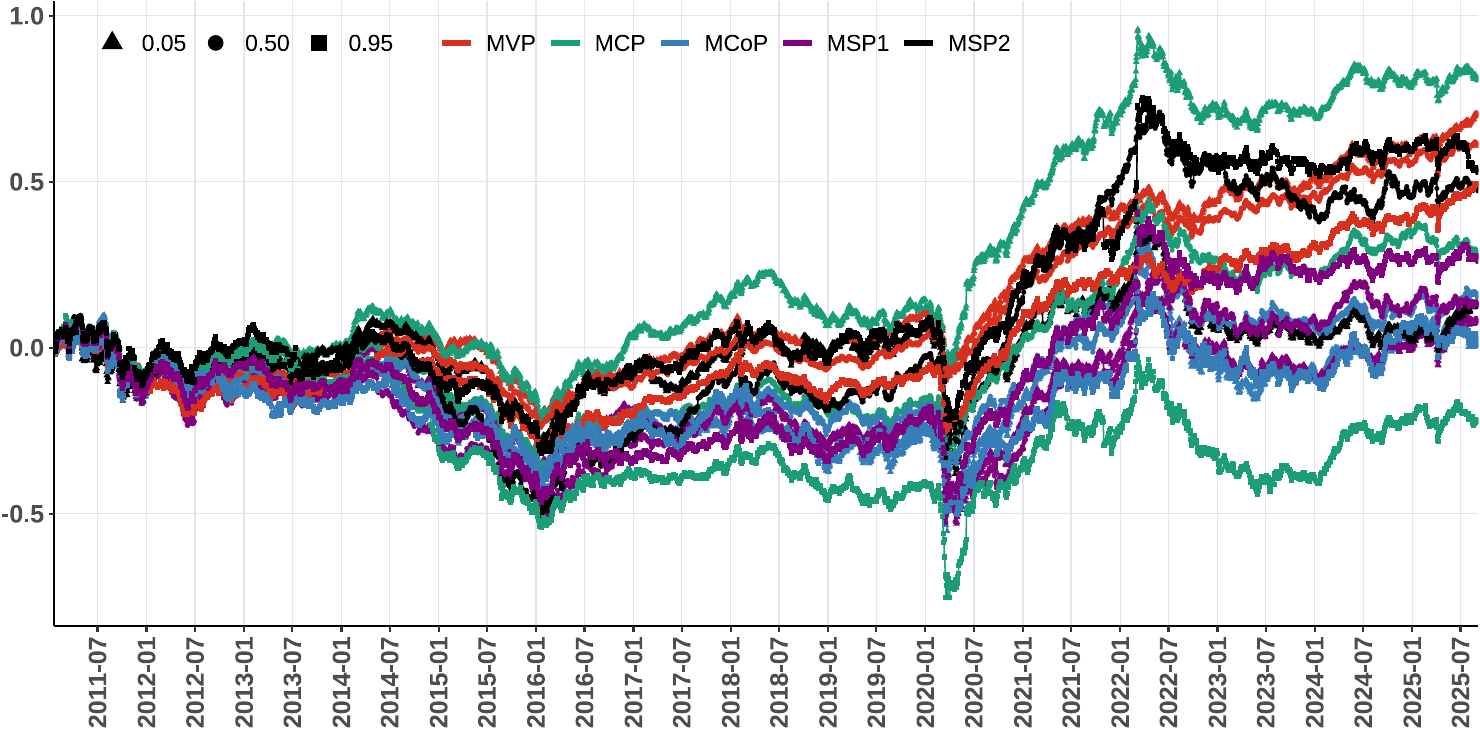}
	\caption{Cumulative returns of the five portfolio strategies across 
		three quantile levels. Colors distinguish strategies (MVP, MCP, MCoP, 
		MSP1, MSP2); marker shapes distinguish $\tau=0.05$ (triangle), 
		$\tau=0.50$ (circle), and $\tau=0.95$ (square).}
	\label{Fig:cumsumR}
\end{figure}

Table~\ref{Tb:portfolio:performance} and Fig.~\ref{Fig:cumsumR} jointly characterize the performance of the five portfolio strategies. MVP delivers consistently positive Sharpe ratios across all three quantile levels, with its cumulative return paths remaining comparatively stable and among the upper-performing trajectories throughout the sample. MCP achieves the highest terminal cumulative return at $\tau=0.05$ and the highest CVaR Sharpe at that quantile, but its performance deteriorates sharply at $\tau=0.95$, where both return and Sharpe ratio turn negative. MCoP underperforms throughout, with cumulative returns near zero or negative across all quantile levels. Looking at the cumulative return paths as a whole, the trajectories of the different strategies are broadly similar in shape: all experience notable declines during the COVID-19 pandemic in early 2020 and again following the onset of the Russia--Ukraine conflict in 2022, before recovering to varying degrees.

A key finding concerns the sensitivity of the motif-based portfolios to the backbone specification. MSP1, based on the more stringent threshold $\alpha=0.05$, yields limited risk-adjusted gains across most quantiles. MSP2, based on the less restrictive $\alpha=0.10$, performs noticeably better: it maintains positive Sharpe ratios across all three quantile levels, and its cumulative return paths lie systematically above those of MSP1, especially after 2020. At $\tau=0.50$ and $\tau=0.95$, MSP2 also outperforms MCP and MCoP in Sharpe ratio. This gap is not driven by isolated episodes but reflects a persistent difference over time, suggesting that the economic value of motif-position information depends critically on how aggressively the backbone network is filtered. The more restrictive specification may leave the structural similarity matrix $\mathbf{H}_t$ too sparse to generate effective portfolio weights, whereas a moderately relaxed threshold preserves richer cross-asset structural information.

From 2020 onward, the cumulative return paths of MSP2 and MVP largely converge, reaching similar terminal values around 0.5, despite relying on entirely different information sources: MVP on the return covariance matrix and MSP2 on the motif-based structural similarity matrix. This convergence suggests that the local network topology captured by $\mathbf{H}_t$ encodes risk-relevant information that partially overlaps with what the covariance matrix provides. It is also notable that the meaningful divergence among strategies occurs during and after these two major tail events, precisely the type of episodes where motif-structural information is most distinctive. This is consistent with the earlier finding that the association between positional diversity and directional connectedness is strongest at the tail quantiles.

\begin{figure}[H]
	\centering
	\includegraphics[width=0.495\linewidth]{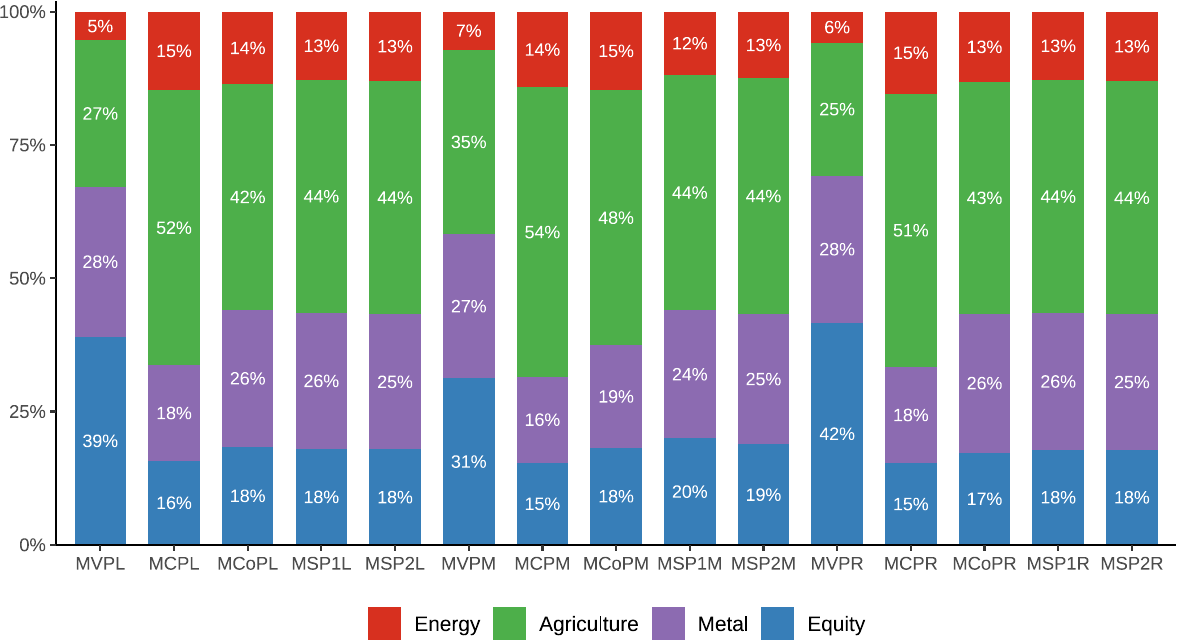}
	\includegraphics[width=0.495\linewidth]{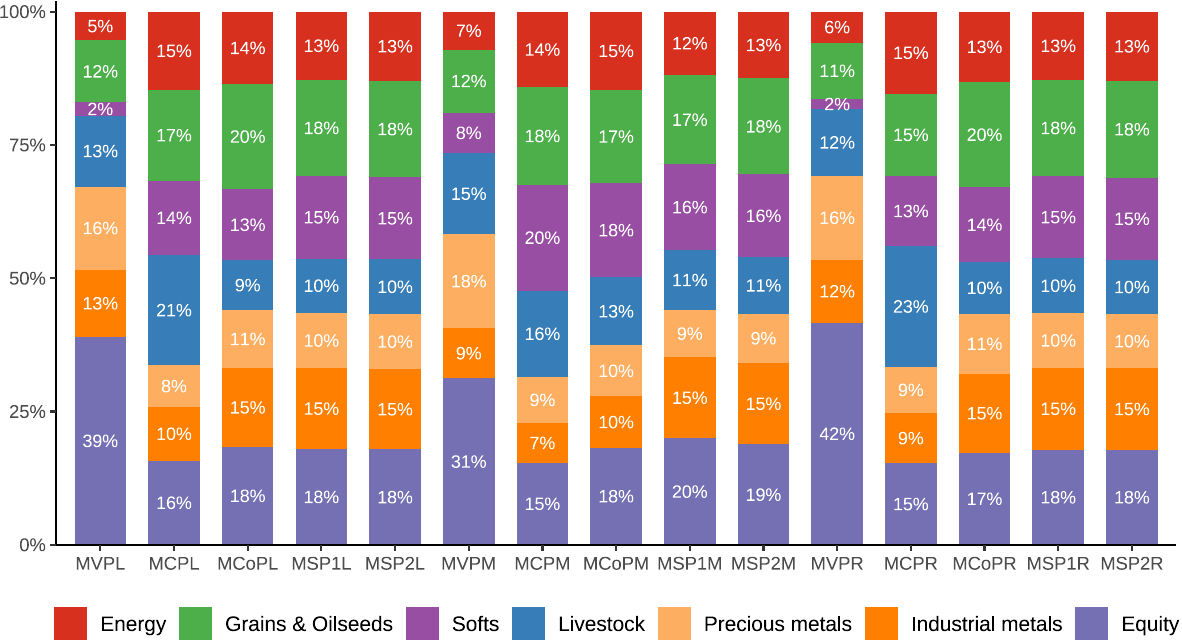}
	\caption{Average portfolio weights aggregated by asset groups. The left 
		panel uses coarse groupings and 
		the right panel uses detailed groupings. 
		Suffixes L, M, and R denote $\tau=0.05$, $0.50$, and $0.95$, 
		respectively. }
	\label{Fig:weights}
\end{figure}

Fig.~\ref{Fig:weights} reveals the allocation structure behind these performance differences. MVP tilts more toward equity and metals, with its equity share especially elevated at the two tail quantiles (39\% and 42\% at $\tau = 0.05$ and $\tau = 0.95$, versus 31\% at $\tau = 0.50$). MCP, in contrast, is consistently dominated by agricultural exposures, which account for roughly half of total capital across all three quantile levels, echoing the agriculture-dominated allocations reported by \cite{FJ-Shahzad-Bouri-Karim-Sadorsky-2025-EnergyEcon} under connectedness-based optimization. MCoP sits between MVP and MCP, with an agriculture share of 42\% to 48\% and an equity allocation in the 17\% to 18\% range. MSP1 and MSP2 show very similar group-level patterns, with agriculture taking the largest share, followed by metals and equity, and energy receiving the smallest allocation. This similarity holds in both the coarse and detailed groupings and is largely stable across quantile levels, indicating that motif-based portfolios exhibit more consistent 
sectoral composition across quantile regimes than MVP.

The near-identical group-level weights of MSP1 and MSP2 are particularly informative given their differing performance outcomes. This pattern implies that the advantage of MSP2 does not stem from a fundamentally different sectoral tilt but rather from finer within-group weight differences, consistent with the interpretation that the less restrictive backbone ($\alpha = 0.10$) preserves more granular structural information within the similarity matrix $\mathbf{H}_t$. The resulting asset-level reweighting inside each sector translates into the performance gains documented in Table~\ref{Tb:portfolio:performance}, even when the aggregate sectoral exposure is held almost constant.

\section{Conclusion}
\label{S7:conclusion}

Connectedness measures have become a standard tool for quantifying risk
spillovers, yet the structural information embedded in the resulting
networks has been exploited almost exclusively at the level of
individual nodes or system-wide aggregates. This paper develops an
integrated framework that joins multiscale backbone extraction,
directed triadic motif significance testing, a colored motif taxonomy
encoding asset heterogeneity, and orbit-informed portfolio
construction. By shifting the analytical unit from individual nodes to
triadic subgraphs and the positions assets occupy within them, the
framework recovers mesoscopic structural information that node-level
or aggregate connectedness statistics cannot isolate.

The empirical application to 39 commodity and equity futures from 2010
to 2025 reveals a pervasive duality between normal and extreme market
states that manifests at every layer of analysis. At the aggregate
level, the total connectedness index traces a pronounced U-shaped
profile across conditional quantiles, reaching near-saturation at both
tails while remaining moderate and crisis-sensitive at the median. Cross-sector transmission consistently
dominates within-sector transmission across the full quantile
spectrum, and the exclusive commodity network operates at a higher
baseline than the equity-commodity channel, so equity futures act as a
meaningful but secondary risk bridge rather than as the backbone of
the system.

The structural picture inverts when we move from aggregate to local.
Under normal conditions, the filtered backbone is sparse, persistent,
and organized into sector-specific cores, and motif testing identifies
reciprocal triads as the statistically preferred building blocks, with
motifs 238, 110, 78, and 46 all strongly over-represented and purely
unidirectional configurations systematically suppressed. Under extreme
conditions, the same filtering yields structurally unstable and
largely disconnected backbones, and motif z-scores compress toward
zero despite very high raw triad counts. High aggregate connectedness and stable local architecture are thus not only distinct but inversely related: the former characterizes tail states, whereas the latter is a property of normal market conditions.

The colored motif analysis sharpens this contrast by adding a sectoral
dimension. At the median, the most strongly over-represented
configurations are overwhelmingly monochromatic reciprocal triads, led
by the fully reciprocal triangle within precious metals, followed
closely by its industrial-metals, equity, and grains-and-oilseeds
counterparts. Reciprocal motif dominance is therefore not a
system-wide property but a sector-specific one, concentrated in a
handful of tightly interconnected clusters. Cross-sector transmission
operates mainly through asymmetric configurations in which a
reciprocal within-sector core broadcasts risk unidirectionally to
nodes in other sectors, as seen in the significantly over-represented
motif 14 variants that span metal, equity, and
agricultural color combinations. At both tails, this layered
architecture dissolves. Even when specific colored motifs appear
frequently in raw counts, they are no longer statistically
distinguishable from the degree-preserving null, and sectoral identity
loses its organizing role in local transmission.

Orbit-position diversity behaves as a tail-specific structural marker.
At both tails, cross-sectional correlations show that assets with more
even orbit participation are predominantly net transmitters rather
than receivers of spillovers, and the pattern persists day by day
rather than being an artifact of time averaging. Near the median this
association attenuates and becomes regime-dependent. Positional
diversity thus identifies assets whose embedding in many local roles
coincides with an outsized role in propagating risk, and does so
precisely in the states where node-level directional indices are least
informative. Building on this structural signal, the minimum
structural similarity portfolio constructed from the motif-based
similarity matrix delivers positive risk-adjusted returns across all
three quantiles under a moderately relaxed backbone threshold, and
outperforms both minimum correlation and minimum connectedness
benchmarks at the median and at the upper tail. The sensitivity to
the filtering level is itself informative. Since the two MSP variants
carry nearly identical group-level weights, the performance gap stems
from finer within-sector reweighting rather than from a different
sectoral tilt, showing that the portfolio-relevant signal operates at
a granular within-sector level that aggregate connectedness measures
do not capture. The cumulative path of MSP2 also converges toward that
of the minimum variance portfolio from 2020 onward despite the two
strategies drawing on entirely different information sources, which
suggests that orbit-based structural similarity captures risk-relevant
information that partially overlaps with but does not duplicate what
return covariances convey.

Taken together, these results argue for treating risk transmission as
a topological object whose local organization carries information
beyond that contained in pairwise or aggregate network statistics. The
contrast between a stable sector-cored reciprocal architecture under
normal conditions and a saturated but structurally unstable
architecture under extremes has direct implications for systemic risk
monitoring. Periods of very high aggregate connectedness should not be
read as periods of reliable structural signal, because the same forces
that push spillovers toward saturation also erode the stability of
local building blocks. Sectoral identity shapes the local organization
of risk transmission in economically interpretable ways, with some
sectors forming tightly integrated reciprocal cores that then project
shocks across sector boundaries. For portfolio management, the
evidence supports using motif-based structural similarity as a
complement to covariance and connectedness inputs, especially for
diversifying away from sector-cored reciprocal clusters that would
otherwise go undetected.

Several limitations and extensions deserve mention. The colored motif
taxonomy treats sector labels as fixed throughout the sample, whereas
financialization dynamics and shifting commodity-equity linkages may
cause effective sector boundaries to evolve over time, and replacing
these labels with data-driven partitions derived from community
detection on the backbone itself would relax this assumption. The
current framework also assesses motif significance on a day-by-day
basis without modeling how specific triadic configurations emerge,
persist, or dissolve across consecutive trading sessions. Moving from
triadic to higher-order or temporal motifs could enrich the picture
of how risk propagates dynamically and reveal whether the appearance
or disappearance of particular subgraph patterns serves as an
early-warning signal of regime transitions between the structured
median state and the diffuse tail state documented here. Examining
whether these motif-level signals forecast regime shifts or systemic
stress episodes out of sample represents a natural next step.
Incorporating macroeconomic or policy variables as additional nodes
would allow the framework to capture interactions between
cross-market spillovers and broader economic conditions. Another promising extension is to link the motif-based approach with higher-order moment risk analysis and investment optimization, thereby examining whether local subgraph structures contain information about skewness, kurtosis, downside risk, and portfolio allocation under non-normal return distributions.
Finally,
applying the framework to sovereign credit default swap networks,
cryptocurrency markets, or cross-border banking exposures would
clarify how general the divergence between local architecture and
aggregate connectedness actually is.

\appendix
\newpage
\section{Risk spillovers and network backbones at the tail quantiles}

\setcounter{figure}{0}
\setcounter{table}{0}

\begin{figure}[H]
	\centering
	\includegraphics[width=0.49\textwidth]{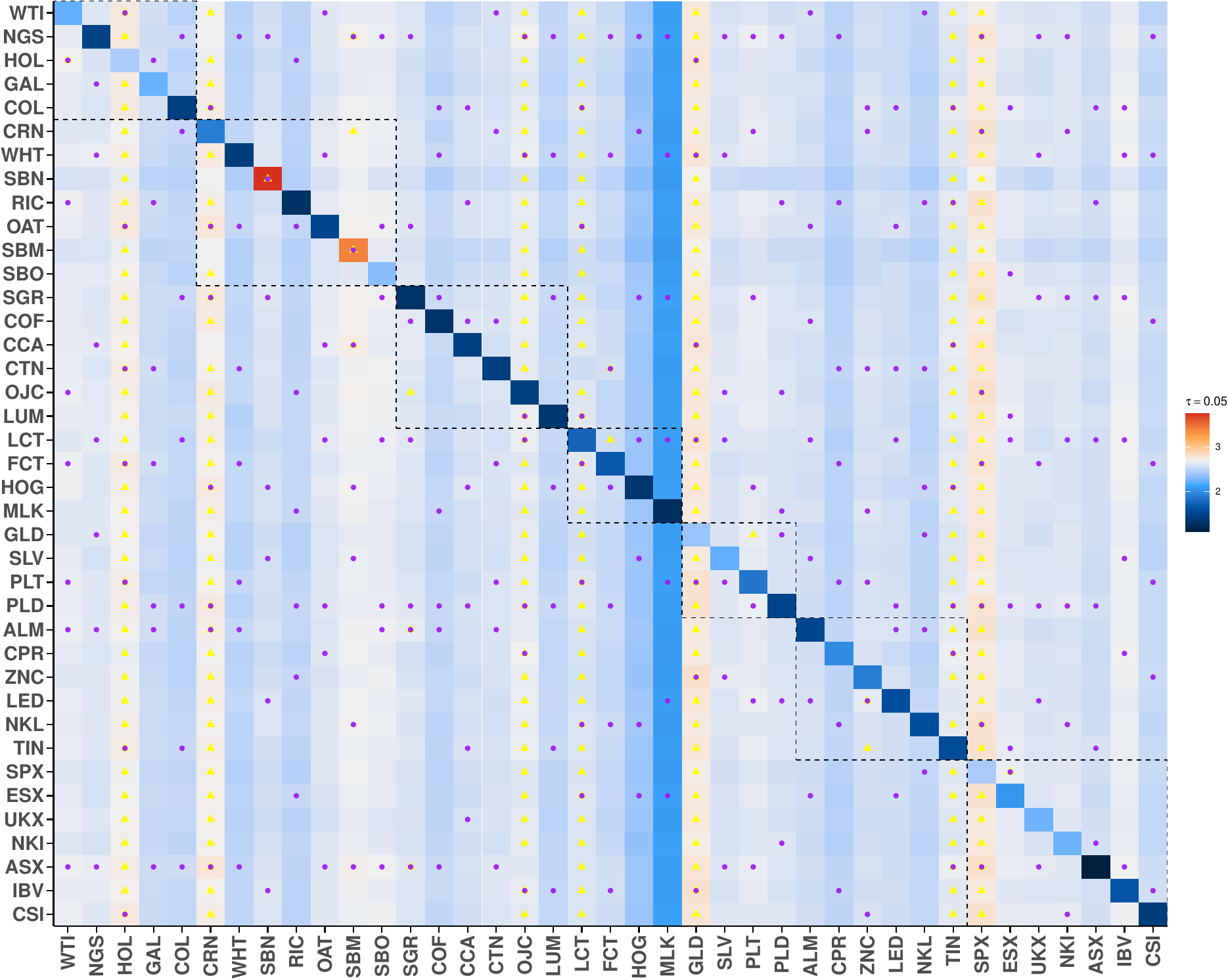}
	\includegraphics[width=0.49\textwidth]{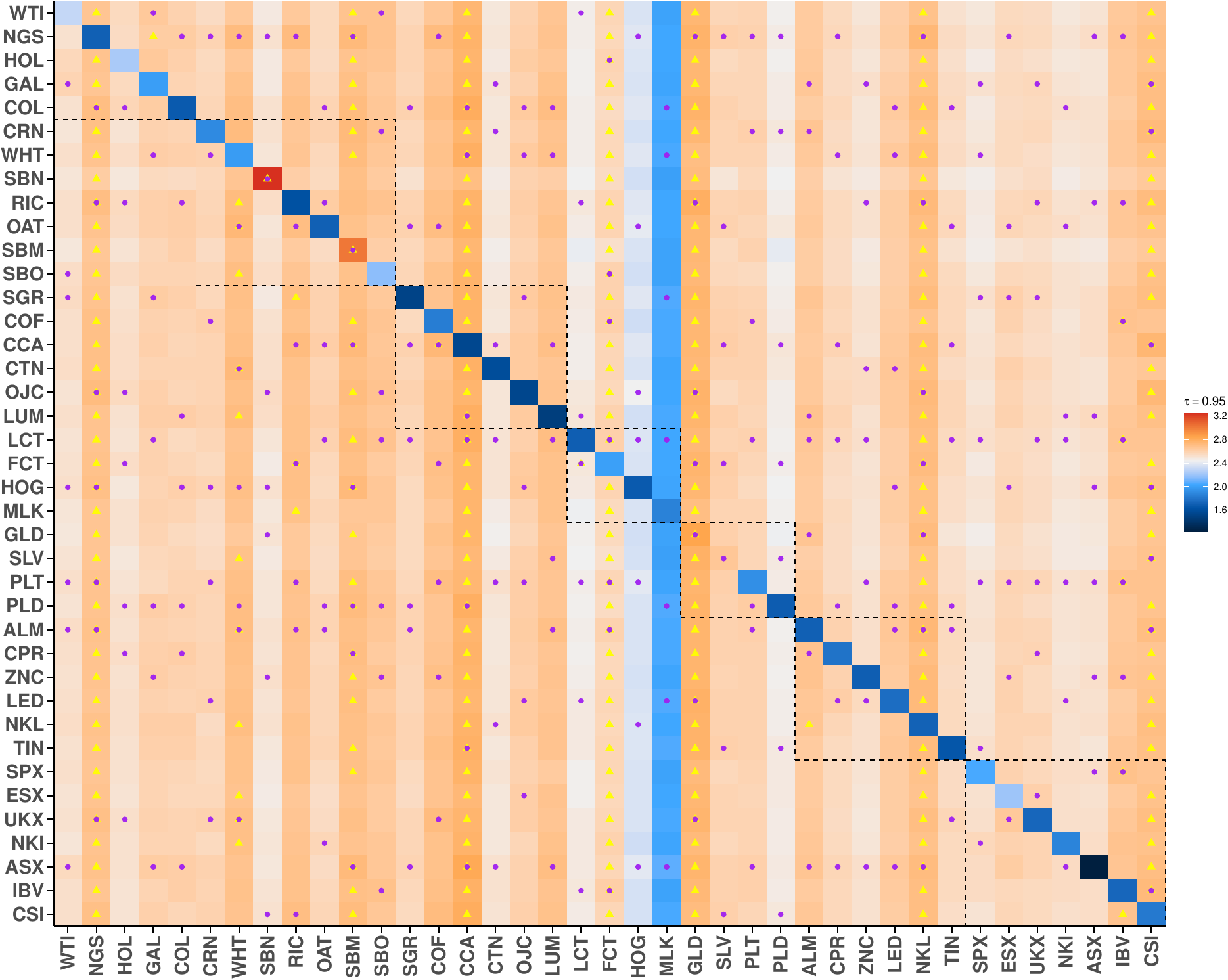}
	\caption{Averaged dynamic risk spillover heatmaps at the tail 
		quantiles. Left panel: $\tau=0.05$. Right panel: $\tau=0.95$. 
		Markers follow the same convention as Fig.~\ref{Fig:Static:Connectedness:M}(a).}
	\label{Fig:Static:Connectedness:LR}
\end{figure}

\begin{figure}[H]
	\centering
 	\includegraphics[width=0.49\textwidth]{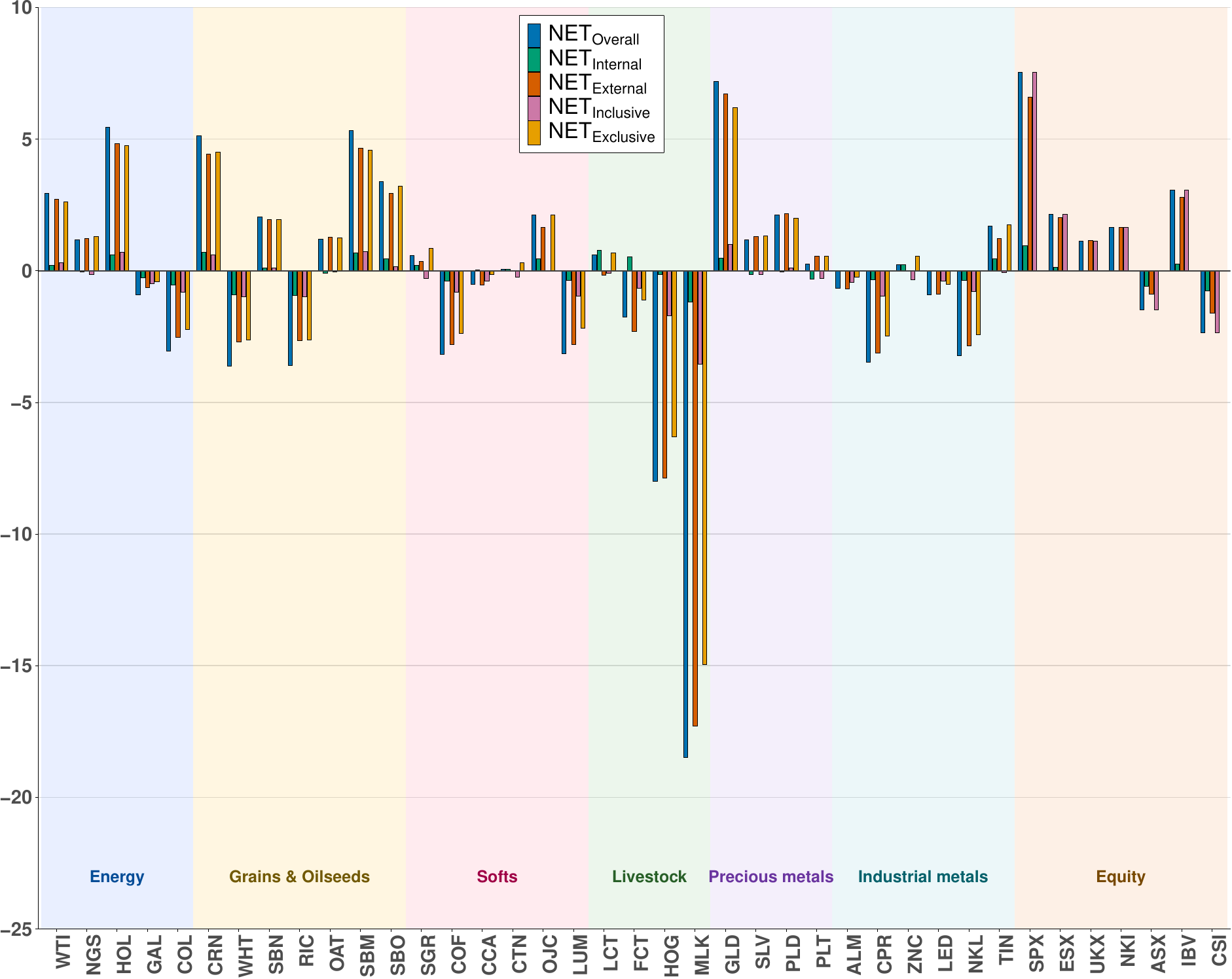}
	\includegraphics[width=0.49\textwidth]{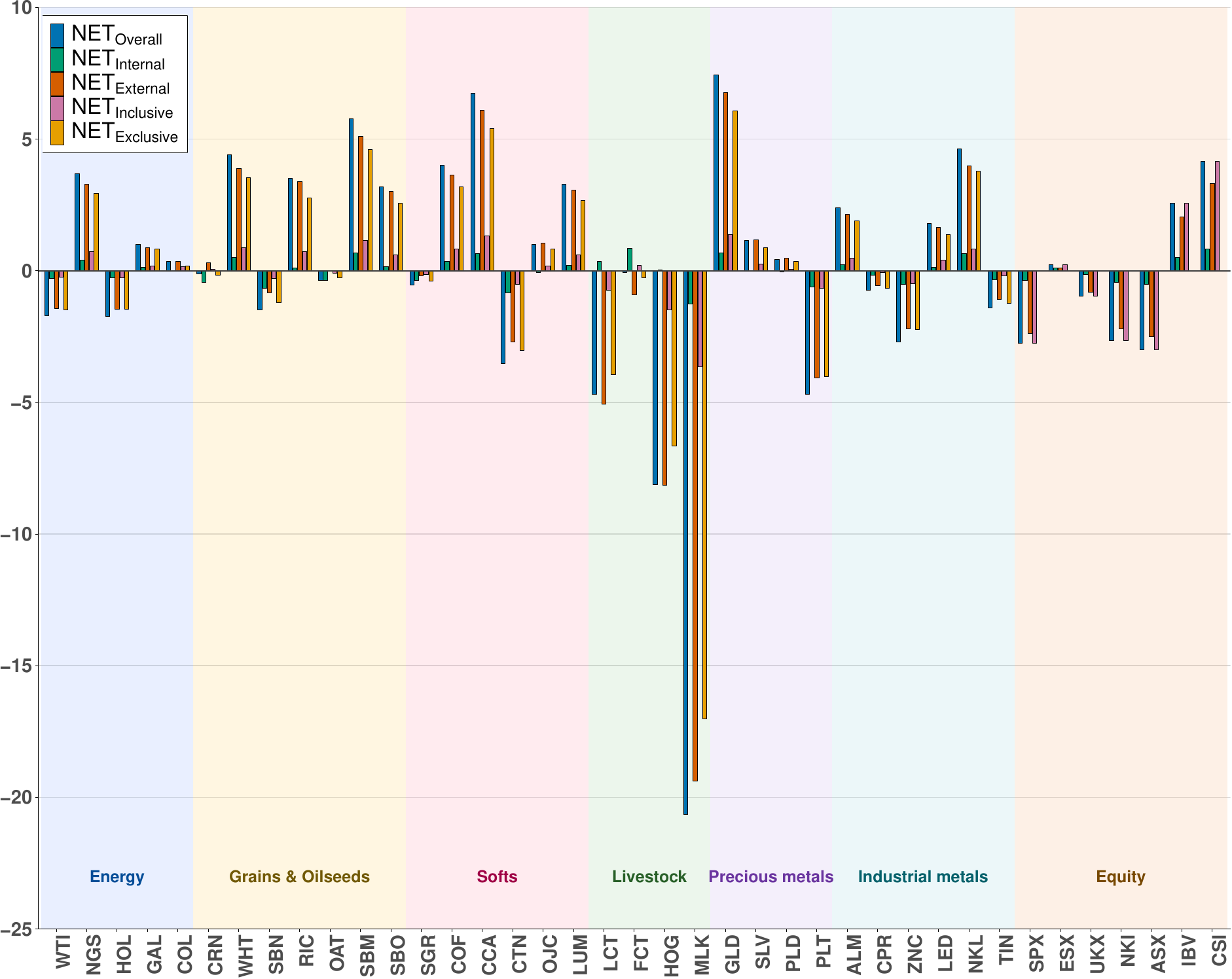}
	\caption{Net spillovers at the tail quantiles, decomposed into 
		overall, internal, external, inclusive, and exclusive components. 
		Left panel: $\tau=0.05$. Right panel: $\tau=0.95$.}
	\label{Fig:Static:Connectedness:int:ext:NET:LR}
\end{figure}

\begin{figure}[H]
	\centering
	\includegraphics[width=8cm]{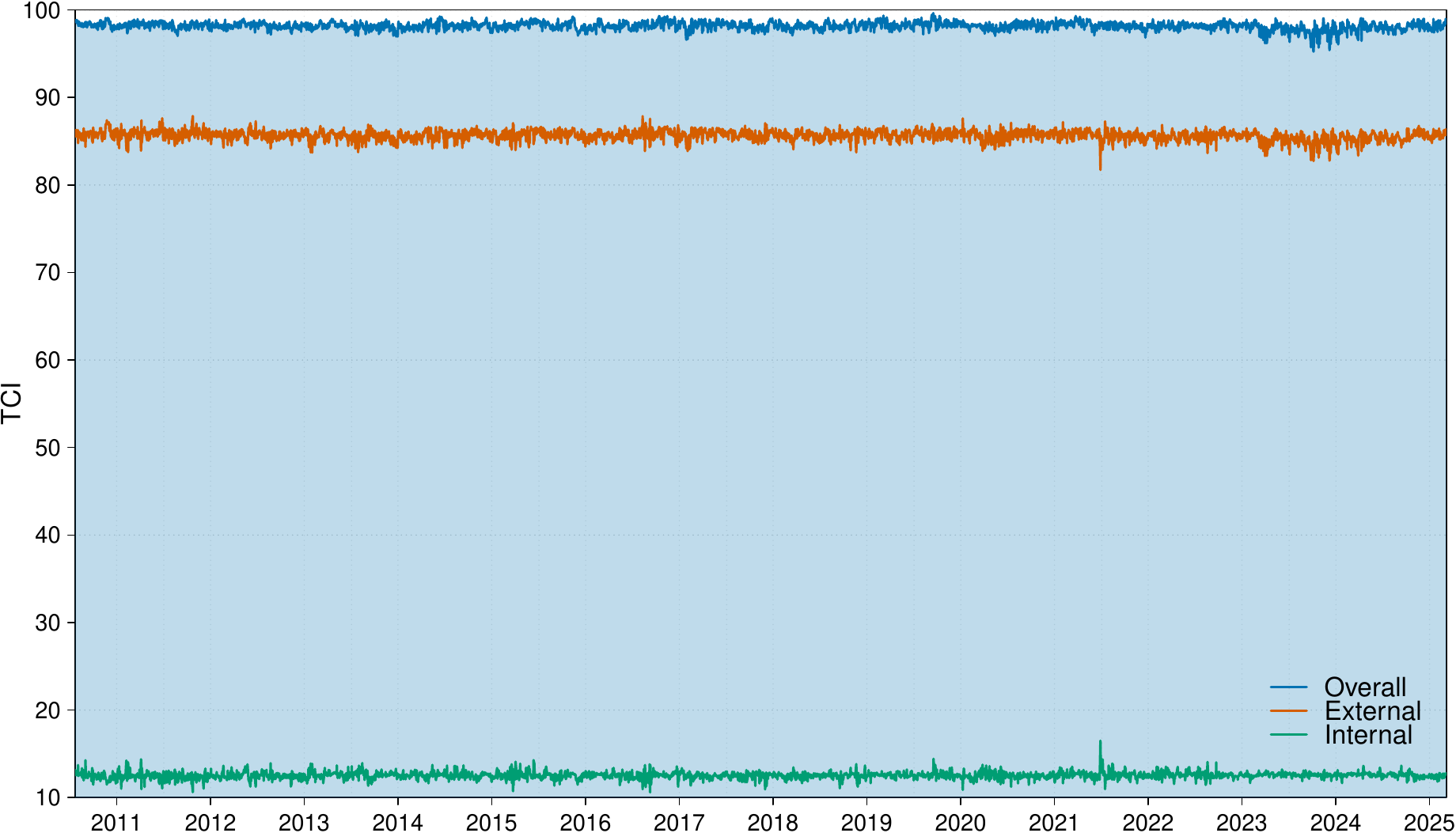}
	\includegraphics[width=8cm]{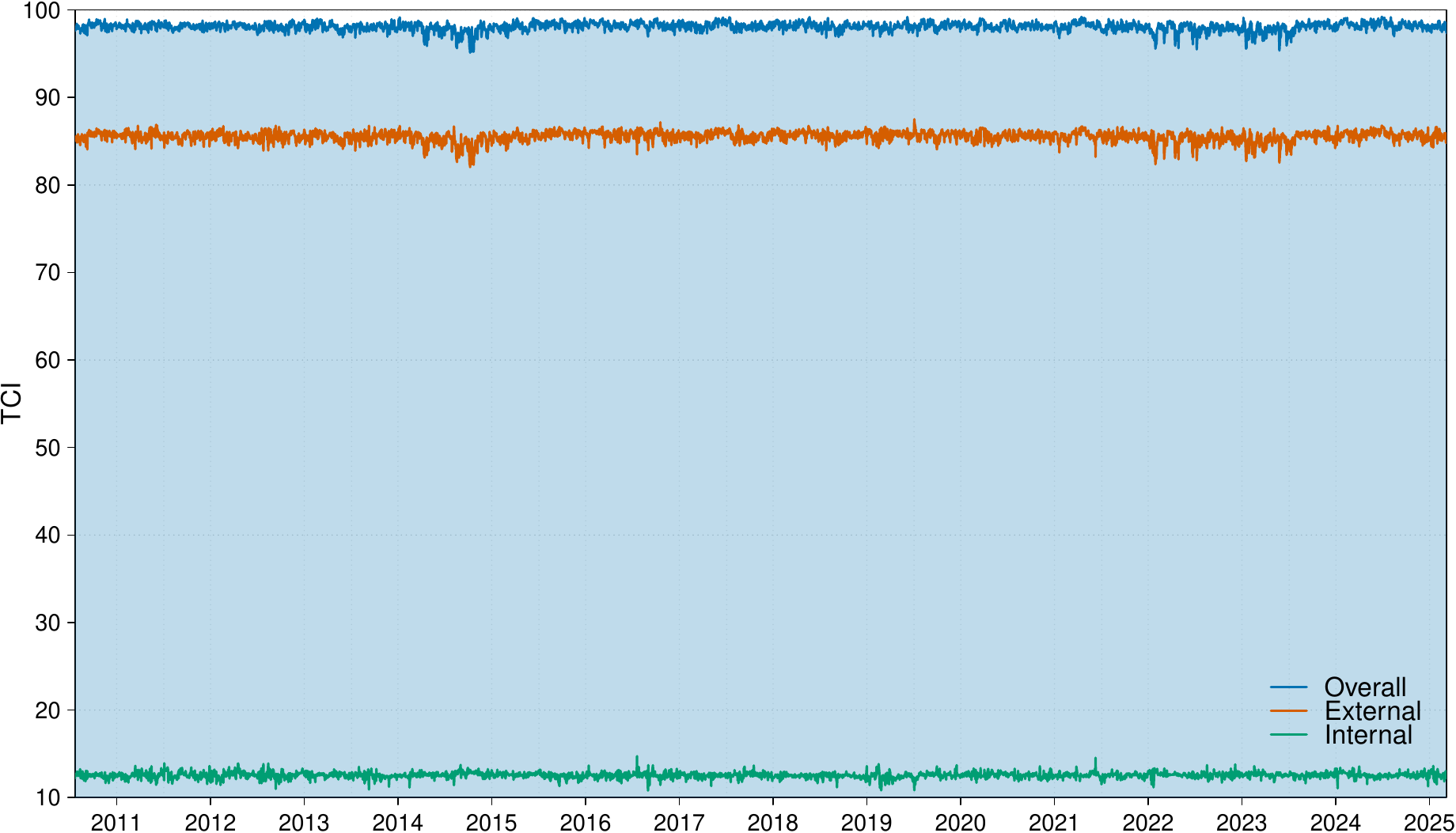}
	\includegraphics[width=8cm]{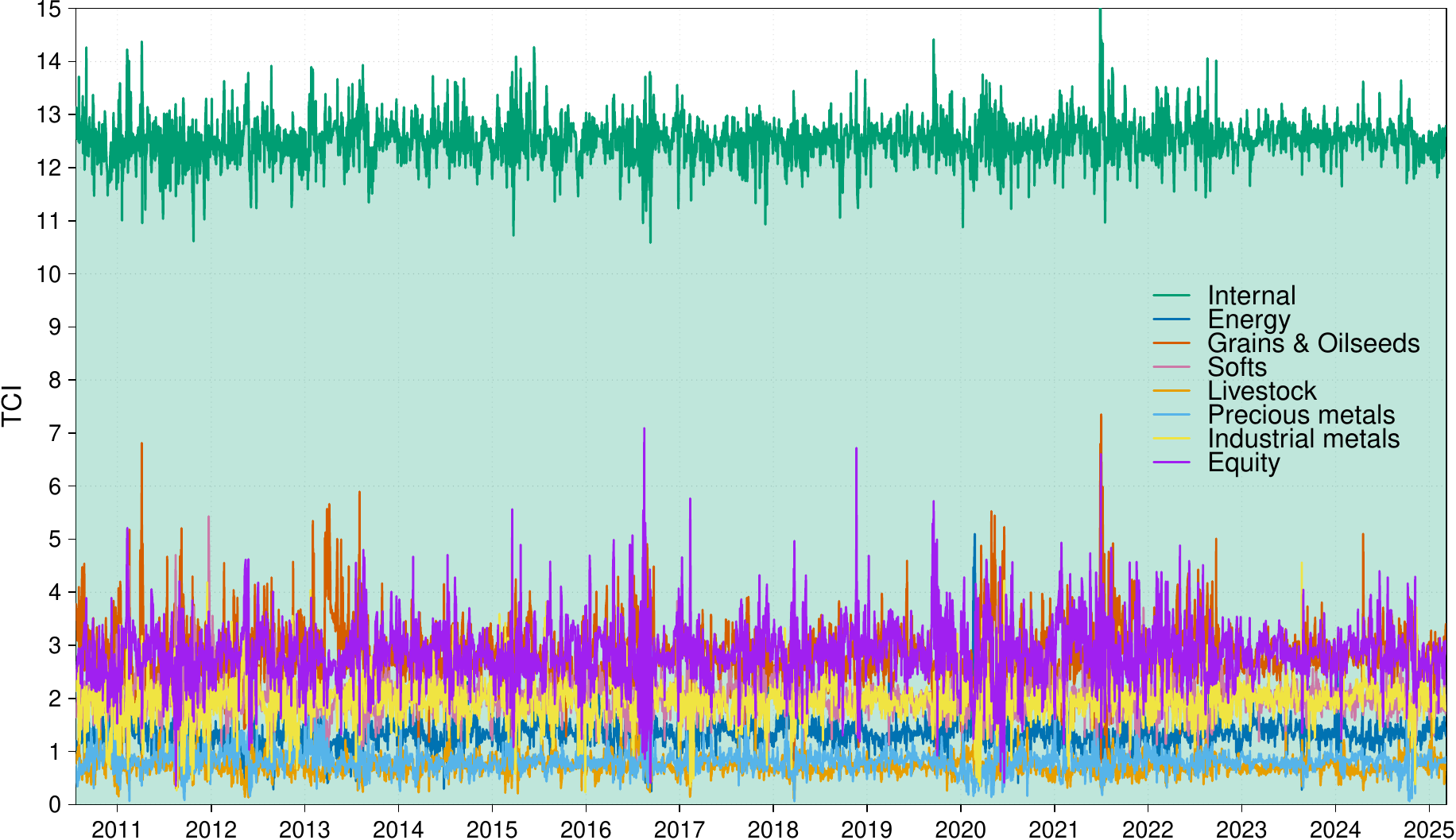}
	\includegraphics[width=8cm]{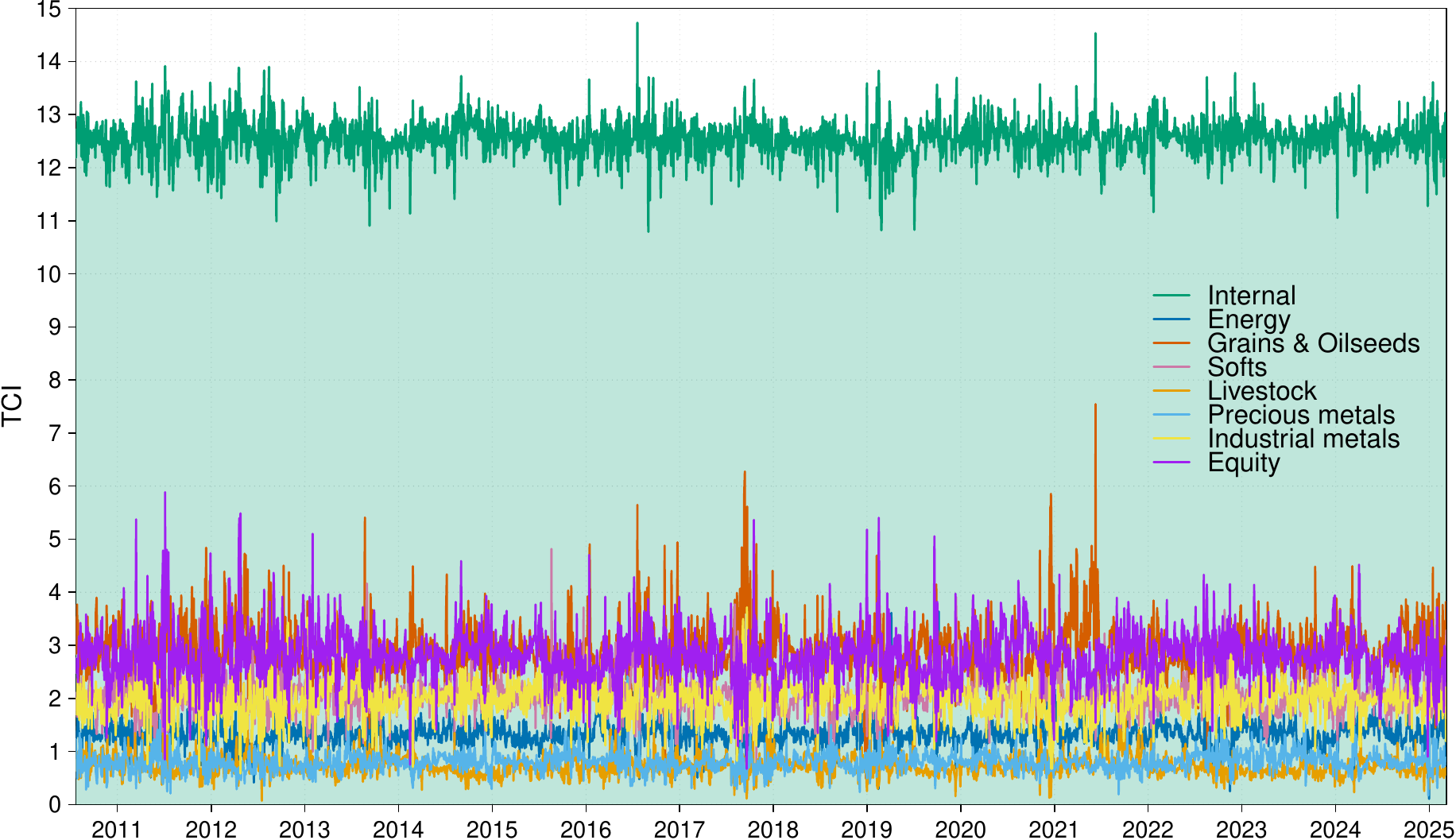}
	\includegraphics[width=8cm]{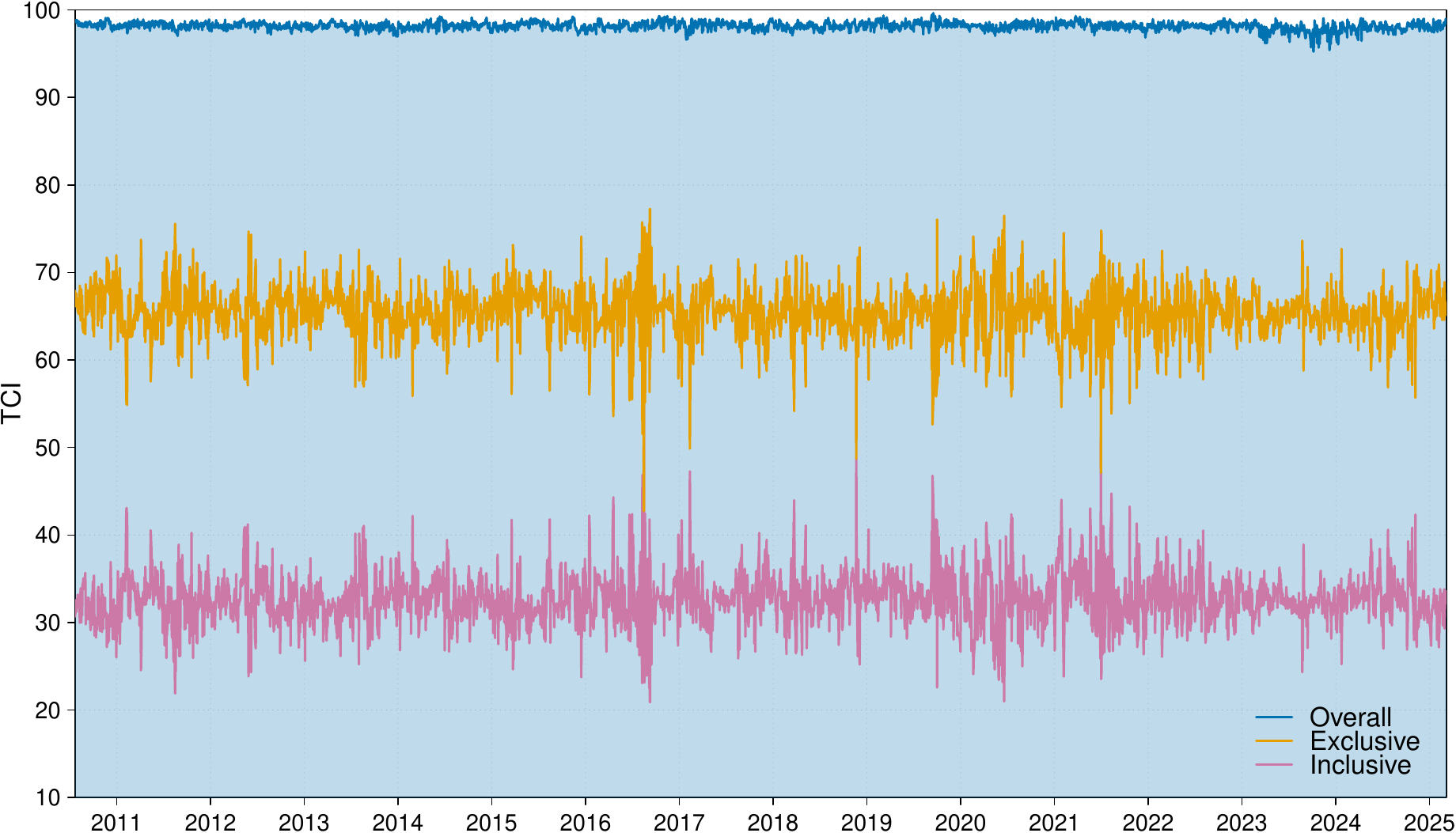}
	\includegraphics[width=8cm]{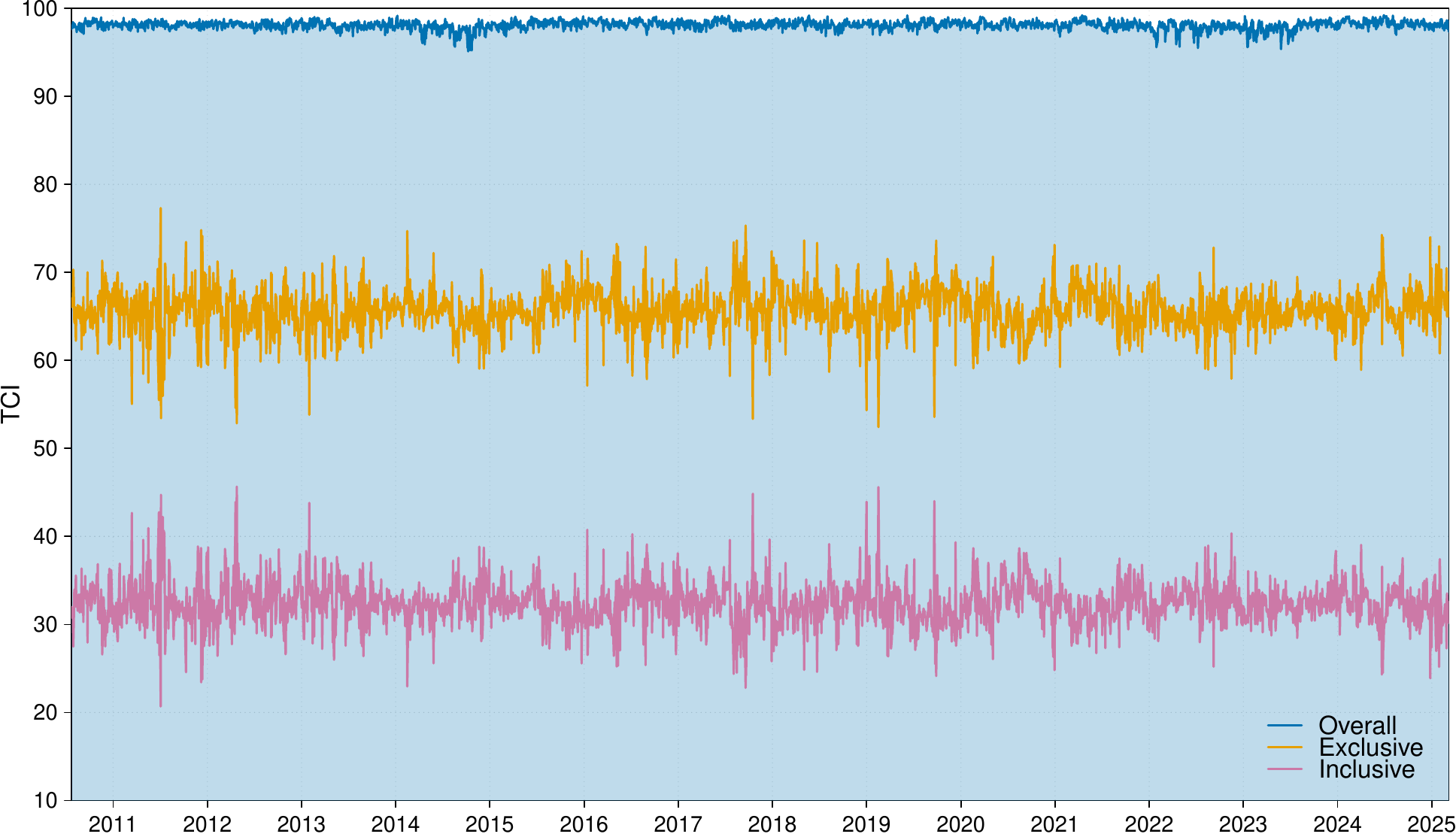}
	\caption{Time-varying TCI at the tail quantiles. 
		Left column: $\tau=0.05$. Right column: $\tau=0.95$. 
		Top row: overall, external, and internal. 
		Middle row: sector-level decomposition of internal connectedness. 
		Bottom row: overall, exclusive, and inclusive.}
	\label{Fig:Dynamic:TCI:LR}
\end{figure}

\begin{figure}[H]
	\centering
	\begin{overpic}[width=0.45\textwidth]{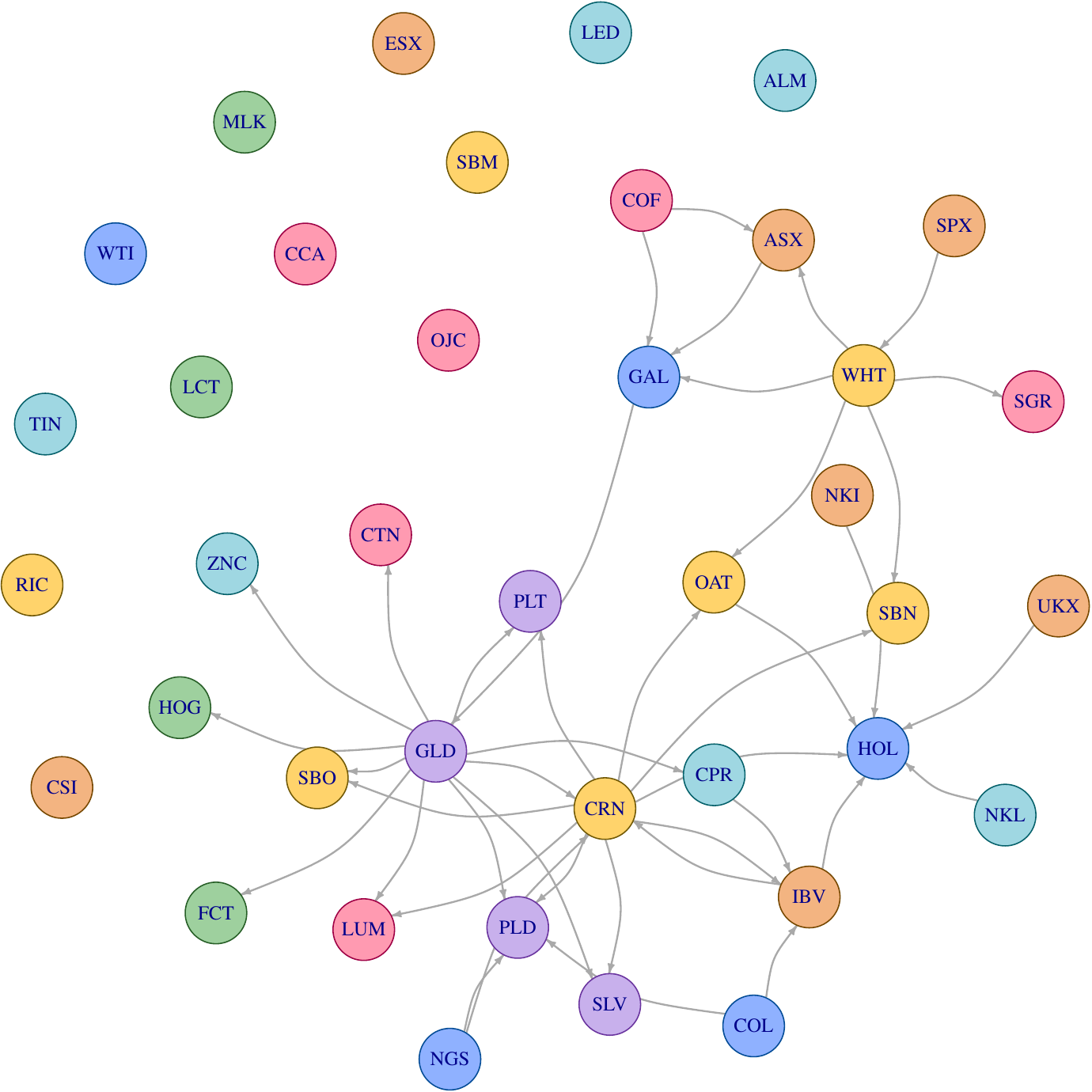}
		\put(1,90.5){\small\bfseries (a)} 
	\end{overpic}
	\hfill
	\begin{overpic}[width=0.45\textwidth]{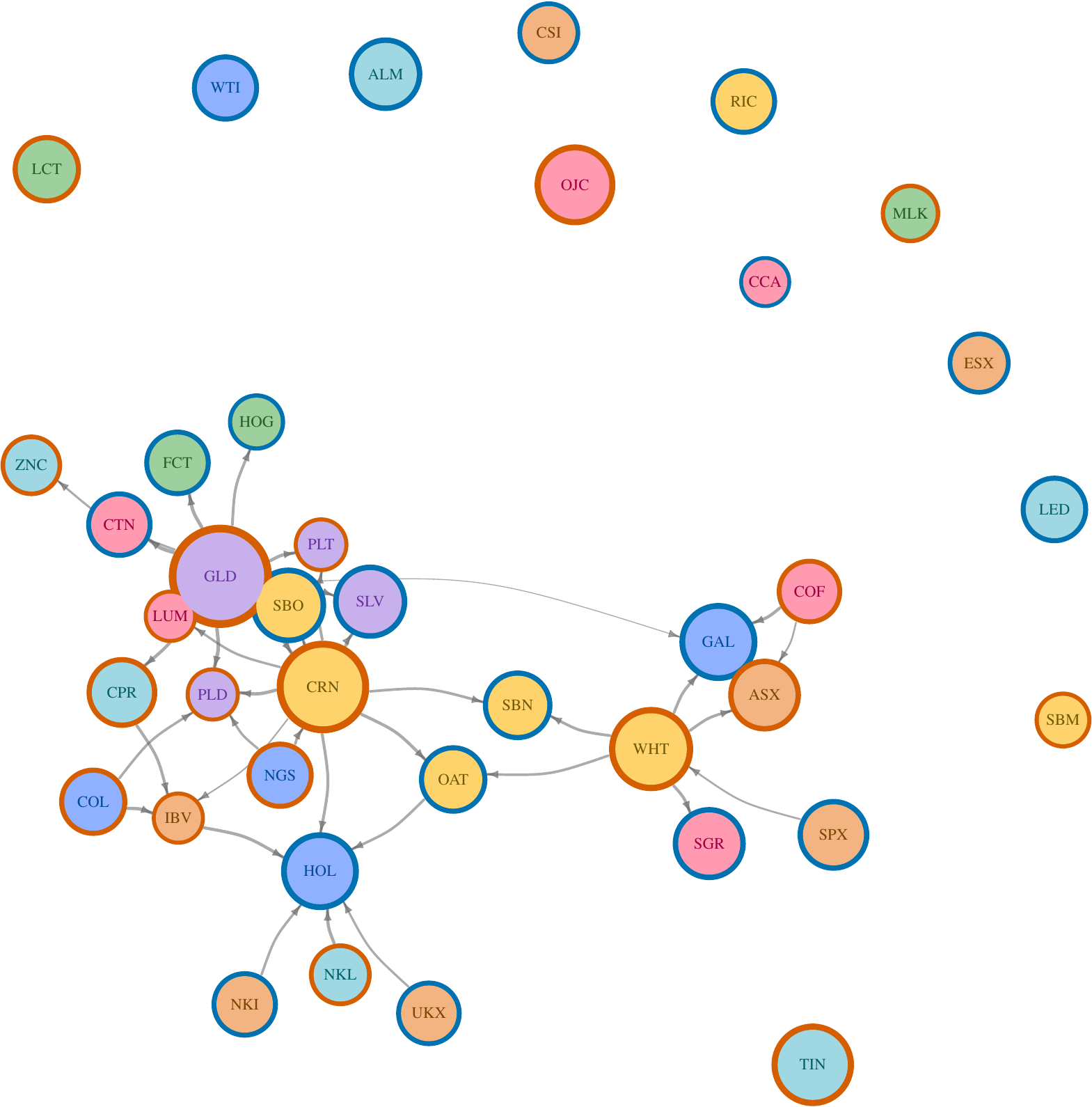}
		\put(1,90.5){\small\bfseries (b)}
	\end{overpic}
	\begin{overpic}[width=0.45\textwidth]{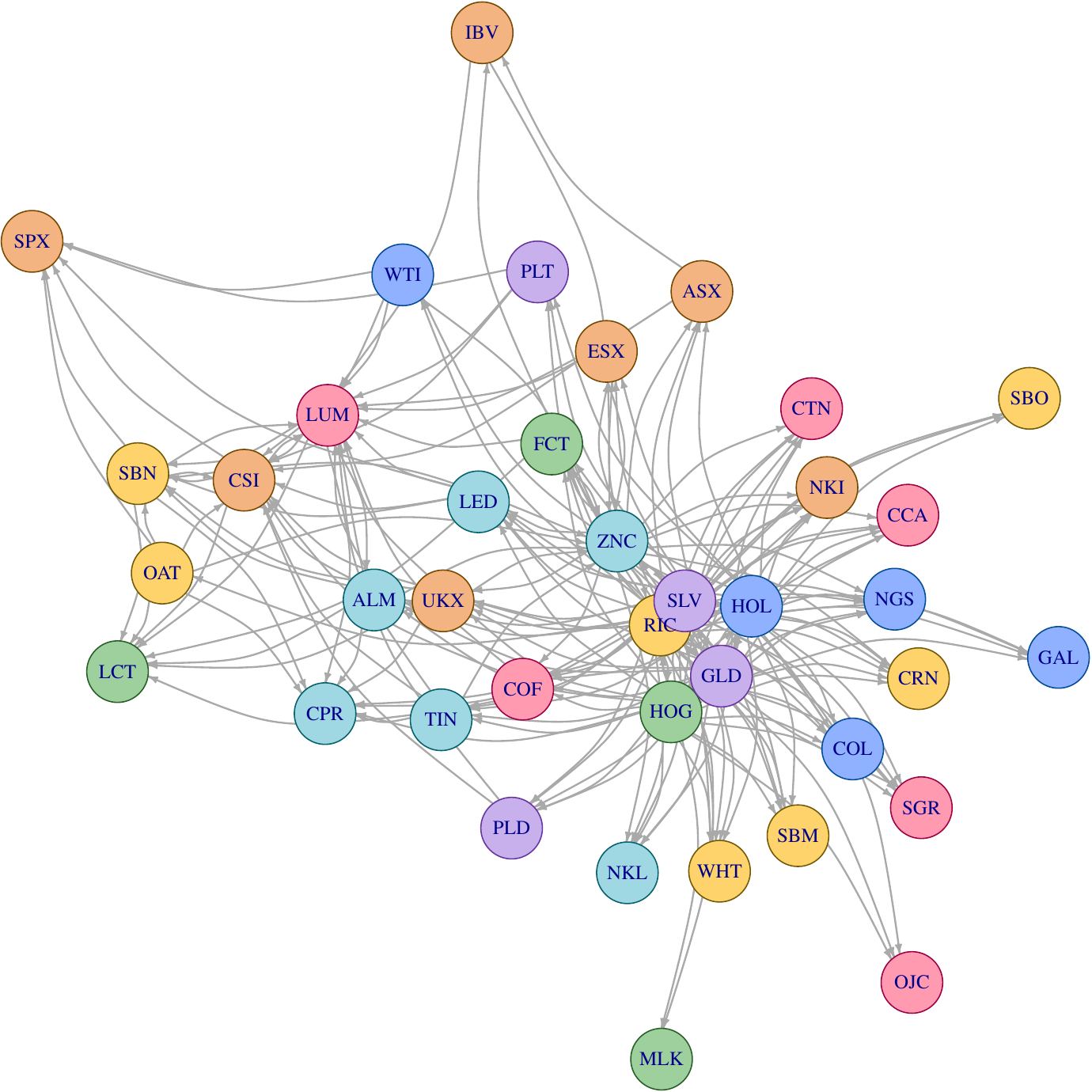}
		\put(1,90.5){\small\bfseries (c)} 
	\end{overpic}
	\hfill
	\begin{overpic}[width=0.45\textwidth]{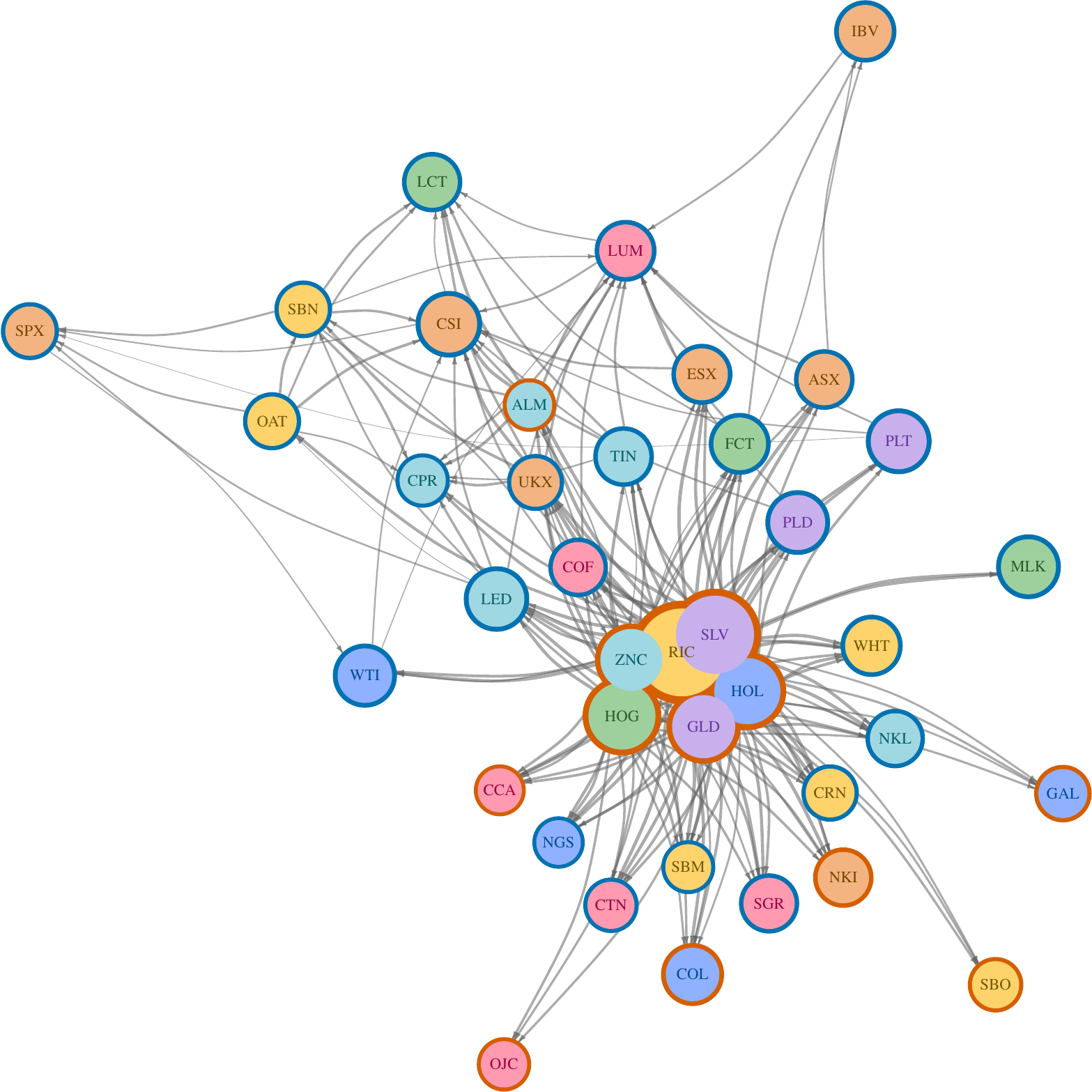}
		\put(1,90.5){\small\bfseries (d)}
	\end{overpic}
	\caption{Network backbones of a representative daily connectedness structure at extreme quantiles ($\alpha = 0.1$). Panels (a) and (b) correspond to $\tau=0.05$, while panels (c) and (d) correspond to $\tau=0.95$. Regarding the topology, panels (a) and (c) display the CT-based topology, whereas panels (b) and (d) display the NPDC-based topology. In the NPDC plots, outer rings denote net transmitters (orange) and net receivers (blue).}
	\label{Fig:CT:NPDC:Backbone:20120120:LR}
\end{figure}

\begin{figure}[H]
  \centering
  \begin{overpic}[width=0.45\textwidth]{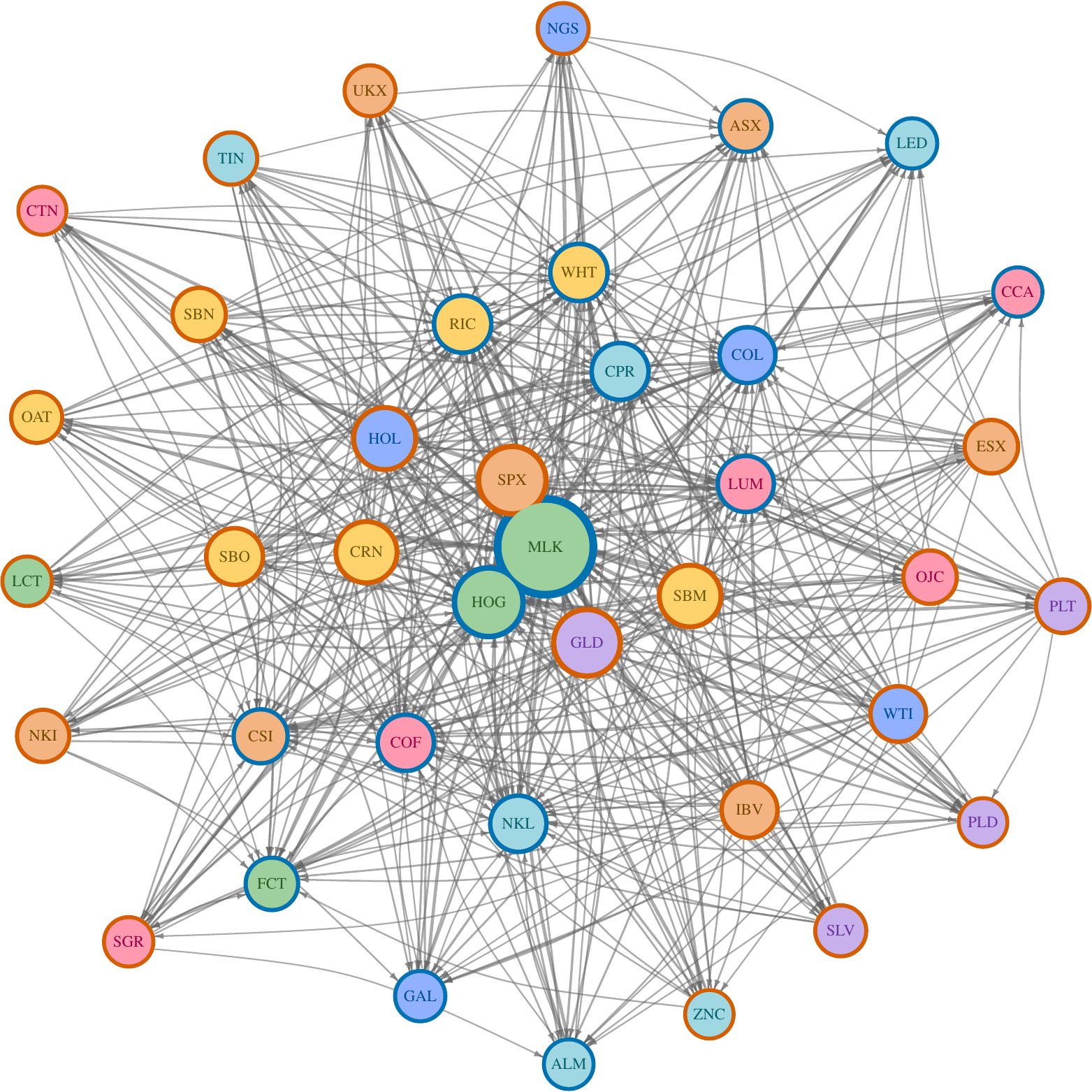}
    \put(1,90.5){\small\bfseries (a)} 
  \end{overpic}
  \hfill
  \begin{overpic}[width=0.45\textwidth]{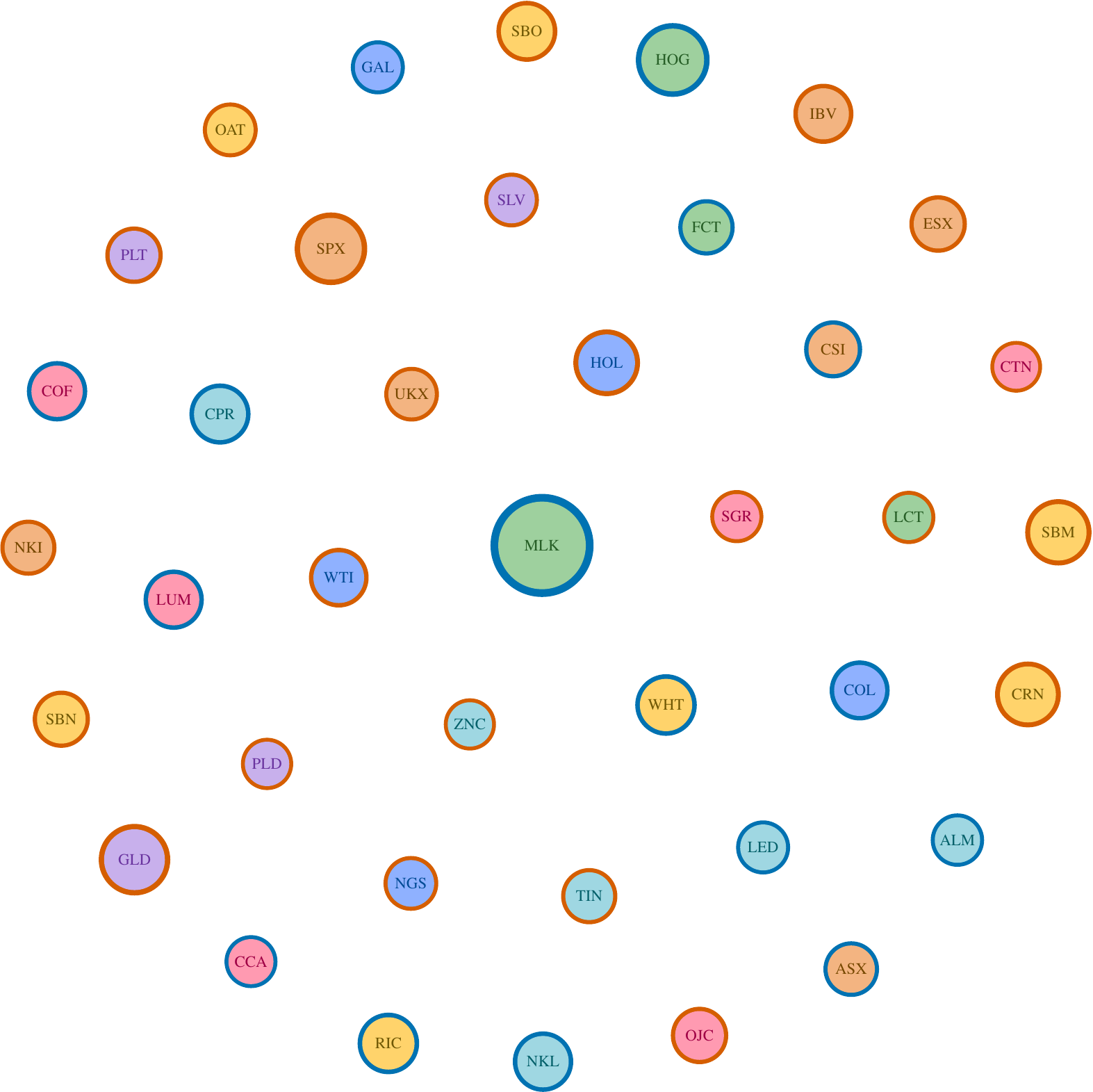}
    \put(1,90.5){\small\bfseries (b)}
  \end{overpic}
   \hfill
  \begin{overpic}[width=0.45\textwidth]{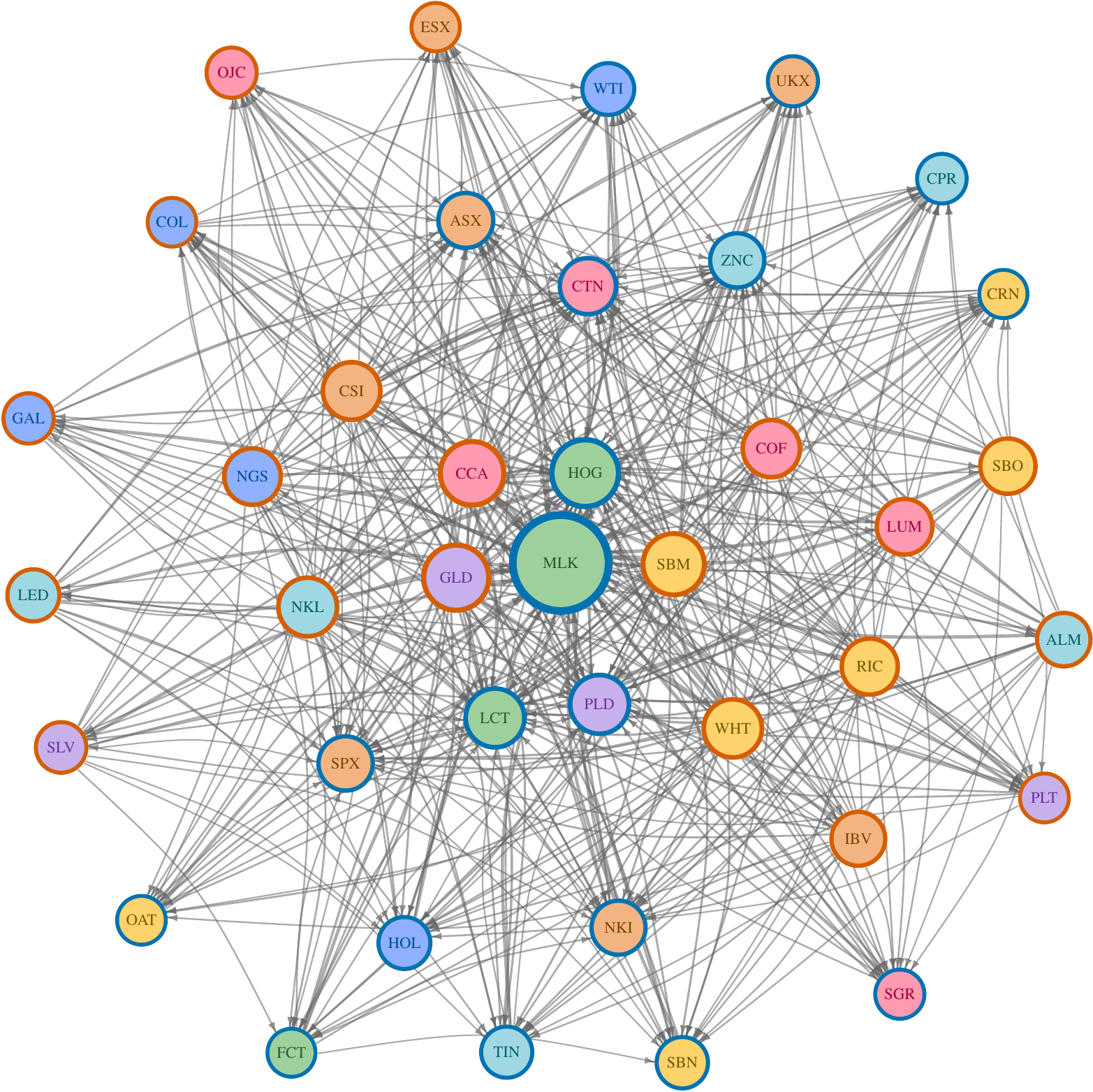}
    \put(1,90.5){\small\bfseries (c)}
  \end{overpic}
   \hfill
  \begin{overpic}[width=0.45\textwidth]{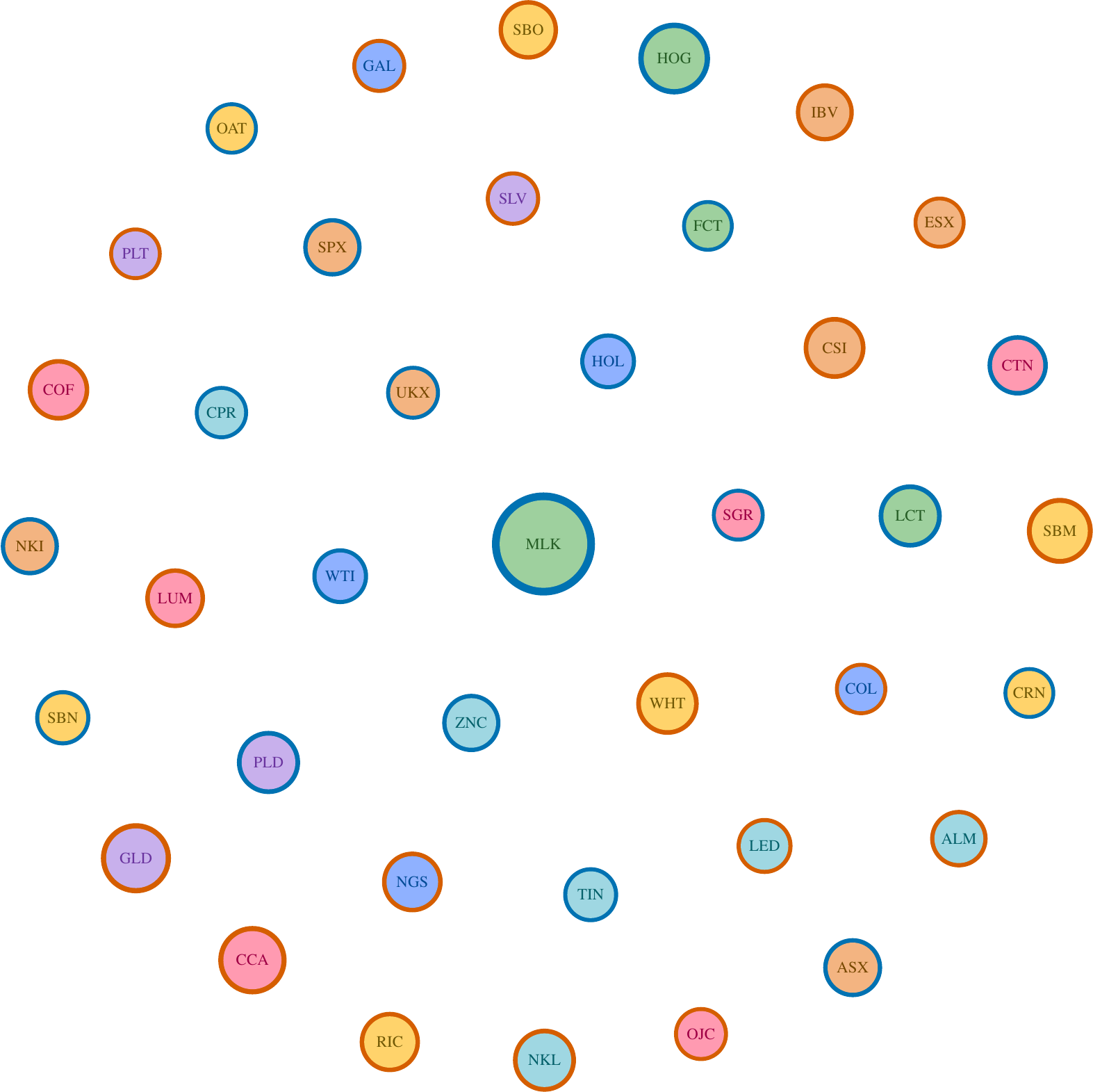}
    \put(1,90.5){\small\bfseries (d)}
  \end{overpic}
  \caption{Visualization of overall spillover risk networks.
Panels (a) and (b) display the networks at $\tau=0.05$, while panels (c) and (d) display them at $\tau=0.95$. Regarding the construction method, (a) and (c) employ the threshold-based method (cutoff = 0.1), whereas (b) and (d) employ the backbone-based method ($\alpha = 0.1$). Outer rings distinguish net transmitters (orange) from net receivers (blue). Node size scales with net total directional connectedness; edge thickness represents pairwise spillover strength.}
  \label{Fig:NPDC:Backbone:Network:LR}
\end{figure}

\newpage
\section{Directed triad motif statistics at the tail quantiles}

\begin{table}[H]
	\centering
	\setlength{\abovecaptionskip}{0pt}
	\setlength{\belowcaptionskip}{10pt}
	\caption{Daily directed motif statistics at the tail quantiles.}
	\label{Tab:motif:LR}
	\resizebox{\textwidth}{!}{
\begin{tabular}{llrrrrr|llrrrrr}
			\toprule
			\multicolumn{7}{c|}{$\alpha=0.05$} & \multicolumn{7}{c}{$\alpha=0.10$}\\
			\cmidrule(lr){1-7} \cmidrule(lr){8-14}
			id & motif & $\mu$ & $\sigma$ & $\mu_{\text{rnd}}$ & $\sigma_{\text{rnd}}$ & $z$ &
			id & motif & $\mu$ & $\sigma$ & $\mu_{\text{rnd}}$ & $\sigma_{\text{rnd}}$ & $z$\\
			\midrule
			\multicolumn{14}{l}{\textit{left tail:} $\tau=0.05$}\\
			36 & \motif{36} & 10.214 & 34.976 & 7.676 & 1.931 & 1.315 &
			36 & \motif{36} & 25.621 & 64.863 & 19.337 & 3.761 & 1.671\\
			98 & \motif{98} & 0.117 & 1.483 & 0.023 & 0.188 & 0.498 &
			6 & \motif{6} & 284.362 & 534.789 & 277.598 & 9.057 & 0.747\\
			6 & \motif{6} & 134.696 & 347.482 & 132.243 & 4.942 & 0.496 &
			238 & \motif{238} & 0.987 & 4.107 & 0.586 & 0.624 & 0.643\\
			12 & \motif{12} & 8.190 & 27.421 & 5.715 & 5.378 & 0.460 &
			12 & \motif{12} & 28.500 & 68.345 & 22.027 & 10.145 & 0.638\\
			238 & \motif{238} & 0.181 & 1.111 & 0.087 & 0.209 & 0.448 &
			46 & \motif{46} & 22.703 & 69.989 & 19.364 & 5.685 & 0.587\\
\bottomrule
	\end{tabular}}
\end{table}

\newpage
\section{Colored triad motifs at the tail quantiles}

\begin{table}[H]
	\centering
	\setlength{\abovecaptionskip}{0pt}
	\setlength{\belowcaptionskip}{10pt}
	\caption{Colored triad motifs at the left quantile (two classes). }
	\label{tab:colored_triads_2class_L}
	\resizebox{\textwidth}{!}{
		\begin{tabular}{llrrrrr|llrrrrr}
			\toprule
			\multicolumn{7}{c|}{$\alpha=0.05$} & \multicolumn{7}{c}{$\alpha=0.10$}\\
			\cmidrule(lr){1-7} \cmidrule(lr){8-14}
			id & motif & $\mu$ & $\sigma$ & $\mu_{\text{rnd}}$ & $\sigma_{\text{rnd}}$ & $z$ &
			id & motif & $\mu$ & $\sigma$ & $\mu_{\text{rnd}}$ & $\sigma_{\text{rnd}}$ & $z$\\
			\midrule
			 6 & \begin{xy}
				\POS (0,3) *{\textcolor{Commodity}{\medbullet}} ="a",
				\POS (-3.5,-1.5) *{\textcolor{Commodity}{\medbullet}} ="b",
				\POS (3.5,-1.5) *{\textcolor{Equity}{\medbullet}} ="c"
				\POS "a" \ar @{->} "b"
				\POS "a" \ar @{->} "c"
			\end{xy} & 119.782 & 186.855 & 80.002 & 45.069 & 0.883 & 6 & \begin{xy}
				\POS (0,3) *{\textcolor{Commodity}{\medbullet}} ="a",
				\POS (-3.5,-1.5) *{\textcolor{Commodity}{\medbullet}} ="b",
				\POS (3.5,-1.5) *{\textcolor{Equity}{\medbullet}} ="c"
				\POS "a" \ar @{->} "b"
				\POS "a" \ar @{->} "c"
			\end{xy} & 165.296 & 238.372 & 109.721 & 52.277 & 1.063\\
			46 & \begin{xy}
				\POS (0,3) *{\textcolor{Commodity}{\medbullet}} ="a",
				\POS (-3.5,-1.5) *{\textcolor{Commodity}{\medbullet}} ="b",
				\POS (3.5,-1.5) *{\textcolor{Equity}{\medbullet}} ="c"
				\POS "a" \ar @{<->} "b"
				\POS "a" \ar @{->} "c"
				\POS "b" \ar @{->} "c"
			\end{xy} & 16.476 & 27.037 & 11.808 & 7.622 & 0.613 & 46 & \begin{xy}
				\POS (0,3) *{\textcolor{Commodity}{\medbullet}} ="a",
				\POS (-3.5,-1.5) *{\textcolor{Commodity}{\medbullet}} ="b",
				\POS (3.5,-1.5) *{\textcolor{Equity}{\medbullet}} ="c"
				\POS "a" \ar @{<->} "b"
				\POS "a" \ar @{->} "c"
				\POS "b" \ar @{->} "c"
			\end{xy} & 30.047 & 48.148 & 20.477 & 10.335 & 0.926\\
\bottomrule
	\end{tabular}}
	\begin{flushleft}
		\footnotesize
		\textit{Legend: }
		\textcolor{Equity}{{\normalsize\textbullet}}\, Equity;
		\textcolor{Commodity}{{\normalsize\textbullet}}\,  Commodity.
	\end{flushleft} 
\end{table}


\begin{table}[H]
	\centering
	\setlength{\abovecaptionskip}{0pt}
	\setlength{\belowcaptionskip}{10pt}
	\caption{Colored triad motifs at the left quantile (four classes).}
	\label{tab:colored_triads_4class_L}
	\resizebox{\textwidth}{!}{
		\begin{tabular}{llrrrrr|llrrrrr}
			\toprule
			\multicolumn{7}{c|}{$\alpha=0.05$} & \multicolumn{7}{c}{$\alpha=0.10$}\\
			\cmidrule(lr){1-7} \cmidrule(lr){8-14}
			id & motif & $\mu$ & $\sigma$ & $\mu_{\text{rnd}}$ & $\sigma_{\text{rnd}}$ & $z$ &
			id & motif & $\mu$ & $\sigma$ & $\mu_{\text{rnd}}$ & $\sigma_{\text{rnd}}$ & $z$\\
			\midrule
			      	6 & \begin{xy}
							\POS (0,3) *{\textcolor{Agriculture}{\medbullet}} ="a",
							\POS (-3.5,-1.5) *{\textcolor{Energy}{\medbullet}} ="b",
							\POS (3.5,-1.5) *{\textcolor{Metal}{\medbullet}} ="c"
							\POS "a" \ar @{->} "b"
							\POS "a" \ar @{->} "c"
						\end{xy} & 18.983 & 33.650 & 9.856 & 6.981 & 1.307 & 6 & \begin{xy}
							\POS (0,3) *{\textcolor{Agriculture}{\medbullet}} ="a",
							\POS (-3.5,-1.5) *{\textcolor{Energy}{\medbullet}} ="b",
							\POS (3.5,-1.5) *{\textcolor{Metal}{\medbullet}} ="c"
							\POS "a" \ar @{->} "b"
							\POS "a" \ar @{->} "c"
						\end{xy} & 26.506 & 43.456 & 13.422 & 8.701 & 1.504\\
						6 & \begin{xy}
							\POS (0,3) *{\textcolor{Agriculture}{\medbullet}} ="a",
							\POS (-3.5,-1.5) *{\textcolor{Energy}{\medbullet}} ="b",
							\POS (3.5,-1.5) *{\textcolor{Equity}{\medbullet}} ="c"
							\POS "a" \ar @{->} "b"
							\POS "a" \ar @{->} "c"
						\end{xy} & 14.058 & 26.311 & 6.900 & 5.878 & 1.218 & 6 & \begin{xy}
							\POS (0,3) *{\textcolor{Agriculture}{\medbullet}} ="a",
							\POS (-3.5,-1.5) *{\textcolor{Energy}{\medbullet}} ="b",
							\POS (3.5,-1.5) *{\textcolor{Equity}{\medbullet}} ="c"
							\POS "a" \ar @{->} "b"
							\POS "a" \ar @{->} "c"
						\end{xy} & 19.486 & 33.560 & 9.494 & 7.218 & 1.384\\
\bottomrule
	\end{tabular}}
	\begin{flushleft}
		\footnotesize
		\textit{Legend: }
		\textcolor{Equity}{{\normalsize\textbullet}}\, Equity;
		\textcolor{Metal}{{\normalsize\textbullet}}\, Metal;
		\textcolor{Agriculture}{{\normalsize\textbullet}}\, Agriculture;
		\textcolor{Energy}{{\normalsize\textbullet}}\,  Energy.
	\end{flushleft} 
\end{table}


\begin{table}[htp]
	\centering
	\setlength{\abovecaptionskip}{0pt}
	\setlength{\belowcaptionskip}{10pt}
	\caption{Colored triad motifs at the left quantile (seven classes).}
	\label{tab:colored_triads_7class_L}
	\resizebox{\textwidth}{!}{
		\begin{tabular}{llrrrrr|llrrrrr}
			\toprule
			\multicolumn{7}{c|}{$\alpha=0.05$} & \multicolumn{7}{c}{$\alpha=0.10$}\\
			\cmidrule(lr){1-7} \cmidrule(lr){8-14}
			id & motif & $\mu$ & $\sigma$ & $\mu_{\text{rnd}}$ & $\sigma_{\text{rnd}}$ & $z$ &
			id & motif & $\mu$ & $\sigma$ & $\mu_{\text{rnd}}$ & $\sigma_{\text{rnd}}$ & $z$\\
			\midrule
			
			6 & \begin{xy}
				\POS (0,3) *{\textcolor{Equity}{\medbullet}} ="a",
				\POS (-3.5,-1.5) *{\textcolor{Energy}{\medbullet}} ="b",
				\POS (3.5,-1.5) *{\textcolor{GrainsOilseeds}{\medbullet}} ="c"
				\POS "a" \ar @{->} "b"
				\POS "a" \ar @{->} "c"
			\end{xy} & 6.850 & 13.766 & 2.955 & 3.327 & 1.171 & 6 & \begin{xy}
				\POS (0,3) *{\textcolor{Equity}{\medbullet}} ="a",
				\POS (-3.5,-1.5) *{\textcolor{Energy}{\medbullet}} ="b",
				\POS (3.5,-1.5) *{\textcolor{GrainsOilseeds}{\medbullet}} ="c"
				\POS "a" \ar @{->} "b"
				\POS "a" \ar @{->} "c"
			\end{xy} & 9.264 & 16.802 & 3.974 & 3.886 & 1.361\\
			6 & \begin{xy}
				\POS (0,3) *{\textcolor{GrainsOilseeds}{\medbullet}} ="a",
				\POS (-3.5,-1.5) *{\textcolor{Energy}{\medbullet}} ="b",
				\POS (3.5,-1.5) *{\textcolor{Preciousmetals}{\medbullet}} ="c"
				\POS "a" \ar @{->} "b"
				\POS "a" \ar @{->} "c"
			\end{xy} & 4.559 & 10.037 & 1.795 & 2.490 & 1.110 & 6 & \begin{xy}
				\POS (0,3) *{\textcolor{GrainsOilseeds}{\medbullet}} ="a",
				\POS (-3.5,-1.5) *{\textcolor{Energy}{\medbullet}} ="b",
				\POS (3.5,-1.5) *{\textcolor{Preciousmetals}{\medbullet}} ="c"
				\POS "a" \ar @{->} "b"
				\POS "a" \ar @{->} "c"
			\end{xy} & 5.808 & 11.182 & 2.376 & 2.851 & 1.204\\
			108 & \begin{xy}
				\POS (0,3) *{\textcolor{Softs}{\medbullet}} ="a",
				\POS (-3.5,-1.5) *{\textcolor{Industrialmetals}{\medbullet}} ="b",
				\POS (3.5,-1.5) *{\textcolor{Preciousmetals}{\medbullet}} ="c"
				\POS "b" \ar @{->} "a"
				\POS "a" \ar @{<->} "c"
				\POS "b" \ar @{->} "c"
			\end{xy} & 0.244 & 0.820 & 0.038 & 0.200 & 1.028 & 6 & \begin{xy}
				\POS (0,3) *{\textcolor{GrainsOilseeds}{\medbullet}} ="a",
				\POS (-3.5,-1.5) *{\textcolor{Energy}{\medbullet}} ="b",
				\POS (3.5,-1.5) *{\textcolor{GrainsOilseeds}{\medbullet}} ="c"
				\POS "a" \ar @{->} "b"
				\POS "a" \ar @{->} "c"
			\end{xy} & 4.650 & 9.734 & 1.796 & 2.600 & 1.098\\
			6 & \begin{xy}
				\POS (0,3) *{\textcolor{GrainsOilseeds}{\medbullet}} ="a",
				\POS (-3.5,-1.5) *{\textcolor{Energy}{\medbullet}} ="b",
				\POS (3.5,-1.5) *{\textcolor{Livestock}{\medbullet}} ="c"
				\POS "a" \ar @{->} "b"
				\POS "a" \ar @{->} "c"
			\end{xy} & 4.119 & 8.662 & 1.749 & 2.429 & 0.976 & 6 & \begin{xy}
				\POS (0,3) *{\textcolor{GrainsOilseeds}{\medbullet}} ="a",
				\POS (-3.5,-1.5) *{\textcolor{Energy}{\medbullet}} ="b",
				\POS (3.5,-1.5) *{\textcolor{Softs}{\medbullet}} ="c"
				\POS "a" \ar @{->} "b"
				\POS "a" \ar @{->} "c"
			\end{xy} & 7.369 & 13.771 & 3.473 & 3.581 & 1.088\\
			6 & \begin{xy}
				\POS (0,3) *{\textcolor{GrainsOilseeds}{\medbullet}} ="a",
				\POS (-3.5,-1.5) *{\textcolor{Energy}{\medbullet}} ="b",
				\POS (3.5,-1.5) *{\textcolor{GrainsOilseeds}{\medbullet}} ="c"
				\POS "a" \ar @{->} "b"
				\POS "a" \ar @{->} "c"
			\end{xy} & 3.671 & 8.532 & 1.395 & 2.350 & 0.969 & 6 & \begin{xy}
				\POS (0,3) *{\textcolor{GrainsOilseeds}{\medbullet}} ="a",
				\POS (-3.5,-1.5) *{\textcolor{Energy}{\medbullet}} ="b",
				\POS (3.5,-1.5) *{\textcolor{Livestock}{\medbullet}} ="c"
				\POS "a" \ar @{->} "b"
				\POS "a" \ar @{->} "c"
			\end{xy} & 5.237 & 10.225 & 2.370 & 2.840 & 1.010\\
			238 & \begin{xy}
				\POS (0,3) *{\textcolor{Energy}{\medbullet}} ="a",
				\POS (-3.5,-1.5) *{\textcolor{Energy}{\medbullet}} ="b",
				\POS (3.5,-1.5) *{\textcolor{Livestock}{\medbullet}} ="c"
				\POS "a" \ar @{<->} "b"
				\POS "a" \ar @{<->} "c"
				\POS "b" \ar @{<->} "c"
			\end{xy} & 0.171 & 0.399 & 0.024 & 0.153 & 0.963 & 6 & \begin{xy}
				\POS (0,3) *{\textcolor{GrainsOilseeds}{\medbullet}} ="a",
				\POS (-3.5,-1.5) *{\textcolor{Energy}{\medbullet}} ="b",
				\POS (3.5,-1.5) *{\textcolor{Industrialmetals}{\medbullet}} ="c"
				\POS "a" \ar @{->} "b"
				\POS "a" \ar @{->} "c"
			\end{xy} & 7.093 & 13.140 & 3.485 & 3.605 & 1.001\\
			6 & \begin{xy}
				\POS (0,3) *{\textcolor{GrainsOilseeds}{\medbullet}} ="a",
				\POS (-3.5,-1.5) *{\textcolor{Energy}{\medbullet}} ="b",
				\POS (3.5,-1.5) *{\textcolor{Softs}{\medbullet}} ="c"
				\POS "a" \ar @{->} "b"
				\POS "a" \ar @{->} "c"
			\end{xy} & 5.433 & 11.342 & 2.586 & 3.078 & 0.925 & 6 & \begin{xy}
				\POS (0,3) *{\textcolor{Equity}{\medbullet}} ="a",
				\POS (-3.5,-1.5) *{\textcolor{Energy}{\medbullet}} ="b",
				\POS (3.5,-1.5) *{\textcolor{Preciousmetals}{\medbullet}} ="c"
				\POS "a" \ar @{->} "b"
				\POS "a" \ar @{->} "c"
			\end{xy} & 5.139 & 10.792 & 2.355 & 2.824 & 0.986\\
			6 & \begin{xy}
				\POS (0,3) *{\textcolor{Equity}{\medbullet}} ="a",
				\POS (-3.5,-1.5) *{\textcolor{Energy}{\medbullet}} ="b",
				\POS (3.5,-1.5) *{\textcolor{Preciousmetals}{\medbullet}} ="c"
				\POS "a" \ar @{->} "b"
				\POS "a" \ar @{->} "c"
			\end{xy} & 3.817 & 9.467 & 1.768 & 2.456 & 0.835 & 6 & \begin{xy}
				\POS (0,3) *{\textcolor{Equity}{\medbullet}} ="a",
				\POS (-3.5,-1.5) *{\textcolor{GrainsOilseeds}{\medbullet}} ="b",
				\POS (3.5,-1.5) *{\textcolor{Preciousmetals}{\medbullet}} ="c"
				\POS "a" \ar @{->} "b"
				\POS "a" \ar @{->} "c"
			\end{xy} & 6.864 & 14.151 & 3.258 & 3.710 & 0.972\\
			6 & \begin{xy}
				\POS (0,3) *{\textcolor{Equity}{\medbullet}} ="a",
				\POS (-3.5,-1.5) *{\textcolor{GrainsOilseeds}{\medbullet}} ="b",
				\POS (3.5,-1.5) *{\textcolor{Preciousmetals}{\medbullet}} ="c"
				\POS "a" \ar @{->} "b"
				\POS "a" \ar @{->} "c"
			\end{xy} & 5.071 & 12.360 & 2.419 & 3.213 & 0.825 & 6 & \begin{xy}
				\POS (0,3) *{\textcolor{Equity}{\medbullet}} ="a",
				\POS (-3.5,-1.5) *{\textcolor{Softs}{\medbullet}} ="b",
				\POS (3.5,-1.5) *{\textcolor{Preciousmetals}{\medbullet}} ="c"
				\POS "a" \ar @{->} "b"
				\POS "a" \ar @{->} "c"
			\end{xy} & 5.938 & 12.316 & 2.813 & 3.264 & 0.957\\
			6 & \begin{xy}
				\POS (0,3) *{\textcolor{Equity}{\medbullet}} ="a",
				\POS (-3.5,-1.5) *{\textcolor{Softs}{\medbullet}} ="b",
				\POS (3.5,-1.5) *{\textcolor{Preciousmetals}{\medbullet}} ="c"
				\POS "a" \ar @{->} "b"
				\POS "a" \ar @{->} "c"
			\end{xy} & 4.426 & 10.902 & 2.095 & 2.833 & 0.823 & 6 & \begin{xy}
				\POS (0,3) *{\textcolor{Equity}{\medbullet}} ="a",
				\POS (-3.5,-1.5) *{\textcolor{GrainsOilseeds}{\medbullet}} ="b",
				\POS (3.5,-1.5) *{\textcolor{GrainsOilseeds}{\medbullet}} ="c"
				\POS "a" \ar @{->} "b"
				\POS "a" \ar @{->} "c"
			\end{xy} & 8.678 & 14.940 & 4.813 & 4.151 & 0.931\\
			6 & \begin{xy}
				\POS (0,3) *{\textcolor{GrainsOilseeds}{\medbullet}} ="a",
				\POS (-3.5,-1.5) *{\textcolor{Energy}{\medbullet}} ="b",
				\POS (3.5,-1.5) *{\textcolor{Industrialmetals}{\medbullet}} ="c"
				\POS "a" \ar @{->} "b"
				\POS "a" \ar @{->} "c"
			\end{xy} & 5.040 & 10.810 & 2.570 & 3.060 & 0.807 & 6 & \begin{xy}
				\POS (0,3) *{\textcolor{Equity}{\medbullet}} ="a",
				\POS (-3.5,-1.5) *{\textcolor{Livestock}{\medbullet}} ="b",
				\POS (3.5,-1.5) *{\textcolor{Preciousmetals}{\medbullet}} ="c"
				\POS "a" \ar @{->} "b"
				\POS "a" \ar @{->} "c"
			\end{xy} & 4.064 & 8.497 & 1.907 & 2.372 & 0.910\\
			6 & \begin{xy}
				\POS (0,3) *{\textcolor{Equity}{\medbullet}} ="a",
				\POS (-3.5,-1.5) *{\textcolor{GrainsOilseeds}{\medbullet}} ="b",
				\POS (3.5,-1.5) *{\textcolor{GrainsOilseeds}{\medbullet}} ="c"
				\POS "a" \ar @{->} "b"
				\POS "a" \ar @{->} "c"
			\end{xy} & 6.418 & 12.599 & 3.522 & 3.604 & 0.804 & 6 & \begin{xy}
				\POS (0,3) *{\textcolor{Livestock}{\medbullet}} ="a",
				\POS (-3.5,-1.5) *{\textcolor{Softs}{\medbullet}} ="b",
				\POS (3.5,-1.5) *{\textcolor{Preciousmetals}{\medbullet}} ="c"
				\POS "a" \ar @{->} "b"
				\POS "a" \ar @{->} "c"
			\end{xy} & 3.665 & 7.761 & 1.673 & 2.312 & 0.861\\
			6 & \begin{xy}
				\POS (0,3) *{\textcolor{Equity}{\medbullet}} ="a",
				\POS (-3.5,-1.5) *{\textcolor{Livestock}{\medbullet}} ="b",
				\POS (3.5,-1.5) *{\textcolor{Preciousmetals}{\medbullet}} ="c"
				\POS "a" \ar @{->} "b"
				\POS "a" \ar @{->} "c"
			\end{xy} & 3.106 & 7.599 & 1.447 & 2.086 & 0.796 & 6 & \begin{xy}
				\POS (0,3) *{\textcolor{Livestock}{\medbullet}} ="a",
				\POS (-3.5,-1.5) *{\textcolor{Energy}{\medbullet}} ="b",
				\POS (3.5,-1.5) *{\textcolor{Preciousmetals}{\medbullet}} ="c"
				\POS "a" \ar @{->} "b"
				\POS "a" \ar @{->} "c"
			\end{xy} & 3.101 & 6.654 & 1.405 & 1.998 & 0.849\\
			6 & \begin{xy}
				\POS (0,3) *{\textcolor{Livestock}{\medbullet}} ="a",
				\POS (-3.5,-1.5) *{\textcolor{Softs}{\medbullet}} ="b",
				\POS (3.5,-1.5) *{\textcolor{Preciousmetals}{\medbullet}} ="c"
				\POS "a" \ar @{->} "b"
				\POS "a" \ar @{->} "c"
			\end{xy} & 2.811 & 6.893 & 1.247 & 2.006 & 0.780 & 6 & \begin{xy}
				\POS (0,3) *{\textcolor{Livestock}{\medbullet}} ="a",
				\POS (-3.5,-1.5) *{\textcolor{GrainsOilseeds}{\medbullet}} ="b",
				\POS (3.5,-1.5) *{\textcolor{Preciousmetals}{\medbullet}} ="c"
				\POS "a" \ar @{->} "b"
				\POS "a" \ar @{->} "c"
			\end{xy} & 4.158 & 8.676 & 1.933 & 2.626 & 0.847\\
			6 & \begin{xy}
				\POS (0,3) *{\textcolor{Livestock}{\medbullet}} ="a",
				\POS (-3.5,-1.5) *{\textcolor{GrainsOilseeds}{\medbullet}} ="b",
				\POS (3.5,-1.5) *{\textcolor{Preciousmetals}{\medbullet}} ="c"
				\POS "a" \ar @{->} "b"
				\POS "a" \ar @{->} "c"
			\end{xy} & 3.172 & 7.600 & 1.436 & 2.265 & 0.766 & 6 & \begin{xy}
				\POS (0,3) *{\textcolor{Equity}{\medbullet}} ="a",
				\POS (-3.5,-1.5) *{\textcolor{Softs}{\medbullet}} ="b",
				\POS (3.5,-1.5) *{\textcolor{Livestock}{\medbullet}} ="c"
				\POS "a" \ar @{->} "b"
				\POS "a" \ar @{->} "c"
			\end{xy} & 5.531 & 11.383 & 2.799 & 3.247 & 0.841\\
			98 & \begin{xy}
				\POS (0,3) *{\textcolor{Preciousmetals}{\medbullet}} ="a",
				\POS (-3.5,-1.5) *{\textcolor{Preciousmetals}{\medbullet}} ="b",
				\POS (3.5,-1.5) *{\textcolor{Preciousmetals}{\medbullet}} ="c"
				\POS "a" \ar @{->} "b"
				\POS "c" \ar @{->} "a"
				\POS "b" \ar @{->} "c"
			\end{xy} & 0.125 & 0.331 & 0.020 & 0.139 & 0.757 & 6 & \begin{xy}
				\POS (0,3) *{\textcolor{Equity}{\medbullet}} ="a",
				\POS (-3.5,-1.5) *{\textcolor{GrainsOilseeds}{\medbullet}} ="b",
				\POS (3.5,-1.5) *{\textcolor{Livestock}{\medbullet}} ="c"
				\POS "a" \ar @{->} "b"
				\POS "a" \ar @{->} "c"
			\end{xy} & 6.395 & 13.030 & 3.281 & 3.741 & 0.832\\
			6 & \begin{xy}
				\POS (0,3) *{\textcolor{Equity}{\medbullet}} ="a",
				\POS (-3.5,-1.5) *{\textcolor{Softs}{\medbullet}} ="b",
				\POS (3.5,-1.5) *{\textcolor{Livestock}{\medbullet}} ="c"
				\POS "a" \ar @{->} "b"
				\POS "a" \ar @{->} "c"
			\end{xy} & 4.193 & 9.754 & 2.068 & 2.812 & 0.756 & 46 & \begin{xy}
				\POS (0,3) *{\textcolor{GrainsOilseeds}{\medbullet}} ="a",
				\POS (-3.5,-1.5) *{\textcolor{Livestock}{\medbullet}} ="b",
				\POS (3.5,-1.5) *{\textcolor{Preciousmetals}{\medbullet}} ="c"
				\POS "a" \ar @{<->} "b"
				\POS "a" \ar @{->} "c"
				\POS "b" \ar @{->} "c"
			\end{xy} & 1.056 & 2.503 & 0.359 & 0.859 & 0.811\\
			6 & \begin{xy}
				\POS (0,3) *{\textcolor{Livestock}{\medbullet}} ="a",
				\POS (-3.5,-1.5) *{\textcolor{Energy}{\medbullet}} ="b",
				\POS (3.5,-1.5) *{\textcolor{Preciousmetals}{\medbullet}} ="c"
				\POS "a" \ar @{->} "b"
				\POS "a" \ar @{->} "c"
			\end{xy} & 2.345 & 5.789 & 1.046 & 1.729 & 0.751 & 6 & \begin{xy}
				\POS (0,3) *{\textcolor{Equity}{\medbullet}} ="a",
				\POS (-3.5,-1.5) *{\textcolor{Energy}{\medbullet}} ="b",
				\POS (3.5,-1.5) *{\textcolor{Livestock}{\medbullet}} ="c"
				\POS "a" \ar @{->} "b"
				\POS "a" \ar @{->} "c"
			\end{xy} & 4.611 & 9.621 & 2.350 & 2.816 & 0.803\\
			6 & \begin{xy}
				\POS (0,3) *{\textcolor{Equity}{\medbullet}} ="a",
				\POS (-3.5,-1.5) *{\textcolor{GrainsOilseeds}{\medbullet}} ="b",
				\POS (3.5,-1.5) *{\textcolor{Livestock}{\medbullet}} ="c"
				\POS "a" \ar @{->} "b"
				\POS "a" \ar @{->} "c"
			\end{xy} & 4.788 & 11.252 & 2.394 & 3.188 & 0.751 & 6 & \begin{xy}
				\POS (0,3) *{\textcolor{Equity}{\medbullet}} ="a",
				\POS (-3.5,-1.5) *{\textcolor{GrainsOilseeds}{\medbullet}} ="b",
				\POS (3.5,-1.5) *{\textcolor{Softs}{\medbullet}} ="c"
				\POS "a" \ar @{->} "b"
				\POS "a" \ar @{->} "c"
			\end{xy} & 8.609 & 17.208 & 4.831 & 4.715 & 0.801\\
			46 & \begin{xy}
				\POS (0,3) *{\textcolor{Energy}{\medbullet}} ="a",
				\POS (-3.5,-1.5) *{\textcolor{Livestock}{\medbullet}} ="b",
				\POS (3.5,-1.5) *{\textcolor{Livestock}{\medbullet}} ="c"
				\POS "a" \ar @{<->} "b"
				\POS "a" \ar @{->} "c"
				\POS "b" \ar @{->} "c"
			\end{xy} & 0.329 & 1.150 & 0.073 & 0.347 & 0.737 & 6 & \begin{xy}
				\POS (0,3) *{\textcolor{GrainsOilseeds}{\medbullet}} ="a",
				\POS (-3.5,-1.5) *{\textcolor{GrainsOilseeds}{\medbullet}} ="b",
				\POS (3.5,-1.5) *{\textcolor{Softs}{\medbullet}} ="c"
				\POS "a" \ar @{->} "b"
				\POS "a" \ar @{->} "c"
			\end{xy} & 7.128 & 12.699 & 4.139 & 3.786 & 0.789\\
			6 & \begin{xy}
				\POS (0,3) *{\textcolor{Equity}{\medbullet}} ="a",
				\POS (-3.5,-1.5) *{\textcolor{Energy}{\medbullet}} ="b",
				\POS (3.5,-1.5) *{\textcolor{Livestock}{\medbullet}} ="c"
				\POS "a" \ar @{->} "b"
				\POS "a" \ar @{->} "c"
			\end{xy} & 3.524 & 8.147 & 1.752 & 2.439 & 0.727 & 6 & \begin{xy}
				\POS (0,3) *{\textcolor{Equity}{\medbullet}} ="a",
				\POS (-3.5,-1.5) *{\textcolor{Energy}{\medbullet}} ="b",
				\POS (3.5,-1.5) *{\textcolor{Softs}{\medbullet}} ="c"
				\POS "a" \ar @{->} "b"
				\POS "a" \ar @{->} "c"
			\end{xy} & 6.211 & 12.568 & 3.453 & 3.555 & 0.776\\
			6 & \begin{xy}
				\POS (0,3) *{\textcolor{GrainsOilseeds}{\medbullet}} ="a",
				\POS (-3.5,-1.5) *{\textcolor{GrainsOilseeds}{\medbullet}} ="b",
				\POS (3.5,-1.5) *{\textcolor{Preciousmetals}{\medbullet}} ="c"
				\POS "a" \ar @{->} "b"
				\POS "a" \ar @{->} "c"
			\end{xy} & 4.060 & 8.229 & 2.104 & 2.705 & 0.723 & 6 & \begin{xy}
				\POS (0,3) *{\textcolor{Preciousmetals}{\medbullet}} ="a",
				\POS (-3.5,-1.5) *{\textcolor{Energy}{\medbullet}} ="b",
				\POS (3.5,-1.5) *{\textcolor{Softs}{\medbullet}} ="c"
				\POS "a" \ar @{->} "b"
				\POS "a" \ar @{->} "c"
			\end{xy} & 4.033 & 8.981 & 2.053 & 2.563 & 0.773\\
			6 & \begin{xy}
				\POS (0,3) *{\textcolor{Preciousmetals}{\medbullet}} ="a",
				\POS (-3.5,-1.5) *{\textcolor{GrainsOilseeds}{\medbullet}} ="b",
				\POS (3.5,-1.5) *{\textcolor{Softs}{\medbullet}} ="c"
				\POS "a" \ar @{->} "b"
				\POS "a" \ar @{->} "c"
			\end{xy} & 4.155 & 10.637 & 2.075 & 2.891 & 0.719 & 6 & \begin{xy}
				\POS (0,3) *{\textcolor{Preciousmetals}{\medbullet}} ="a",
				\POS (-3.5,-1.5) *{\textcolor{GrainsOilseeds}{\medbullet}} ="b",
				\POS (3.5,-1.5) *{\textcolor{Softs}{\medbullet}} ="c"
				\POS "a" \ar @{->} "b"
				\POS "a" \ar @{->} "c"
			\end{xy} & 5.417 & 11.781 & 2.830 & 3.357 & 0.771\\
			102 & \begin{xy}
				\POS (0,3) *{\textcolor{Energy}{\medbullet}} ="a",
				\POS (-3.5,-1.5) *{\textcolor{GrainsOilseeds}{\medbullet}} ="b",
				\POS (3.5,-1.5) *{\textcolor{Softs}{\medbullet}} ="c"
				\POS "a" \ar @{->} "b"
				\POS "a" \ar @{<->} "c"
				\POS "b" \ar @{->} "c"
			\end{xy} & 0.167 & 0.709 & 0.033 & 0.187 & 0.714 & 6 & \begin{xy}
				\POS (0,3) *{\textcolor{Industrialmetals}{\medbullet}} ="a",
				\POS (-3.5,-1.5) *{\textcolor{GrainsOilseeds}{\medbullet}} ="b",
				\POS (3.5,-1.5) *{\textcolor{Preciousmetals}{\medbullet}} ="c"
				\POS "a" \ar @{->} "b"
				\POS "a" \ar @{->} "c"
			\end{xy} & 5.428 & 11.709 & 2.835 & 3.372 & 0.769\\
			6 & \begin{xy}
				\POS (0,3) *{\textcolor{GrainsOilseeds}{\medbullet}} ="a",
				\POS (-3.5,-1.5) *{\textcolor{GrainsOilseeds}{\medbullet}} ="b",
				\POS (3.5,-1.5) *{\textcolor{Softs}{\medbullet}} ="c"
				\POS "a" \ar @{->} "b"
				\POS "a" \ar @{->} "c"
			\end{xy} & 5.376 & 10.847 & 3.042 & 3.315 & 0.704 & 6 & \begin{xy}
				\POS (0,3) *{\textcolor{Equity}{\medbullet}} ="a",
				\POS (-3.5,-1.5) *{\textcolor{Softs}{\medbullet}} ="b",
				\POS (3.5,-1.5) *{\textcolor{Industrialmetals}{\medbullet}} ="c"
				\POS "a" \ar @{->} "b"
				\POS "a" \ar @{->} "c"
			\end{xy} & 7.315 & 14.406 & 4.144 & 4.137 & 0.766\\
			46 & \begin{xy}
				\POS (0,3) *{\textcolor{Energy}{\medbullet}} ="a",
				\POS (-3.5,-1.5) *{\textcolor{GrainsOilseeds}{\medbullet}} ="b",
				\POS (3.5,-1.5) *{\textcolor{Livestock}{\medbullet}} ="c"
				\POS "a" \ar @{<->} "b"
				\POS "a" \ar @{->} "c"
				\POS "b" \ar @{->} "c"
			\end{xy} & 0.824 & 2.475 & 0.255 & 0.820 & 0.693 & 6 & \begin{xy}
				\POS (0,3) *{\textcolor{Industrialmetals}{\medbullet}} ="a",
				\POS (-3.5,-1.5) *{\textcolor{Softs}{\medbullet}} ="b",
				\POS (3.5,-1.5) *{\textcolor{Preciousmetals}{\medbullet}} ="c"
				\POS "a" \ar @{->} "b"
				\POS "a" \ar @{->} "c"
			\end{xy} & 4.710 & 10.349 & 2.444 & 2.960 & 0.766\\
			6 & \begin{xy}
				\POS (0,3) *{\textcolor{Preciousmetals}{\medbullet}} ="a",
				\POS (-3.5,-1.5) *{\textcolor{Energy}{\medbullet}} ="b",
				\POS (3.5,-1.5) *{\textcolor{Softs}{\medbullet}} ="c"
				\POS "a" \ar @{->} "b"
				\POS "a" \ar @{->} "c"
			\end{xy} & 3.068 & 7.967 & 1.530 & 2.227 & 0.691 & 6 & \begin{xy}
				\POS (0,3) *{\textcolor{GrainsOilseeds}{\medbullet}} ="a",
				\POS (-3.5,-1.5) *{\textcolor{GrainsOilseeds}{\medbullet}} ="b",
				\POS (3.5,-1.5) *{\textcolor{Preciousmetals}{\medbullet}} ="c"
				\POS "a" \ar @{->} "b"
				\POS "a" \ar @{->} "c"
			\end{xy} & 5.171 & 9.275 & 2.824 & 3.073 & 0.764\\
			102 & \begin{xy}
				\POS (0,3) *{\textcolor{Energy}{\medbullet}} ="a",
				\POS (-3.5,-1.5) *{\textcolor{Energy}{\medbullet}} ="b",
				\POS (3.5,-1.5) *{\textcolor{Industrialmetals}{\medbullet}} ="c"
				\POS "a" \ar @{->} "b"
				\POS "a" \ar @{<->} "c"
				\POS "b" \ar @{->} "c"
			\end{xy} & 0.135 & 0.440 & 0.024 & 0.164 & 0.675 & 6 & \begin{xy}
				\POS (0,3) *{\textcolor{Equity}{\medbullet}} ="a",
				\POS (-3.5,-1.5) *{\textcolor{GrainsOilseeds}{\medbullet}} ="b",
				\POS (3.5,-1.5) *{\textcolor{Industrialmetals}{\medbullet}} ="c"
				\POS "a" \ar @{->} "b"
				\POS "a" \ar @{->} "c"
			\end{xy} & 8.348 & 16.311 & 4.796 & 4.672 & 0.760\\
			6 & \begin{xy}
				\POS (0,3) *{\textcolor{GrainsOilseeds}{\medbullet}} ="a",
				\POS (-3.5,-1.5) *{\textcolor{GrainsOilseeds}{\medbullet}} ="b",
				\POS (3.5,-1.5) *{\textcolor{Livestock}{\medbullet}} ="c"
				\POS "a" \ar @{->} "b"
				\POS "a" \ar @{->} "c"
			\end{xy} & 3.849 & 7.748 & 2.071 & 2.678 & 0.664 & 6 & \begin{xy}
				\POS (0,3) *{\textcolor{Industrialmetals}{\medbullet}} ="a",
				\POS (-3.5,-1.5) *{\textcolor{Energy}{\medbullet}} ="b",
				\POS (3.5,-1.5) *{\textcolor{Preciousmetals}{\medbullet}} ="c"
				\POS "a" \ar @{->} "b"
				\POS "a" \ar @{->} "c"
			\end{xy} & 3.998 & 8.940 & 2.053 & 2.575 & 0.755\\
			6 & \begin{xy}
				\POS (0,3) *{\textcolor{Industrialmetals}{\medbullet}} ="a",
				\POS (-3.5,-1.5) *{\textcolor{Softs}{\medbullet}} ="b",
				\POS (3.5,-1.5) *{\textcolor{Preciousmetals}{\medbullet}} ="c"
				\POS "a" \ar @{->} "b"
				\POS "a" \ar @{->} "c"
			\end{xy} & 3.523 & 9.157 & 1.820 & 2.581 & 0.660 & 6 & \begin{xy}
				\POS (0,3) *{\textcolor{Equity}{\medbullet}} ="a",
				\POS (-3.5,-1.5) *{\textcolor{Energy}{\medbullet}} ="b",
				\POS (3.5,-1.5) *{\textcolor{Industrialmetals}{\medbullet}} ="c"
				\POS "a" \ar @{->} "b"
				\POS "a" \ar @{->} "c"
			\end{xy} & 6.075 & 12.199 & 3.449 & 3.562 & 0.737\\
			\bottomrule
	\end{tabular}}
	\begin{flushleft}
		\footnotesize
		\textit{Legend: }
		\textcolor{Equity}{{\normalsize\textbullet}}\, Equity;
		\textcolor{Industrialmetals}{{\normalsize\textbullet}}\, Industrial metals;
		\textcolor{Preciousmetals}{{\normalsize\textbullet}}\, Precious metals;
		\textcolor{Energy}{{\normalsize\textbullet}}\,  Energy;
		\textcolor{GrainsOilseeds}{{\normalsize\textbullet}}\, Grains \& oilseeds;
		\textcolor{Softs}{{\normalsize\textbullet}}\, Softs;
		\textcolor{Livestock}{{\normalsize\textbullet}}\, Livestock.
	\end{flushleft} 
\end{table}

\begin{table}[H]
	\centering
	\setlength{\abovecaptionskip}{0pt}
	\setlength{\belowcaptionskip}{10pt}
	\caption{Colored triad motifs at the right quantile (two classes). }
	\label{tab:colored_triads_2class_R}
	\resizebox{\textwidth}{!}{
		\begin{tabular}{llrrrrr|llrrrrr}
			\toprule
			\multicolumn{7}{c|}{$\alpha=0.05$} & \multicolumn{7}{c}{$\alpha=0.10$}\\
			\cmidrule(lr){1-7} \cmidrule(lr){8-14}
			id & motif & $\mu$ & $\sigma$ & $\mu_{\text{rnd}}$ & $\sigma_{\text{rnd}}$ & $z$ &
			id & motif & $\mu$ & $\sigma$ & $\mu_{\text{rnd}}$ & $\sigma_{\text{rnd}}$ & $z$\\
			\midrule
			            78 & \begin{xy}
				\POS (0,3) *{\textcolor{Equity}{\medbullet}} ="a",
				\POS (-3.5,-1.5) *{\textcolor{Commodity}{\medbullet}} ="b",
				\POS (3.5,-1.5) *{\textcolor{Equity}{\medbullet}} ="c"
				\POS "a" \ar @{<->} "b"
				\POS "a" \ar @{<->} "c"
			\end{xy} & 0.786 & 3.564 & 0.171 & 0.566 & 1.085 & 6 & \begin{xy}
				\POS (0,3) *{\textcolor{Commodity}{\medbullet}} ="a",
				\POS (-3.5,-1.5) *{\textcolor{Commodity}{\medbullet}} ="b",
				\POS (3.5,-1.5) *{\textcolor{Equity}{\medbullet}} ="c"
				\POS "a" \ar @{->} "b"
				\POS "a" \ar @{->} "c"
			\end{xy} & 135.058 & 199.111 & 91.000 & 47.371 & 0.930\\
			6 & \begin{xy}
				\POS (0,3) *{\textcolor{Commodity}{\medbullet}} ="a",
				\POS (-3.5,-1.5) *{\textcolor{Commodity}{\medbullet}} ="b",
				\POS (3.5,-1.5) *{\textcolor{Equity}{\medbullet}} ="c"
				\POS "a" \ar @{->} "b"
				\POS "a" \ar @{->} "c"
			\end{xy} & 93.447 & 158.291 & 62.786 & 37.636 & 0.815 & 46 & \begin{xy}
				\POS (0,3) *{\textcolor{Commodity}{\medbullet}} ="a",
				\POS (-3.5,-1.5) *{\textcolor{Commodity}{\medbullet}} ="b",
				\POS (3.5,-1.5) *{\textcolor{Equity}{\medbullet}} ="c"
				\POS "a" \ar @{<->} "b"
				\POS "a" \ar @{->} "c"
				\POS "b" \ar @{->} "c"
			\end{xy} & 23.935 & 36.474 & 16.663 & 9.110 & 0.798\\
			46 & \begin{xy}
				\POS (0,3) *{\textcolor{Commodity}{\medbullet}} ="a",
				\POS (-3.5,-1.5) *{\textcolor{Commodity}{\medbullet}} ="b",
				\POS (3.5,-1.5) *{\textcolor{Equity}{\medbullet}} ="c"
				\POS "a" \ar @{<->} "b"
				\POS "a" \ar @{->} "c"
				\POS "b" \ar @{->} "c"
			\end{xy} & 15.418 & 20.861 & 10.129 & 7.111 & 0.744 & 6 & \begin{xy}
				\POS (0,3) *{\textcolor{Equity}{\medbullet}} ="a",
				\POS (-3.5,-1.5) *{\textcolor{Commodity}{\medbullet}} ="b",
				\POS (3.5,-1.5) *{\textcolor{Equity}{\medbullet}} ="c"
				\POS "a" \ar @{->} "b"
				\POS "a" \ar @{->} "c"
			\end{xy} & 16.720 & 40.996 & 8.850 & 12.189 & 0.646\\
			108 & \begin{xy}
				\POS (0,3) *{\textcolor{Commodity}{\medbullet}} ="a",
				\POS (-3.5,-1.5) *{\textcolor{Equity}{\medbullet}} ="b",
				\POS (3.5,-1.5) *{\textcolor{Equity}{\medbullet}} ="c"
				\POS "b" \ar @{->} "a"
				\POS "a" \ar @{<->} "c"
				\POS "b" \ar @{->} "c"
			\end{xy} & 0.358 & 1.075 & 0.113 & 0.363 & 0.673 & 46 & \begin{xy}
				\POS (0,3) *{\textcolor{Commodity}{\medbullet}} ="a",
				\POS (-3.5,-1.5) *{\textcolor{Equity}{\medbullet}} ="b",
				\POS (3.5,-1.5) *{\textcolor{Equity}{\medbullet}} ="c"
				\POS "a" \ar @{<->} "b"
				\POS "a" \ar @{->} "c"
				\POS "b" \ar @{->} "c"
			\end{xy} & 3.013 & 6.104 & 1.619 & 2.303 & 0.605\\
			78 & \begin{xy}
				\POS (0,3) *{\textcolor{Equity}{\medbullet}} ="a",
				\POS (-3.5,-1.5) *{\textcolor{Equity}{\medbullet}} ="b",
				\POS (3.5,-1.5) *{\textcolor{Equity}{\medbullet}} ="c"
				\POS "a" \ar @{<->} "b"
				\POS "a" \ar @{<->} "c"
			\end{xy} & 0.233 & 0.761 & 0.051 & 0.275 & 0.663 & 78 & \begin{xy}
				\POS (0,3) *{\textcolor{Commodity}{\medbullet}} ="a",
				\POS (-3.5,-1.5) *{\textcolor{Commodity}{\medbullet}} ="b",
				\POS (3.5,-1.5) *{\textcolor{Equity}{\medbullet}} ="c"
				\POS "a" \ar @{<->} "b"
				\POS "a" \ar @{<->} "c"
			\end{xy} & 3.231 & 8.130 & 2.132 & 2.041 & 0.538\\
			46 & \begin{xy}
				\POS (0,3) *{\textcolor{Commodity}{\medbullet}} ="a",
				\POS (-3.5,-1.5) *{\textcolor{Equity}{\medbullet}} ="b",
				\POS (3.5,-1.5) *{\textcolor{Equity}{\medbullet}} ="c"
				\POS "a" \ar @{<->} "b"
				\POS "a" \ar @{->} "c"
				\POS "b" \ar @{->} "c"
			\end{xy} & 2.007 & 3.883 & 0.983 & 1.762 & 0.581 & 108 & \begin{xy}
				\POS (0,3) *{\textcolor{Commodity}{\medbullet}} ="a",
				\POS (-3.5,-1.5) *{\textcolor{Equity}{\medbullet}} ="b",
				\POS (3.5,-1.5) *{\textcolor{Equity}{\medbullet}} ="c"
				\POS "b" \ar @{->} "a"
				\POS "a" \ar @{<->} "c"
				\POS "b" \ar @{->} "c"
			\end{xy} & 0.311 & 1.118 & 0.147 & 0.417 & 0.393\\
			6 & \begin{xy}
				\POS (0,3) *{\textcolor{Equity}{\medbullet}} ="a",
				\POS (-3.5,-1.5) *{\textcolor{Commodity}{\medbullet}} ="b",
				\POS (3.5,-1.5) *{\textcolor{Equity}{\medbullet}} ="c"
				\POS "a" \ar @{->} "b"
				\POS "a" \ar @{->} "c"
			\end{xy} & 11.556 & 36.858 & 6.124 & 9.766 & 0.556 & 78 & \begin{xy}
				\POS (0,3) *{\textcolor{Equity}{\medbullet}} ="a",
				\POS (-3.5,-1.5) *{\textcolor{Commodity}{\medbullet}} ="b",
				\POS (3.5,-1.5) *{\textcolor{Equity}{\medbullet}} ="c"
				\POS "a" \ar @{<->} "b"
				\POS "a" \ar @{<->} "c"
			\end{xy} & 0.436 & 2.262 & 0.208 & 0.614 & 0.371\\
			102 & \begin{xy}
				\POS (0,3) *{\textcolor{Commodity}{\medbullet}} ="a",
				\POS (-3.5,-1.5) *{\textcolor{Equity}{\medbullet}} ="b",
				\POS (3.5,-1.5) *{\textcolor{Commodity}{\medbullet}} ="c"
				\POS "a" \ar @{->} "b"
				\POS "a" \ar @{<->} "c"
				\POS "b" \ar @{->} "c"
			\end{xy} & 1.281 & 2.966 & 0.744 & 0.971 & 0.553 & 108 & \begin{xy}
				\POS (0,3) *{\textcolor{Commodity}{\medbullet}} ="a",
				\POS (-3.5,-1.5) *{\textcolor{Commodity}{\medbullet}} ="b",
				\POS (3.5,-1.5) *{\textcolor{Equity}{\medbullet}} ="c"
				\POS "b" \ar @{->} "a"
				\POS "a" \ar @{<->} "c"
				\POS "b" \ar @{->} "c"
			\end{xy} & 2.020 & 4.050 & 1.528 & 1.399 & 0.351\\
			78 & \begin{xy}
				\POS (0,3) *{\textcolor{Commodity}{\medbullet}} ="a",
				\POS (-3.5,-1.5) *{\textcolor{Commodity}{\medbullet}} ="b",
				\POS (3.5,-1.5) *{\textcolor{Equity}{\medbullet}} ="c"
				\POS "a" \ar @{<->} "b"
				\POS "a" \ar @{<->} "c"
			\end{xy} & 2.737 & 5.857 & 1.799 & 1.885 & 0.498 & 102 & \begin{xy}
				\POS (0,3) *{\textcolor{Commodity}{\medbullet}} ="a",
				\POS (-3.5,-1.5) *{\textcolor{Equity}{\medbullet}} ="b",
				\POS (3.5,-1.5) *{\textcolor{Commodity}{\medbullet}} ="c"
				\POS "a" \ar @{->} "b"
				\POS "a" \ar @{<->} "c"
				\POS "b" \ar @{->} "c"
			\end{xy} & 1.367 & 3.587 & 1.074 & 1.219 & 0.240\\
			102 & \begin{xy}
				\POS (0,3) *{\textcolor{Equity}{\medbullet}} ="a",
				\POS (-3.5,-1.5) *{\textcolor{Equity}{\medbullet}} ="b",
				\POS (3.5,-1.5) *{\textcolor{Commodity}{\medbullet}} ="c"
				\POS "a" \ar @{->} "b"
				\POS "a" \ar @{<->} "c"
				\POS "b" \ar @{->} "c"
			\end{xy} & 0.333 & 1.563 & 0.130 & 0.485 & 0.421 & 108 & \begin{xy}
				\POS (0,3) *{\textcolor{Equity}{\medbullet}} ="a",
				\POS (-3.5,-1.5) *{\textcolor{Equity}{\medbullet}} ="b",
				\POS (3.5,-1.5) *{\textcolor{Equity}{\medbullet}} ="c"
				\POS "b" \ar @{->} "a"
				\POS "a" \ar @{<->} "c"
				\POS "b" \ar @{->} "c"
			\end{xy} & 0.071 & 0.346 & 0.036 & 0.196 & 0.181\\
			108 & \begin{xy}
				\POS (0,3) *{\textcolor{Commodity}{\medbullet}} ="a",
				\POS (-3.5,-1.5) *{\textcolor{Commodity}{\medbullet}} ="b",
				\POS (3.5,-1.5) *{\textcolor{Equity}{\medbullet}} ="c"
				\POS "b" \ar @{->} "a"
				\POS "a" \ar @{<->} "c"
				\POS "b" \ar @{->} "c"
			\end{xy} & 1.639 & 3.094 & 1.200 & 1.255 & 0.350 & 238 & \begin{xy}
				\POS (0,3) *{\textcolor{Commodity}{\medbullet}} ="a",
				\POS (-3.5,-1.5) *{\textcolor{Commodity}{\medbullet}} ="b",
				\POS (3.5,-1.5) *{\textcolor{Commodity}{\medbullet}} ="c"
				\POS "a" \ar @{<->} "b"
				\POS "a" \ar @{<->} "c"
				\POS "b" \ar @{<->} "c"
			\end{xy} & 3.418 & 4.748 & 3.095 & 1.990 & 0.162\\
			238 & \begin{xy}
				\POS (0,3) *{\textcolor{Commodity}{\medbullet}} ="a",
				\POS (-3.5,-1.5) *{\textcolor{Commodity}{\medbullet}} ="b",
				\POS (3.5,-1.5) *{\textcolor{Commodity}{\medbullet}} ="c"
				\POS "a" \ar @{<->} "b"
				\POS "a" \ar @{<->} "c"
				\POS "b" \ar @{<->} "c"
			\end{xy} & 1.642 & 1.771 & 1.456 & 1.102 & 0.168 & 74 & \begin{xy}
				\POS (0,3) *{\textcolor{Commodity}{\medbullet}} ="a",
				\POS (-3.5,-1.5) *{\textcolor{Equity}{\medbullet}} ="b",
				\POS (3.5,-1.5) *{\textcolor{Equity}{\medbullet}} ="c"
				\POS "a" \ar @{<->} "b"
				\POS "c" \ar @{->} "a"
			\end{xy} & 0.287 & 0.896 & 0.214 & 0.485 & 0.152\\
			38 & \begin{xy}
				\POS (0,3) *{\textcolor{Commodity}{\medbullet}} ="a",
				\POS (-3.5,-1.5) *{\textcolor{Commodity}{\medbullet}} ="b",
				\POS (3.5,-1.5) *{\textcolor{Commodity}{\medbullet}} ="c"
				\POS "a" \ar @{->} "b"
				\POS "a" \ar @{->} "c"
				\POS "b" \ar @{->} "c"
			\end{xy} & 9.404 & 13.660 & 8.685 & 4.373 & 0.164 & 108 & \begin{xy}
				\POS (0,3) *{\textcolor{Commodity}{\medbullet}} ="a",
				\POS (-3.5,-1.5) *{\textcolor{Commodity}{\medbullet}} ="b",
				\POS (3.5,-1.5) *{\textcolor{Commodity}{\medbullet}} ="c"
				\POS "b" \ar @{->} "a"
				\POS "a" \ar @{<->} "c"
				\POS "b" \ar @{->} "c"
			\end{xy} & 3.542 & 5.743 & 3.277 & 1.908 & 0.139\\
			74 & \begin{xy}
				\POS (0,3) *{\textcolor{Equity}{\medbullet}} ="a",
				\POS (-3.5,-1.5) *{\textcolor{Equity}{\medbullet}} ="b",
				\POS (3.5,-1.5) *{\textcolor{Commodity}{\medbullet}} ="c"
				\POS "a" \ar @{<->} "b"
				\POS "c" \ar @{->} "a"
			\end{xy} & 0.196 & 1.209 & 0.120 & 0.518 & 0.147 & 46 & \begin{xy}
				\POS (0,3) *{\textcolor{Commodity}{\medbullet}} ="a",
				\POS (-3.5,-1.5) *{\textcolor{Commodity}{\medbullet}} ="b",
				\POS (3.5,-1.5) *{\textcolor{Commodity}{\medbullet}} ="c"
				\POS "a" \ar @{<->} "b"
				\POS "a" \ar @{->} "c"
				\POS "b" \ar @{->} "c"
			\end{xy} & 37.915 & 56.271 & 35.829 & 16.026 & 0.130\\
			74 & \begin{xy}
				\POS (0,3) *{\textcolor{Equity}{\medbullet}} ="a",
				\POS (-3.5,-1.5) *{\textcolor{Commodity}{\medbullet}} ="b",
				\POS (3.5,-1.5) *{\textcolor{Commodity}{\medbullet}} ="c"
				\POS "a" \ar @{<->} "b"
				\POS "c" \ar @{->} "a"
			\end{xy} & 0.723 & 2.798 & 0.561 & 1.110 & 0.146 & 36 & \begin{xy}
				\POS (0,3) *{\textcolor{Commodity}{\medbullet}} ="a",
				\POS (-3.5,-1.5) *{\textcolor{Equity}{\medbullet}} ="b",
				\POS (3.5,-1.5) *{\textcolor{Equity}{\medbullet}} ="c"
				\POS "a" \ar @{->} "c"
				\POS "b" \ar @{->} "c"
			\end{xy} & 0.948 & 3.227 & 0.795 & 1.383 & 0.111\\
			98 & \begin{xy}
				\POS (0,3) *{\textcolor{Commodity}{\medbullet}} ="a",
				\POS (-3.5,-1.5) *{\textcolor{Commodity}{\medbullet}} ="b",
				\POS (3.5,-1.5) *{\textcolor{Equity}{\medbullet}} ="c"
				\POS "a" \ar @{->} "b"
				\POS "c" \ar @{->} "a"
				\POS "b" \ar @{->} "c"
			\end{xy} & 1.256 & 2.047 & 1.119 & 0.966 & 0.143 & 78 & \begin{xy}
				\POS (0,3) *{\textcolor{Equity}{\medbullet}} ="a",
				\POS (-3.5,-1.5) *{\textcolor{Equity}{\medbullet}} ="b",
				\POS (3.5,-1.5) *{\textcolor{Equity}{\medbullet}} ="c"
				\POS "a" \ar @{<->} "b"
				\POS "a" \ar @{<->} "c"
			\end{xy} & 0.076 & 0.404 & 0.050 & 0.232 & 0.108\\
			102 & \begin{xy}
				\POS (0,3) *{\textcolor{Equity}{\medbullet}} ="a",
				\POS (-3.5,-1.5) *{\textcolor{Equity}{\medbullet}} ="b",
				\POS (3.5,-1.5) *{\textcolor{Equity}{\medbullet}} ="c"
				\POS "a" \ar @{->} "b"
				\POS "a" \ar @{<->} "c"
				\POS "b" \ar @{->} "c"
			\end{xy} & 0.057 & 0.232 & 0.035 & 0.192 & 0.118 & 110 & \begin{xy}
				\POS (0,3) *{\textcolor{Equity}{\medbullet}} ="a",
				\POS (-3.5,-1.5) *{\textcolor{Equity}{\medbullet}} ="b",
				\POS (3.5,-1.5) *{\textcolor{Equity}{\medbullet}} ="c"
				\POS "a" \ar @{<->} "b"
				\POS "a" \ar @{<->} "c"
				\POS "b" \ar @{->} "c"
			\end{xy} & 0.049 & 0.266 & 0.031 & 0.180 & 0.100\\
			102 & \begin{xy}
				\POS (0,3) *{\textcolor{Equity}{\medbullet}} ="a",
				\POS (-3.5,-1.5) *{\textcolor{Commodity}{\medbullet}} ="b",
				\POS (3.5,-1.5) *{\textcolor{Commodity}{\medbullet}} ="c"
				\POS "a" \ar @{->} "b"
				\POS "a" \ar @{<->} "c"
				\POS "b" \ar @{->} "c"
			\end{xy} & 0.891 & 2.513 & 0.756 & 1.192 & 0.113 & 12 & \begin{xy}
				\POS (0,3) *{\textcolor{Commodity}{\medbullet}} ="a",
				\POS (-3.5,-1.5) *{\textcolor{Commodity}{\medbullet}} ="b",
				\POS (3.5,-1.5) *{\textcolor{Commodity}{\medbullet}} ="c"
				\POS "b" \ar @{->} "a"
				\POS "a" \ar @{->} "c"
			\end{xy} & 25.450 & 40.963 & 24.616 & 8.364 & 0.100\\
			14 & \begin{xy}
				\POS (0,3) *{\textcolor{Commodity}{\medbullet}} ="a",
				\POS (-3.5,-1.5) *{\textcolor{Commodity}{\medbullet}} ="b",
				\POS (3.5,-1.5) *{\textcolor{Commodity}{\medbullet}} ="c"
				\POS "a" \ar @{<->} "b"
				\POS "a" \ar @{->} "c"
			\end{xy} & 23.108 & 32.357 & 21.876 & 10.961 & 0.112 & 38 & \begin{xy}
				\POS (0,3) *{\textcolor{Commodity}{\medbullet}} ="a",
				\POS (-3.5,-1.5) *{\textcolor{Commodity}{\medbullet}} ="b",
				\POS (3.5,-1.5) *{\textcolor{Commodity}{\medbullet}} ="c"
				\POS "a" \ar @{->} "b"
				\POS "a" \ar @{->} "c"
				\POS "b" \ar @{->} "c"
			\end{xy} & 17.093 & 24.747 & 16.447 & 6.772 & 0.095\\
			12 & \begin{xy}
				\POS (0,3) *{\textcolor{Commodity}{\medbullet}} ="a",
				\POS (-3.5,-1.5) *{\textcolor{Commodity}{\medbullet}} ="b",
				\POS (3.5,-1.5) *{\textcolor{Commodity}{\medbullet}} ="c"
				\POS "b" \ar @{->} "a"
				\POS "a" \ar @{->} "c"
			\end{xy} & 14.804 & 22.121 & 14.127 & 6.083 & 0.111 & 110 & \begin{xy}
				\POS (0,3) *{\textcolor{Commodity}{\medbullet}} ="a",
				\POS (-3.5,-1.5) *{\textcolor{Equity}{\medbullet}} ="b",
				\POS (3.5,-1.5) *{\textcolor{Commodity}{\medbullet}} ="c"
				\POS "a" \ar @{<->} "b"
				\POS "a" \ar @{<->} "c"
				\POS "b" \ar @{->} "c"
			\end{xy} & 0.748 & 1.785 & 0.668 & 0.943 & 0.085\\
			110 & \begin{xy}
				\POS (0,3) *{\textcolor{Commodity}{\medbullet}} ="a",
				\POS (-3.5,-1.5) *{\textcolor{Commodity}{\medbullet}} ="b",
				\POS (3.5,-1.5) *{\textcolor{Commodity}{\medbullet}} ="c"
				\POS "a" \ar @{<->} "b"
				\POS "a" \ar @{<->} "c"
				\POS "b" \ar @{->} "c"
			\end{xy} & 1.544 & 1.846 & 1.443 & 1.070 & 0.094 & 110 & \begin{xy}
				\POS (0,3) *{\textcolor{Commodity}{\medbullet}} ="a",
				\POS (-3.5,-1.5) *{\textcolor{Commodity}{\medbullet}} ="b",
				\POS (3.5,-1.5) *{\textcolor{Commodity}{\medbullet}} ="c"
				\POS "a" \ar @{<->} "b"
				\POS "a" \ar @{<->} "c"
				\POS "b" \ar @{->} "c"
			\end{xy} & 2.995 & 4.514 & 2.857 & 1.775 & 0.078\\
			110 & \begin{xy}
				\POS (0,3) *{\textcolor{Equity}{\medbullet}} ="a",
				\POS (-3.5,-1.5) *{\textcolor{Equity}{\medbullet}} ="b",
				\POS (3.5,-1.5) *{\textcolor{Equity}{\medbullet}} ="c"
				\POS "a" \ar @{<->} "b"
				\POS "a" \ar @{<->} "c"
				\POS "b" \ar @{->} "c"
			\end{xy} & 0.034 & 0.182 & 0.021 & 0.145 & 0.092 & 110 & \begin{xy}
				\POS (0,3) *{\textcolor{Equity}{\medbullet}} ="a",
				\POS (-3.5,-1.5) *{\textcolor{Equity}{\medbullet}} ="b",
				\POS (3.5,-1.5) *{\textcolor{Commodity}{\medbullet}} ="c"
				\POS "a" \ar @{<->} "b"
				\POS "a" \ar @{<->} "c"
				\POS "b" \ar @{->} "c"
			\end{xy} & 0.168 & 0.697 & 0.136 & 0.425 & 0.074\\
			14 & \begin{xy}
				\POS (0,3) *{\textcolor{Commodity}{\medbullet}} ="a",
				\POS (-3.5,-1.5) *{\textcolor{Commodity}{\medbullet}} ="b",
				\POS (3.5,-1.5) *{\textcolor{Equity}{\medbullet}} ="c"
				\POS "a" \ar @{<->} "b"
				\POS "a" \ar @{->} "c"
			\end{xy} & 5.383 & 8.246 & 5.097 & 3.310 & 0.086 & 74 & \begin{xy}
				\POS (0,3) *{\textcolor{Equity}{\medbullet}} ="a",
				\POS (-3.5,-1.5) *{\textcolor{Equity}{\medbullet}} ="b",
				\POS (3.5,-1.5) *{\textcolor{Commodity}{\medbullet}} ="c"
				\POS "a" \ar @{<->} "b"
				\POS "c" \ar @{->} "a"
			\end{xy} & 0.272 & 1.204 & 0.218 & 0.736 & 0.074\\
			36 & \begin{xy}
				\POS (0,3) *{\textcolor{Commodity}{\medbullet}} ="a",
				\POS (-3.5,-1.5) *{\textcolor{Equity}{\medbullet}} ="b",
				\POS (3.5,-1.5) *{\textcolor{Equity}{\medbullet}} ="c"
				\POS "a" \ar @{->} "c"
				\POS "b" \ar @{->} "c"
			\end{xy} & 0.661 & 2.809 & 0.568 & 1.076 & 0.086 & 98 & \begin{xy}
				\POS (0,3) *{\textcolor{Commodity}{\medbullet}} ="a",
				\POS (-3.5,-1.5) *{\textcolor{Commodity}{\medbullet}} ="b",
				\POS (3.5,-1.5) *{\textcolor{Commodity}{\medbullet}} ="c"
				\POS "a" \ar @{->} "b"
				\POS "c" \ar @{->} "a"
				\POS "b" \ar @{->} "c"
			\end{xy} & 3.164 & 5.015 & 3.032 & 1.840 & 0.072\\
			36 & \begin{xy}
				\POS (0,3) *{\textcolor{Commodity}{\medbullet}} ="a",
				\POS (-3.5,-1.5) *{\textcolor{Commodity}{\medbullet}} ="b",
				\POS (3.5,-1.5) *{\textcolor{Commodity}{\medbullet}} ="c"
				\POS "a" \ar @{->} "c"
				\POS "b" \ar @{->} "c"
			\end{xy} & 12.712 & 28.917 & 12.277 & 5.069 & 0.086 & 38 & \begin{xy}
				\POS (0,3) *{\textcolor{Equity}{\medbullet}} ="a",
				\POS (-3.5,-1.5) *{\textcolor{Commodity}{\medbullet}} ="b",
				\POS (3.5,-1.5) *{\textcolor{Commodity}{\medbullet}} ="c"
				\POS "a" \ar @{->} "b"
				\POS "a" \ar @{->} "c"
				\POS "b" \ar @{->} "c"
			\end{xy} & 4.065 & 7.258 & 3.857 & 3.106 & 0.067\\
			38 & \begin{xy}
				\POS (0,3) *{\textcolor{Equity}{\medbullet}} ="a",
				\POS (-3.5,-1.5) *{\textcolor{Commodity}{\medbullet}} ="b",
				\POS (3.5,-1.5) *{\textcolor{Commodity}{\medbullet}} ="c"
				\POS "a" \ar @{->} "b"
				\POS "a" \ar @{->} "c"
				\POS "b" \ar @{->} "c"
			\end{xy} & 2.201 & 4.244 & 2.031 & 2.053 & 0.082 & 36 & \begin{xy}
				\POS (0,3) *{\textcolor{Equity}{\medbullet}} ="a",
				\POS (-3.5,-1.5) *{\textcolor{Equity}{\medbullet}} ="b",
				\POS (3.5,-1.5) *{\textcolor{Equity}{\medbullet}} ="c"
				\POS "a" \ar @{->} "c"
				\POS "b" \ar @{->} "c"
			\end{xy} & 0.180 & 0.710 & 0.153 & 0.429 & 0.064\\
			74 & \begin{xy}
				\POS (0,3) *{\textcolor{Equity}{\medbullet}} ="a",
				\POS (-3.5,-1.5) *{\textcolor{Equity}{\medbullet}} ="b",
				\POS (3.5,-1.5) *{\textcolor{Equity}{\medbullet}} ="c"
				\POS "a" \ar @{<->} "b"
				\POS "c" \ar @{->} "a"
			\end{xy} & 0.044 & 0.205 & 0.030 & 0.184 & 0.078 & 102 & \begin{xy}
				\POS (0,3) *{\textcolor{Equity}{\medbullet}} ="a",
				\POS (-3.5,-1.5) *{\textcolor{Equity}{\medbullet}} ="b",
				\POS (3.5,-1.5) *{\textcolor{Commodity}{\medbullet}} ="c"
				\POS "a" \ar @{->} "b"
				\POS "a" \ar @{<->} "c"
				\POS "b" \ar @{->} "c"
			\end{xy} & 0.247 & 1.462 & 0.206 & 0.644 & 0.064\\
			110 & \begin{xy}
				\POS (0,3) *{\textcolor{Commodity}{\medbullet}} ="a",
				\POS (-3.5,-1.5) *{\textcolor{Commodity}{\medbullet}} ="b",
				\POS (3.5,-1.5) *{\textcolor{Equity}{\medbullet}} ="c"
				\POS "a" \ar @{<->} "b"
				\POS "a" \ar @{<->} "c"
				\POS "b" \ar @{->} "c"
			\end{xy} & 0.395 & 0.942 & 0.348 & 0.633 & 0.073 & 36 & \begin{xy}
				\POS (0,3) *{\textcolor{Commodity}{\medbullet}} ="a",
				\POS (-3.5,-1.5) *{\textcolor{Commodity}{\medbullet}} ="b",
				\POS (3.5,-1.5) *{\textcolor{Commodity}{\medbullet}} ="c"
				\POS "a" \ar @{->} "c"
				\POS "b" \ar @{->} "c"
			\end{xy} & 17.581 & 37.648 & 17.183 & 6.287 & 0.063\\
			110 & \begin{xy}
				\POS (0,3) *{\textcolor{Equity}{\medbullet}} ="a",
				\POS (-3.5,-1.5) *{\textcolor{Commodity}{\medbullet}} ="b",
				\POS (3.5,-1.5) *{\textcolor{Commodity}{\medbullet}} ="c"
				\POS "a" \ar @{<->} "b"
				\POS "a" \ar @{<->} "c"
				\POS "b" \ar @{->} "c"
			\end{xy} & 0.395 & 1.014 & 0.349 & 0.695 & 0.066 & 102 & \begin{xy}
				\POS (0,3) *{\textcolor{Commodity}{\medbullet}} ="a",
				\POS (-3.5,-1.5) *{\textcolor{Commodity}{\medbullet}} ="b",
				\POS (3.5,-1.5) *{\textcolor{Commodity}{\medbullet}} ="c"
				\POS "a" \ar @{->} "b"
				\POS "a" \ar @{<->} "c"
				\POS "b" \ar @{->} "c"
			\end{xy} & 4.724 & 9.882 & 4.588 & 2.479 & 0.055\\
			108 & \begin{xy}
				\POS (0,3) *{\textcolor{Commodity}{\medbullet}} ="a",
				\POS (-3.5,-1.5) *{\textcolor{Commodity}{\medbullet}} ="b",
				\POS (3.5,-1.5) *{\textcolor{Commodity}{\medbullet}} ="c"
				\POS "b" \ar @{->} "a"
				\POS "a" \ar @{<->} "c"
				\POS "b" \ar @{->} "c"
			\end{xy} & 2.742 & 4.100 & 2.670 & 1.699 & 0.042 & 14 & \begin{xy}
				\POS (0,3) *{\textcolor{Commodity}{\medbullet}} ="a",
				\POS (-3.5,-1.5) *{\textcolor{Commodity}{\medbullet}} ="b",
				\POS (3.5,-1.5) *{\textcolor{Commodity}{\medbullet}} ="c"
				\POS "a" \ar @{<->} "b"
				\POS "a" \ar @{->} "c"
			\end{xy} & 35.359 & 45.626 & 34.566 & 14.655 & 0.054\\
			\bottomrule
	\end{tabular}}
	\begin{flushleft}
		\footnotesize
		\textit{Legend: }
		\textcolor{Equity}{{\normalsize\textbullet}}\, Equity;
		\textcolor{Commodity}{{\normalsize\textbullet}}\,  Commodity.
	\end{flushleft} 
\end{table}


\begin{table}[H]
	\centering
	\setlength{\abovecaptionskip}{0pt}
	\setlength{\belowcaptionskip}{10pt}
	\caption{Colored triad motifs at the right quantile (four classes).}
	\label{tab:colored_triads_4class_R}
	\resizebox{\textwidth}{!}{
		\begin{tabular}{llrrrrr|llrrrrr}
			\toprule
			\multicolumn{7}{c|}{$\alpha=0.05$} & \multicolumn{7}{c}{$\alpha=0.10$}\\
			\cmidrule(lr){1-7} \cmidrule(lr){8-14}
			id & motif & $\mu$ & $\sigma$ & $\mu_{\text{rnd}}$ & $\sigma_{\text{rnd}}$ & $z$ &
			id & motif & $\mu$ & $\sigma$ & $\mu_{\text{rnd}}$ & $\sigma_{\text{rnd}}$ & $z$\\
			\midrule
			            78 & \begin{xy}
				\POS (0,3) *{\textcolor{Equity}{\medbullet}} ="a",
				\POS (-3.5,-1.5) *{\textcolor{Energy}{\medbullet}} ="b",
				\POS (3.5,-1.5) *{\textcolor{Equity}{\medbullet}} ="c"
				\POS "a" \ar @{<->} "b"
				\POS "a" \ar @{<->} "c"
			\end{xy} & 0.630 & 2.003 & 0.052 & 0.253 & 2.282 & 6 & \begin{xy}
				\POS (0,3) *{\textcolor{Agriculture}{\medbullet}} ="a",
				\POS (-3.5,-1.5) *{\textcolor{Energy}{\medbullet}} ="b",
				\POS (3.5,-1.5) *{\textcolor{Metal}{\medbullet}} ="c"
				\POS "a" \ar @{->} "b"
				\POS "a" \ar @{->} "c"
			\end{xy} & 21.921 & 37.523 & 11.165 & 7.697 & 1.397\\
			6 & \begin{xy}
				\POS (0,3) *{\textcolor{Agriculture}{\medbullet}} ="a",
				\POS (-3.5,-1.5) *{\textcolor{Energy}{\medbullet}} ="b",
				\POS (3.5,-1.5) *{\textcolor{Metal}{\medbullet}} ="c"
				\POS "a" \ar @{->} "b"
				\POS "a" \ar @{->} "c"
			\end{xy} & 15.384 & 29.025 & 7.720 & 5.835 & 1.314 & 6 & \begin{xy}
				\POS (0,3) *{\textcolor{Agriculture}{\medbullet}} ="a",
				\POS (-3.5,-1.5) *{\textcolor{Energy}{\medbullet}} ="b",
				\POS (3.5,-1.5) *{\textcolor{Equity}{\medbullet}} ="c"
				\POS "a" \ar @{->} "b"
				\POS "a" \ar @{->} "c"
			\end{xy} & 15.515 & 26.297 & 7.859 & 6.454 & 1.186\\
			6 & \begin{xy}
				\POS (0,3) *{\textcolor{Agriculture}{\medbullet}} ="a",
				\POS (-3.5,-1.5) *{\textcolor{Energy}{\medbullet}} ="b",
				\POS (3.5,-1.5) *{\textcolor{Equity}{\medbullet}} ="c"
				\POS "a" \ar @{->} "b"
				\POS "a" \ar @{->} "c"
			\end{xy} & 10.761 & 21.479 & 5.440 & 4.948 & 1.076 & 6 & \begin{xy}
				\POS (0,3) *{\textcolor{Agriculture}{\medbullet}} ="a",
				\POS (-3.5,-1.5) *{\textcolor{Energy}{\medbullet}} ="b",
				\POS (3.5,-1.5) *{\textcolor{Agriculture}{\medbullet}} ="c"
				\POS "a" \ar @{->} "b"
				\POS "a" \ar @{->} "c"
			\end{xy} & 18.340 & 32.805 & 8.971 & 7.946 & 1.179\\
			6 & \begin{xy}
				\POS (0,3) *{\textcolor{Agriculture}{\medbullet}} ="a",
				\POS (-3.5,-1.5) *{\textcolor{Energy}{\medbullet}} ="b",
				\POS (3.5,-1.5) *{\textcolor{Agriculture}{\medbullet}} ="c"
				\POS "a" \ar @{->} "b"
				\POS "a" \ar @{->} "c"
			\end{xy} & 12.712 & 25.412 & 6.239 & 6.173 & 1.049 & 6 & \begin{xy}
				\POS (0,3) *{\textcolor{Equity}{\medbullet}} ="a",
				\POS (-3.5,-1.5) *{\textcolor{Agriculture}{\medbullet}} ="b",
				\POS (3.5,-1.5) *{\textcolor{Metal}{\medbullet}} ="c"
				\POS "a" \ar @{->} "b"
				\POS "a" \ar @{->} "c"
			\end{xy} & 28.839 & 55.334 & 15.566 & 12.515 & 1.061\\
			6 & \begin{xy}
				\POS (0,3) *{\textcolor{Equity}{\medbullet}} ="a",
				\POS (-3.5,-1.5) *{\textcolor{Agriculture}{\medbullet}} ="b",
				\POS (3.5,-1.5) *{\textcolor{Metal}{\medbullet}} ="c"
				\POS "a" \ar @{->} "b"
				\POS "a" \ar @{->} "c"
			\end{xy} & 19.843 & 46.008 & 10.740 & 9.914 & 0.918 & 6 & \begin{xy}
				\POS (0,3) *{\textcolor{Agriculture}{\medbullet}} ="a",
				\POS (-3.5,-1.5) *{\textcolor{Agriculture}{\medbullet}} ="b",
				\POS (3.5,-1.5) *{\textcolor{Metal}{\medbullet}} ="c"
				\POS "a" \ar @{->} "b"
				\POS "a" \ar @{->} "c"
			\end{xy} & 53.562 & 82.958 & 35.688 & 17.900 & 0.999\\
			6 & \begin{xy}
				\POS (0,3) *{\textcolor{Equity}{\medbullet}} ="a",
				\POS (-3.5,-1.5) *{\textcolor{Energy}{\medbullet}} ="b",
				\POS (3.5,-1.5) *{\textcolor{Metal}{\medbullet}} ="c"
				\POS "a" \ar @{->} "b"
				\POS "a" \ar @{->} "c"
			\end{xy} & 6.315 & 14.199 & 3.254 & 3.488 & 0.877 & 6 & \begin{xy}
				\POS (0,3) *{\textcolor{Equity}{\medbullet}} ="a",
				\POS (-3.5,-1.5) *{\textcolor{Energy}{\medbullet}} ="b",
				\POS (3.5,-1.5) *{\textcolor{Metal}{\medbullet}} ="c"
				\POS "a" \ar @{->} "b"
				\POS "a" \ar @{->} "c"
			\end{xy} & 8.780 & 16.825 & 4.702 & 4.475 & 0.911\\
			6 & \begin{xy}
				\POS (0,3) *{\textcolor{Agriculture}{\medbullet}} ="a",
				\POS (-3.5,-1.5) *{\textcolor{Agriculture}{\medbullet}} ="b",
				\POS (3.5,-1.5) *{\textcolor{Metal}{\medbullet}} ="c"
				\POS "a" \ar @{->} "b"
				\POS "a" \ar @{->} "c"
			\end{xy} & 36.894 & 64.366 & 24.616 & 14.198 & 0.865 & 46 & \begin{xy}
				\POS (0,3) *{\textcolor{Agriculture}{\medbullet}} ="a",
				\POS (-3.5,-1.5) *{\textcolor{Equity}{\medbullet}} ="b",
				\POS (3.5,-1.5) *{\textcolor{Metal}{\medbullet}} ="c"
				\POS "a" \ar @{<->} "b"
				\POS "a" \ar @{->} "c"
				\POS "b" \ar @{->} "c"
			\end{xy} & 5.701 & 9.898 & 2.876 & 3.164 & 0.893\\
			46 & \begin{xy}
				\POS (0,3) *{\textcolor{Agriculture}{\medbullet}} ="a",
				\POS (-3.5,-1.5) *{\textcolor{Equity}{\medbullet}} ="b",
				\POS (3.5,-1.5) *{\textcolor{Metal}{\medbullet}} ="c"
				\POS "a" \ar @{<->} "b"
				\POS "a" \ar @{->} "c"
				\POS "b" \ar @{->} "c"
			\end{xy} & 3.839 & 6.012 & 1.730 & 2.489 & 0.847 & 46 & \begin{xy}
				\POS (0,3) *{\textcolor{Agriculture}{\medbullet}} ="a",
				\POS (-3.5,-1.5) *{\textcolor{Metal}{\medbullet}} ="b",
				\POS (3.5,-1.5) *{\textcolor{Metal}{\medbullet}} ="c"
				\POS "a" \ar @{<->} "b"
				\POS "a" \ar @{->} "c"
				\POS "b" \ar @{->} "c"
			\end{xy} & 4.044 & 8.048 & 1.868 & 2.716 & 0.801\\
			78 & \begin{xy}
				\POS (0,3) *{\textcolor{Equity}{\medbullet}} ="a",
				\POS (-3.5,-1.5) *{\textcolor{Agriculture}{\medbullet}} ="b",
				\POS (3.5,-1.5) *{\textcolor{Equity}{\medbullet}} ="c"
				\POS "a" \ar @{<->} "b"
				\POS "a" \ar @{<->} "c"
			\end{xy} & 0.396 & 1.617 & 0.104 & 0.384 & 0.761 & 6 & \begin{xy}
				\POS (0,3) *{\textcolor{Agriculture}{\medbullet}} ="a",
				\POS (-3.5,-1.5) *{\textcolor{Agriculture}{\medbullet}} ="b",
				\POS (3.5,-1.5) *{\textcolor{Equity}{\medbullet}} ="c"
				\POS "a" \ar @{->} "b"
				\POS "a" \ar @{->} "c"
			\end{xy} & 37.569 & 58.286 & 24.914 & 16.047 & 0.789\\
			46 & \begin{xy}
				\POS (0,3) *{\textcolor{Agriculture}{\medbullet}} ="a",
				\POS (-3.5,-1.5) *{\textcolor{Metal}{\medbullet}} ="b",
				\POS (3.5,-1.5) *{\textcolor{Metal}{\medbullet}} ="c"
				\POS "a" \ar @{<->} "b"
				\POS "a" \ar @{->} "c"
				\POS "b" \ar @{->} "c"
			\end{xy} & 2.616 & 4.836 & 1.098 & 2.106 & 0.721 & 6 & \begin{xy}
				\POS (0,3) *{\textcolor{Metal}{\medbullet}} ="a",
				\POS (-3.5,-1.5) *{\textcolor{Agriculture}{\medbullet}} ="b",
				\POS (3.5,-1.5) *{\textcolor{Metal}{\medbullet}} ="c"
				\POS "a" \ar @{->} "b"
				\POS "a" \ar @{->} "c"
			\end{xy} & 18.373 & 42.849 & 10.041 & 11.088 & 0.751\\
			6 & \begin{xy}
				\POS (0,3) *{\textcolor{Agriculture}{\medbullet}} ="a",
				\POS (-3.5,-1.5) *{\textcolor{Agriculture}{\medbullet}} ="b",
				\POS (3.5,-1.5) *{\textcolor{Equity}{\medbullet}} ="c"
				\POS "a" \ar @{->} "b"
				\POS "a" \ar @{->} "c"
			\end{xy} & 26.102 & 46.948 & 17.235 & 12.863 & 0.689 & 6 & \begin{xy}
				\POS (0,3) *{\textcolor{Agriculture}{\medbullet}} ="a",
				\POS (-3.5,-1.5) *{\textcolor{Energy}{\medbullet}} ="b",
				\POS (3.5,-1.5) *{\textcolor{Energy}{\medbullet}} ="c"
				\POS "a" \ar @{->} "b"
				\POS "a" \ar @{->} "c"
			\end{xy} & 7.315 & 14.272 & 4.535 & 4.057 & 0.685\\
			108 & \begin{xy}
				\POS (0,3) *{\textcolor{Agriculture}{\medbullet}} ="a",
				\POS (-3.5,-1.5) *{\textcolor{Agriculture}{\medbullet}} ="b",
				\POS (3.5,-1.5) *{\textcolor{Metal}{\medbullet}} ="c"
				\POS "b" \ar @{->} "a"
				\POS "a" \ar @{<->} "c"
				\POS "b" \ar @{->} "c"
			\end{xy} & 1.010 & 2.068 & 0.466 & 0.794 & 0.686 & 78 & \begin{xy}
				\POS (0,3) *{\textcolor{Equity}{\medbullet}} ="a",
				\POS (-3.5,-1.5) *{\textcolor{Agriculture}{\medbullet}} ="b",
				\POS (3.5,-1.5) *{\textcolor{Metal}{\medbullet}} ="c"
				\POS "a" \ar @{<->} "b"
				\POS "a" \ar @{<->} "c"
			\end{xy} & 0.809 & 3.031 & 0.353 & 0.669 & 0.681\\
			6 & \begin{xy}
				\POS (0,3) *{\textcolor{Metal}{\medbullet}} ="a",
				\POS (-3.5,-1.5) *{\textcolor{Energy}{\medbullet}} ="b",
				\POS (3.5,-1.5) *{\textcolor{Metal}{\medbullet}} ="c"
				\POS "a" \ar @{->} "b"
				\POS "a" \ar @{->} "c"
			\end{xy} & 4.241 & 11.435 & 2.163 & 3.134 & 0.663 & 46 & \begin{xy}
				\POS (0,3) *{\textcolor{Agriculture}{\medbullet}} ="a",
				\POS (-3.5,-1.5) *{\textcolor{Agriculture}{\medbullet}} ="b",
				\POS (3.5,-1.5) *{\textcolor{Metal}{\medbullet}} ="c"
				\POS "a" \ar @{<->} "b"
				\POS "a" \ar @{->} "c"
				\POS "b" \ar @{->} "c"
			\end{xy} & 10.698 & 18.123 & 6.598 & 6.076 & 0.675\\
			6 & \begin{xy}
				\POS (0,3) *{\textcolor{Metal}{\medbullet}} ="a",
				\POS (-3.5,-1.5) *{\textcolor{Agriculture}{\medbullet}} ="b",
				\POS (3.5,-1.5) *{\textcolor{Metal}{\medbullet}} ="c"
				\POS "a" \ar @{->} "b"
				\POS "a" \ar @{->} "c"
			\end{xy} & 12.782 & 34.908 & 6.948 & 8.932 & 0.653 & 6 & \begin{xy}
				\POS (0,3) *{\textcolor{Metal}{\medbullet}} ="a",
				\POS (-3.5,-1.5) *{\textcolor{Energy}{\medbullet}} ="b",
				\POS (3.5,-1.5) *{\textcolor{Metal}{\medbullet}} ="c"
				\POS "a" \ar @{->} "b"
				\POS "a" \ar @{->} "c"
			\end{xy} & 5.625 & 13.189 & 3.070 & 3.865 & 0.661\\
			108 & \begin{xy}
				\POS (0,3) *{\textcolor{Agriculture}{\medbullet}} ="a",
				\POS (-3.5,-1.5) *{\textcolor{Equity}{\medbullet}} ="b",
				\POS (3.5,-1.5) *{\textcolor{Equity}{\medbullet}} ="c"
				\POS "b" \ar @{->} "a"
				\POS "a" \ar @{<->} "c"
				\POS "b" \ar @{->} "c"
			\end{xy} & 0.256 & 0.879 & 0.073 & 0.284 & 0.645 & 78 & \begin{xy}
				\POS (0,3) *{\textcolor{Metal}{\medbullet}} ="a",
				\POS (-3.5,-1.5) *{\textcolor{Agriculture}{\medbullet}} ="b",
				\POS (3.5,-1.5) *{\textcolor{Metal}{\medbullet}} ="c"
				\POS "a" \ar @{<->} "b"
				\POS "a" \ar @{<->} "c"
			\end{xy} & 0.606 & 2.938 & 0.239 & 0.584 & 0.628\\
			78 & \begin{xy}
				\POS (0,3) *{\textcolor{Metal}{\medbullet}} ="a",
				\POS (-3.5,-1.5) *{\textcolor{Agriculture}{\medbullet}} ="b",
				\POS (3.5,-1.5) *{\textcolor{Metal}{\medbullet}} ="c"
				\POS "a" \ar @{<->} "b"
				\POS "a" \ar @{<->} "c"
			\end{xy} & 0.571 & 1.801 & 0.207 & 0.567 & 0.642 & 6 & \begin{xy}
				\POS (0,3) *{\textcolor{Equity}{\medbullet}} ="a",
				\POS (-3.5,-1.5) *{\textcolor{Agriculture}{\medbullet}} ="b",
				\POS (3.5,-1.5) *{\textcolor{Equity}{\medbullet}} ="c"
				\POS "a" \ar @{->} "b"
				\POS "a" \ar @{->} "c"
			\end{xy} & 8.958 & 22.061 & 4.750 & 6.738 & 0.625\\
			46 & \begin{xy}
				\POS (0,3) *{\textcolor{Agriculture}{\medbullet}} ="a",
				\POS (-3.5,-1.5) *{\textcolor{Agriculture}{\medbullet}} ="b",
				\POS (3.5,-1.5) *{\textcolor{Metal}{\medbullet}} ="c"
				\POS "a" \ar @{<->} "b"
				\POS "a" \ar @{->} "c"
				\POS "b" \ar @{->} "c"
			\end{xy} & 6.544 & 10.347 & 3.951 & 4.574 & 0.567 & 6 & \begin{xy}
				\POS (0,3) *{\textcolor{Equity}{\medbullet}} ="a",
				\POS (-3.5,-1.5) *{\textcolor{Metal}{\medbullet}} ="b",
				\POS (3.5,-1.5) *{\textcolor{Equity}{\medbullet}} ="c"
				\POS "a" \ar @{->} "b"
				\POS "a" \ar @{->} "c"
			\end{xy} & 5.502 & 13.349 & 2.895 & 4.253 & 0.613\\
			78 & \begin{xy}
				\POS (0,3) *{\textcolor{Agriculture}{\medbullet}} ="a",
				\POS (-3.5,-1.5) *{\textcolor{Energy}{\medbullet}} ="b",
				\POS (3.5,-1.5) *{\textcolor{Equity}{\medbullet}} ="c"
				\POS "a" \ar @{<->} "b"
				\POS "a" \ar @{<->} "c"
			\end{xy} & 0.389 & 1.420 & 0.156 & 0.412 & 0.566 & 78 & \begin{xy}
				\POS (0,3) *{\textcolor{Equity}{\medbullet}} ="a",
				\POS (-3.5,-1.5) *{\textcolor{Energy}{\medbullet}} ="b",
				\POS (3.5,-1.5) *{\textcolor{Metal}{\medbullet}} ="c"
				\POS "a" \ar @{<->} "b"
				\POS "a" \ar @{<->} "c"
			\end{xy} & 0.317 & 1.274 & 0.120 & 0.339 & 0.581\\
			78 & \begin{xy}
				\POS (0,3) *{\textcolor{Energy}{\medbullet}} ="a",
				\POS (-3.5,-1.5) *{\textcolor{Energy}{\medbullet}} ="b",
				\POS (3.5,-1.5) *{\textcolor{Equity}{\medbullet}} ="c"
				\POS "a" \ar @{<->} "b"
				\POS "a" \ar @{<->} "c"
			\end{xy} & 0.206 & 0.631 & 0.061 & 0.259 & 0.559 & 78 & \begin{xy}
				\POS (0,3) *{\textcolor{Agriculture}{\medbullet}} ="a",
				\POS (-3.5,-1.5) *{\textcolor{Energy}{\medbullet}} ="b",
				\POS (3.5,-1.5) *{\textcolor{Metal}{\medbullet}} ="c"
				\POS "a" \ar @{<->} "b"
				\POS "a" \ar @{<->} "c"
			\end{xy} & 0.576 & 1.863 & 0.278 & 0.529 & 0.564\\
			6 & \begin{xy}
				\POS (0,3) *{\textcolor{Equity}{\medbullet}} ="a",
				\POS (-3.5,-1.5) *{\textcolor{Agriculture}{\medbullet}} ="b",
				\POS (3.5,-1.5) *{\textcolor{Equity}{\medbullet}} ="c"
				\POS "a" \ar @{->} "b"
				\POS "a" \ar @{->} "c"
			\end{xy} & 6.201 & 19.899 & 3.283 & 5.381 & 0.542 & 6 & \begin{xy}
				\POS (0,3) *{\textcolor{Equity}{\medbullet}} ="a",
				\POS (-3.5,-1.5) *{\textcolor{Energy}{\medbullet}} ="b",
				\POS (3.5,-1.5) *{\textcolor{Equity}{\medbullet}} ="c"
				\POS "a" \ar @{->} "b"
				\POS "a" \ar @{->} "c"
			\end{xy} & 2.833 & 6.810 & 1.498 & 2.387 & 0.559\\
			6 & \begin{xy}
				\POS (0,3) *{\textcolor{Equity}{\medbullet}} ="a",
				\POS (-3.5,-1.5) *{\textcolor{Energy}{\medbullet}} ="b",
				\POS (3.5,-1.5) *{\textcolor{Equity}{\medbullet}} ="c"
				\POS "a" \ar @{->} "b"
				\POS "a" \ar @{->} "c"
			\end{xy} & 2.117 & 6.312 & 1.082 & 1.975 & 0.524 & 6 & \begin{xy}
				\POS (0,3) *{\textcolor{Energy}{\medbullet}} ="a",
				\POS (-3.5,-1.5) *{\textcolor{Energy}{\medbullet}} ="b",
				\POS (3.5,-1.5) *{\textcolor{Metal}{\medbullet}} ="c"
				\POS "a" \ar @{->} "b"
				\POS "a" \ar @{->} "c"
			\end{xy} & 4.264 & 8.695 & 2.725 & 2.808 & 0.548\\
			102 & \begin{xy}
				\POS (0,3) *{\textcolor{Equity}{\medbullet}} ="a",
				\POS (-3.5,-1.5) *{\textcolor{Energy}{\medbullet}} ="b",
				\POS (3.5,-1.5) *{\textcolor{Metal}{\medbullet}} ="c"
				\POS "a" \ar @{->} "b"
				\POS "a" \ar @{<->} "c"
				\POS "b" \ar @{->} "c"
			\end{xy} & 0.152 & 0.509 & 0.041 & 0.214 & 0.519 & 78 & \begin{xy}
				\POS (0,3) *{\textcolor{Equity}{\medbullet}} ="a",
				\POS (-3.5,-1.5) *{\textcolor{Energy}{\medbullet}} ="b",
				\POS (3.5,-1.5) *{\textcolor{Equity}{\medbullet}} ="c"
				\POS "a" \ar @{<->} "b"
				\POS "a" \ar @{<->} "c"
			\end{xy} & 0.171 & 1.022 & 0.048 & 0.228 & 0.542\\
			78 & \begin{xy}
				\POS (0,3) *{\textcolor{Equity}{\medbullet}} ="a",
				\POS (-3.5,-1.5) *{\textcolor{Equity}{\medbullet}} ="b",
				\POS (3.5,-1.5) *{\textcolor{Equity}{\medbullet}} ="c"
				\POS "a" \ar @{<->} "b"
				\POS "a" \ar @{<->} "c"
			\end{xy} & 0.189 & 0.691 & 0.050 & 0.269 & 0.517 & 108 & \begin{xy}
				\POS (0,3) *{\textcolor{Energy}{\medbullet}} ="a",
				\POS (-3.5,-1.5) *{\textcolor{Agriculture}{\medbullet}} ="b",
				\POS (3.5,-1.5) *{\textcolor{Agriculture}{\medbullet}} ="c"
				\POS "b" \ar @{->} "a"
				\POS "a" \ar @{<->} "c"
				\POS "b" \ar @{->} "c"
			\end{xy} & 0.412 & 1.230 & 0.153 & 0.478 & 0.541\\
			6 & \begin{xy}
				\POS (0,3) *{\textcolor{Agriculture}{\medbullet}} ="a",
				\POS (-3.5,-1.5) *{\textcolor{Energy}{\medbullet}} ="b",
				\POS (3.5,-1.5) *{\textcolor{Energy}{\medbullet}} ="c"
				\POS "a" \ar @{->} "b"
				\POS "a" \ar @{->} "c"
			\end{xy} & 4.955 & 11.547 & 3.221 & 3.357 & 0.516 & 46 & \begin{xy}
				\POS (0,3) *{\textcolor{Energy}{\medbullet}} ="a",
				\POS (-3.5,-1.5) *{\textcolor{Agriculture}{\medbullet}} ="b",
				\POS (3.5,-1.5) *{\textcolor{Metal}{\medbullet}} ="c"
				\POS "a" \ar @{<->} "b"
				\POS "a" \ar @{->} "c"
				\POS "b" \ar @{->} "c"
			\end{xy} & 3.885 & 9.568 & 2.020 & 3.665 & 0.509\\
			6 & \begin{xy}
				\POS (0,3) *{\textcolor{Equity}{\medbullet}} ="a",
				\POS (-3.5,-1.5) *{\textcolor{Metal}{\medbullet}} ="b",
				\POS (3.5,-1.5) *{\textcolor{Equity}{\medbullet}} ="c"
				\POS "a" \ar @{->} "b"
				\POS "a" \ar @{->} "c"
			\end{xy} & 3.801 & 11.863 & 2.023 & 3.445 & 0.516 & 78 & \begin{xy}
				\POS (0,3) *{\textcolor{Equity}{\medbullet}} ="a",
				\POS (-3.5,-1.5) *{\textcolor{Metal}{\medbullet}} ="b",
				\POS (3.5,-1.5) *{\textcolor{Metal}{\medbullet}} ="c"
				\POS "a" \ar @{<->} "b"
				\POS "a" \ar @{<->} "c"
			\end{xy} & 0.428 & 2.336 & 0.195 & 0.470 & 0.496\\
			102 & \begin{xy}
				\POS (0,3) *{\textcolor{Agriculture}{\medbullet}} ="a",
				\POS (-3.5,-1.5) *{\textcolor{Equity}{\medbullet}} ="b",
				\POS (3.5,-1.5) *{\textcolor{Energy}{\medbullet}} ="c"
				\POS "a" \ar @{->} "b"
				\POS "a" \ar @{<->} "c"
				\POS "b" \ar @{->} "c"
			\end{xy} & 0.207 & 0.825 & 0.069 & 0.272 & 0.509 & 6 & \begin{xy}
				\POS (0,3) *{\textcolor{Equity}{\medbullet}} ="a",
				\POS (-3.5,-1.5) *{\textcolor{Metal}{\medbullet}} ="b",
				\POS (3.5,-1.5) *{\textcolor{Metal}{\medbullet}} ="c"
				\POS "a" \ar @{->} "b"
				\POS "a" \ar @{->} "c"
			\end{xy} & 11.611 & 22.370 & 8.303 & 6.670 & 0.496\\
			6 & \begin{xy}
				\POS (0,3) *{\textcolor{Energy}{\medbullet}} ="a",
				\POS (-3.5,-1.5) *{\textcolor{Energy}{\medbullet}} ="b",
				\POS (3.5,-1.5) *{\textcolor{Metal}{\medbullet}} ="c"
				\POS "a" \ar @{->} "b"
				\POS "a" \ar @{->} "c"
			\end{xy} & 3.140 & 7.468 & 1.981 & 2.397 & 0.483 & 78 & \begin{xy}
				\POS (0,3) *{\textcolor{Metal}{\medbullet}} ="a",
				\POS (-3.5,-1.5) *{\textcolor{Energy}{\medbullet}} ="b",
				\POS (3.5,-1.5) *{\textcolor{Metal}{\medbullet}} ="c"
				\POS "a" \ar @{<->} "b"
				\POS "a" \ar @{<->} "c"
			\end{xy} & 0.239 & 1.448 & 0.091 & 0.307 & 0.484\\
			6 & \begin{xy}
				\POS (0,3) *{\textcolor{Equity}{\medbullet}} ="a",
				\POS (-3.5,-1.5) *{\textcolor{Metal}{\medbullet}} ="b",
				\POS (3.5,-1.5) *{\textcolor{Metal}{\medbullet}} ="c"
				\POS "a" \ar @{->} "b"
				\POS "a" \ar @{->} "c"
			\end{xy} & 8.230 & 18.719 & 5.745 & 5.401 & 0.460 & 108 & \begin{xy}
				\POS (0,3) *{\textcolor{Agriculture}{\medbullet}} ="a",
				\POS (-3.5,-1.5) *{\textcolor{Agriculture}{\medbullet}} ="b",
				\POS (3.5,-1.5) *{\textcolor{Metal}{\medbullet}} ="c"
				\POS "b" \ar @{->} "a"
				\POS "a" \ar @{<->} "c"
				\POS "b" \ar @{->} "c"
			\end{xy} & 1.011 & 2.148 & 0.590 & 0.875 & 0.482\\
			78 & \begin{xy}
				\POS (0,3) *{\textcolor{Equity}{\medbullet}} ="a",
				\POS (-3.5,-1.5) *{\textcolor{Agriculture}{\medbullet}} ="b",
				\POS (3.5,-1.5) *{\textcolor{Metal}{\medbullet}} ="c"
				\POS "a" \ar @{<->} "b"
				\POS "a" \ar @{<->} "c"
			\end{xy} & 0.561 & 1.364 & 0.295 & 0.615 & 0.434 & 6 & \begin{xy}
				\POS (0,3) *{\textcolor{Energy}{\medbullet}} ="a",
				\POS (-3.5,-1.5) *{\textcolor{Energy}{\medbullet}} ="b",
				\POS (3.5,-1.5) *{\textcolor{Equity}{\medbullet}} ="c"
				\POS "a" \ar @{->} "b"
				\POS "a" \ar @{->} "c"
			\end{xy} & 3.045 & 6.282 & 1.965 & 2.275 & 0.475\\
			46 & \begin{xy}
				\POS (0,3) *{\textcolor{Agriculture}{\medbullet}} ="a",
				\POS (-3.5,-1.5) *{\textcolor{Equity}{\medbullet}} ="b",
				\POS (3.5,-1.5) *{\textcolor{Equity}{\medbullet}} ="c"
				\POS "a" \ar @{<->} "b"
				\POS "a" \ar @{->} "c"
				\POS "b" \ar @{->} "c"
			\end{xy} & 1.079 & 2.575 & 0.534 & 1.262 & 0.431 & 46 & \begin{xy}
				\POS (0,3) *{\textcolor{Energy}{\medbullet}} ="a",
				\POS (-3.5,-1.5) *{\textcolor{Equity}{\medbullet}} ="b",
				\POS (3.5,-1.5) *{\textcolor{Metal}{\medbullet}} ="c"
				\POS "a" \ar @{<->} "b"
				\POS "a" \ar @{->} "c"
				\POS "b" \ar @{->} "c"
			\end{xy} & 1.689 & 4.739 & 0.855 & 1.845 & 0.452\\
			\bottomrule
	\end{tabular}}
	\begin{flushleft}
		\footnotesize
		\textit{Legend: }
		\textcolor{Equity}{{\normalsize\textbullet}}\, Equity;
		\textcolor{Metal}{{\normalsize\textbullet}}\, Metal;
		\textcolor{Agriculture}{{\normalsize\textbullet}}\, Agriculture;
		\textcolor{Energy}{{\normalsize\textbullet}}\,  Energy.
	\end{flushleft} 
\end{table}


\begin{table}[htp]
	\centering
	\setlength{\abovecaptionskip}{0pt}
	\setlength{\belowcaptionskip}{10pt}
	\caption{Colored triad motifs at the right quantile (seven classes).}
	\label{tab:colored_triads_7class_R}
	\resizebox{\textwidth}{!}{
		\begin{tabular}{llrrrrr|llrrrrr}
			\toprule
			\multicolumn{7}{c|}{$\alpha=0.05$} & \multicolumn{7}{c}{$\alpha=0.10$}\\
			\cmidrule(lr){1-7} \cmidrule(lr){8-14}
			id & motif & $\mu$ & $\sigma$ & $\mu_{\text{rnd}}$ & $\sigma_{\text{rnd}}$ & $z$ &
			id & motif & $\mu$ & $\sigma$ & $\mu_{\text{rnd}}$ & $\sigma_{\text{rnd}}$ & $z$\\
			\midrule
			  78 & \begin{xy}
				\POS (0,3) *{\textcolor{Equity}{\medbullet}} ="a",
				\POS (-3.5,-1.5) *{\textcolor{Energy}{\medbullet}} ="b",
				\POS (3.5,-1.5) *{\textcolor{Equity}{\medbullet}} ="c"
				\POS "a" \ar @{<->} "b"
				\POS "a" \ar @{<->} "c"
			\end{xy} & 0.567 & 1.909 & 0.046 & 0.227 & 2.295 & 6 & \begin{xy}
				\POS (0,3) *{\textcolor{Equity}{\medbullet}} ="a",
				\POS (-3.5,-1.5) *{\textcolor{Energy}{\medbullet}} ="b",
				\POS (3.5,-1.5) *{\textcolor{GrainsOilseeds}{\medbullet}} ="c"
				\POS "a" \ar @{->} "b"
				\POS "a" \ar @{->} "c"
			\end{xy} & 6.894 & 13.121 & 3.314 & 3.547 & 1.009\\
			110 & \begin{xy}
				\POS (0,3) *{\textcolor{Preciousmetals}{\medbullet}} ="a",
				\POS (-3.5,-1.5) *{\textcolor{Livestock}{\medbullet}} ="b",
				\POS (3.5,-1.5) *{\textcolor{Livestock}{\medbullet}} ="c"
				\POS "a" \ar @{<->} "b"
				\POS "a" \ar @{<->} "c"
				\POS "b" \ar @{->} "c"
			\end{xy} & 0.182 & 0.575 & 0.014 & 0.126 & 1.339 & 6 & \begin{xy}
				\POS (0,3) *{\textcolor{GrainsOilseeds}{\medbullet}} ="a",
				\POS (-3.5,-1.5) *{\textcolor{Energy}{\medbullet}} ="b",
				\POS (3.5,-1.5) *{\textcolor{Softs}{\medbullet}} ="c"
				\POS "a" \ar @{->} "b"
				\POS "a" \ar @{->} "c"
			\end{xy} & 6.110 & 11.882 & 2.869 & 3.233 & 1.003\\
			102 & \begin{xy}
				\POS (0,3) *{\textcolor{Equity}{\medbullet}} ="a",
				\POS (-3.5,-1.5) *{\textcolor{Preciousmetals}{\medbullet}} ="b",
				\POS (3.5,-1.5) *{\textcolor{GrainsOilseeds}{\medbullet}} ="c"
				\POS "a" \ar @{->} "b"
				\POS "a" \ar @{<->} "c"
				\POS "b" \ar @{->} "c"
			\end{xy} & 0.250 & 0.722 & 0.034 & 0.186 & 1.161 & 6 & \begin{xy}
				\POS (0,3) *{\textcolor{GrainsOilseeds}{\medbullet}} ="a",
				\POS (-3.5,-1.5) *{\textcolor{Energy}{\medbullet}} ="b",
				\POS (3.5,-1.5) *{\textcolor{Preciousmetals}{\medbullet}} ="c"
				\POS "a" \ar @{->} "b"
				\POS "a" \ar @{->} "c"
			\end{xy} & 4.301 & 8.601 & 1.968 & 2.579 & 0.905\\
			6 & \begin{xy}
				\POS (0,3) *{\textcolor{Equity}{\medbullet}} ="a",
				\POS (-3.5,-1.5) *{\textcolor{Energy}{\medbullet}} ="b",
				\POS (3.5,-1.5) *{\textcolor{GrainsOilseeds}{\medbullet}} ="c"
				\POS "a" \ar @{->} "b"
				\POS "a" \ar @{->} "c"
			\end{xy} & 5.097 & 11.083 & 2.313 & 2.823 & 0.986 & 6 & \begin{xy}
				\POS (0,3) *{\textcolor{GrainsOilseeds}{\medbullet}} ="a",
				\POS (-3.5,-1.5) *{\textcolor{Energy}{\medbullet}} ="b",
				\POS (3.5,-1.5) *{\textcolor{Industrialmetals}{\medbullet}} ="c"
				\POS "a" \ar @{->} "b"
				\POS "a" \ar @{->} "c"
			\end{xy} & 5.598 & 11.553 & 2.837 & 3.194 & 0.864\\
			110 & \begin{xy}
				\POS (0,3) *{\textcolor{GrainsOilseeds}{\medbullet}} ="a",
				\POS (-3.5,-1.5) *{\textcolor{Preciousmetals}{\medbullet}} ="b",
				\POS (3.5,-1.5) *{\textcolor{Preciousmetals}{\medbullet}} ="c"
				\POS "a" \ar @{<->} "b"
				\POS "a" \ar @{<->} "c"
				\POS "b" \ar @{->} "c"
			\end{xy} & 0.121 & 0.409 & 0.013 & 0.113 & 0.958 & 6 & \begin{xy}
				\POS (0,3) *{\textcolor{Equity}{\medbullet}} ="a",
				\POS (-3.5,-1.5) *{\textcolor{GrainsOilseeds}{\medbullet}} ="b",
				\POS (3.5,-1.5) *{\textcolor{Softs}{\medbullet}} ="c"
				\POS "a" \ar @{->} "b"
				\POS "a" \ar @{->} "c"
			\end{xy} & 7.595 & 14.775 & 3.958 & 4.224 & 0.861\\
			6 & \begin{xy}
				\POS (0,3) *{\textcolor{GrainsOilseeds}{\medbullet}} ="a",
				\POS (-3.5,-1.5) *{\textcolor{Energy}{\medbullet}} ="b",
				\POS (3.5,-1.5) *{\textcolor{Softs}{\medbullet}} ="c"
				\POS "a" \ar @{->} "b"
				\POS "a" \ar @{->} "c"
			\end{xy} & 4.505 & 9.597 & 2.042 & 2.615 & 0.942 & 78 & \begin{xy}
				\POS (0,3) *{\textcolor{Industrialmetals}{\medbullet}} ="a",
				\POS (-3.5,-1.5) *{\textcolor{Softs}{\medbullet}} ="b",
				\POS (3.5,-1.5) *{\textcolor{Industrialmetals}{\medbullet}} ="c"
				\POS "a" \ar @{<->} "b"
				\POS "a" \ar @{<->} "c"
			\end{xy} & 0.252 & 1.394 & 0.052 & 0.239 & 0.837\\
			102 & \begin{xy}
				\POS (0,3) *{\textcolor{Equity}{\medbullet}} ="a",
				\POS (-3.5,-1.5) *{\textcolor{Equity}{\medbullet}} ="b",
				\POS (3.5,-1.5) *{\textcolor{Preciousmetals}{\medbullet}} ="c"
				\POS "a" \ar @{->} "b"
				\POS "a" \ar @{<->} "c"
				\POS "b" \ar @{->} "c"
			\end{xy} & 0.219 & 0.739 & 0.034 & 0.203 & 0.912 & 6 & \begin{xy}
				\POS (0,3) *{\textcolor{Equity}{\medbullet}} ="a",
				\POS (-3.5,-1.5) *{\textcolor{Energy}{\medbullet}} ="b",
				\POS (3.5,-1.5) *{\textcolor{Softs}{\medbullet}} ="c"
				\POS "a" \ar @{->} "b"
				\POS "a" \ar @{->} "c"
			\end{xy} & 5.570 & 11.010 & 2.865 & 3.232 & 0.837\\
			6 & \begin{xy}
				\POS (0,3) *{\textcolor{GrainsOilseeds}{\medbullet}} ="a",
				\POS (-3.5,-1.5) *{\textcolor{Energy}{\medbullet}} ="b",
				\POS (3.5,-1.5) *{\textcolor{Preciousmetals}{\medbullet}} ="c"
				\POS "a" \ar @{->} "b"
				\POS "a" \ar @{->} "c"
			\end{xy} & 3.337 & 7.262 & 1.414 & 2.120 & 0.907 & 6 & \begin{xy}
				\POS (0,3) *{\textcolor{Preciousmetals}{\medbullet}} ="a",
				\POS (-3.5,-1.5) *{\textcolor{GrainsOilseeds}{\medbullet}} ="b",
				\POS (3.5,-1.5) *{\textcolor{Softs}{\medbullet}} ="c"
				\POS "a" \ar @{->} "b"
				\POS "a" \ar @{->} "c"
			\end{xy} & 4.811 & 9.896 & 2.329 & 3.045 & 0.815\\
			102 & \begin{xy}
				\POS (0,3) *{\textcolor{Equity}{\medbullet}} ="a",
				\POS (-3.5,-1.5) *{\textcolor{Energy}{\medbullet}} ="b",
				\POS (3.5,-1.5) *{\textcolor{Preciousmetals}{\medbullet}} ="c"
				\POS "a" \ar @{->} "b"
				\POS "a" \ar @{<->} "c"
				\POS "b" \ar @{->} "c"
			\end{xy} & 0.214 & 0.558 & 0.035 & 0.203 & 0.886 & 6 & \begin{xy}
				\POS (0,3) *{\textcolor{Preciousmetals}{\medbullet}} ="a",
				\POS (-3.5,-1.5) *{\textcolor{Energy}{\medbullet}} ="b",
				\POS (3.5,-1.5) *{\textcolor{Softs}{\medbullet}} ="c"
				\POS "a" \ar @{->} "b"
				\POS "a" \ar @{->} "c"
			\end{xy} & 3.539 & 7.498 & 1.690 & 2.314 & 0.799\\
			102 & \begin{xy}
				\POS (0,3) *{\textcolor{Industrialmetals}{\medbullet}} ="a",
				\POS (-3.5,-1.5) *{\textcolor{Preciousmetals}{\medbullet}} ="b",
				\POS (3.5,-1.5) *{\textcolor{GrainsOilseeds}{\medbullet}} ="c"
				\POS "a" \ar @{->} "b"
				\POS "a" \ar @{<->} "c"
				\POS "b" \ar @{->} "c"
			\end{xy} & 0.188 & 0.726 & 0.029 & 0.181 & 0.877 & 6 & \begin{xy}
				\POS (0,3) *{\textcolor{GrainsOilseeds}{\medbullet}} ="a",
				\POS (-3.5,-1.5) *{\textcolor{Energy}{\medbullet}} ="b",
				\POS (3.5,-1.5) *{\textcolor{GrainsOilseeds}{\medbullet}} ="c"
				\POS "a" \ar @{->} "b"
				\POS "a" \ar @{->} "c"
			\end{xy} & 3.355 & 7.708 & 1.486 & 2.371 & 0.788\\
			78 & \begin{xy}
				\POS (0,3) *{\textcolor{Equity}{\medbullet}} ="a",
				\POS (-3.5,-1.5) *{\textcolor{Preciousmetals}{\medbullet}} ="b",
				\POS (3.5,-1.5) *{\textcolor{Equity}{\medbullet}} ="c"
				\POS "a" \ar @{<->} "b"
				\POS "a" \ar @{<->} "c"
			\end{xy} & 0.250 & 0.829 & 0.042 & 0.239 & 0.871 & 6 & \begin{xy}
				\POS (0,3) *{\textcolor{GrainsOilseeds}{\medbullet}} ="a",
				\POS (-3.5,-1.5) *{\textcolor{Energy}{\medbullet}} ="b",
				\POS (3.5,-1.5) *{\textcolor{Livestock}{\medbullet}} ="c"
				\POS "a" \ar @{->} "b"
				\POS "a" \ar @{->} "c"
			\end{xy} & 3.918 & 8.155 & 1.961 & 2.569 & 0.762\\
			102 & \begin{xy}
				\POS (0,3) *{\textcolor{Industrialmetals}{\medbullet}} ="a",
				\POS (-3.5,-1.5) *{\textcolor{Industrialmetals}{\medbullet}} ="b",
				\POS (3.5,-1.5) *{\textcolor{Livestock}{\medbullet}} ="c"
				\POS "a" \ar @{->} "b"
				\POS "a" \ar @{<->} "c"
				\POS "b" \ar @{->} "c"
			\end{xy} & 0.179 & 0.658 & 0.027 & 0.176 & 0.861 & 6 & \begin{xy}
				\POS (0,3) *{\textcolor{Equity}{\medbullet}} ="a",
				\POS (-3.5,-1.5) *{\textcolor{Softs}{\medbullet}} ="b",
				\POS (3.5,-1.5) *{\textcolor{Preciousmetals}{\medbullet}} ="c"
				\POS "a" \ar @{->} "b"
				\POS "a" \ar @{->} "c"
			\end{xy} & 4.582 & 9.653 & 2.324 & 2.977 & 0.759\\
			108 & \begin{xy}
				\POS (0,3) *{\textcolor{Livestock}{\medbullet}} ="a",
				\POS (-3.5,-1.5) *{\textcolor{Preciousmetals}{\medbullet}} ="b",
				\POS (3.5,-1.5) *{\textcolor{Preciousmetals}{\medbullet}} ="c"
				\POS "b" \ar @{->} "a"
				\POS "a" \ar @{<->} "c"
				\POS "b" \ar @{->} "c"
			\end{xy} & 0.094 & 0.291 & 0.011 & 0.097 & 0.860 & 6 & \begin{xy}
				\POS (0,3) *{\textcolor{Equity}{\medbullet}} ="a",
				\POS (-3.5,-1.5) *{\textcolor{GrainsOilseeds}{\medbullet}} ="b",
				\POS (3.5,-1.5) *{\textcolor{Preciousmetals}{\medbullet}} ="c"
				\POS "a" \ar @{->} "b"
				\POS "a" \ar @{->} "c"
			\end{xy} & 5.268 & 10.825 & 2.708 & 3.380 & 0.758\\
			78 & \begin{xy}
				\POS (0,3) *{\textcolor{Industrialmetals}{\medbullet}} ="a",
				\POS (-3.5,-1.5) *{\textcolor{GrainsOilseeds}{\medbullet}} ="b",
				\POS (3.5,-1.5) *{\textcolor{Industrialmetals}{\medbullet}} ="c"
				\POS "a" \ar @{<->} "b"
				\POS "a" \ar @{<->} "c"
			\end{xy} & 0.222 & 0.786 & 0.039 & 0.213 & 0.859 & 6 & \begin{xy}
				\POS (0,3) *{\textcolor{Equity}{\medbullet}} ="a",
				\POS (-3.5,-1.5) *{\textcolor{Energy}{\medbullet}} ="b",
				\POS (3.5,-1.5) *{\textcolor{Preciousmetals}{\medbullet}} ="c"
				\POS "a" \ar @{->} "b"
				\POS "a" \ar @{->} "c"
			\end{xy} & 3.891 & 8.222 & 1.955 & 2.566 & 0.754\\
			6 & \begin{xy}
				\POS (0,3) *{\textcolor{GrainsOilseeds}{\medbullet}} ="a",
				\POS (-3.5,-1.5) *{\textcolor{Energy}{\medbullet}} ="b",
				\POS (3.5,-1.5) *{\textcolor{Industrialmetals}{\medbullet}} ="c"
				\POS "a" \ar @{->} "b"
				\POS "a" \ar @{->} "c"
			\end{xy} & 4.290 & 9.822 & 2.047 & 2.627 & 0.853 & 6 & \begin{xy}
				\POS (0,3) *{\textcolor{Equity}{\medbullet}} ="a",
				\POS (-3.5,-1.5) *{\textcolor{Livestock}{\medbullet}} ="b",
				\POS (3.5,-1.5) *{\textcolor{Preciousmetals}{\medbullet}} ="c"
				\POS "a" \ar @{->} "b"
				\POS "a" \ar @{->} "c"
			\end{xy} & 3.135 & 6.689 & 1.572 & 2.135 & 0.732\\
			6 & \begin{xy}
				\POS (0,3) *{\textcolor{GrainsOilseeds}{\medbullet}} ="a",
				\POS (-3.5,-1.5) *{\textcolor{Energy}{\medbullet}} ="b",
				\POS (3.5,-1.5) *{\textcolor{GrainsOilseeds}{\medbullet}} ="c"
				\POS "a" \ar @{->} "b"
				\POS "a" \ar @{->} "c"
			\end{xy} & 2.706 & 6.796 & 1.083 & 1.996 & 0.813 & 6 & \begin{xy}
				\POS (0,3) *{\textcolor{Industrialmetals}{\medbullet}} ="a",
				\POS (-3.5,-1.5) *{\textcolor{GrainsOilseeds}{\medbullet}} ="b",
				\POS (3.5,-1.5) *{\textcolor{Softs}{\medbullet}} ="c"
				\POS "a" \ar @{->} "b"
				\POS "a" \ar @{->} "c"
			\end{xy} & 6.187 & 13.251 & 3.443 & 3.872 & 0.709\\
			98 & \begin{xy}
				\POS (0,3) *{\textcolor{GrainsOilseeds}{\medbullet}} ="a",
				\POS (-3.5,-1.5) *{\textcolor{Industrialmetals}{\medbullet}} ="b",
				\POS (3.5,-1.5) *{\textcolor{Industrialmetals}{\medbullet}} ="c"
				\POS "a" \ar @{->} "b"
				\POS "c" \ar @{->} "a"
				\POS "b" \ar @{->} "c"
			\end{xy} & 0.222 & 0.831 & 0.047 & 0.217 & 0.804 & 6 & \begin{xy}
				\POS (0,3) *{\textcolor{Equity}{\medbullet}} ="a",
				\POS (-3.5,-1.5) *{\textcolor{Softs}{\medbullet}} ="b",
				\POS (3.5,-1.5) *{\textcolor{Industrialmetals}{\medbullet}} ="c"
				\POS "a" \ar @{->} "b"
				\POS "a" \ar @{->} "c"
			\end{xy} & 5.996 & 13.035 & 3.401 & 3.716 & 0.698\\
			108 & \begin{xy}
				\POS (0,3) *{\textcolor{Softs}{\medbullet}} ="a",
				\POS (-3.5,-1.5) *{\textcolor{Equity}{\medbullet}} ="b",
				\POS (3.5,-1.5) *{\textcolor{Equity}{\medbullet}} ="c"
				\POS "b" \ar @{->} "a"
				\POS "a" \ar @{<->} "c"
				\POS "b" \ar @{->} "c"
			\end{xy} & 0.220 & 0.865 & 0.040 & 0.225 & 0.803 & 6 & \begin{xy}
				\POS (0,3) *{\textcolor{Equity}{\medbullet}} ="a",
				\POS (-3.5,-1.5) *{\textcolor{GrainsOilseeds}{\medbullet}} ="b",
				\POS (3.5,-1.5) *{\textcolor{Industrialmetals}{\medbullet}} ="c"
				\POS "a" \ar @{->} "b"
				\POS "a" \ar @{->} "c"
			\end{xy} & 6.925 & 14.833 & 3.972 & 4.245 & 0.696\\
			108 & \begin{xy}
				\POS (0,3) *{\textcolor{Softs}{\medbullet}} ="a",
				\POS (-3.5,-1.5) *{\textcolor{Equity}{\medbullet}} ="b",
				\POS (3.5,-1.5) *{\textcolor{Industrialmetals}{\medbullet}} ="c"
				\POS "b" \ar @{->} "a"
				\POS "a" \ar @{<->} "c"
				\POS "b" \ar @{->} "c"
			\end{xy} & 0.244 & 0.790 & 0.056 & 0.241 & 0.781 & 6 & \begin{xy}
				\POS (0,3) *{\textcolor{Industrialmetals}{\medbullet}} ="a",
				\POS (-3.5,-1.5) *{\textcolor{Softs}{\medbullet}} ="b",
				\POS (3.5,-1.5) *{\textcolor{Preciousmetals}{\medbullet}} ="c"
				\POS "a" \ar @{->} "b"
				\POS "a" \ar @{->} "c"
			\end{xy} & 3.846 & 8.880 & 1.999 & 2.666 & 0.693\\
			110 & \begin{xy}
				\POS (0,3) *{\textcolor{GrainsOilseeds}{\medbullet}} ="a",
				\POS (-3.5,-1.5) *{\textcolor{Preciousmetals}{\medbullet}} ="b",
				\POS (3.5,-1.5) *{\textcolor{Softs}{\medbullet}} ="c"
				\POS "a" \ar @{<->} "b"
				\POS "a" \ar @{<->} "c"
				\POS "b" \ar @{->} "c"
			\end{xy} & 0.125 & 0.389 & 0.018 & 0.137 & 0.776 & 6 & \begin{xy}
				\POS (0,3) *{\textcolor{Equity}{\medbullet}} ="a",
				\POS (-3.5,-1.5) *{\textcolor{Energy}{\medbullet}} ="b",
				\POS (3.5,-1.5) *{\textcolor{Industrialmetals}{\medbullet}} ="c"
				\POS "a" \ar @{->} "b"
				\POS "a" \ar @{->} "c"
			\end{xy} & 5.094 & 10.815 & 2.862 & 3.228 & 0.691\\
			238 & \begin{xy}
				\POS (0,3) *{\textcolor{Softs}{\medbullet}} ="a",
				\POS (-3.5,-1.5) *{\textcolor{Softs}{\medbullet}} ="b",
				\POS (3.5,-1.5) *{\textcolor{Softs}{\medbullet}} ="c"
				\POS "a" \ar @{<->} "b"
				\POS "a" \ar @{<->} "c"
				\POS "b" \ar @{<->} "c"
			\end{xy} & 0.111 & 0.314 & 0.016 & 0.126 & 0.754 & 46 & \begin{xy}
				\POS (0,3) *{\textcolor{GrainsOilseeds}{\medbullet}} ="a",
				\POS (-3.5,-1.5) *{\textcolor{Industrialmetals}{\medbullet}} ="b",
				\POS (3.5,-1.5) *{\textcolor{Preciousmetals}{\medbullet}} ="c"
				\POS "a" \ar @{<->} "b"
				\POS "a" \ar @{->} "c"
				\POS "b" \ar @{->} "c"
			\end{xy} & 1.102 & 2.938 & 0.419 & 0.991 & 0.689\\
			78 & \begin{xy}
				\POS (0,3) *{\textcolor{Industrialmetals}{\medbullet}} ="a",
				\POS (-3.5,-1.5) *{\textcolor{Energy}{\medbullet}} ="b",
				\POS (3.5,-1.5) *{\textcolor{Industrialmetals}{\medbullet}} ="c"
				\POS "a" \ar @{<->} "b"
				\POS "a" \ar @{<->} "c"
			\end{xy} & 0.217 & 0.657 & 0.044 & 0.230 & 0.754 & 6 & \begin{xy}
				\POS (0,3) *{\textcolor{Equity}{\medbullet}} ="a",
				\POS (-3.5,-1.5) *{\textcolor{Preciousmetals}{\medbullet}} ="b",
				\POS (3.5,-1.5) *{\textcolor{Industrialmetals}{\medbullet}} ="c"
				\POS "a" \ar @{->} "b"
				\POS "a" \ar @{->} "c"
			\end{xy} & 4.216 & 9.018 & 2.334 & 2.734 & 0.689\\
			78 & \begin{xy}
				\POS (0,3) *{\textcolor{Equity}{\medbullet}} ="a",
				\POS (-3.5,-1.5) *{\textcolor{Equity}{\medbullet}} ="b",
				\POS (3.5,-1.5) *{\textcolor{Equity}{\medbullet}} ="c"
				\POS "a" \ar @{<->} "b"
				\POS "a" \ar @{<->} "c"
			\end{xy} & 0.226 & 0.750 & 0.049 & 0.236 & 0.752 & 6 & \begin{xy}
				\POS (0,3) *{\textcolor{Equity}{\medbullet}} ="a",
				\POS (-3.5,-1.5) *{\textcolor{Livestock}{\medbullet}} ="b",
				\POS (3.5,-1.5) *{\textcolor{Industrialmetals}{\medbullet}} ="c"
				\POS "a" \ar @{->} "b"
				\POS "a" \ar @{->} "c"
			\end{xy} & 4.212 & 9.224 & 2.333 & 2.735 & 0.687\\
			6 & \begin{xy}
				\POS (0,3) *{\textcolor{Equity}{\medbullet}} ="a",
				\POS (-3.5,-1.5) *{\textcolor{GrainsOilseeds}{\medbullet}} ="b",
				\POS (3.5,-1.5) *{\textcolor{Softs}{\medbullet}} ="c"
				\POS "a" \ar @{->} "b"
				\POS "a" \ar @{->} "c"
			\end{xy} & 5.359 & 12.263 & 2.816 & 3.419 & 0.744 & 6 & \begin{xy}
				\POS (0,3) *{\textcolor{Industrialmetals}{\medbullet}} ="a",
				\POS (-3.5,-1.5) *{\textcolor{Energy}{\medbullet}} ="b",
				\POS (3.5,-1.5) *{\textcolor{Softs}{\medbullet}} ="c"
				\POS "a" \ar @{->} "b"
				\POS "a" \ar @{->} "c"
			\end{xy} & 4.464 & 9.646 & 2.487 & 2.958 & 0.668\\
			78 & \begin{xy}
				\POS (0,3) *{\textcolor{Industrialmetals}{\medbullet}} ="a",
				\POS (-3.5,-1.5) *{\textcolor{Softs}{\medbullet}} ="b",
				\POS (3.5,-1.5) *{\textcolor{Industrialmetals}{\medbullet}} ="c"
				\POS "a" \ar @{<->} "b"
				\POS "a" \ar @{<->} "c"
			\end{xy} & 0.250 & 1.199 & 0.054 & 0.269 & 0.729 & 6 & \begin{xy}
				\POS (0,3) *{\textcolor{Industrialmetals}{\medbullet}} ="a",
				\POS (-3.5,-1.5) *{\textcolor{GrainsOilseeds}{\medbullet}} ="b",
				\POS (3.5,-1.5) *{\textcolor{Preciousmetals}{\medbullet}} ="c"
				\POS "a" \ar @{->} "b"
				\POS "a" \ar @{->} "c"
			\end{xy} & 4.363 & 10.026 & 2.339 & 3.049 & 0.664\\
			78 & \begin{xy}
				\POS (0,3) *{\textcolor{Energy}{\medbullet}} ="a",
				\POS (-3.5,-1.5) *{\textcolor{Softs}{\medbullet}} ="b",
				\POS (3.5,-1.5) *{\textcolor{Industrialmetals}{\medbullet}} ="c"
				\POS "a" \ar @{<->} "b"
				\POS "a" \ar @{<->} "c"
			\end{xy} & 0.263 & 1.445 & 0.065 & 0.272 & 0.729 & 6 & \begin{xy}
				\POS (0,3) *{\textcolor{Livestock}{\medbullet}} ="a",
				\POS (-3.5,-1.5) *{\textcolor{Energy}{\medbullet}} ="b",
				\POS (3.5,-1.5) *{\textcolor{Softs}{\medbullet}} ="c"
				\POS "a" \ar @{->} "b"
				\POS "a" \ar @{->} "c"
			\end{xy} & 3.252 & 6.871 & 1.710 & 2.344 & 0.658\\
			6 & \begin{xy}
				\POS (0,3) *{\textcolor{Equity}{\medbullet}} ="a",
				\POS (-3.5,-1.5) *{\textcolor{Energy}{\medbullet}} ="b",
				\POS (3.5,-1.5) *{\textcolor{Softs}{\medbullet}} ="c"
				\POS "a" \ar @{->} "b"
				\POS "a" \ar @{->} "c"
			\end{xy} & 3.908 & 9.022 & 2.032 & 2.587 & 0.725 & 6 & \begin{xy}
				\POS (0,3) *{\textcolor{Industrialmetals}{\medbullet}} ="a",
				\POS (-3.5,-1.5) *{\textcolor{Energy}{\medbullet}} ="b",
				\POS (3.5,-1.5) *{\textcolor{Preciousmetals}{\medbullet}} ="c"
				\POS "a" \ar @{->} "b"
				\POS "a" \ar @{->} "c"
			\end{xy} & 3.216 & 7.507 & 1.693 & 2.317 & 0.657\\
			6 & \begin{xy}
				\POS (0,3) *{\textcolor{Equity}{\medbullet}} ="a",
				\POS (-3.5,-1.5) *{\textcolor{Energy}{\medbullet}} ="b",
				\POS (3.5,-1.5) *{\textcolor{Preciousmetals}{\medbullet}} ="c"
				\POS "a" \ar @{->} "b"
				\POS "a" \ar @{->} "c"
			\end{xy} & 2.947 & 7.271 & 1.412 & 2.127 & 0.722 & 6 & \begin{xy}
				\POS (0,3) *{\textcolor{Equity}{\medbullet}} ="a",
				\POS (-3.5,-1.5) *{\textcolor{Softs}{\medbullet}} ="b",
				\POS (3.5,-1.5) *{\textcolor{Livestock}{\medbullet}} ="c"
				\POS "a" \ar @{->} "b"
				\POS "a" \ar @{->} "c"
			\end{xy} & 4.288 & 8.854 & 2.336 & 2.982 & 0.654\\
			78 & \begin{xy}
				\POS (0,3) *{\textcolor{Equity}{\medbullet}} ="a",
				\POS (-3.5,-1.5) *{\textcolor{Energy}{\medbullet}} ="b",
				\POS (3.5,-1.5) *{\textcolor{Livestock}{\medbullet}} ="c"
				\POS "a" \ar @{<->} "b"
				\POS "a" \ar @{<->} "c"
			\end{xy} & 0.229 & 0.636 & 0.058 & 0.240 & 0.710 & 6 & \begin{xy}
				\POS (0,3) *{\textcolor{Industrialmetals}{\medbullet}} ="a",
				\POS (-3.5,-1.5) *{\textcolor{Livestock}{\medbullet}} ="b",
				\POS (3.5,-1.5) *{\textcolor{Preciousmetals}{\medbullet}} ="c"
				\POS "a" \ar @{->} "b"
				\POS "a" \ar @{->} "c"
			\end{xy} & 2.638 & 6.261 & 1.368 & 1.946 & 0.653\\
			74 & \begin{xy}
				\POS (0,3) *{\textcolor{Industrialmetals}{\medbullet}} ="a",
				\POS (-3.5,-1.5) *{\textcolor{Preciousmetals}{\medbullet}} ="b",
				\POS (3.5,-1.5) *{\textcolor{Livestock}{\medbullet}} ="c"
				\POS "a" \ar @{<->} "b"
				\POS "c" \ar @{->} "a"
			\end{xy} & 0.111 & 0.605 & 0.018 & 0.133 & 0.705 & 6 & \begin{xy}
				\POS (0,3) *{\textcolor{Livestock}{\medbullet}} ="a",
				\POS (-3.5,-1.5) *{\textcolor{Softs}{\medbullet}} ="b",
				\POS (3.5,-1.5) *{\textcolor{Preciousmetals}{\medbullet}} ="c"
				\POS "a" \ar @{->} "b"
				\POS "a" \ar @{->} "c"
			\end{xy} & 2.752 & 6.039 & 1.383 & 2.099 & 0.652\\
			\bottomrule
	\end{tabular}}
	\begin{flushleft}
		\footnotesize
		\textit{Legend: }
		\textcolor{Equity}{{\normalsize\textbullet}}\, Equity;
		\textcolor{Industrialmetals}{{\normalsize\textbullet}}\, Industrial metals;
		\textcolor{Preciousmetals}{{\normalsize\textbullet}}\, Precious metals;
		\textcolor{Energy}{{\normalsize\textbullet}}\,  Energy;
		\textcolor{GrainsOilseeds}{{\normalsize\textbullet}}\, Grains \& oilseeds;
		\textcolor{Softs}{{\normalsize\textbullet}}\, Softs;
		\textcolor{Livestock}{{\normalsize\textbullet}}\, Livestock.
	\end{flushleft} 
\end{table}

\end{document}